\documentclass[aps,10pt,prd,twocolumn,superscriptaddress,preprintnumbers,eqsecnum,%
showkeys,nofootinbib,showpacs]{revtex4-1}
\usepackage{bm}
\usepackage{graphicx}
\usepackage[hypertexnames=false]{hyperref}
\usepackage{amsmath,amssymb,amsfonts,latexsym}
\newcommand{\lambdadp}{{\lambda'}_{\!\! d}}
\newcommand{\lambdanp}{{\lambda'}_{\!\! n}}
\newcommand{\lambdapp}{{\lambda'}_{\!\! p}}
\newcommand{\be}{\begin{eqnarray}}
\newcommand{\ee}{\end{eqnarray}}
\newcommand{\beq}{\begin{equation}}
\newcommand{\eeq}{\end{equation}}
\begin{document}
\title{Polarized electron-deuteron deep-inelastic scattering with spectator nucleon tagging}
\author{W.~Cosyn}
\email[ E-mail: ]{wcosyn@fiu.edu}
\affiliation{Department of Physics, Florida International University, Miami, Florida 33199, USA}
\affiliation{Department of Physics and Astronomy, Ghent University, B9000 Ghent, Belgium}
\author{C.~Weiss}
\email[ E-mail: ]{weiss@jlab.org}
\affiliation{Theory Center, Jefferson Lab, Newport News, VA 23606, USA}
\begin{abstract}
\begin{description}
\item[Background] DIS on the polarized deuteron with detection of a proton in the nuclear breakup
region (spectator tagging) represents a unique method for extracting the neutron spin structure
functions and studying nuclear modifications. The tagged proton momentum controls the nuclear
configuration during the DIS process and enables a differential analysis of nuclear effects.
Such measurements could be performed with the future electron-ion collider (EIC) and forward
proton detectors if deuteron beam polarization could be achieved.
\item[Purpose] Develop theoretical framework for polarized deuteron DIS with spectator tagging.
Formulate procedures for neutron spin structure extraction.
\item[Methods] A covariant spin density matrix formalism is used to describe general deuteron
polarization in collider experiments (vector/tensor, pure/mixed). Light-front (LF) quantum mechanics
is employed to factorize nuclear and nucleonic structure in the DIS process. A 4-dimensional
representation of LF spin structure is used to construct the polarized deuteron LF wave function
and efficiently evaluate the spin sums. Free neutron structure is extracted using the
impulse approximation and analyticity in the tagged proton momentum (pole extrapolation).
\item[Results] General expressions of the polarized tagged DIS observables in collider experiments.
Analytic and numerical study of the polarized deuteron LF spectral function and nucleon
momentum distributions. Practical procedures for neutron spin structure extraction from the tagged
deuteron spin asymmetries.
\item[Conclusions] Spectator tagging provides new tools for precise neutron spin structure measurements.
D-wave depolarization and nuclear binding effects can be eliminated through the tagged proton
momentum dependence. The methods can be extended to tensor-polarized observables,
spin-orbit effects, and diffractive processes.
\end{description}
\end{abstract}
\preprint{JLAB-THY-20-3203}
\maketitle
\tableofcontents
\section{Introduction}
\label{sec:intro}
Exploring the spin-dependent partonic structure of the nucleon and the numerous polarization-induced
phenomena in QCD processes is a principal objective of modern nuclear physics; see
Refs.~\cite{Anselmino:1994gn,Burkardt:2008jw,Kuhn:2008sy,Aidala:2012mv} for a review. This program
requires measurements of deep-inelastic lepton scattering (DIS) on both the proton and the neutron.
Proton and neutron data are needed to separate the isovector and isoscalar combinations of the
nucleon spin structure functions, which are subject to different short--distance dynamics
(QCD evolution, higher--twist effects, small--$x$ asymptotics) and give access to different
combinations of the parton densities (non-singlet quarks vs.\ gluons and singlet quarks)
\cite{deFlorian:2009vb,Leader:2010rb,Blumlein:2010rn,Nocera:2014gqa,Sato:2016tuz}.

The isovector structure function $g_{1p} - g_{1n}$ exhibits pure non-singlet QCD evolution
and provides direct access to the flavor-nonsinglet polarized quark
densities. Its moment (integral over $x$) can be used for precision studies of perturbative QCD and extraction
of the strong coupling constant, see Refs.~\cite{Altarelli:1996nm,Pasechnik:2009yc,
Baikov:2010je,Deur:2014vea,Cvetic:2016rot,Kotlorz:2018bxp,Ayala:2018ulm,Deur:2016tte}
and references therein, and is asymptotically constrained by the Bjorken sum rule, a fundamental prediction
of current algebra that can be tested experimentally \cite{Bjorken:1969mm}.
The isoscalar structure function $g_{1p} + g_{1n}$ exhibits singlet evolution and can be used to
extract the polarized gluon density and the flavor-singlet
polarized quark densities.
Both isospin combinations are needed to determine the flavor decomposition of the polarized
quark densities and their contributions to the nucleon spin;
see Refs.~\cite{deFlorian:2009vb,Leader:2010rb,Blumlein:2010rn,Nocera:2014gqa,Sato:2016tuz}
and references therein. The subasymptotic power corrections to the spin structure functions give
access to higher-twist matrix elements describing non-perturbative quark-gluon correlations in the nucleon,
for which theoretical calculations predict a significant isospin dependence
\cite{Braun:1986ty,Balitsky:1989jb,Stein:1995si,Balla:1997hf}, consistent with empirical
extractions \cite{Meziani:2004ne,Sidorov:2006vu}. Neutron and proton data together are also needed
to explore the dynamical mechanisms causing single-spin asymmetries in semi-inclusive DIS,
where there are signs of large isovector structures, see Refs.~\cite{Anselmino:2009pn,Anselmino:2014eza}
and references therein.

Neutron spin structure functions are measured in DIS on polarized light nuclei, principally the
deuteron $d \equiv {}^2$H and ${}^3$He. Experiments have been performed at SLAC
\cite{Anthony:1996mw,Abe:1998wq,Abe:1997cx,Abe:1997qk,Anthony:1999py,
Anthony:1999rm,Anthony:2000fn,Anthony:2002hy} , DESY HERMES
\cite{Ackerstaff:1997ws,Airapetian:2006vy}, CERN SMC \cite{Adeva:1998vv}, CERN COMPASS
\cite{Alexakhin:2006oza,Alekseev:2009ac,Adolph:2016myg} and JLab 6 GeV
\cite{Deur:2014vea,Zheng:2004ce,Posik:2014usi,Prok:2014ltt,Chen:2011zzp}, and will be extended
further with the JLab 12 GeV Upgrade \cite{Malace:2014uea}. The extraction of the neutron structure
functions from the nuclear DIS data faces considerable theoretical challenges; see
\cite{Frankfurt:1988nt,Arneodo:1992wf,CiofidegliAtti:1993zs,Melnitchouk:1994tx,Kulagin:1994cj,Piller:1995mf,%
Frankfurt:1996nf,Bissey:2001cw,Ethier:2013hna} and references therein. The DIS process can happen on the
protons or neutrons in the nucleus, causing dilution of the neutron signal.
Spin depolarization occurs due to higher partial waves in the nuclear wave function.
Nuclear binding modifies the apparent neutron structure functions through the Fermi motion and dynamical effects
(EMC effect at $x > 0.3$; antishadowing at $x \sim 0.1$; nuclear shadowing at $x < 0.01$). These modifications
reveal different aspects of nucleon interactions in QCD and are themselves objects of study. In spin
structure measurements with polarized ${}^3$He, the presence of $\Delta$ isobars in the nuclear wave
function (non-nucleonic degrees of freedom) \cite{Frankfurt:1996nf,Bissey:2001cw} modifies the
effective neutron polarization compared to non-relativistic nuclear structure calculations
\cite{Friar:1990vx,CiofidegliAtti:1993zs}.

The main difficulty in the theoretical treatment of the nuclear modifications lies in the fact
that they strongly depend on the nuclear configurations present during the DIS process.
Both the state of motion and the strength of interaction of the active nucleon depend on the
configuration and exhibit considerable variation, as determined by the quantum-mechanical motion
of the interacting system. In inclusive measurements one attempts to account for these effects
by modeling their dependence on the nuclear configuration and summing over all configurations.
The resulting theoretical uncertainty represents a significant source of the systematic error
in neutron spin structure function extraction. Efforts should be made to reduce this theoretical
uncertainty as the experimental data are becoming more precise. Another strategy is to consider
alternative measurements that permit control of the nuclear configuration during the DIS process.

DIS on the deuteron with detection of a proton in the nuclear fragmentation region,
$e + d \rightarrow e' + X + p$, represents a unique method for performing neutron
structure measurements in controlled nuclear configurations \cite{Frankfurt:1981mk}.
The proton is detected with momenta $p_p \lesssim$ 300 MeV in the deuteron rest frame.
At such momenta the deuteron can be described with good accuracy in terms of nucleonic
degrees of freedom ($pn$), and its wave function in nucleonic degrees of freedom is known
well both in non-relativistic and in LF quantum mechanics (see below). Configurations with
$\Delta$ isobars are suppressed in the isospin-0 system \cite{Frankfurt:1981mk}.
The detection of the proton identifies DIS events with active neutron and eliminates dilution.
The measurement of the proton momentum controls the nuclear configuration during the DIS process
and enables a differential treatment of nuclear effects. By extrapolating the proton momentum
dependence to the unphysical region one can reach $pn$ configurations where the neutron is
effectively free, and nuclear binding effects and final-state interactions disappear,
which makes possible the extraction of free neutron structure (pole extrapolation, or on-shell
extrapolation) \cite{Sargsian:2005rm}. The technique is theoretically appealing and practically feasible.
If it could be applied to polarized electron-deuteron scattering with proton tagging,
one could use it to extract the free neutron spin structure function.

Measurements of DIS on the deuteron with proton tagging have been performed in fixed-target
experiments at JLab with 6 GeV beam energy with the CLAS spectrometer and the BoNuS
proton detector \cite{Baillie:2011za,Tkachenko:2014byy}. The data provide constraints on
the $F_{2n}/F_{2d}$ structure function ratio and the $d/u$ quark density ratio at large $x$.
The measurements will be extended to 11 GeV energy with the BoNuS and ALERT
detectors \cite{Bonus12,Armstrong:2017zqr}. The BoNuS setup detects only protons
with momenta $p_p \gtrsim$ 70 MeV, as slower protons cannot escape the target;
the measurements therefore use only a small part of the deuteron momentum distribution
and preclude accurate on-shell extrapolation. The setup is also limited to unpolarized targets.
Other DIS experiments with proton and neutron tagging at larger momenta $p_{p, n} \sim$ few 100 MeV
explore the EMC effect and its connection with nucleon
short-range correlations \cite{Klimenko:2005zz,Hen:2011,Hen:2014vua}.

The future Electron-Ion Collider (EIC) at BNL will greatly expand the capabilities for DIS
measurements on the proton and on light and heavy ions \cite{Boer:2011fh,Accardi:2012qut}.
The proposed design will enable electron-proton collisions at center-of-mass (CM) energies
$\sqrt{s_{ep}} \sim$ 20--140 GeV and luminosities $\sim 10^{33}$--$10^{34}$ cm$^{-2}$ s$^{-1}$,
and electron-deuteron collisions at electron-nucleon CM energies $\sqrt{s_{eN}} \sim$ 20--100 GeV
and similar luminosities per nucleon \cite{Aschenauer:2014cki,EICpCDR}.\footnote{For a given ion/proton
storage ring design (ring radius, magnetic field, etc.), the energy per nucleon in a relativistic ion beam
with $A > 1$ is generally lower than that of a proton beam by a factor $Z/A$ (the nuclear charge to mass
number ratio). When the ion/proton beams collide with an electron beam of fixed energy, the squared CM 
energies per nucleon of the collision are therefore related as $s_{eN} \approx (Z/A) \, s_{ep}$.}
In DIS on the deuteron at the collider, the spectator nucleon moves forward with approximately
half the deuteron beam momentum and can be detected with forward detectors integrated into the
interaction region. The development and integration of such forward detectors have been a priority
of the machine design effort and have made major progress. The current conceptual design includes a
magnetic dipole spectrometer with multiple active elements for forward proton detection, and a
zero-degree calorimeter for forward neutron detection. The apparatus can detect protons with
transverse momenta from zero to $\sim$ few 100 MeV with a resolution $\lesssim 30$ MeV,
and longitudinal momenta from $\sim 0.5$--$1.5$ times the nominal spectator momentum;
for details and neutron detection see Ref.~\cite{Jentsch:2020}. The EIC thus provides
excellent capabilities for deuteron DIS with proton tagging. The collider offers many
advantages over the fixed-target setup: protons can be detected down to zero momentum
in the deuteron rest frame; the magnetic spectrometer provides good momentum resolution;
and tagged measurements can be performed with polarized deuteron beams. 

Polarization of the deuteron beams in the EIC is regarded as technically possible and considered
as a future option \cite{EICpCDR}. Maintaining deuteron polarization in a storage ring is more
challenging than proton polarization, as the small magnetic moment of the deuteron renders
spin manipulation more difficult. Possibilities for realizing deuteron polarization in the EIC
circular storage ring are under investigation; see e.g.\ Ref.~\cite{Higinbotham:2020}.
(A unique solution to the problem of deuteron polarization is a figure-8 layout of the ion ring
that compensates the spin precession within one turn, as was proposed in an earlier
alternative EIC design \cite{Abeyratne:2012ah}.) 
Together with the forward detection capabilities, deuteron polarization raises the prospect of using
polarized deuteron DIS with proton tagging for precision measurements of neutron spin structure at EIC.
The setup could also be used for measurements of bound proton spin structure with neutron tagging,
of spin-dependent diffractive processes on the deuteron with proton and neutron tagging,
and of tensor-polarized deuteron structure. The physics potential such measurements at EIC
has been explored in an R\&D project \cite{LD1506,Guzey:2014jva,Cosyn:2014zfa}.

In this article we develop the theoretical framework for DIS on the polarized deuteron with spectator tagging. 
The development proceeds in three steps. In the first step, we derive the general expressions of the differential 
cross section of polarized electron-deuteron DIS with spectator tagging for the case of arbitrary deuteron 
polarization (vector and tensor), including the spin asymmetry observables corresponding to specific polarization 
states in colliding-beam experiments (depolarization factors). In the second step we separate the high-energy DIS 
process from the low-energy nuclear structure using methods of light-front quantization, calculate the deuteron 
structure elements entering in the description of tagged DIS in the impulse approximation (IA), and study their 
properties (limiting cases, sum rules). In the third step we evaluate the longitudinal spin asymmetries 
in polarized tagged DIS, study their dependence on the tagged proton momentum analytically and numerically,
and formulate the procedures for neutron spin structure extraction, including pole extrapolation.
We find that the neutron spin structure function $g_{1n}$ can be extracted efficiently from the tagged 
longitudinal spin asymmetry formed with the deuteron's $\pm 1$ spin states only (without the $0$ state,
involving effective tensor polarization). We comment on possible extensions of the methods to the
study of tensor-polarized observables, spin-orbit effects in deuteron breakup, and exclusive scattering processes.
Many of the applications considered here were originally discussed in the work of Ref.~\cite{Frankfurt:1983qs}.
Some preliminary results of our study were reported earlier in Ref.~\cite{Cosyn:2019hem}.

Our theoretical treatment of deuteron structure in tagged DIS employs the methods of LF quantization of 
nuclear systems developed in Refs.~\cite{Frankfurt:1981mk,Frankfurt:1988nt} and summarized in 
Ref.~\cite{Strikman:2017koc} (for a general review of LF quantization, see
Refs.~\cite{Coester:1992cg,Brodsky:1997de,Heinzl:1998kz}).
High-energy processes such as DIS probe the nucleus at fixed 
LF time $x^+ = x^0 + x^3$. A description of the nucleus in terms of nucleonic degrees of freedom at fixed 
$x^+$ permits a smooth matching of nuclear and nucleonic structure, preserves the partonic sum rules 
for the nucleus (baryon number, LF momentum, spin), and exhibits a close correspondence with 
nonrelativistic nuclear structure. The LF wave function of the deuteron in nucleon degrees of freedom
can be obtained by solving the dynamical equation with realistic $NN$ interactions or 
constructed approximately from the nonrelativistic wave function. The deuteron and nucleon spin states
are introduced as LF helicity states, or boost-invariant extensions of the rest-frame spin states,
and the deuteron LF spin structure is obtained in direct analogy to the nonrelativistic system (S and D waves). 
In the traditional ``3-dimensional'' formulation of LF spin structure one describes the nucleons by LF 2-spinors 
that are related to the canonical 2-spinors by the Melosh rotation, and constructs the deuteron LF wave function 
from the 3-dimensional wave function in the center-of-mass frame. In the present work we employ a ``4-dimensional'' 
formulation \cite{Kondratyuk:1983kq,Frankfurt:1983qs}, in which the nucleons are described by LF bispinors and the 
coupling to the deuteron is implemented through a 4-dimensional vertex function (the equivalence of the two 
formulations is demonstrated in Appendix~\ref{app:wave_function}). It permits efficient evaluation of the sums over
the nucleon LF helicities and leads to LF formulas in close analogy with those of relativistically covariant 
quantum field theory (Feynman diagrams). In particular, in the 4-dimensional formulation the effective polarization 
of the neutron in the deuteron (at a given LF momentum) can be described by a spin density matrix
of the same form as that in covariant field theory, with the entire deuteron structure information 
condensed in an axial 4-vector (polarization vector).

The article is organized as follows. In Sec.~\ref{sec:spin_density_matrices} we review the formalism of
relativistic spin density matrices of the spin-1/2 and spin-1 system, which will be used throughout this work. 
In Sec.~\ref{sec:scattering} we present the general expressions of the cross section and structure functions
of tagged DIS on the polarized deuteron with vector and tensor polarization, including the spin asymmetries 
measured in colliding-beam experiments.
We express the kinematic factors (effective polarizations, depolarization factors) in manifestly relativistically 
invariant form as suitable for colliding-beam experiments. In Sec.~\ref{sec:deuteron} we summarize the
elements of LF quantization and describe the deuteron LF wave function in the 4-dimensional formulation
of the spin structure, including its correspondence with the nonrelativistic wave function.
In Sec.~\ref{sec:nucleon_operators} we develop the formalism for the evaluation of nucleon one-body operators 
in the polarized deuteron at fixed LF momentum of the spectator. We derive the effective spin density
matrix of the neutron, calculate the LF momentum distribution of the neutron in the polarized deuteron,
and study its momentum and spin dependence. In Sec.~\ref{sec:ia} we calculate the polarized tagged DIS 
cross section and the spin asymmetries in the IA, separating nuclear and nucleonic structure,
and study the dependence on the tagged proton momentum. In Sec.~\ref{sec:neutron_spin_structure}
we study the analytic properties in the tagged proton momentum and discuss the strategies for 
neutron spin structure extraction through pole extrapolation. In Sec.~\ref{sec:summary} we 
summarize the methodological and practical results and discuss possible extensions of the methods
to other scattering processes and to nuclei with $A > 2$. Appendix~\ref{app:spinors} summarizes the 
definition and properties of the LF helicity spinors used in the calculations. Appendix~\ref{app:wave_function} 
describes the 3-dimensional formulation of the deuteron spin structure in LF quantization and demonstrates
the equivalence to the 4-dimensional formulation used in the calculations.

Some explanations are in order regarding the limitations of the present study.
In the polarized tagged DIS cross section we consider only the structures after integration over the azimuthal angle 
of the proton (in the frame where the deuteron and the virtual photon momenta are collinear); these structures
correspond to those measured in ``untagged'' DIS and are used for neutron spin structure function extraction.
When the azimuthal angle dependence is included, the number of independent structures in the polarized tagged 
DIS cross section for the spin-1 target becomes very considerable, especially in the case of tensor polarization;
this case will be considered in a separate study \cite{Cosyn:inprep}.

In the treatment of nuclear structure effects we limit ourselves
to the IA, which is sufficient for studying the tagged spin asymmetries used for neutron structure extraction
and their analytic properties at small proton momenta. FSI in unpolarized tagged DIS at intermediate $x$ 
($\sim$ 0.1--0.5) were calculated in Ref.~\cite{Strikman:2017koc} and found to be moderate at proton
momenta $\lesssim$100 MeV (in the deuteron rest frame); the calculations could be extended to the polarized case.
In the practical applications we consider the leading-twist longitudinal spin asymmetries used for tagged measurements 
of the neutron spin structure function $g_{1n}$; our general expressions cover also the power-suppressed transverse 
spin asymmetry and the contributions of $g_{2n}$, and the IA calculations could easily be extended to 
these observables.
\section{Spin density matrices}
\label{sec:spin_density_matrices}
\subsection{Spin-1/2 particle}
We begin by reviewing the formalism of spin density matrices for ensembles of spin states (mixed polarization states)
of spin-1/2 and spin-1 particles. We focus on the relativistically covariant representation of the 
density matrices in terms of 4-vectors and tensors, which will be used throughout the subsequent calculations.

Consider a relativistic spin-1/2 particle with spin states labeled by the quantum number 
$\lambda = \pm \frac{1}{2}$; the exact definition of the spin states is not needed here and will be specified later.
An ensemble of spin states is described by the density matrix in spin quantum numbers,
\beq
\rho(\lambda, \lambda'), \hspace{2em} \sum_{\lambda} \rho(\lambda, \lambda) = 1.
\eeq
Each spin state of the particle corresponds to a bispinor wave function $u(p, \lambda)$, 
normalized such that $\bar u u = 2m$, where $p$ is the 4-momentum and $m$ the mass. 
The spin density matrix in bispinor representation is defined as
\beq
\rho \; \equiv \; \sum_{\lambda, \lambda'} \rho(\lambda, \lambda') \; u(p, \lambda) \bar u(p, \lambda'), 
\hspace{2em} \textrm{tr}[\rho] \; = \; 2 m. 
\label{spin_density_1/2_bispinor}
\eeq
A general spin observable is given by a matrix in spin quantum numbers
$O(\lambda', \lambda)$. In the bispinor representation it corresponds to a bilinear form
\beq
O(\lambda', \lambda) \; \equiv \;  \bar u(\lambda', p) \Gamma u(\lambda, p) ,
\eeq
where the specific form of the matrix $\Gamma$ depends on the observable. The expectation value of
the observable in the spin ensemble is then obtained as
\beq
\langle O \rangle \; \equiv \; \sum_{\lambda, \lambda'} \rho(\lambda, \lambda') O(\lambda', \lambda)
\; = \; \textrm{tr}[\rho \Gamma] .
\eeq
The spin density matrix Eq.~(\ref{spin_density_1/2_bispinor}) transforms covariantly under Lorentz 
transformations. It can be decomposed into an unpolarized and a polarized part,
\beq
\rho \; = \; \rho [\textrm{unpol}] + \rho [\textrm{pol}] .
\label{density_matrix_spin12}
\eeq
The unpolarized part depends only on the particle 4-momentum and is given by
\beq
\rho [\textrm{unpol}] \; = \; {\textstyle\frac{1}{2}}(p\gamma + m) 
\hspace{2em} [p\gamma\equiv p^\mu \gamma_\mu].
\eeq
The polarized part is parameterized in terms of a real axial 4-vector $s$ (``polarization 4-vector'')
\begin{subequations}
\begin{align}
\rho [\textrm{pol}] \; &= \; {\textstyle\frac{1}{2}}(p\gamma + m) (s \gamma) \gamma^5, 
\hspace{2em}
sp \; = \; 0, 
\\
\hspace{2em} s^\mu \; &= \; -\frac{1}{2m} \, \textrm{tr}[\rho \gamma^\mu \gamma^5] ,
\label{density_matrix_11/2_pol}
\end{align}
\end{subequations}
where we follow the conventions of Ref.~\cite{LLIV},\footnote{In this convention
the sign of $\gamma^5$ is opposite to the Bjorken-Drell convention.}
\begin{subequations}
\begin{align}
& \gamma^5 \; \equiv \; -i \gamma^0 \gamma^1 \gamma^2 \gamma^3 ,
\\
& \text{tr} \left[ \gamma^\alpha \gamma^\beta \gamma^\gamma \gamma^\delta 
\gamma^5 \right] \;\; = \;\; 4 i \epsilon^{\alpha\beta\gamma\delta} ,
\hspace{2em}
\epsilon^{0123} = 1 .
\end{align}
\end{subequations}
The polarization 4-vector is, up to a factor, equal to the axial current of the particle.
Specifically, in the particle's rest frame, $p[\textrm{RF}] = (m, \bm{0})$, the components
of the polarization 4-vector are
\begin{subequations}
\begin{align}
s[\textrm{RF}] \; &= \; (0, \bm{S}), \hspace{2em} 0 \leq |\bm{S}|^2 \leq 1 ,
\\[1ex]
S^i \; &= \;  
\sum_{\lambda, \lambda'} \rho(\lambda, \lambda') \sigma^i (\lambda', \lambda) .
\end{align}
\end{subequations}
$\bm{S}$ is the polarization 3-vector in the rest frame, and 
$\sigma^i (\lambda', \lambda) \equiv \langle \lambda' | \sigma^i | \lambda \rangle
(i = 1,2,3)$ are the matrix elements of the spin operator between states with 
spin projections $\lambda$ and $\lambda'$ [if the spin is quantized along the $z$-axis, 
these matrix elements are the Pauli matrices: 
$\sigma^i (\lambda', \lambda) = (\sigma^i)_{\lambda'\lambda}$].
It follows that the 4-vector in any frame satisfies
\be
s^2 \; < \; 0, \hspace{2em} 0 \; \leq \; |s^2| \; \leq \; 1.
\ee
\subsection{Spin-1 particle}
\label{subsec:spin_density_spin_1}
The spin-1 particle can be treated in analogy with the spin-1/2 case;
see Ref.~\cite{Hoodbhoy:1988am,Leader:2001gr} for a general discussion. The spin states of the spin-1
particle are labeled by the quantum number $\lambda = (-1, 0, 1)$. An ensemble of
spin states is described by the density matrix
\beq
\rho(\lambda, \lambda'), \hspace{2em} \sum_{\lambda} \rho(\lambda, \lambda) = 1.
\label{density_matrix_spin1_orig}
\eeq
Each spin state corresponds to a 4-vector wave function
\beq
\epsilon^\alpha (p, \lambda), \hspace{2em}
\epsilon^2 \; = \; -1, \hspace{2em} \epsilon p \; = \; 0 .
\eeq
The spin density matrix in 4-tensor representation is defined as
\begin{subequations}
\begin{align}
\rho^{\alpha\beta} 
\; &\equiv \; \sum_{\lambda, \lambda'} \rho(\lambda, \lambda') \; \epsilon^\alpha (p, \lambda) 
\epsilon^{\beta\ast} (p, \lambda') 
\label{density_matrix_spin1}
\\[1ex]
p_\alpha \rho^{\alpha\beta} \; &= \; \rho^{\alpha\beta}  p_\beta \; = \; 0,
\hspace{2em} \rho^{\alpha}_{\;\;\alpha} \; = \; -1.
\end{align}
\end{subequations}
A general spin observable is given by a matrix in spin quantum numbers $O(\lambda', \lambda)$.
In the 4-tensor representation it corresponds to a bilinear form
\beq
O(\lambda', \lambda) \; \equiv \;  \epsilon^{\beta\ast}(\lambda') R_{\beta\alpha} \epsilon^\alpha (\lambda) ,
\eeq
and the expectation value in the ensemble is obtained as
\beq
\langle O \rangle \; \equiv \; \sum_{\lambda, \lambda'} \rho(\lambda, \lambda') 
O(\lambda', \lambda)
\; = \; \rho^{\alpha\beta} R_{\beta\alpha} .
\label{spin_average_tensor}
\eeq
The spin density matrix Eq.~(\ref{density_matrix_spin1}) can be decomposed into
an unpolarized, a vector-polarized, and a tensor-polarized part,
\beq
\rho^{\alpha\beta} \; = \; \rho^{\alpha\beta}[\textrm{unpol}]
+ \rho^{\alpha\beta}[\textrm{vector}] + \rho^{\alpha\beta}[\textrm{tensor}] .
\label{density_matrix_spin1_decomposition}
\eeq
The unpolarized part is given by
\beq
\rho^{\alpha\beta}[\textrm{unpol}]
\; = \; \frac{1}{3} \left( -g^{\alpha\beta} + \frac{p^\alpha p^\beta}{p^2} \right) .
\label{rho_deuteron_unpol}
\eeq
The vector-polarized part is parameterized in terms of a real axial 4-vector $s$
[cf.\ Eq.~(\ref{density_matrix_11/2_pol}) for the spin-1/2 particle]
\begin{subequations}
\begin{align}
& \rho^{\alpha\beta}[\textrm{vector}] \; = \; \frac{i}{2M} \epsilon^{\alpha\beta\gamma\delta}
p_\gamma s_\delta,
\hspace{2em}
sp \; = \; 0,
\\[1ex]
& s^\mu \; = \; \rho^{\alpha\beta} (L^\mu)_{\beta\alpha},
\\[1ex]
& (L^\mu)^{\beta\alpha} \; \equiv \; \frac{i}{M} \epsilon^{\mu\nu\beta\alpha} p_\nu ,
\\[1ex]
& (L^\mu)^{\alpha\beta} \; = \; -(L^\mu)^{\beta\alpha}, \hspace{2em} (L^\mu)^{\alpha\beta} p_\mu = 0 .
\end{align}
\end{subequations}
Here $M$ is the particle mass, and the antisymmetric matrices $(L^\mu)^{\alpha\beta}$
are the 4-dimensional representation of the generators of spatial rotations.
In the particle's rest frame $p = (M, \bm{0})$ the components of the axial vector are
\begin{subequations}
\begin{align}
s[\textrm{RF}] \; &= \; (0, \bm{S}), \hspace{2em} 0 < |\bm{S}|^2 < 1,
\label{a_restframe}
\\[1ex]
S^i \; &= \;  
\sum_{\lambda, \lambda'} \rho(\lambda, \lambda') L^i (\lambda', \lambda) .
\end{align}
\end{subequations}
$\bm{S}$ is the polarization 3-vector in the rest frame, and 
$L^i (\lambda', \lambda) = \langle \lambda' | L^i | \lambda \rangle$ is the
angular momentum operator, represented as a matrix in the spin quantum numbers $\lambda'$ and $\lambda$. 
Again it follows that in any frame
\beq
s^2 < 0, \hspace{2em} 0 < |s^2| < 1 .
\eeq
The description of vector polarization of the spin-1 particle is thus completely analogous to
that of the spin-1/2 particle. 

The tensor-polarized part of the density matrix Eq.~(\ref{density_matrix_spin1}) is specific
to the spin-1 system. It can be parameterized in terms of a real, symmetric, traceless 4-tensor $t^{\mu\nu}$,
\begin{subequations}
\begin{align}
& \rho^{\alpha\beta}[\textrm{tensor}] \; = \; -t^{\alpha\beta},
\label{tensor_def}
\\[1ex]
& t^{\alpha\beta} \; = \; t^{\beta\alpha},
\hspace{2em} 
t^\alpha_{\;\;\alpha} \; = \; 0,
\hspace{2em} 
p_\alpha t^{\alpha\beta} \; = \; t^{\alpha\beta} p_\beta \; = \; 0.
\end{align}
\end{subequations}
In the rest frame the 4-tensor components are
\begin{subequations}
\begin{align}
& t^{0\beta}[\textrm{RF}] \; = \; t^{\alpha 0}[\textrm{RF}] \; = \; 0, 
\hspace{2em} t^{ij}[\textrm{RF}] \; \equiv \; T^{ij},
\\
& T^{ij} \; = \; T^{ji}, \hspace{2em} T^{ii} \; = \; 0.
\end{align}
\end{subequations}
$T^{ij}$ is the conventional 3-dimensional polarization tensor in the rest frame. Its general decomposition,
positivity conditions, and other properties, are described in Ref.~\cite{Leader:2001gr}.\footnote{The prefactor 
accompanying the tensor in Eq.~(\ref{tensor_def}) is conventional.
We choose $-1$ in order to simplify the subsequent covariant expressions. Reference~\cite{Leader:2001gr} 
uses $-\sqrt{2/3}$.} In the present study we need only a special tensor structure, which can 
be constructed directly in 4-dimensional form (see below); we therefore do not need to consider
the general properties of the 3-dimensional tensor.

In applications we need the spin density matrices of pure states polarized along some given direction.
They can be constructed in 4-dimensional form, as a superposition of the unpolarized, vector-polarized 
and tensor-polarized parts with certain parameters. The vector-polarized part is expressed in terms 
of the special axial 4-vector
\begin{align}
& s^\mu(N, \Lambda) \; \equiv \; \Lambda N^\mu, \hspace{1em} pN \; = \; 0, \hspace{1em} N^2 \; = \; -1, 
\label{vector_special}
\end{align}
where the 4-vector $N$ defines the direction of polarization and $\Lambda = (-1, 0, +1)$ is the 
spin projection along that direction. Note that $s^\mu = 0$ in the state with $\Lambda = 0$. 
The tensor-polarized part is expressed in terms of the special tensor
\begin{align}
t^{\alpha\beta}(N, \Lambda) \; & \equiv \; \frac{1}{6}
\left( g^{\alpha\beta} - \frac{p^\alpha p^\beta}{p^2} + 3 N^\alpha N^\beta \right)
\nonumber \\
&\times \; 
\left\{ 
\begin{array}{rl} 1, & \Lambda = \pm 1 \\
(-2), & \Lambda = 0
\end{array}
\right\} .
\label{tensor_special}
\end{align}
The density matrices of the pure states polarized in the direction $N$ are then given by
\begin{subequations}
\begin{align}
\rho^{\alpha\beta}(N, \Lambda)
\; &= \; \rho^{\alpha\beta}[\textrm{unpol}] 
\nonumber
\\[1ex]
&+ \; \rho^{\alpha\beta} [\textrm{vector}, s(N, \Lambda)]
\nonumber
\\[1ex]
&+ \; \rho^{\alpha\beta} [\textrm{tensor}, t(N, \Lambda)] ,
\label{density_pure}
\\[2ex]
\rho^{\alpha\beta}(N, \pm 1)
\; &= \; \frac{1}{2} \left( -g^{\alpha\beta} + \frac{p^\alpha p^\beta}{p^2} - N^\alpha N^\beta \right)
\nonumber
\\
& \pm \; \frac{i}{2M} \epsilon^{\alpha\beta\gamma\delta} p_\gamma N_\delta ,
\label{density_pure_pm1}
\\[2ex]
\rho^{\alpha\beta}(N, 0)
\; &= \; N^\alpha N^\beta .
\label{density_pure_0}
\end{align}
\end{subequations}
Eqs.~(\ref{density_pure_pm1}) and (\ref{density_pure_0}) can be verified in the rest frame, by
considering the case of polarization along the $z$-axis, $N = (0, \bm{e}_z)$, and comparing with the
explicit expression of the density matrix in terms of the polarization vectors; the general relation
then follows from relativistic covariance.  

Notice that the pure states with projections $\pm 1$
involve the unpolarized, vector-polarized and tensor-polarized parts of the density matrix, while
the state with projection $0$ involves only the unpolarized and tensor-polarized parts.
Conversely, the unpolarized, vector-polarized and tensor-polarized parts are expressed
as combinations of pure spin states as
\begin{subequations}
\begin{align}
& \rho^{\alpha\beta}[\textrm{unpol}]
\nonumber
\\
&= \; \frac{1}{3} \left[ \rho^{\alpha\beta}(N, +1) 
+ \rho^{\alpha\beta}(N, -1) + \rho^{\alpha\beta}(N, 0) \right] ,
\\[2ex]
& \rho^{\alpha\beta}[\textrm{vector}, s(N, \pm 1)]
\nonumber
\\
&= \; \pm \frac{1}{2} \left[ \rho^{\alpha\beta}(N, +1)
- \rho^{\alpha\beta}(N, -1)
\right] ,
\\[2ex]
& \rho^{\alpha\beta}[\textrm{vector}, s(N, 0)]
\; = \; 0 ,
\\[2ex]
& \rho^{\alpha\beta} [\textrm{tensor}, t(N, \pm 1)] 
\nonumber
\\
&= \frac{1}{6} \left[ \rho^{\alpha\beta}(N, +1)
+ \rho^{\alpha\beta}(N, -1) - 2 \rho^{\alpha\beta}(N, 0)
\right] ,
\\[2ex]
& \rho^{\alpha\beta} [\textrm{tensor}, t(N, 0)] 
\nonumber \\
&= -\frac{1}{3} \left[ \rho^{\alpha\beta}(N, +1) 
+ \rho^{\alpha\beta}(N, -1) - 2 \rho^{\alpha\beta}(N, 0)
\right] .
\end{align}
\end{subequations}
The relations demonstrate how vector and tensor polarization can be prepared as a superposition of
pure polarization states. The vector-polarized part can be prepared by taking the difference
of the two pure states with $\pm 1$. To prepare vector and tensor polarization, one needs a
superposition of all three polarization states with $\pm 1$ and $0$.

In spin asymmetry measurements we encounter the difference and sum of pure states with 
projection $\pm 1$,
\begin{subequations}
\begin{align}
& \frac{1}{2} \left[ \rho^{\alpha\beta}(N, +1)
\; - \; \rho^{\alpha\beta}(N, -1) \right]
\nonumber
\\[1ex]
&= \; \rho^{\alpha\beta} [\textrm{vector}, s(N, +1)]
\nonumber
\\[1ex]
&= \frac{i}{2M} \epsilon^{\alpha\beta\gamma\delta} p_\gamma N_\delta ,
\label{density_diff_pm1}
\\[2ex]
& \frac{1}{2} \left[ \rho^{\alpha\beta}(N, +1)
\; + \; \rho^{\alpha\beta}(N, -1) \right]
\nonumber
\\[1ex]
&= \; \rho^{\alpha\beta}[\textrm{unpol}] \; + \; \rho^{\alpha\beta} [\textrm{tensor}, t(N, +1)]
\nonumber
\\[1ex]
&= \frac{1}{2} \left( -g^{\alpha\beta} + \frac{p^\alpha p^\beta}{p^2} - N^\alpha N^\beta \right) .
\label{density_sum_pm1}
\end{align}
\end{subequations}
The difference Eq.~(\ref{density_diff_pm1}) involves only the vector-polarized part of the density matrix;
the sum Eq.~(\ref{density_sum_pm1}), involves both the unpolarized and the tensor-polarized parts.
The sum appears in the denominator of spin asymmetry measurements (see below). Inclusion of the 
tensor-polarized part of the density matrix is therefore necessary in the calculation of spin asymmetries
of the spin-1 system. Notice one important difference between spin-1/2 and spin-1 systems: 
In the spin-1/2 system both the polarized and the unpolarized density are formed from the ``maximum-spin'' 
$\pm 1/2$ states, while in the spin-1 system only the polarized density is formed from the 
``maximum-spin'' $\pm 1$ states, and the unpolarized density requires the $0$ spin state in addition.
This has consequences for the number of experimental spin asymmetries that can be formed in the
spin-1 case (see Sec.~\ref{subsec:spin_asymmetries}).
\section{Polarized tagged electron-deuteron scattering}
\label{sec:scattering}
\subsection{Kinematic variables and cross section}
We now discuss the general form of the cross section and structure functions of tagged DIS on the polarized deuteron. 
The structural decomposition and kinematic factors are presented in invariant form. The information on the polarization 
of the deuteron enters through invariants formed out of the 4-vector $s^\mu$ and the 4-tensor $t^{\mu\nu}$
(cf.\ Sec.~\ref{sec:spin_density_matrices}), which can be evaluated in any frame (``effective polarizations''). 
The cross section formulas presented here are general and make no assumption regarding composite 
nuclear structure; results of specific dynamical calculations will be described in Sec.~\ref{sec:ia}. 
The expressions are given with exact kinematic factors including $1/Q^2$ suppressed terms; simplifications pertaining 
to the DIS limit will be made only in the dynamical calculations in Sec.~\ref{sec:ia}.
Our notation and conventions follow those of Ref.~\cite{Strikman:2017koc} unless stated otherwise.

%
%
\begin{figure}
\includegraphics[width=.16\textwidth]{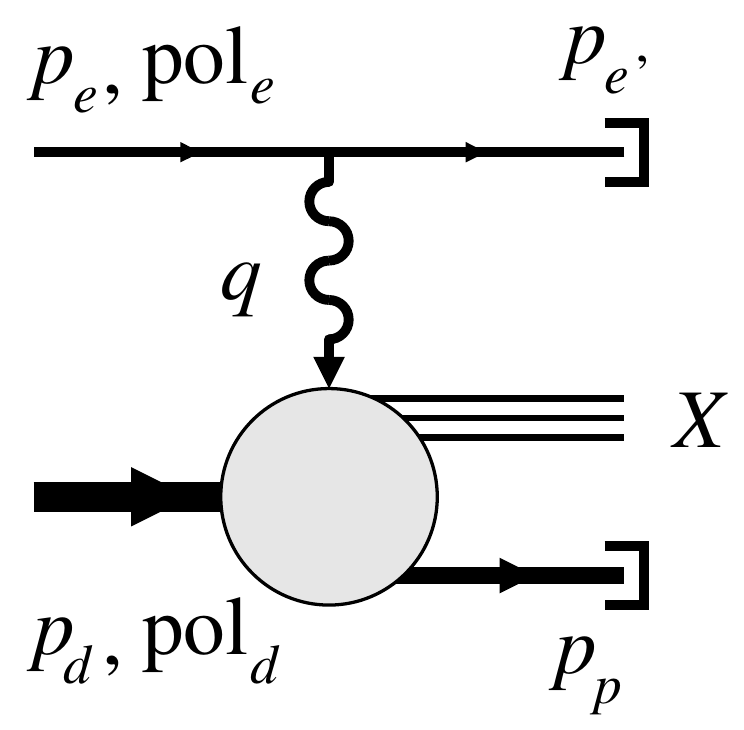}
\caption[]{Polarized tagged DIS on the deuteron, Eq.~(\ref{tagged_dis}).}
\label{fig:deut_tagged_pol}
\end{figure}
We consider polarized deep-inelastic electron scattering on the deuteron with detection of an identified
proton in the deuteron fragmentation region (``tagged DIS,'' see Fig.~\ref{fig:deut_tagged_pol}),
\be
e(p_e, \textrm{pol}_e) + d(p_d, \textrm{pol}_d) &\rightarrow & e'(p_{e'}) + X + p(p_p) .
\label{tagged_dis}
\ee
$p_e$ and $p_d$ are the 4-momenta of the electron and deuteron in the initial state; $\textrm{pol}_e$
and $\textrm{pol}_d$ indicate the variables characterizing their experimental polarization, which will
be specified below. $p_{e'}$ and $p_p$ are the 4-momenta of the scattered electron and the
detected proton in the final state.
The 4-momentum transfer is defined as the difference of the initial and final electron 4-momenta,
\beq
q \; \equiv \; p_e - p_{e'} .
\eeq
The kinematics is described by the invariants
\begin{align}
& (p_d p_e) \; > \; 0, \hspace{1.5em} (p_d q) \; > \; 0, \hspace{1.5em} Q^2 \; \equiv \; - q^2 \; > \; 0 .
\end{align}
The conventional scaling variables are defined as
\begin{subequations}
\begin{align}
x_d \; &\equiv \; \displaystyle \frac{-q^2}{2 (p_d q)}, 
\hspace{2em} 0 < x_d < 1,
\\[1ex]
y \; &\equiv \; \displaystyle \frac{(p_d q)}{(p_d p_e)},
\hspace{2em} 0 < y < 1.
\end{align}
\end{subequations}
The variable $x_d$ is the conventional Bjorken variable for scattering on the deuteron. The rescaled variable
\beq
x \; \equiv \; 2 x_d, \hspace{2em} 0 < x < 2,
\label{x_def}
\eeq
corresponds to the effective Bjorken variable for scattering from a nucleon in the deuteron in the absence of 
nuclear binding. We use $x_d$ in the general kinematic formulas in this section (for easier comparison with 
standard formulas), and $x$ in the dynamical calculations (for simpler matching with nucleon structure functions).
The invariant variables involving the tagged proton momentum are described in detail in Ref.~\cite{Strikman:2017koc}
and will be quoted below.

The differential cross section of polarized tagged DIS, Eq.~(\ref{tagged_dis}), in leading order
of the electromagnetic interaction is given by
\be
d\sigma [ed \rightarrow e'Xp] \; &=& \; \frac{2\pi \alpha_{\rm em}^2 y^2}{Q^6}
\; dx_d \, dQ^2 \, \frac{d\phi_{e'}}{2\pi}
\nonumber \\
&\times& \; \left[ w_e^{\mu\nu} (W_d)_{\mu\nu} \right] \; d\Gamma_p ,
\label{dsigma_tagged}
\ee
where $\alpha_{\rm em} \equiv e^2/(4\pi) \approx 1/137$ is the fine structure constant, 
$d\phi_{e'}$ is the differential of the azimuthal angle of the scattered electron around the incident
electron direction, and $d\Gamma_p$ is the invariant phase space element in the tagged proton 
momentum \cite{Strikman:2017koc}. The expression in brackets is the contraction between the electron and
deuteron scattering tensors. The initial electron is in a pure spin state described by its helicity 
$\Lambda_e = \pm \frac{1}{2}$; we neglect the electron mass ($m_e \rightarrow 0$) and assume helicity 
conservation ${\Lambda'}_{\!\! e} = \Lambda_e$. The electron tensor is given by
\begin{align}
w_e^{\mu\nu} \; &\equiv \; w_e^{\mu\nu}(p_{e'}, p_e, \Lambda_e)
\nonumber \\
&= \; \langle e' (p_e', \Lambda_e) | J^\mu | e (p_e, \Lambda_e) \rangle^\ast 
\nonumber \\
&\times \; \langle e' (p_e', \Lambda_e) | J^\nu | e (p_e, \Lambda_e) \rangle ,
\end{align}
where $J^{\mu}$ is the electromagnetic current operator at space-time point $x = 0$. 
The tensor consists of an unpolarized (helicity-independent) and a polarized (helicity-dependent) part,
\begin{subequations}
\begin{align}
w_e^{\mu\nu} &= w_e^{\mu\nu}[\textrm{unpol}] + w_e^{\mu\nu}[\textrm{pol}],
\\[1ex]
w_e^{\mu\nu}[\textrm{unpol}] &= 4 p_{e}^\mu p_{e}^\nu 
- 2 (q^\mu p_{e}^\nu + p_{e}^\mu q^\nu) + q^2 g^{\mu\nu} ,
\label{electron_tensor_unpol}
\\[1ex]
w_e^{\mu\nu}[\textrm{pol}] &= (2\Lambda_e) 2 i \epsilon^{\mu\nu\alpha\beta} q_\alpha p_{e, \beta} .
\end{align}
\end{subequations}
The deuteron is in an ensemble of spin states described by a general density matrix 
in spin quantum numbers, $\rho_d$, cf.\ Eq.~(\ref{density_matrix_spin1_orig}).
The deuteron tensor is given by the ensemble average
\begin{subequations}
\begin{align}
& W_d^{\mu\nu} \; \equiv \;
W_d^{\mu\nu} (p_d, q, p_p| \rho_d)
\nonumber
\\
&= \;
\sum_{\lambdadp, \lambda_d} \rho_d (\lambda_d, \lambdadp) \;
W_d^{\mu\nu} (p_d, q, p_p| \lambdadp, \lambda_d) ,
\label{deuteron_tensor_ensemble}
\\[2ex]
& W_d^{\mu\nu} (p_d, q, p_p| \lambdadp, \lambda_d)
\nonumber
\\[1ex]
&\equiv \; (4\pi)^{-1} \sum_X \; (2\pi)^4 \; 
\delta^{(4)} (q + p_d - p_p - p_X)
\nonumber \\
&\times \langle p(p_p), X| J^{\mu} |D(p_d, \lambdadp) \rangle^\ast \,
\langle p(p_p), X| J^{\nu} |D(p_d, \lambda_d) \rangle .
\label{w_mu_nu_current}
\end{align}
\end{subequations}
The last expression is a generalized scattering tensor defined as a matrix between pure spin states, 
$\lambdadp$ and $\lambda_d$, which generally involves non-diagonal elements $\lambdadp \neq \lambda_d$.
The deuteron density matrix can be expressed in covariant form and parameterized by an axial 
4-vector and a 4-tensor, cf.\ Eqs.~(\ref{density_matrix_spin1_decomposition}) et seq.,
\beq
\rho_d \; \leftrightarrow \; s_d^\mu, \, t_d^{\mu\nu} .
\label{deuteron_polarization_vector_tensor}
\eeq
The averaged deuteron tensor Eq.~(\ref{deuteron_tensor_ensemble}) can therefore be organized into an 
unpolarized part, a vector-polarized part linear in $s_d$, and a tensor-polarized part 
linear in $t_d$, 
\begin{align}
& W_d^{\mu\nu} (p_d, q, p_p | \rho_d)
\nonumber
\\[1ex]
&= W_d^{\mu\nu}[\textrm{unpol}] \; + \; W_d^{\mu\nu}[\textrm{vector}] \; + \; W_d^{\mu\nu}[\textrm{tensor}] .
\label{deuteron_tensor_decomposition}
\end{align}
The further structural decomposition of these terms can be performed by using the polarization parameters 
$s_d$ and $t_d$ as building blocks in the construction of independent tensor structures.
This technique permits a simple derivation of the spin structure of the polarized cross section
and represents the main motivation for working with the covariant form of the spin 
density matrix (Sec.~\ref{sec:spin_density_matrices}).
The deuteron tensor satisfies the transversality conditions
\beq
q_\mu W_d^{\mu\nu} \; = \; 0, \hspace{2em} W_d^{\mu\nu} q_\nu \; = \; 0,
\label{transversality}
\eeq
which express the conservation of the electromagnetic current. Because they hold for any polarization state, 
the conditions must be satisfied by the individual terms in the decomposition Eq.~(\ref{deuteron_tensor_decomposition})
and constrains their tensor structure.

For constructing the independent tensor structures, we introduce a set of orthonormal basis vectors 
in the subspace spanned by the 4-momenta $p_d$ and $q$ (``longitudinal subspace''). With
\begin{subequations}
\begin{align}
L^\mu 
\; &\equiv \; p_d^\mu - \frac{(p_d q) q^\mu}{q^2}, 
\hspace{1em} (qL) \; = \; 0, \hspace{1em} L^2 \; > \; 0, 
\label{L_def}
\\[1ex]
L^2 
\; &= \;
\frac{(p_d q)^2}{Q^2} (1 + \gamma^2) 
\;\; = \;\;
\frac{Q^2}{4 x_d^2} (1 + \gamma^2) ,
\\[1ex]
\gamma^2 
\; &\equiv \; 
\frac{M_d^2 Q^2}{(p_d q)^2} \; = \; \frac{4 x_d^2 M_d^2}{Q^2} ,
\label{gamma_def}
\end{align}
\end{subequations}
where $M_d$ is the deuteron mass, we define the unit vectors
\begin{align}
& e_L^\mu \; \equiv \; \frac{L^\mu}{\sqrt{L^2}}, 
\hspace{2em}
e_q^\mu \; \equiv \; \frac{q^\mu}{\sqrt{-q^2}}, 
\nonumber 
\\[1ex]
& e_L^2 \; = \; 1, 
\hspace{2em}
e_q^2 \; = \; -1,
\hspace{2em}
(e_L e_q) \; = \; 0 .
\label{unit_vectors_L_q}
\end{align}
By constructing normalized tensors out of the unit vectors, we separate ``geometry'' from ``structure''
and obtain the invariant structure functions with a natural normalization.
\subsection{Unpolarized part}
\label{subsec:cross_section_unpol}
The unpolarized part of the deuteron tensor Eq.~(\ref{deuteron_tensor_decomposition}) is symmetric in 
the indices $\mu\nu$. Its decomposition is of the form (this is different from Ref.~\cite{Strikman:2017koc};
see below)
\begin{align}
W_d^{\mu\nu}[\textrm{unpol}]
&= - \frac{1}{2} g_T^{\mu\nu} \, F_{[UU,T]d}
\; + \; 
\frac{1}{2} g_L^{\mu\nu} \, F_{[UU,L]d}
\nonumber
\\[1ex]
&+ \textrm{($p_{pT}$-dependent structures)},
\label{deuteron_tensor_unpol}
\\[2ex]
g_T^{\mu\nu} \; &\equiv \; g^{\mu\nu} + e_q^\mu e_q^\nu - e_L^\mu e_L^\nu ,
\\[2ex]
g_L^{\mu\nu} \; &\equiv \; e_L^\mu e_L^\nu .
\end{align}
$g_T^{\mu\nu}$ is the projector on the ``transverse'' subspace, orthogonal to the longitudinal subspace.
In Eq.~(\ref{deuteron_tensor_unpol}) we omit tensor structures that depend on the transverse part of the 
proton momentum; these structures correspond to terms in the 
cross sections that depend on the azimuthal angle of the tagged proton momentum in the collinear 
frame and vanish upon integration over the latter (cf.\ Sec.~\ref{subsec:collinear_frames}). The invariant
structure functions multiplying the tensors depend on $x_d$ and $Q^2$, as well as on variables 
specifying the momentum of the final-state proton (to be described in Sec.~\ref{subsec:spectator_momentum}),
\beq
F_{[UU,T] d} \;\; \equiv \;\; F_{[UU,T]d} (x, Q^2, \{p_p\}), \hspace{2em} \text{etc.}
\label{structure_function_arguments}
\eeq
Here and in the following, we use a notation analogous to that of Refs.~\cite{Bacchetta:2006tn,Cosyn:inprep}
to identify the structure functions corresponding to different electron, deuteron, and virtual photon polarization:
\begin{align}
& F_{[\text{electron--deuteron}, \; \text{photon}]d},
\nonumber
\\[2ex]
& \left\{
\begin{array}{lcl}
\text{electron} &=& U, L \\
\text{deuteron} &=& U, S_L, S_T, T_{LL}, T_{LT}, T_{TT} \\
\text{photon} &=& L, T
\end{array}
\right.
\end{align}
(the precise meaning of the labels will become clear in the following). While more burdensome than the conventional 
notation in simple cases, the new notation is physically meaningful and greatly helps with managing more complex 
expressions involving vector and tensor polarization. The unpolarized deuteron structure functions in 
Eq.~(\ref{deuteron_tensor_unpol}) are related to those of Ref.~\cite{Strikman:2017koc} as
\begin{subequations}
\begin{align}
F_{Ld}(\textrm{Ref.~\cite{Strikman:2017koc}}) &= F_{[UU,L]d},
\\[1ex]
F_{Td}(\textrm{Ref.~\cite{Strikman:2017koc}}) &= F_{[UU,T]d} + F_{[UU,L]d}
\label{structure_functions_SW}
\end{align}
\end{subequations}
They are related to the conventional unpolarized structure functions 
$F_{1d}$ and $F_{2d}$ as
\begin{subequations}
\label{F1_F2}
\begin{align}
F_{[UU, T] d} \; &= \; 2 F_{1d},
\\[1ex]
F_{[UU,L]d} \; &= \; ( 1 + \gamma^2 ) \frac{F_{2d}}{x_d} - 2 F_{1d} ,
\\[1ex]
F_{1d} \; &= \; \frac{1}{2} F_{[UU,T]d},
\\[1ex]
F_{2d} \; &= \; \frac{x_d \left( F_{[UU, T]d} + F_{[UU, L]d} \right)}{1 + \gamma^2} .
\end{align}
\end{subequations}
The longitudinal-transverse ($L/T$) ratio is defined as
\begin{align}
R_d \; \equiv \; \frac{F_{[UU,L]d}}{F_{[UU,T]d}} \; = \; \frac{(1 + \gamma^2) F_{2d}}{2 x_d F_{1d}} - 1 .
\end{align}

For computing the contraction of the unpolarized electron tensor Eq.~(\ref{electron_tensor_unpol})
with the unpolarized deuteron tensor Eq.~(\ref{deuteron_tensor_unpol}), one introduces the virtual 
photon polarization parameter
\begin{align}
\epsilon \; &\equiv \; \frac{\displaystyle \phantom{-} g_L^{\mu\nu} \, (w_e)_{\mu\nu}[\textrm{unpol}]}
{\displaystyle -g_T^{\mu\nu} \, (w_e)_{\mu\nu}[\textrm{unpol}]}
\nonumber
\\[1ex]
\; &= \; \frac{\displaystyle 1 - y - \gamma^2 y^2/4}
{\displaystyle 1 - y + y^2/2 + \gamma^2 y^2/4} ,
\end{align}
which satisfies the relation
\beq
\frac{(2 - y)^2}{y^2} \; \frac{1 - \epsilon}{1 + \epsilon}
\;\; = \;\; 1 + \gamma^2 .
\eeq
The contractions of the electron momentum $p_e$ with the momenta $p_d$ and $q$ can then be expressed in terms 
of either of the variables $y$ or $\epsilon$. Specifically, the contractions of $p_e$ with the longitudinal 
basis vectors Eq.~(\ref{unit_vectors_L_q}) are
\begin{subequations}
\begin{align}
(e_L p_e) \; &= \; 
\frac{Q}{2} \sqrt{\frac{1 + \epsilon}{1 - \epsilon}}
\; = \; \frac{Q (1 - y/2)}{y \sqrt{1 + \gamma^2}},
\\[1ex]
(e_q p_e) \; &= \; -\frac{Q}{2} .
\end{align}
\end{subequations}
The contraction of the unpolarized electron tensor Eq.~(\ref{electron_tensor_unpol})
with the unpolarized deuteron tensor Eq.~(\ref{deuteron_tensor_unpol}) is obtained as
\begin{subequations}
\begin{align}
& (w_e)_{\mu\nu}[\textrm{unpol}] \, W_d^{\mu\nu}[\textrm{unpol}]
\nonumber \\
&= \; \frac{Q^2}{1 - \epsilon} \left( F_{[UU,T]d} + \epsilon F_{[UU,L]d} \right)
\nonumber \\[1ex]
&+ \textrm{($p_{pT}$-dependent structures)}
\\
&= \; \frac{Q^2}{1 - \epsilon} F_{[UU,T]d} (1 + \epsilon R_d)
\nonumber \\[1ex]
&+ \textrm{($p_{pT}$-dependent structures)}.
\end{align}
\end{subequations}
The expression does not include structures that explicitly depend on the proton
transverse momentum. We note that, if such structures were included, the unpolarized
deuteron tensor would no longer be symmetric and have a non-zero contraction with the
polarized electron tensor, resulting in an electron single-spin dependent term
in the cross section \cite{Cosyn:inprep}.
\subsection{Vector-polarized part}
\label{subsec:cross_section_vector}
The vector-polarized part of the deuteron tensor Eq.~(\ref{deuteron_tensor_decomposition}) 
depends linearly on the axial 4-vector $s_d$ and contains terms antisymmetric and symmetric in 
$\mu\nu$ (the symmetric term corresponds to a single-spin dependence of the unpolarized
electron scattering cross section that is forbidden in strictly inclusive DIS but
allowed in tagged DIS; see below).
In constructing the independent tensor structures we must take into account that the
axial 4-vector is orthogonal to the deuteron momentum, $(p_d s_d) = 0$. It is convenient 
to introduce an alternative set of longitudinal basis vectors aligned with $p_d$ rather than $q$. 
With 
\begin{subequations}
\begin{align}
L_\ast^\mu \; &= \; q^\mu - \frac{(p_d q)}{p_d^2} p_d^\mu,
\hspace{2em} (p_d L_\ast) \; = \; 0, 
\\[1ex]
L_\ast^2 \; &= \; - \frac{(p_d q)^2}{p_d^2} (1 + \gamma^2) \; = \; -\frac{Q^2}{M_d^2} L^2 
\end{align}
\end{subequations}
we define unit vectors
\begin{align}
& e_d^\mu \; \equiv \; \frac{p_d^\mu}{\sqrt{p_d^2}}, 
\hspace{2em}
e_{L\ast}^\mu \; \equiv \; \frac{L_\ast^\mu}{\sqrt{-L_\ast^2}}, 
\nonumber
\\[1ex]
& e_d^2 \; = \; 1, 
\hspace{2em}
e_{L\ast}^2 \; = \; -1,
\hspace{2em}
(e_d e_{L\ast}) \; = \; 0 
\label{unit_longitudinal_alt}
\end{align}
The relation between the two sets of unit vectors, Eqs.~(\ref{unit_vectors_L_q}) and 
(\ref{unit_longitudinal_alt}), is
\beq
\left( \begin{array}{l}
e_d \\[1ex] e_{L\ast}
\end{array} \right)
\; = \;
\frac{1}{\gamma}
\left( \begin{array}{cc}
\sqrt{1 + \gamma^2} & -1 \\[1ex] -1 & \sqrt{1 + \gamma^2}
\end{array} \right)
\left( \begin{array}{l}
e_L \\[1ex] e_q
\end{array} \right)
\label{basis_relation}
\eeq
The antisymmetric term in the vector-polarized deuteron tensor is then decomposed as
\begin{align}
W_d^{\mu\nu}[\textrm{vector}]
\; &= \; \frac{i}{2} \epsilon^{\mu\nu\rho\sigma} e_{q, \rho}
\left\{ e_{L\ast, \sigma} \, e_{L\ast, \tau} \, \gamma F_{[LS_L]d}
\right.
\nonumber
\\[1ex]
&+ \; \left. \left( e_{L\ast, \sigma} \, e_{L\ast, \tau} + g_{\sigma\tau} \right) F_{[LS_T]d} \right\} s_d^\tau
\nonumber
\\[2ex]
&+ \textrm{($p_{pT}$-dependent structures).}
\label{deuteron_tensor_polarized}
\end{align}
Again we omit terms corresponding to azimuthal-angle dependent structures.
The factor $\gamma$ in first term ensures proper normalization of the tensor,
\beq
\gamma \; = \; \frac{1}{\sqrt{(e_q e_{L\ast})^2 - 1}}
\eeq
The polarized deuteron structure functions in Eq.~(\ref{deuteron_tensor_polarized})
are related to the conventional structure functions $g_{1d}$ and $g_{2d}$ as
\begin{subequations}
\begin{align}
F_{[LS_L]d} \; &= \; 2 (g_{1d} - \gamma^2 g_{2d}) ,
\\[2ex]
F_{[LS_T]d} \; & = \; - 2\gamma (g_{1d} + g_{2d}) ,
\\[1ex]
g_{1d} \; &= \; \frac{F_{[LS_L]d} - \gamma F_{[LS_T]d}}{2(1 + \gamma^2)} ,
\\[1ex]
g_{2d} \; &= \; \frac{-\gamma F_{[LS_L]d} - F_{[LS_T]d}}{2 \gamma (1 + \gamma^2)} .
\end{align}
\end{subequations}
The symmetric term of the vector-polarized deuteron tensor is parameterized as
\begin{subequations}
\begin{align}
W_d^{\mu\nu}[\textrm{vector}]
\; &= \; -\frac{1}{2} (e_L^\mu X^\nu + X^\mu e_L^\nu) \, F_{[US_T]}
\nonumber
\\[1ex]
&+ \; \textrm{($p_{pT}$-dependent structures)} ,
\\[1ex]
X^\mu \; &\equiv \; \epsilon^{\mu\alpha\beta\gamma} e_{L, \alpha} \, e_{q, \beta} \, s_{d, \gamma}
\label{deuteron_tensor_polarized_symm}
\end{align}
\end{subequations}
$X$ is a true 4-vector constructed from the axial 4-vector $s_d$.

To compute the contraction with electron tensor, we expand the electron 4-momentum
in the basis vectors Eq.~(\ref{unit_longitudinal_alt}),
\be
p_e^\mu &=& (e_d p_e) \, e_d^\mu \; - \; (e_{L\ast} p_e) \, e_{L\ast}^\mu \; + \; p_{eT}^\mu ,
\ee
where
\begin{subequations}
\begin{align}
(e_d p_e) \; &= \; \frac{Q}{2\gamma} \left( \sqrt{1 + \gamma^2} \sqrt{\frac{1 + \epsilon}{1 - \epsilon}} + 1 \right)
\; = \; \frac{Q}{\gamma y} ,
\\[2ex]
(e_{L\ast} p_e) \; &= \; - \frac{Q}{2\gamma} \left( \sqrt{\frac{1 + \epsilon}{1 - \epsilon}} 
+ \sqrt{1 + \gamma^2}  \right)
\nonumber
\\[2ex]
&= \; -\frac{Q (1 + \gamma^2 y/2)}{\gamma y \sqrt{1 + \gamma^2}} ,
\label{contraction_eL*_pe}
\\[2ex]
p_{eT}^2 \; &= \; - \frac{Q^2 \epsilon}{2 (1 - \epsilon)}
\nonumber
\\[2ex]
&= \; -\frac{Q^2 (1 - y - \gamma^2 y^2/4)}{(1 + \gamma^2) y^2} .
\end{align}
\end{subequations}
The spacelike 4-vector $p_{eT}$ is the ``transverse'' part of the electron 4-momentum, i.e., the component 
orthogonal to the longitudinal subspace spanned by $p_d$ and $q$ or the related unit vectors.
We define transverse unit vectors as
\begin{align}
e_{T1}^\mu \; &\equiv \; \frac{p_{eT}^\mu}{\sqrt{-p_{eT}^2}},
\nonumber
\\[1ex]
e_{T2}^\mu \; &\equiv \; \epsilon^{\mu\alpha\beta\gamma} e_{d, \alpha} \, e_{L\ast, \beta} \, e_{T1, \gamma}
\; = \; \epsilon^{\mu\alpha\beta\gamma} e_{L, \alpha} \, e_{q, \beta} \, e_{T1, \gamma},
\nonumber
\\[2ex]
e_{T1}^2 \; &= \; e_{T2}^2 \; = \; -1 .
\label{unit_transverse}
\end{align}
$e_{T1}$ is along the direction of $p_{eT}$ in transverse space, while $e_{T2}$ is orthogonal to it.
With these definitions, the set
\beq
\{ e_d, \, e_{L\ast}, \, e_{T1}, \, e_{T2} \}
\label{basis_set}
\eeq
provides a complete orthonormal basis of the 4-dimensional space and can be used to expand
other kinematic vectors [the relation to the other basis set with $e_q$ and $e_L$ 
is given by Eq.~(\ref{basis_relation})].
We expand the deuteron polarization 4-vector $s_d$ in the second basis set. The contraction 
of the electron tensor with the vector-polarized deuteron tensor is obtained as
\begin{subequations}
\begin{align}
&(w_e)_{\mu\nu}[\textrm{pol}] \; W_d^{\mu\nu}[\textrm{vector}]
\nonumber
\\
&= \; (2\Lambda_e) \, \frac{Q^2}{1 - \epsilon} \left\{ \sqrt{1 - \epsilon^2} \; S_L \; F_{[LS_L]d}
\right.
\nonumber
\\
&+ \; \sqrt{2 \epsilon (1 - \epsilon)} \; S_T \cos\phi_S \;
F_{[L S_T]d}
\nonumber
\\[1ex]
&+ \left. \textrm{($p_{pT}$-dependent structures)} \phantom{\sqrt{0^2}} \hspace{-1.6em} \right\} ,
\\[2ex]
& (w_e)_{\mu\nu}[\textrm{unpol}] \; W_d^{\mu\nu}[\textrm{vector}]
\nonumber
\\
&= \; \frac{Q^2}{1 - \epsilon} \sqrt{2 \epsilon (1 + \epsilon)} \; S_T \sin\phi_S \; F_{[US_T]d}
\nonumber
\\[1ex]
&+ \textrm{($p_{pT}$-dependent structures)},
\label{contraction_unpol_vector}
\end{align}
\end{subequations}
where the effective vector polarizations are defined as
\begin{subequations}
\label{S_invariant}
\begin{align}
S_L \; &\equiv \; \phantom{-} (e_{L\ast} s_d),
\label{S_L_invariant}
\\[2ex]
S_T \cos\phi_S \; &\equiv \; -(e_{T1} s_d),
\label{S_T_cos_invariant}
\\[2ex]
S_T \sin\phi_S \; &\equiv \; -(e_{T2} s_d) .
\label{S_T_sin_invariant}
\end{align}
\end{subequations}
They are given in invariant form, as contractions of the deuteron polarization 4-vector $s_d$ with the
kinematic vectors of the scattering process, and can be evaluated in any frame, depending 
on the experimental setup.\footnote{In Eqs.~(\ref{S_T_cos_invariant}) and (\ref{S_T_sin_invariant})
we express the contractions of $s_d$ with the transverse basis vectors, $e_{T1}$ and $e_{T2}$,
in terms of a magnitude $S_T > 0$ and an angle $\phi_S$. This does not imply reference to 
any particular frame, as both parameters are unambiguously defined in terms of the invariant 
4-vector contractions. In the collinear frames of Sec.~\ref{subsec:collinear_frames}, $S_T$ 
and $\phi_S$ do indeed correspond to the magnitude and azimuthal angle of the transverse
component of the spin vector. The same applies to the effective tensor polarizations
introduced in Sec.~\ref{subsec:cross_section_tensor}.} 
In Sec.~\ref{subsec:effective_polarization} we derive their
specific values in colliding-beam experiments with polarized beams. Note that the
effective polarizations satisfy the relation (``sum rule'')
\beq
S_L^2 + S_T^2 \; = \; -s_d^2 \; = \; |\bm{S}_d|^2,
\label{effective_polarization_total}
\eeq
where $|\bm{S}_d|^2$ is the squared modulus of the deuteron polarization vector 
in the rest frame, cf.\ Eq.~(\ref{a_restframe}). 

Some comments are in order regarding the symmetric term of the vector-polarized deuteron tensor 
Eq.~(\ref{deuteron_tensor_polarized_symm}) and the resulting deuteron spin dependence 
in unpolarized electron scattering Eq.~(\ref{contraction_unpol_vector}).
This term describes a dependence of the unpolarized electron scattering cross section on the 
deuteron spin perpendicular to the electron scattering plane (normal single-spin asymmetry).
In strictly inclusive electron scattering such a single-spin dependence is forbidden in leading
order of the electromagnetic interaction (one-photon exchange) and can appear only in higher 
orders (two-photon exchange) \cite{Christ:1966zz,Afanasev:2007ii}. Because tagged DIS
is semi-inclusive scattering, in which one places conditions on the hadronic final state,
the standard argument prohibiting a single-spin dependence in leading order is not applicable.
We therefore cannot rule out a single-spin dependence of the tagged DIS cross section,
even after integration over the azimuthal angle of the tagged proton momentum
in the collinear frame (see below). There certainly are non-zero single-spin dependent terms 
in the azimuthal-angle dependent tagged DIS cross section \cite{Cosyn:inprep}.
\subsection{Tensor-polarized part}
\label{subsec:cross_section_tensor}
The tensor-polarized part of the deuteron tensor Eq.~(\ref{deuteron_tensor_decomposition}) 
depends linearly on the 4-tensor $t_d$, Eq.~(\ref{deuteron_polarization_vector_tensor}).
Its decomposition in independent structures can be derived using the same methods as 
for the vector-polarized part in Sec.~\ref{subsec:cross_section_vector}. We expand
the tensor $t_d$ in the basis Eq.~(\ref{basis_set}) and construct all
independent structures satisfying the transversality condition Eq.~(\ref{transversality}). 
In this way we obtain the decomposition
\begin{align}
& W_d^{\mu\nu}[\textrm{tensor}]
\nonumber
\\
&=  \frac{1}{2} \, e_{L\ast}^\rho e_{L\ast}^\sigma (t_d)_{\rho\sigma} 
\left( -g_T^{\mu\nu} F_{[UT_{LL},T] d} + e_L^\mu e_L^\nu F_{[UT_{LL},L] d} \right)
\nonumber \\[1ex]
&-\frac{1}{2} \, e_{L\ast}^\rho \, 
(g_T^{\sigma\mu} \, e_L^\nu +  g_T^{\sigma\nu} e_L^\mu ) \, (t_d)_{\rho\sigma} \; F_{[UT_{LT}]d} 
\nonumber \\[1ex]
&+
\frac{1}{2} \, ( g_T^{\mu\rho} \, g_T^{\sigma\nu} - \epsilon_T^{\mu\rho} \, \epsilon_T^{\sigma\nu})
\, (t_d)_{\rho\sigma} \, F_{[UT_{TT}]d} 
\nonumber \\[1ex]
&- \frac{i}{2} \, \epsilon^{\mu\nu\rho\sigma} \, e_{q,\rho} \,
(\epsilon_T)_{\sigma\tau} \, e_{L\ast,\omega} \, (t_d)^{\tau\omega} \, F_{[LT_{LT}]d}
\nonumber \\[2ex]
&+ \textrm{($p_{pT}$-dependent structures),}
\label{decomposition_tensor}
\end{align}
where
\begin{subequations}
\begin{align}
g_T^{\mu\nu} \; &= \; g^{\mu\nu} + e_q^\mu e_q^\nu -e_L^\mu e_L^\nu 
\; = \; g^{\mu\nu} + e_{L*}^\mu e_{L*}^\nu -e_d^\mu e_d^\nu
\nonumber \\
&= \; -e_{T1}^\mu e_{T1}^\nu - e_{T2}^\mu e_{T2}^\nu \,,
\\[1ex]
\epsilon_T^{\mu\nu} \; &= \; \epsilon^{\mu\nu\rho\sigma} e_{L,\rho} \, e_{q,\sigma}\; =\; 
\epsilon^{\mu\nu\rho\sigma} e_{d, \rho} \, e_{L*,\sigma}
\nonumber \\
&= \; e_{T1}^\mu e_{T2}^\nu - e_{T2}^\mu e_{T1}^\nu\,.
\label{deuteron_tensor_tensor}
\end{align}
\end{subequations}
Again we omit terms corresponding to azimuthal-angle dependent structures. The first two terms in 
Eq.~(\ref{decomposition_tensor}) are symmetric in $\mu\nu$ and have the same structure as the unpolarized 
deuteron tensor Eq.~(\ref{deuteron_tensor_unpol}). The third and fourth term are likewise symmetric 
in $\mu\nu$. These terms contribute to the cross section of unpolarized electron scattering from
the tensor-polarized deuteron. For reference we note that our symmetric tensor-polarized structure functions 
in Eq.~(\ref{deuteron_tensor_tensor}) are related to the $b_{1d},... b_{4d}$ structure functions
of Ref.~\cite{Hoodbhoy:1988am} by\footnote{Ref.~\cite{Hoodbhoy:1988am} considers inclusive DIS
on the tensor-polarized deuteron, while we consider tagged DIS. The correspondence pertains to the 
tagged structure functions that survive integration over the proton momentum, which are
the ones listed in Eq.~(\ref{decomposition_tensor}).}
\begin{subequations}
\begin{align}
& F_{[UT_{LL},L]d}
\nonumber \\
&= \;   \frac{1}{x_d}\left[2(1+\gamma^2)x_d b_{1d}-(1+\gamma^2)^2
\left( \frac {1}{3} b_{2d} + b_{3d} + b_{4d} \right)\right.
\nonumber
\\[1ex] 
& \left. -(1+\gamma^2)\left(\frac{1}{3}b_{2d}-
b_{4d}\right)-\left(\frac{1}{3}b_{2d}-b_{3d} \right) \right] ,
\\[2ex]
& F_{[UT_{LL},T]d}
\nonumber
\\[1ex] 
&= \; -\left[2(1+\gamma^2) b_{1d}
-\frac{\gamma^2}{x_d}\left(\frac{1}{6} b_{2d} - \frac{1}{2} b_{3d} \right)\right ] ,
\\[2ex]
& F_{[UT_{LT}]d}
\nonumber
\\[1ex] 
 &= \; -\frac{\gamma}{2x_d}\left[(1+\gamma^2)
\left(\frac{1}{3}
b_{2d} - b_{4d}\right) + \left(\frac{2}{3}b_{2d} - 2b_{3d} \right)\right] ,
\\[1ex]
& F_{[UT_{TT}]d}
\; = \; - \frac{\gamma^2}{x_d}
\left( \frac{1}{6} b_{2d} - \frac{1}{2} b_{3d} \right) .
\end{align}
\end{subequations}
The tensor-polarized part of the deuteron tensor also contains a structure antisymmetric in $\mu\nu$, 
analogous to that appearing in the vector-polarized part Eq.~(\ref{deuteron_tensor_polarized_symm}).
This structure is absent in inclusive DIS but may be non-zero in tagged DIS (cf.\ the discussion in 
Sec.~\ref{subsec:cross_section_vector}).

The contraction of the electron tensor with the tensor-polarized part of the deuteron tensor 
Eq.~(\ref{decomposition_tensor}) is computed in the same way as for the vector-polarized part.
We obtain
\begin{subequations}
\begin{align}
& (w_e)_{\mu\nu}[\textrm{unpol}] \, W_d^{\mu\nu}[\textrm{tensor}]
\nonumber
\\
&= \; \frac{Q^2}{1 - \epsilon} \; \left\{ T_{LL} \,
\left( F_{[UT_{LL},T]d} + \epsilon F_{[UT_{LL},L]d} \right) \phantom{\sqrt{0}} \right.
\nonumber \\[1ex]
&+ \; \sqrt{2\epsilon(1+\epsilon)}\; T_{LT} \cos \phi_{T_{L}} \;
F_{[UT_{LT}]d}
\nonumber\\[1ex]
&+ \; \epsilon \; T_{TT} \cos 2\phi_{T_T} \; F_{[UT_{TT}]d}  
\nonumber \\[1ex]
&+ \; (2\Lambda_e) \sqrt{2\epsilon(1-\epsilon)} \; T_{LT}
\sin \phi_{T_{L}} \; F_{[LT_{LT}]d}
\nonumber
\\[1ex]
&+ \left. \textrm{($p_{pT}$-dependent structures)} \phantom{\sqrt{0^2}} \hspace{-1.6em} \right\} ,
\end{align}
\end{subequations}
where the effective tensor polarizations are defined as
\begin{subequations}
\label{T_invariant}
\begin{align}
T_{LL} \; &\equiv \; (e_{L\ast} \, t_d \, e_{L\ast})
\; \equiv \; e_{L\ast}^\rho \, (t_d)_{\rho\sigma} \, e_{L\ast}^\sigma ,
\\[1ex]
T_{LT}\cos \phi_{T_{L}} \; &\equiv \; -(e_{T1} \, t_d \, e_{L\ast}),
\\[1ex]
T_{LT}\sin \phi_{T_{L}} \; &\equiv \; -(e_{T2} \, t_d \, e_{L\ast}),
\\[1ex]
T_{TT}\cos 2\phi_{T_{T}} \; &\equiv \; (e_{T1} \, t_d \, e_{T1}) - (e_{T2} \, t_d \, e_{T2}) .
\end{align}
\end{subequations}
They are given in invariant form, as contractions of the deuteron polarization 4-tensor $t_d$ 
with the kinematic vectors of the scattering process. Regarding the representation of the transverse
contractions in terms of magnitudes and angles, the same comments apply as in the
vector-polarized case.
\subsection{Cross section summary}
Combining the results of Secs.~\ref{subsec:cross_section_unpol}, \ref{subsec:cross_section_vector}
and \ref{subsec:cross_section_tensor}, and using Eq.(\ref{dsigma_tagged}), we can now assemble the
general expression of the cross section of polarized tagged electron-deuteron scattering.
We separate the terms independent of the electron helicity ($U$) and proportional 
to the electron helicity ($L$).
\begin{subequations}
\label{dsigma_tagged_explicit}
\begin{align}
d\sigma [ed \rightarrow e'Xp] \; &= \; \frac{2\pi \alpha_{\rm em}^2 y^2}{Q^4 (1 - \epsilon)}
\; dx_d \, dQ^2 \, \frac{d\phi_{e'}}{2\pi}
\nonumber \\
&\times \; \left( \mathcal{F}_{[U]d} \; + \; \mathcal{F}_{[L]d} \right) \; d\Gamma_p ,
\end{align}
\begin{align}
\mathcal{F}_{[U]d} \; &= \; F_{[UU,T]d} \; + \; \epsilon F_{[UU,L]d}
\nonumber \\[2ex]
&+ \; T_{LL} \, \left( F_{[UT_{LL},T]d} + \epsilon F_{[UT_{LL},L]d} \right)
\nonumber
\\[2ex]
&+ \; \sqrt{2 \epsilon (1 + \epsilon)} \; S_T \sin\phi_S \; F_{[US_T]d}
\nonumber
\\[2ex]
&+ \; \sqrt{2\epsilon(1+\epsilon)}\; T_{LT} \cos \phi_{T_{L}} \; F_{[UT_{LT}]d}
\nonumber
\\[2ex]
&+ \; \epsilon \; T_{TT} \cos 2\phi_{T_T} \; F_{[UT_{TT}]d} 
\nonumber
\\[2ex]
&+ \; \textrm{($p_{pT}$-dependent structures)} ,
\\[2ex]
\mathcal{F}_{[L]d} \; &= \; (2\Lambda_e) \left\{ \sqrt{1 - \epsilon^2} \; S_L \; F_{[LS_L]d}
\right.
\nonumber
\\[2ex]
&+ \; \sqrt{2 \epsilon (1 - \epsilon)} \; S_T \cos\phi_S \; 
F_{[L S_T]d}
\nonumber
\\[2ex]
&+ \; \sqrt{2\epsilon(1-\epsilon)} \; T_{LT} \sin \phi_{T_{L}} \; F_{[LT_{LT}]d}
\nonumber
\\[2ex]
&+ \left. \textrm{($p_{pT}$-dependent structures)} \phantom{\sqrt{0^2}} \hspace{-1.6em} \right\} .
\end{align}
\end{subequations}
The expression includes all terms that do not depend on the azimuthal angle of the tagged proton
momentum in the collinear frame and do not vanish upon integration over that variable. The
effective polarization parameters are defined in Eqs.~(\ref{S_invariant})
and (\ref{T_invariant}). The invariant structure functions depend on 
$x$ and $Q^2$, as well as on variables specifying the tagged proton momentum (to be described
in Sec.~\ref{subsec:spectator_momentum}), cf.\ Eq.~(\ref{structure_function_arguments}).
\subsection{Effective polarizations}
\label{subsec:effective_polarization}
\begin{figure}
\includegraphics[width=.4\textwidth]{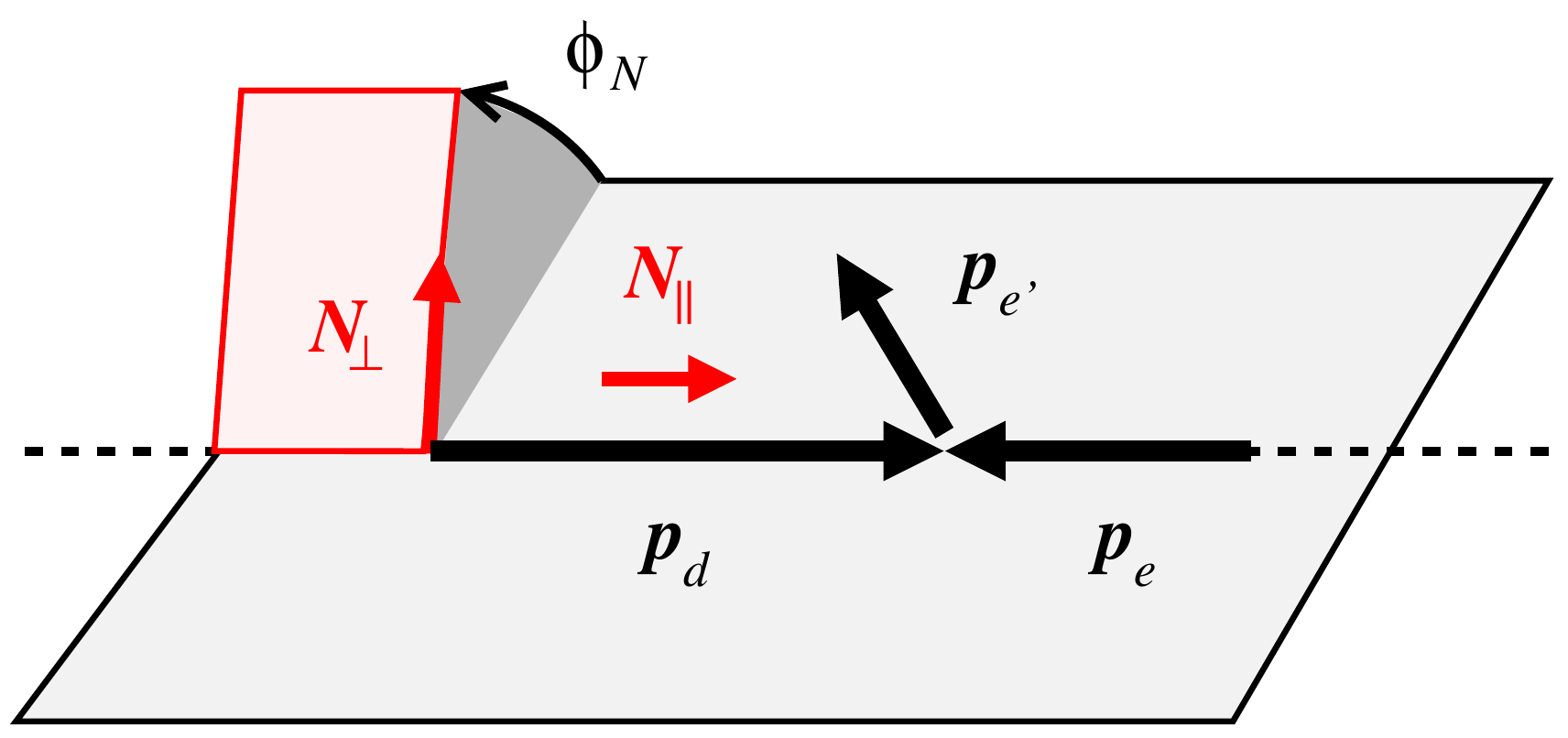}
\caption[]{The experimental deuteron polarization in a frame where the deuteron and electron momenta 
are collinear (beam axis). The vectors $\bm{N}_\parallel$ and $\bm{N}_\perp$ indicate the
directions of parallel and perpendicular polarization. The angle $\phi_N$ of $\bm{N}_\perp$
is measured relative to the plane defined by the beam axis and the scattered electron momentum.}
\label{fig:electron_deuteron_frame}
\end{figure}

In the cross section Eq.~(\ref{dsigma_tagged_explicit}) the information about deuteron polarization is
contained in the invariant effective polarizations Eqs.~(\ref{S_invariant})
and (\ref{T_invariant}), which are defined in terms of contractions of the 
deuteron polarization 4-vector and 4-tensor with kinematic vectors of the scattering process. 
In experiments the deuteron polarization is prepared with respect to some fixed axes determined
by the experimental setup. In order to evaluate the cross section and spin asymmetries one has to express 
the invariant effective polarizations in terms of the experimental polarizations specific to that setup.
Here we consider the situation that the experimental polarizations are specified in a reference frame 
in which the electron and deuteron 3-momenta are collinear and define an axis (see Fig.~\ref{fig:electron_deuteron_frame}),
\beq
\bm{p}_{e} \parallel \bm{p}_d .
\label{electron_deuteron_frame}
\eeq
This covers two cases of interest: 
(a) Fixed-target experiments ($\bm{p}_d = 0$), in which the deuteron polarization is specified
relative to the electron beam axis; 
(b) Colliding-beam experiments, in which the beams collide head-on (zero crossing angle) and the
deuteron polarization is specified relative to the common beam axis.
We refer to the common axis as the ``beam axis'' and denote the directions parallel and 
perpendicular to it by $\parallel$ and $\perp$. We consider 
pure deuteron spin states polarized along a fixed axis (parallel or perpendicular to the beam axis),
and denote the spin projection along this axis by $\Lambda_d = \{\pm 1, 0\}$.\footnote{It is 
important to distinguish between the frames 
in which the electron and deuteron momentum are collinear, Eq.~(\ref{electron_deuteron_frame})
(in which the experimental polarization is prepared), and the frames in which the virtual photon
and the deuteron momentum are collinear, Sec.~\ref{subsec:collinear_frames} (in which the 
theoretical analysis of the cross section is performed). We use ``parallel'' and ``perpendicular''
to refer to the directions in electron-deuteron frame, and ``longitudinal'' and ``transverse'' to 
refer to the directions in virtual photon-deuteron frame. The term ``collinear frame'' per se
always refers to the virtual photon-deuteron frame.}
The covariant deuteron density matrix for these pure polarization states is obtained
from the general expressions in Sec.~\ref{subsec:spin_density_spin_1}, Eqs.(\ref{vector_special}) et seq.
%
%

In pure deuteron spin states polarized parallel to the beam axis, the polarization 4-vector 
is of the form Eq.~(\ref{vector_special}),
\begin{align}
s_d^\mu \; = \; \Lambda_d N^\mu, 
\hspace{2em} p_d N \; = \; 0, \hspace{2em} N^2 \; = \; -1.
\label{a_parallel}
\end{align}
The 4-vector $N$ can be expanded covariantly in the electron and deuteron 4-momentum as
\be
N^\mu &=& \frac{1}{M_d} \left[ p_d^\mu - \frac{M_d^2}{(p_e p_d)} p_e^\mu \right] .
\label{n_parallel}
\ee
In the deuteron rest frame its components are
\beq
N^\mu [\textrm{RF}] = \left( 0, -\frac{\bm{p}_e}{|\bm{p}_e|} \right),
\eeq
and one sees that $\Lambda_d = +1 \; (-1)$ corresponds to deuteron polarization 
opposite to the direction (in the direction) of the electron momentum. The effective vector 
polarizations Eq.~(\ref{S_invariant}) are calculated as contractions
of Eq.~(\ref{a_parallel}), using Eq.~(\ref{contraction_eL*_pe}) and (\ref{effective_polarization_total}).
We obtain
\begin{subequations}
\begin{align}
S_L \; &= \; \frac{1 + \gamma^2 y/2}{\sqrt{1 + \gamma^2}} \; \Lambda_d ,
\\[1ex]
S_T \cos\phi_S \; &= \;  -\frac{\gamma \sqrt{1 - y - \gamma^2 y^2/4}}{\sqrt{1 + \gamma^2}} \; \Lambda_d ,
\\[2ex]
S_T \sin\phi_S \; &= \; 0 .
\end{align}
\label{S_parallel}
\end{subequations}
The polarization 4-tensor in the same pure states is given by the general formula
Eq.~(\ref{tensor_special}), with the unit vector $N$ of Eq.~(\ref{n_parallel})
and the spin projection $\Lambda_d$. The effective tensor polarizations, Eq.~(\ref{T_invariant}), 
are calculated as contractions of that 4-tensor. In the case $\Lambda_d = \pm 1$ they
can be obtained from the vector polarizations Eq.~(\ref{S_parallel}) as
\begin{subequations}
\label{T_from_S}
\begin{align}
T_{LL} 
&= \frac{1}{6} \left( 3 S_L^2 - 1\right) ,
\\[2ex]
T_{LT}\cos\phi_{T_L} 
&= \frac{1}{2} S_LS_T\cos\phi_S ,
\\[2ex]
T_{LT}\sin\phi_{T_L} 
&= \frac{1}{2} S_LS_T\sin\phi_S ,
\\[2ex]
T_{TT}\cos2\phi_{T_T} 
&= \frac{1}{2} S_T^2 (\cos^2\phi_S - \sin^2\phi_S)
\\[2ex]
& \hspace{1em}  [\textrm{$\Lambda_d = \pm 1$ only}] .
\nonumber
\end{align}
\end{subequations}
The explicit expressions are
\begin{subequations}
\label{T_parallel}
\begin{align}
T_{LL} 
&= \frac{1}{6} \left[ \frac{3 (1 + \gamma^2 y/2)^2}{1 + \gamma^2} - 1 \right] ,
\\[2ex]
T_{LT}\cos\phi_{T_L} 
&= \frac{1}{2} \, \frac{\gamma  (1 + \gamma^2 y/2) \sqrt{1 - y - \gamma^2 y^2/4}}{1 + \gamma^2} ,
\\[2ex]
T_{LT}\sin\phi_{T_L} 
&= 0 ,
\\[2ex]
T_{TT}\cos2\phi_{T_T} 
&=
\frac{1}{2} \; \frac{\gamma^2 \left(1 - y - \gamma^2 y^2/4\right)}{1 + \gamma^2} 
\\[2ex]
& \hspace{1em}  [\textrm{$\Lambda_d = \pm 1$}].
\end{align}
\end{subequations}
In the case $\Lambda_d = 0$ the effective tensor polarizations are given by $(-2)$ times the 
expressions in Eq.~(\ref{T_parallel}), cf.\ Eq.~(\ref{tensor_special}).

In pure deuteron spin states polarized perpendicular to the beam direction, the deuteron polarization 
vector is again of the form of Eq.~(\ref{vector_special}) and Eq.~(\ref{a_parallel}),
but with a different 4-vector $N$, satisfying the conditions
\begin{align}
p_e N \; = \; 0, \hspace{2em} p_d N \; = \; 0, \hspace{2em} N^2 = -1
\label{condition_perpendicular}
\end{align}
(in a frame where $\bm{p}_e$ and $\bm{p}_d$ are collinear, the first two conditions require that
$\bm{p}_e\bm{N} = 0$). An explicit representation can be found by expanding the 4-vector in 
the basis Eq.~(\ref{basis_set}) and imposing the conditions Eq.~(\ref{condition_perpendicular}),
\begin{align}
N^\mu \; &= \; \cos\phi_N \;
\frac{(e_{T1} p_e) \, e_{L\ast}^\mu - (e_{L\ast} p_e) \, e_{T1}^\mu}{
\sqrt{(e_{T1} p_e)^2 + (e_{L\ast} p_e)^2}}
\nonumber \\
&+ \; \sin \phi_N \, e_{T2}^\mu \;,
\label{n_perpendicular}
\end{align}
where the angle $\phi_N$ is a free parameter. In the deuteron rest frame the vector components become
\begin{subequations}
\begin{align}
N^\mu [\textrm{RF}] \; &= \; (0, \, \bm{N}), \hspace{2em} \bm{p}_e \bm{N} \; = \; 0,
\\[1ex]
\bm{N} \; &= \; \cos\phi_N \, \frac{\bm{p}_{e'} - (\bm{p}_{e}\bm{p}_{e'}) \bm{p}_{e} / |\bm{p}_{e}|^2}
{\sqrt{|\bm{p}_{e'}|^2 - (\bm{p}_{e}\bm{p}_{e'})^2 / |\bm{p}_{e}|^2}}
\nonumber
\\[1ex]
&+ \; \sin\phi_N \, \frac{\bm{p}_{e'} \times \bm{p}_{e}}{|\bm{p}_{e'} \times \bm{p}_{e}|} ,
\end{align}
\end{subequations}
and one identifies $\phi_N$ as the angle of the polarization direction relative to the plane spanned 
by the vectors $\bm{p}_e$ and $\bm{p}_{e'}$ (electron scattering plane). With the polarization 4-vector
given by Eqs.~(\ref{a_parallel}) and (\ref{n_perpendicular}), the effective vector polarizations
Eq.~(\ref{S_invariant}) for perpendicular polarization are evaluated in the 
same manner as in the case of parallel polarization, and we obtain
\begin{subequations}
\begin{align}
S_L &= \frac{\gamma\sqrt{1-y-\gamma^2y^2/4}}{\sqrt{1+\gamma^2}} \; \cos \phi_N \; \Lambda_d ,
\\[1ex]
S_T\cos\phi_S &= \frac{1+\gamma^2y/2}{\sqrt{1+\gamma^2}}\; \cos\phi_N \; \Lambda_d ,
\\[2ex]
S_T\sin\phi_S &= \sin\phi_N \; \Lambda_d .
\end{align}
\label{S_perpendicular}
\end{subequations}
The deuteron polarization tensor in the perpendicular polarized states is again given by
the general formula Eq.~(\ref{tensor_special}), with the unit vector $N$ given by Eq.~(\ref{n_perpendicular}).
The effective tensor polarizations for perpendicular polarization are evaluated in the
same manner as for parallel polarization. In the case $\Lambda_d = \pm 1$ they again can 
be obtained from the perpendicular vector polarizations Eq.~(\ref{S_perpendicular}) 
through the relations Eq.~(\ref{T_from_S}). The explicit expressions are
\begin{subequations}
\label{T_perpendicular}
\begin{align}
T_{LL} 
&= \frac{1}{6} \left[ \frac{3 \gamma^2(1-y-\gamma^2y^2/4)}{1+\gamma^2} \cos^2\phi_N - 1 \right] ,
\\[2ex]
T_{LT}\cos\phi_{T_L} 
&= \frac{1}{2} \frac{\gamma (1+\gamma^2y/2) \sqrt{1-y-\gamma^2y^2/4}}{1+\gamma^2}
\nonumber \\[1ex]
&\times \; \cos^2\phi_N ,
\\[2ex]
T_{LT}\sin\phi_{T_L} 
&= \frac{1}{2} \; \frac{\gamma \sqrt{1-y-\gamma^2y^2/4}}{\sqrt{1+\gamma^2}} \; \cos\phi_N \sin \phi_N ,
\\[2ex]
T_{TT}\cos2\phi_{T_T} 
&= \frac{1}{2} \left[\frac{1+\gamma^2y+\gamma^4y^2/4}{1+\gamma^2}\cos^2\phi_N - \sin^2\phi_N \right] 
\\[2ex]
& \hspace{1em}  [\textrm{$\Lambda_d = \pm 1$}] .
\nonumber
\end{align}
\end{subequations}
In the case $\Lambda_d = 0$ the effective tensor polarizations are given by $(-2)$ times the 
expressions in Eq.~(\ref{T_perpendicular}).

The effective polarizations depend on the kinematic variables $x_d, y$ and $Q^2$. For a fixed $ed$ 
collision energy, $y$ is determined by $x_d$ and $Q^2$, and the polarizations are functions of 
$x_d$ and $Q^2$ only. For the following applications it is useful to study their scaling behavior
in the DIS limit, $Q^2 \rightarrow \infty$ with $x_d$ fixed. The effective vector polarizations 
Eqs.~(\ref{S_parallel}) and (\ref{S_perpendicular}) scale as
\begin{subequations}
\label{S_scaling}
\begin{align}
\{ S_L, \, S_T \cos\phi_S \}  
&=
{\mathcal O} \{ 1, \, \gamma \} & \text{($\parallel$ pol.),} 
\\[1ex]
\{ S_L, \, S_T \cos\phi_S, \, S_T\sin\phi_S \}  
&=
{\mathcal O} \{ \gamma, \, 1, \, 1 \} & \text{($\perp$ pol.),}
\end{align}
\end{subequations}
where $\gamma = 2 x_d M_d/Q$ is the parameter governing kinematic power corrections, cf.\ Eq.~(\ref{gamma_def}).
Thus $\parallel$ polarization along the beam axis induces mostly $L$ polarization, while $\perp$
polarization induces mostly $T$ polarization, as expected. The scaling behavior of the effective
tensor polarizations follows from that of the vector polarizations, cf.\ Eqs.~(\ref{T_parallel})
and (\ref{T_perpendicular}),
\label{T_scaling}
\begin{align}
& \{ 
T_{LL}, \,
T_{LT}\cos\phi_{T_L}, \,
T_{LT}\sin\phi_{T_L}, \,
T_{TT}\cos2\phi_{T_T} 
\}
\nonumber \\[1ex]
&= \;
\left\{ 
\begin{array}{ll}
{\mathcal O}\{ 1, \, \gamma, \, 0, \, \gamma^2 \} \hspace{1em} & \text{($\parallel$ pol.)} 
\\[2ex]
{\mathcal O}\{ 1, \, \gamma, \, \gamma, \, 1 \} & \text{($\perp$ pol.)}
\end{array}
\right\} .
\end{align}
$T_{LL}$ is of order unity for both $\parallel$ and $\perp$ deuteron polarization. Other than it, the only
tensor polarization that is not power-suppressed in the DIS limit is $T_{TT}\cos2\phi_{T_T}$
for $\perp$ deuteron polarization. Note that these statements refer to the kinematic scaling of 
the effective polarizations, not to the dynamical scaling of the structure functions.

In this section we have derived the effective polarizations for the two idealized situations of
pure deuteron spin states polarized parallel or perpendicular to the beam direction 
(in a frame where the deuteron and electron momenta are collinear). More complex experimental
situations can be treated as a superposition of the cases considered here: the cross section
is linear in the deuteron density matrix, and more general polarization vectors/tensors can be 
represented as sums of the ones considered here. This includes the case of colliding-beam
experiments with finite crossing angle at an EIC \cite{Aschenauer:2014cki,Abeyratne:2012ah}.
\subsection{Spin asymmetries}
\label{subsec:spin_asymmetries}
Experiments typically measure sums and differences of cross sections in different electron and deuteron 
polarization states and their ratios (spin asymmetries). We now derive the expressions for the spin
asymmetries between pure deuteron spin states polarized parallel or perpendicular to the beam axis 
(see Sec.~\ref{subsec:effective_polarization}). We consider the asymmetries formed with all three
deuteron spin states ($\Lambda_d = \pm 1, 0$) and those formed with the two maximum-spin states 
only ($\Lambda_d = \pm 1$) and compare their properties. In the following we write the dependence of 
the differential cross section Eq.~(\ref{dsigma_tagged_explicit}) on the electron and deuteron 
spin projections as
\beq
d\sigma_\parallel (\Lambda_e, \Lambda_d), \hspace{2em} d\sigma_\perp (\Lambda_e, \Lambda_d),
\eeq
where $\parallel$ and $\perp$ distinguish parallel and perpendicular deuteron polarization with 
respect to the beam axis.

We first consider sums of the cross section over deuteron spin states.
The average of the cross section in all three deuteron spin states is
\begin{align}
& \frac{1}{3} \left[ d\sigma_\parallel (\pm{\textstyle\frac{1}{2}}, +1)
+ d\sigma_\parallel (\pm{\textstyle\frac{1}{2}}, -1) + d\sigma_\parallel (\pm{\textstyle\frac{1}{2}}, 0) \right]
\nonumber \\[2ex]
=& \frac{1}{3} \left[ d\sigma_\perp (\pm{\textstyle\frac{1}{2}}, +1)
+ d\sigma_\perp (\pm{\textstyle\frac{1}{2}}, -1) + d\sigma_\perp (\pm{\textstyle\frac{1}{2}}, 0) \right]
\nonumber \\[2ex]
=& \; [...] \left( F_{[UU,T]d} \; + \; \epsilon F_{[UU,L]d} \right),
\label{sigma_average_3}
\end{align}
where $[...]$ denotes the differential phase space and flux factors of Eq.~(\ref{dsigma_tagged_explicit}).
The average involves only the unpolarized structure functions. The result is the same when averaging over 
$\parallel$ and $\perp$ deuteron spin states. It does not depend on the electron spin [as expressed by 
the notation $\pm \frac{1}{2}$ in Eq.~(\ref{sigma_average_3})], so that no additional averaging over
the electron spin is required in order to isolate the unpolarized structure functions.

The averages of the cross section in the two deuteron spin states with projection $\pm 1$ only are
\begin{align}
& \frac{1}{2} \left[ d\sigma_\parallel (\pm{\textstyle\frac{1}{2}}, +1)
+ d\sigma_\parallel (\pm{\textstyle\frac{1}{2}}, -1) \right]
\nonumber \\[1ex]
&= \;
[...] \left[ F_{[UU,T]d} \; + \; \epsilon F_{[UU,L]d}
\right.
\nonumber
\\[1ex]
&+ \; D_{\parallel [T_{LL}]}
\, \left( F_{[UT_{LL},T]d} + \epsilon F_{[UT_{LL},L]d} \right)
\nonumber
\\[1ex]
&+ \left. D_{\parallel [UT_{LT}]}F_{[UT_{LT}]d} \; + \; D_{\parallel [T_{TT}]}F_{[UT_{TT}]d}
\right] ,
\label{sigma_average_2_parallel}
\\[2ex]
& \frac{1}{2} \left[ d\sigma_\perp (\pm{\textstyle\frac{1}{2}}, +1)
+ d\sigma_\perp (\pm{\textstyle\frac{1}{2}}, -1) \right] 
\nonumber \\[1ex]
&= \;
[...] \left[ F_{[UU,T]d} \; + \; \epsilon F_{[UU,L]d} \right.
\nonumber
\\[1ex]
&+ \; D_{\perp [T_{LL}]}
\, \left( F_{[UT_{LL},T]d} + \epsilon F_{[UT_{LL},L]d} \right)
\nonumber \\[1ex]
&+ \; D_{\perp [UT_{LT}]} F_{[UT_{LT}]d} 
\; + \;  D_{\perp [T_{TT}]} F_{[UT_{TT}]d}
\nonumber
\\[1ex]
& \; \pm \left. D_{\perp [LT_{LT}]} F_{[LT_{LT}]d}
\right] .
\label{sigma_average_2_perpendicular}
\end{align}
They involve the unpolarized and tensor-polarized structure functions. The functions $D_{(...)}$ 
(``depolarization factors'') are given by
\begin{subequations}
\label{depolarization_tensor_parallel}
\begin{align}
& D_{\parallel [T_{LL}]} \; = \; T_{LL}[\Lambda_d = \pm 1]
\nonumber \\[2ex]
&= \frac{1}{6} \left[\frac{3 (1 + \gamma^2 y/2)^2}{1 + \gamma^2} - 1 \right] ,
\label{depolarization_TLL_parallel}
\\[3ex]
& D_{\parallel [UT_{LT}]} \; = \; \sqrt{2\epsilon(1+\epsilon)} \; T_{LT}\cos \phi_{T_L}[\Lambda_d = \pm 1] 
\nonumber \\[2ex]
&= 
\frac{\gamma  (1 + \gamma^2 y/2) (1 - y - \gamma^2 y^2/4) (1 - y/2) }{(1 + \gamma^2)(1-y+y^2/2+\gamma^2y^2/4)} ,
\\[3ex]
& D_{\parallel [T_{TT}]} \; = \; \epsilon \; T_{TT}\cos 2\phi_{T_T}[\Lambda_d = \pm 1]
\nonumber \\[2ex]
&= 
\frac{1}{2} \; \frac{\gamma^2 (1 - y - \gamma^2 y^2/4)^2}{(1 + \gamma^2)(1-y+y^2/2+\gamma^2y^2/4)} ;
\end{align}
\end{subequations}
\begin{subequations}
\label{depolarization_tensor_perpendicular}
\begin{align}
& D_{\perp [T_{LL}]} \; = \; T_{LL} [\Lambda_d = \pm 1]
\nonumber \\[2ex]
&= 
\frac{1}{6} \left[ \frac{3 \gamma^2 (1 - y - \gamma^2 y^2/4)}{1 + \gamma^2}\cos^2\phi_N - 1 \right] ,
\\[3ex]
& D_{\perp [UT_{LT}]} \; = \; \sqrt{2\epsilon(1+\epsilon)} \; T_{LT}\cos \phi_{T_L}[\Lambda_d = \pm 1]
\nonumber \\[2ex]
&= \frac{\gamma (1+\gamma^2y/2) (1-y-\gamma^2y^2/4) (1 - y/2)}{(1+\gamma^2)(1-y+y^2/2+\gamma^2y^2/4)}
\; \cos^2\phi_N ,
\\[3ex]
& D_{\perp [LT_{LT}]} \; = \; \sqrt{2\epsilon(1-\epsilon)} \; T_{LT}\sin \phi_{T_L}[\Lambda_d = \pm 1]
\nonumber \\[2ex]
&= \frac{1}{2} \; \frac{\gamma y (1-y-\gamma^2y^2/4)}{1-y+y^2/2+\gamma^2y^2/4} \; \cos\phi_N \sin \phi_N .
\\[3ex]
& D_{\perp [T_{TT}]} \; = \; \epsilon \; T_{TT}\cos 2\phi_{T_T}[\Lambda_d = \pm 1]
\nonumber \\[2ex]
&= \; \frac{1}{2} \left( \frac{1+\gamma^2y+\gamma^4y^2/4}{1+\gamma^2}\cos^2\phi_N - \sin^2\phi_N \right)
\nonumber \\
& \times \; \frac{1 - y - \gamma^2 y^2/4}{1-y+y^2/2+\gamma^2y^2/4} ,
\end{align}
\end{subequations}
In Eq.~(\ref{depolarization_tensor_parallel}) $T_{LL}$ etc.\ are the effective 
tensor polarizations for $\parallel$ polarization with $\Lambda_d = \pm 1$ as given in Eq.~(\ref{T_parallel});
in Eq.~(\ref{depolarization_tensor_perpendicular}) they are the same quantities
for $\perp$ polarization as given in Eq.~(\ref{T_perpendicular}).
Note that the results are different for $\parallel$ and $\perp$ polarization.
In the case of $\parallel$ polarization, Eq.~(\ref{sigma_average_2_parallel}), the summation
over the two deuteron spin states has canceled all electron-spin dependent terms in Eq.~(\ref{dsigma_tagged_explicit}),
so that the result is independent of the electron spin. It is therefore not necessary to average 
explicitly over the electron spin. In the case of $\perp$ polarization, Eq.~(\ref{sigma_average_2_perpendicular}),
the summation over the two deuteron spin states leaves intact the electron spin-dependent term
in Eq.~(\ref{dsigma_tagged_explicit}),
\beq
\sim (2\Lambda_e) T_{LT} \sin \phi_{T_{L}} \; F_{[LT_{LT}]d} .
\eeq
One must therefore average explicitly over the electron spins if one wants to remove the electron spin dependence.

We now turn to differences of the cross section between deuteron spin states. Here we must take into account
that the tagged DIS cross section Eq.~(\ref{dsigma_tagged_explicit}) contains the term 
\beq
\sim S_T \sin\phi_S \; F_{[US_T]d} ,
\label{single_spin}
\eeq
which depends on the deuteron spin but not on the electron spin. In order to isolate the electron 
spin-dependent structure functions $F_{[LS_L]}$ and $F_{[LS_T]}$ we form double spin differences 
with respect to both the deuteron and electron spin, in which the single-spin dependent 
term Eq.~(\ref{single_spin}) drops out.
The double differences of the cross section with respect to the deuteron spin $\pm 1$
($\parallel$ or $\perp$) and the electron spin are given by
\begin{align}
& \frac{1}{4} 
\left[ 
  d\sigma_\parallel (+{\textstyle\frac{1}{2}}, +1)
- d\sigma_\parallel (-{\textstyle\frac{1}{2}}, +1)
\right.
\nonumber \\
& 
\left.
- d\sigma_\parallel (+{\textstyle\frac{1}{2}}, -1)
+ d\sigma_\parallel (-{\textstyle\frac{1}{2}}, -1)
\right]
\nonumber \\[1ex]
=& \; [...] \left( D_{\parallel [S_L]} \; F_{[LS_L]d} \; + \; D_{\parallel [S_T]} \; F_{[LS_T]d} \right) ,
\label{sigma_diff_parallel}
\\[2ex]
& \frac{1}{4} 
\left[ 
  d\sigma_\perp (+{\textstyle\frac{1}{2}}, +1)
- d\sigma_\perp (-{\textstyle\frac{1}{2}}, +1)
\right.
\nonumber \\
& 
\left.
- d\sigma_\perp (+{\textstyle\frac{1}{2}}, -1)
+ d\sigma_\perp (-{\textstyle\frac{1}{2}}, -1)
\right]
\nonumber \\[1ex]
=& \; [...] \left( D_{\perp [S_L]} \; F_{[LS_L]d} \; + \; D_{\perp [S_T]} \; F_{[LS_T]d} \right) ,
\label{sigma_diff_perp}
\end{align}
where the depolarization factors are
\begin{subequations}
\begin{align}
D_{\parallel [S_L]} &= \sqrt{1 - \epsilon^2} \; S_L [\Lambda_d = 1]
\nonumber \\[2ex]
&= 
\frac{y (1 - y/2) (1 + \gamma^2 y/2)}{1 - y + y^2/2 + \gamma^2 y^2/4} ,
\\[4ex]
D_{\parallel [S_T]} &= \sqrt{2 \epsilon (1 - \epsilon )} \; S_T \cos\phi_S [\Lambda_d = 1]
\nonumber \\[2ex]
&= 
\frac{- \gamma y (1 - y - \gamma^2 y^2/4)}{1 - y + y^2/2 + \gamma^2 y^2/4} ;
\end{align}
\label{depolarization_vector_parallel}
\end{subequations}
\begin{subequations}
\begin{align}
D_{\perp [S_L]} &= \sqrt{1 - \epsilon^2} \; S_L [\Lambda_d = 1]
\nonumber \\[2ex]
&= 
\frac{\gamma y (1 - y/2) \sqrt{1 - y - \gamma^2 y^2/4}}{1 - y + y^2/2 + \gamma^2 y^2/4}\cos\phi_N ,
\\[4ex]
D_{\perp [S_T]} &= \sqrt{2 \epsilon (1 - \epsilon )} \; S_T \cos\phi_S [\Lambda_d = 1]
\nonumber \\[2ex]
&= \frac{y (1 + \gamma^2 y/2) \sqrt{1 - y - \gamma^2 y^2/4}}{1 - y + y^2/2 + \gamma^2 y^2/4}\cos\phi_N .
\end{align}
\label{depolarization_vector_perpendicular}
\end{subequations}
In Eq.~(\ref{depolarization_vector_parallel}) $S_L$ and $S_T \cos\phi_S$ are the effective 
vector polarizations for $\parallel$ polarization with $\Lambda_d = 1$ as given in Eq.~(\ref{S_parallel});
in Eq.~(\ref{depolarization_vector_perpendicular}) they are the same quantities
for $\perp$ polarization as given in Eq.~(\ref{S_perpendicular}).

From the spin sums and differences one can form two different ratios (spin asymmetries).
The ratios of the spin differences Eqs.~(\ref{sigma_diff_parallel}) and (\ref{sigma_diff_perp}) 
to the three-state average of the cross section Eq.~(\ref{sigma_average_3}) are
\begin{widetext}
\be
A_{\parallel (3) d} &\equiv& 
\frac{
\frac{1}{4} 
\left[ 
  d\sigma_\parallel (+{\textstyle\frac{1}{2}}, +1)
- d\sigma_\parallel (-{\textstyle\frac{1}{2}}, +1)
- d\sigma_\parallel (+{\textstyle\frac{1}{2}}, -1)
+ d\sigma_\parallel (-{\textstyle\frac{1}{2}}, -1)
\right]
}{
\frac{1}{6} 
\left[ 
  d\sigma_\parallel (+{\textstyle\frac{1}{2}}, +1)
+ d\sigma_\parallel (-{\textstyle\frac{1}{2}}, +1)
+ d\sigma_\parallel (+{\textstyle\frac{1}{2}}, -1)
+ d\sigma_\parallel (-{\textstyle\frac{1}{2}}, -1)
+ d\sigma_\parallel (+{\textstyle\frac{1}{2}},  0)
+ d\sigma_\parallel (-{\textstyle\frac{1}{2}},  0)
\right]
}
\nonumber
\\[4ex]
&=& \frac{D_{\parallel [S_L]} \; F_{[LS_L]d} \; + \; D_{\parallel [S_T]} \; F_{[LS_T]d}}
{F_{[UU,T]d} \; + \; \epsilon F_{[UU,L]d}} ,
\label{A_parallel_3}
\\[4ex]
A_{\perp (3) d} &\equiv&  
\frac{
\frac{1}{4} 
\left[ 
  d\sigma_\perp (+{\textstyle\frac{1}{2}}, +1)
- d\sigma_\perp (-{\textstyle\frac{1}{2}}, +1)
- d\sigma_\perp (+{\textstyle\frac{1}{2}}, -1)
+ d\sigma_\perp (-{\textstyle\frac{1}{2}}, -1)
\right]
}{
\frac{1}{6} 
\left[ 
  d\sigma_\perp (+{\textstyle\frac{1}{2}}, +1)
+ d\sigma_\perp (-{\textstyle\frac{1}{2}}, +1)
+ d\sigma_\perp (+{\textstyle\frac{1}{2}}, -1)
+ d\sigma_\perp (-{\textstyle\frac{1}{2}}, -1)
+ d\sigma_\perp (+{\textstyle\frac{1}{2}},  0)
+ d\sigma_\perp (-{\textstyle\frac{1}{2}},  0)
\right]
}
\nonumber
\\[4ex]
&=& \frac{D_{\perp [S_L]} \; F_{[LS_L]d} \; + \; D_{\perp [S_T]} \; F_{[LS_T]d}}
{F_{[UU,T]d} \; + \; \epsilon F_{[UU,L]d}} .
\label{A_perpendicular_3}
\ee
[Here we have written the denominator as a sum over the electron spin in order to emphasize the 
similarity with the numerator; because the expression in Eq.~(\ref{sigma_average_3}) is 
independent of the electron spin this sum is optional and we could just as well use the 
expression for a fixed electron spin.] The ratios of the spin differences Eqs.~(\ref{sigma_diff_parallel}) 
and (\ref{sigma_diff_perp}) to the two-state averages of the cross section, 
Eqs.~(\ref{sigma_average_2_parallel}) and Eqs.~(\ref{sigma_average_2_perpendicular}), are
\begin{align}
A_{\parallel (2) d} &\equiv 
\frac{
\frac{1}{4} 
\left[ 
  d\sigma_\parallel (+{\textstyle\frac{1}{2}}, +1)
- d\sigma_\parallel (-{\textstyle\frac{1}{2}}, +1)
- d\sigma_\parallel (+{\textstyle\frac{1}{2}}, -1)
+ d\sigma_\parallel (-{\textstyle\frac{1}{2}}, -1)
\right]
}{
\frac{1}{4} 
\left[ 
  d\sigma_\parallel (+{\textstyle\frac{1}{2}}, +1)
+ d\sigma_\parallel (-{\textstyle\frac{1}{2}}, +1)
+ d\sigma_\parallel (+{\textstyle\frac{1}{2}}, -1)
+ d\sigma_\parallel (-{\textstyle\frac{1}{2}}, -1)
\right]
}
\nonumber
\\[2ex]
&= \frac{D_{\parallel [S_L]} \; F_{[LS_L]d} \; + \; D_{\parallel [S_T]} \; F_{[LS_T]d}}
{F_{[UU,T]d} \; + \; \epsilon F_{[UU,L]d} 
\; + \; D_{\parallel [T_{LL}]} \, (F_{[UT_{LL},T]d} + \epsilon F_{[UT_{LL},L]d})
\; + \; D_{\parallel [UT_{LT}]} \, F_{[UT_{LT}]d} \; + \; D_{\parallel [T_{TT}]}F_{[UT_{TT}]d} }
\label{A_parallel_2}
\\[2ex]
A_{\perp (2) d} &\equiv 
\frac{
\frac{1}{4} 
\left[ 
  d\sigma_\perp (+{\textstyle\frac{1}{2}}, +1)
- d\sigma_\perp (-{\textstyle\frac{1}{2}}, +1)
- d\sigma_\perp (+{\textstyle\frac{1}{2}}, -1)
+ d\sigma_\perp (-{\textstyle\frac{1}{2}}, -1)
\right]
}{
\frac{1}{4} 
\left[ 
  d\sigma_\perp (+{\textstyle\frac{1}{2}}, +1)
+ d\sigma_\perp (-{\textstyle\frac{1}{2}}, +1)
+ d\sigma_\perp (+{\textstyle\frac{1}{2}}, -1)
+ d\sigma_\perp (-{\textstyle\frac{1}{2}}, -1)
\right]
}
\nonumber
\\[2ex]
&= \frac{D_{\perp [S_L]} \; F_{[LS_L]d} \; + \; D_{\perp [S_T]} \; F_{[LS_T]d}}
{F_{[UU,T]d} \; + \; \epsilon F_{[UU,L]d} \; + \; D_{\perp [T_{LL}]} \, ( F_{[UT_{LL},T]d} + \epsilon F_{[UT_{LL},L]d})
\; + \; D_{\perp [UT_{LT}]} \, F_{[UT_{LT}]d} \; + \; D_{\perp [T_{TT}]} \, F_{[UT_{TT}]d}} .
\label{A_perpendicular_2}
\end{align}
Now the tensor-polarized structure functions appear in the denominators with the depolarization 
factors Eqs.~(\ref{depolarization_tensor_parallel}) and (\ref{depolarization_tensor_perpendicular}).
Both the three-state asymmetries, $A_{\parallel (3) d}$ and $A_{\perp (3) d}$, and the two-state asymmetries, 
$A_{\parallel (2) d}$ and $A_{\perp (2) d}$, can be used for neutron structure extraction from tagged DIS and 
other spin physics studies and are calculated below.

The scaling behavior of the depolarization factors in the DIS limit can be inferred from
that of the effective polarizations, Eqs.~(\ref{S_scaling}) and (\ref{T_scaling}), and from
the explicit expressions given above,
\begin{align}
\{ D_{\parallel [S_L]}, D_{\parallel [S_T]} \} &= {\mathcal O} \{ 1, \gamma \} ,
\\[2ex]
\{ D_{\parallel [T_{LL}]}, \, D_{\parallel [UT_{LT}]}, \, D_{\parallel [T_{TT}]} \} 
&= {\mathcal O} \{ 1, \gamma, \gamma^2 \} ,
\\[2ex]
\{ D_{\perp [S_L]}, D_{\perp [S_T]} \} &= {\mathcal O} \{ \gamma, 1 \} ,
\\[2ex]
\{ D_{\perp [T_{LL}]}, D_{\perp [UT_{LT}]}, D_{\perp [LT_{LT}]}, D_{\perp [T_{TT}]} \}
&= {\mathcal O} \{ 1, \, \gamma, \gamma, \, 1 \} .
\end{align}
Up to power corrections ${\mathcal O}(\gamma)$, the three-state and two-state asymmetries therefore simplify to
\begin{align}
A_{\parallel (3) d} &= \frac{D_{\parallel [S_L]} \; F_{[LS_L]d}}{F_{[UU,T]d} \; + \; \epsilon F_{[UU,L]d}} ,
\\[2ex]
A_{\parallel (2) d} &= \frac{D_{\parallel [S_L]} \; F_{[LS_L]d}}
{F_{[UU,T]d} \; + \; \epsilon F_{[UU,L]d} 
\; + \; D_{\parallel [T_{LL}]} \, (F_{[UT_{LL},T]d} + \epsilon F_{[UT_{LL},L]d}) } ,
\\[2ex]
A_{\perp (3) d} &=  \frac{D_{\perp [S_T]} \; F_{[LS_T]d}}{F_{[UU,T]d} \; + \; \epsilon F_{[UU,L]d}} ,
\\[2ex]
A_{\perp (2) d} &=  \frac{D_{\perp [S_T]} \; F_{[LS_T]d}}
{F_{[UU,T]d} \; + \; \epsilon F_{[UU,L]d} 
\; + \; D_{\perp [T_{LL}]} \, ( F_{[UT_{LL},T]d} + \epsilon F_{[UT_{LL},L]d})
\; + \; D_{\perp [T_{TT}]} \, F_{[UT_{TT}]d}} .
\end{align}
\end{widetext}
Note that, with our definition of the structure functions, all asymmetries are ${\mathcal O}(1)$ in the DIS limit.
In the asymmetries for $\parallel$ polarization, the numerators involve the longitudinal spin 
structure function $F_{[LS_L]d}$. The denominator of $A_{\parallel (2) d}$ differs from that of
$A_{\parallel (3) d}$ by the tensor-polarized term $F_{[UT_{LL},T]d} + \epsilon F_{[UT_{LL},L]d}$,
which has a form similar to the unpolarized term $F_{[UU,T]d} + \epsilon F_{[UU,L]d}$
(this will become apparent in the dynamical calculations below).
In the asymmetries for $\perp$ polarization, the numerators involve the transverse spin 
structure function $F_{[LS_T]d}$. The denominator of $A_{\perp (2) d}$ differs from that of
$A_{\perp (3) d}$ by two independent tensor-polarized terms and thus has a more complex structure.

For reference we note that in terms of the conventional structure functions the three-state asymmetries 
Eqs.~(\ref{A_parallel_3}) and (\ref{A_perpendicular_3}) are expressed as
\be
A_{\parallel (3) d} \; &=& \; \frac{D_{\parallel 1} g_{1d} \, + \, D_{\parallel 2} g_{2d}}{2 (1 + \epsilon R_d) F_{1d}},
\\[2ex]
A_{\perp (3) d} \; &=& \; \frac{D_{\perp 1} g_{1d} \, + \, D_{\perp 2} g_{2d}}{2 (1 + \epsilon R_d) F_{1d}},
\ee
where the depolarization factors are now given by
\begin{subequations}
\label{depolarization_12_parallel}
\begin{align}
D_{\parallel 1} 
&= \; 2 (D_{\parallel [S_L]} - \gamma D_{\parallel [S_T]})
\nonumber \\[2ex]
&= \; \frac{2 (1 + \gamma^2) y (1 - y/2 - \gamma^2 y^2/4)}{1 - y + y^2/2 + \gamma^2 y^2/4},
\\[4ex]
D_{\parallel 2} 
&= \; -2 \gamma ( \gamma D_{\parallel [S_L]} + D_{\parallel [S_T]} )
\nonumber \\[2ex]
&= \; \frac{-\gamma^2 (1 + \gamma^2) y^2}{1 - y + y^2/2 + \gamma^2 y^2/4},
\end{align}
\end{subequations}
\begin{subequations}
\label{depolarization_12_perpendicular}
\begin{align}
D_{\perp 1} \; &= \; 
2 (D_{\perp [S_L]} - \gamma D_{\perp [S_T]})
\nonumber \\[2ex]
&= \;
\displaystyle
\frac{-\gamma (1 + \gamma^2) y^2 \sqrt{1 - y - \gamma^2 y^2/4}}
{1 - y + y^2/2 + \gamma^2 y^2/4}\cos\phi_N,
\\[4ex]
D_{\perp 2} \; &= \; 
-2 \gamma ( \gamma D_{\perp [S_L]} + D_{\perp [S_T]} )
\nonumber \\[2ex]
&= \;
\displaystyle
\frac{- 2\gamma (1 + \gamma^2) y \sqrt{1 - y - \gamma^2 y^2/4}}
{1 - y + y^2/2 + \gamma^2 y^2/4}\cos\phi_N,
\end{align}
\end{subequations}
and scale as
\begin{align}
\{ D_{\parallel 1}, D_{\parallel 2} \} &= {\mathcal O} \{ 1, \gamma^2 \} ,
\\[2ex]
\{ D_{\perp 1}, D_{\perp 2} \} &= {\mathcal O} \{ \gamma, \gamma \} .
\end{align}
In $A_{\parallel (3) d}$ the structure function $g_1$ appears in the scaling limit,
while $g_2$ appears as a power correction $\sim 1/Q^2$. In $A_{\perp (3) d}$ the 
structure functions $g_1$ and $g_2$ appear at the same level, but the entire 
asymmetry is power-suppressed $\sim 1/Q$
\subsection{Collinear frames}
\label{subsec:collinear_frames}
\begin{figure}
\includegraphics[width=.43\textwidth]{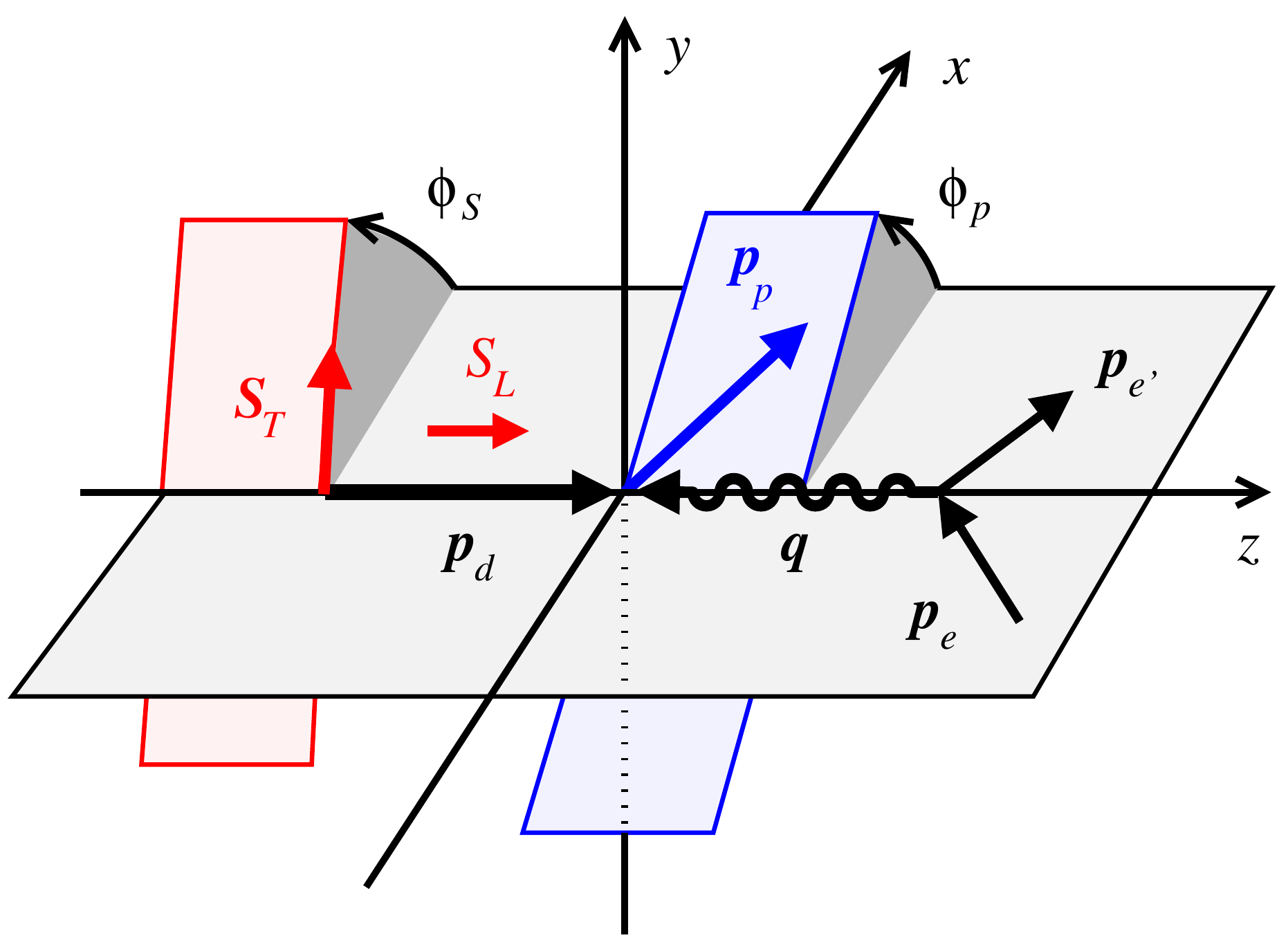}
\caption[]{Tagged deuteron DIS in the collinear frame, Eq.~(\ref{collinear_frame}). 
The deuteron 3-momentum $\bm{p}_d$ and the momentum transfer $\bm{q}$ are collinear and define the
$z$-axis ($\bm{q}$ points in the negative $z$-direction). The initial and final electron 
momenta, $\bm{p}_e$ and $\bm{p}_{e'}$, define the $xz$ plane (both have positive $x$ components). 
The deuteron vector polarization is described by the longitudinal ($z$-) spin $S_L$ and the transverse
($xy$-) spin vector $\bm{S}_T$ (the azimuthal angle $\phi_S$ is measured relative to the positive $x$ axis).
The tagged proton momentum $\bm{p}_p$ has both longitudinal and transverse components (azimuthal angle $\phi_p$).}
\label{fig:collinear_frame_spin}
\end{figure}

In the theoretical description of tagged DIS we consider the process Eq.~(\ref{tagged_dis})
in a frame where the deuteron momentum $\bm{p}_d$ and the momentum transfer $\bm{q}$
are collinear and define the $z$--axis of the coordinate system (see Fig.~\ref{fig:collinear_frame_spin}). 
This condition does not specify a unique frame, but rather a class of 
frames that are related by boosts along the $z$--axis (``collinear frames''). 
We describe the 4-vectors in this frame by their LF components
\beq
v^\pm \;\; \equiv \;\; v^0 \pm v^z, \hspace{2em}
\bm{v}_T \;\; \equiv \;\; (v^x, v^y) ,
\label{lc_def}
\eeq
and use the notation
\beq
v \;\; = \;\; [v^+, v^-, \bm{v}_T] .
\label{notation_lf}
\eeq
The LF components of the 4-momenta $p_d$ and $q$ in the collinear frame are
\begin{subequations}
\label{collinear_frame}
\begin{align}
p_d \; &= \; \left[ p_d^+, \; \frac{M_d^2}{p_d^+}, \; \bm{0}_T \right] ,
\\[1ex]
q \; &= \; \left[ -\xi_d p_d^+, \; \frac{Q^2}{\xi_d p_d^+}, \; \bm{0}_T \right] ,
\end{align}
\end{subequations}
where the parameter $\xi_d$ is related to the scaling variable $x_d$
\beq
\xi_d \;\; = \;\; \frac{2 x_d}{1 + \sqrt{1 + \gamma^2}}
\;\; = \;\; x_d \; + \; {\mathcal O}(\gamma^2).
\eeq
Note that in our convention the 3-vector $\bm{q}$ points in the negative $z$-direction, $q^z = (q^+ - q^-)/2 < 0$.
%
%

The collinear frames are a class of frames related by boosts along the $z$-axis (longitudinal boosts).
The boosts are performed by changing the LF components of the 4-vectors as
\beq
[v^+, v^-, \bm{v}_T] \; \rightarrow \; [e^\eta v^+, e^{-\eta} v^-, \bm{v}_T] ,
\label{vector_lf_notation}
\eeq
where $\eta$ is the rapidity. If the $+$ and $-$ components are expressed as multiples of $p_d^+$ and $1/p_d^+$, 
the boosts can be effected by simply changing the value of $p_d^+$ from the one in the ``old'' frame to the one 
in the ``new'' frame. In this sense $p_d^+$ serves as a parameter that identifies a particular representative
of the class. In particular, the class of collinear frames includes the deuteron rest frame, in which
\beq
p_d^+ = M_d .
\eeq
In this way one can construct a boost-invariant theoretical description that can easily be matched with the 
deuteron rest frame. The deuteron polarization 4-vector in any collinear frame is given by
\be
s_d &=& \left[ \frac{p_d^+}{M_d}  S_d^z, -\frac{M_d}{p_d^+} S_d^z, \bm{S}_{dT} \right] ,
\label{deuteron_polarization_vector_collinear}
\ee
where $S_d^z$ and $\bm{S}_{dT} \equiv (S_d^x, S_d^y)$ are the components of the polarization 
3-vector in the rest frame, cf.\ Eq.~(\ref{a_restframe}),
\begin{subequations}
\begin{align}
\bm{S}_d \; &= \; (\bm{S}_{dT}, S_d^z), 
\\[1ex]
|\bm{S}_d|^2 \; &= \; (S_d^z)^2 + |\bm{S}_{dT}|^2 \; \leq \; 1 .
\end{align}
\end{subequations}
In the collinear frames the longitudinal unit 4-vectors Eq.~(\ref{unit_longitudinal_alt}) have only $+$ and $-$
components,
\begin{subequations}
\label{basis_collinear}
\begin{align}
e_d \; &= \; \left[\frac{p_d^+}{M_d}, \frac{M_d}{p_d^+}, \bm{0}_T \right], 
\\[1ex]
e_{L\ast} \; &= \; \left[ -\frac{p_d^+}{M_d}, \frac{M_d}{p_d^+}, \bm{0}_T \right] . 
\end{align}
\end{subequations}
The transverse unit 4-vectors have only transverse components, which are chosen as the $x$ and $y$ directions,
\beq
e_{T1} \; = \; \left[ 0, 0, \bm{e}_{x} \right],
\hspace{2em}
e_{T2} \; = \; \left[ 0, 0, \bm{e}_{y} \right],
\eeq
such that the electron scattering plane is the $xz$ plane, and the electron has a transverse momentum
along the positive $x$-axis. It is straightforward to compute invariants from these 4-vectors and express 
them in terms of deuteron rest-frame variables. Specifically, one sees that 
the invariant effective vector polarization parameters, defined in Eq.~(\ref{S_invariant}),
coincide with the longitudinal and transverse component of the 3-dimensional deuteron
polarization vector in the rest frame
\begin{subequations}
\begin{align}
S_L \; &\equiv \; S_d^z, 
\\[1ex]
(S_T \cos\phi_S, \, S_T\sin\phi_S) \; &\equiv \; (S_d^x, S_d^y) \; = \; \bm{S}_{dT}.
\end{align}
\end{subequations}
The angle $\phi_S$ can be regarded as the angle of the transverse component of the deuteron spin
in the rest frame (or any collinear frame) relative to electron scattering plane, measured from positive $x$-axis
(see Fig.~\ref{fig:collinear_frame_spin}).

In a similar way one can infer the form of the deuteron polarization 4-tensor in any collinear frame.
For the LF components of a 4-tensor $w^{\mu\nu}$ we use the notation [cf.\ Eq.~(\ref{vector_lf_notation})]
\beq
w 
\; = \; 
\left[\begin{array}{lll}
w^{++} & w^{+-} & w^{+j} \\[2ex]
w^{-+} & w^{--} & w^{-j} \\[2ex]
w^{i+} & w^{i-} & w^{ij}
\end{array} \right] \hspace{2em} (i, j = x, y).
\label{notation_lf_tensor}
\eeq
The LF components of the deuteron polarization 4-tensor in any collinear frame are then given by
\be
t_d \; &=& \; 
\left[\begin{array}{rrr} 
\displaystyle \frac{(p_d^+)^2}{M_d^2} T_d^{zz} & -T_d^{zz} & \displaystyle \frac{p_d^+}{M_d} T_d^{zj} \\[2ex]
-T_d^{zz} & \displaystyle \frac{M_d^2}{(p_d^+)^2} T_d^{zz} & \displaystyle -\frac{M_d}{p_d^+} T_d^{zj} \\[2ex]
\displaystyle\frac{p_d^+}{M_d} T_d^{zi} & \displaystyle -\frac{M_d}{p_d^+} T_d^{zi} & T_d^{ij}
\end{array} \right] ,
\label{deuteron_polarization_tensor_collinear}
\ee
where $T_d^{zz}, T_d^{zi}$ and $T_d^{ij}$ ($i, j = x, y$) 
are the components of is the deuteron polarization 3-tensor in the rest frame,
\beq
T_d \; = \; 
\left(
\begin{array}{rr}
T_d^{ij} & T_d^{iz} \\[2ex]
T_d^{zj} & T_d^{zz} 
\end{array} \right) \hspace{2em} (i, j = x, y).
\eeq
Eq.~(\ref{deuteron_polarization_tensor_collinear}) for the polarization 4-tensor is a 
straightforward generalization of
Eq.~(\ref{deuteron_polarization_vector_collinear}) for the polarization 4-vector.

In experimental or theoretical applications one needs to infer the LF components of a 4-vector in the 
collinear frame from those given in another frame. This can easily be accomplished using the 4-vector
basis Eq.~(\ref{basis_set}). From the timelike and spacelike longitudinal vectors,
$e_d$ and $e_{L\ast}$, we construct the light-like vectors $e_d \pm e_{L\ast}$,
\begin{align}
(e_d \pm e_{L\ast})^2 = 0,
\hspace{2em}  (e_d + e_{L\ast})(e_d - e_{L\ast}) = 2 .
\end{align}
Their components in the collinear frame are [cf.\ Eq.~(\ref{basis_collinear})]
\begin{subequations}
\begin{align}
e_d + e_{L\ast} \; &= \; \left[ 0, \frac{2 M_d}{p_d^+}, \bm{0}_T \right],
\\[1ex]
e_d - e_{L\ast} \; &= \; \left[ \frac{2 p_d^+}{M_d}, 0, \bm{0}_T \right].
\end{align}
\end{subequations}
The scalar products of the vectors with an arbitrary vector $v$ are
\begin{subequations}
\label{projection_lf_+-}
\begin{align}
(e_d + e_{L\ast}) \, v \; &= \; \frac{M_d}{p_d^+} v^+ ,
\\[1ex]
(e_d - e_{L\ast}) \, v \; &= \; \frac{p_d^+}{M_d} v^- .
\end{align}
\end{subequations}
They project out the $+$ and $-$ LF components of the vector. The transverse components 
are projected out as
\beq
v^x = -(e_{T1} v), \hspace{2em} v^y = - (e_{T2} v) .
\label{projection_lf_T}
\eeq
The expressions in Eqs.~(\ref{projection_lf_+-}) and (\ref{projection_lf_T}) are invariant
and can be used to compute the LF components starting from an arbitrary representation of 
the vector $v$ and the basis vectors (e.g., in a frame associated with the experimental setup).
Conversely, the components of $v$ in any frame can be obtained from the LF components in
the collinear frame by expanding $v$ in the basis vectors,
\begin{align}
v^\mu \; &= \; \frac{M_d}{2 p_d^+} v^+ (e_d + e_{L\ast})^\mu 
+ \frac{p_d^+}{2 M_d} v^- (e_d - e_{L\ast})^\mu
\nonumber
\\[1ex]
& + \; v^x e_{T1}^\mu \; + \; v^y e_{T2}^\mu ,
\end{align}
and evaluating the expression with the basis vector components in the desired frame.
\subsection{Spectator momentum variables}
\label{subsec:spectator_momentum}
The invariant structure functions in the tagged DIS cross section Eq.~(\ref{dsigma_tagged_explicit}) depend on 
kinematic variables specifying the final-state proton momentum, cf.\ Eq.~(\ref{structure_function_arguments}).
Several sets of variables are used in experimental analysis and theoretical studies (proton 3-momentum
in deuteron rest frame, LF components in collinear frame, invariant momentum transfer); their relation and 
kinematic limits are summarized in Ref.~\cite{Strikman:2017koc}. In the following calculations the proton 
momentum is specified by its LF components in the collinear frames,
\beq
\alpha_p \; \equiv \; \frac{2 p_p^+}{p_d^+}, \hspace{2em} \bm{p}_{pT} .
\label{recoil_lf}
\eeq
The fraction $\alpha_p$ is boost-invariant (same in all collinear frames) and can be expressed in terms 
of invariants that can be evaluated in any frame, cf.\ Eq.~(\ref{projection_lf_+-}),
\beq
\alpha_p \; = \; \frac{2 (e_d + e_{L\ast}) p_p}{(e_d + e_{L\ast}) p_d} ;
\eeq
the same applies to $\bm{p}_{pT}$, cf.\ Eq.~(\ref{projection_lf_T}). 
In the deuteron rest frame, $\alpha_p$ is given by
\beq
\alpha_p \;\; = \;\; \frac{2(E_p + p_p^z)}{M_d} \; = \; 1 \; + \; \frac{p_p^z}{m} 
\; + \; {\mathcal O}\left(\frac{|\bm{p}_p|^2}{m^2}, \frac{\epsilon_d}{m}\right) ,
\eeq
where $m$ is the nucleon mass and $\epsilon_d$ the deuteron binding energy;
typical values of $\alpha_p$ are therefore of the order
\beq
|\alpha_p - 1| \;\; \sim \;\; \frac{|\bm{p}_p|}{m} \;\; \lesssim \;\; 0.1.
\eeq
The range of $\alpha_p$ is kinematically 
limited by the conservation of 4-momentum in the tagged DIS process, which implies the conservation of LF 
plus momentum, Eq.~(\ref{collinear_frame}),
\beq
\alpha_p/2 \; < \; 1 - \xi_d .
\eeq
The invariant phase space element in the proton momentum is expressed in terms of $\alpha_p$ and $\bm{p}_{pT}$ as
\beq
d\Gamma_p \; \equiv \; [2 (2\pi)^3]^{-1} \; \frac{d^3 p_p}{E_p}
\;\; = \;\; [2 (2\pi)^3]^{-1} \; \frac{d\alpha_p}{\alpha_p} \; d^2p_{pT} .
\label{phase_space_alpha_pt}
\eeq

The structure functions in Eq.~(\ref{dsigma_tagged_explicit}) depend on $\alpha_p$ and the
modulus of the transverse momentum, $|\bm{p}_{pT}|$. The dependence on the azimuthal angle of the
transverse momentum with respect to the $z$ axis, $\phi_p$ (see Fig.~\ref{fig:collinear_frame_spin}), 
is realized explicitly in the decomposition of the cross section, which follows from the 
decomposition of the hadronic tensor; Eq.~(\ref{dsigma_tagged_explicit}) shows only the terms 
that are independent of $\phi_p$ and survive integration over that variable.
\section{Deuteron light-front structure}
\label{sec:deuteron}
\subsection{Light-front nuclear structure}
\label{subsec:light_front_structure}
We now describe the elements of LF nuclear structure used in the theoretical calculation of the tagged DIS 
cross section. The basic method was developed in Refs.~\cite{Frankfurt:1981mk,Frankfurt:1988nt} and is 
summarized in Ref.~\cite{Strikman:2017koc}. In this section we present the formalism for the treatment of 
polarized deuteron structure in the covariant approach of Refs.~\cite{Kondratyuk:1983kq,Frankfurt:1983qs}
and its connection to nonrelativistic deuteron spin structure. The evaluation of nucleonic operators
and the calculation of the tagged DIS cross section are discussed in Secs.~\ref{sec:nucleon_operators}
and \ref{sec:ia}.

High-energy processes such as DIS probe the nucleus with energy transfers much larger than the hadronic mass 
scale and result in hadron production over wide rapidity intervals. One would like to construct a theoretical 
description that starts from the nucleus as a system of protons and neutrons and produces the DIS final state 
by scattering on the nucleons, with eventual corrections due to nuclear binding effects. That such an 
approximation can be obtained is not obvious, as the nuclear initial states relevant for the
process might a priori involve states with energies of the order of the excitation energy, 
which in a relativistic context can no longer be described in terms of nucleonic degrees of freedom. 
The notion of energy, and therefore the relevant states, depend on the relativistic quantization scheme
adopted for the description of the process. LF quantization, which describes the evolution of the process 
in LF time $x^+ = t + z$ (with the $z$-direction along the reaction axis), is unique in that the energies 
of the intermediate states do not grow with the collision energy but remain finite in the high-energy limit.
It therefore permits a composite description of the nuclear initial state in terms of nucleon degrees of freedom, 
which is then matched with the high-energy scattering process on the nucleons, with finite effects 
due to nuclear binding.

\begin{figure}[t]
\includegraphics[width=.32\textwidth]{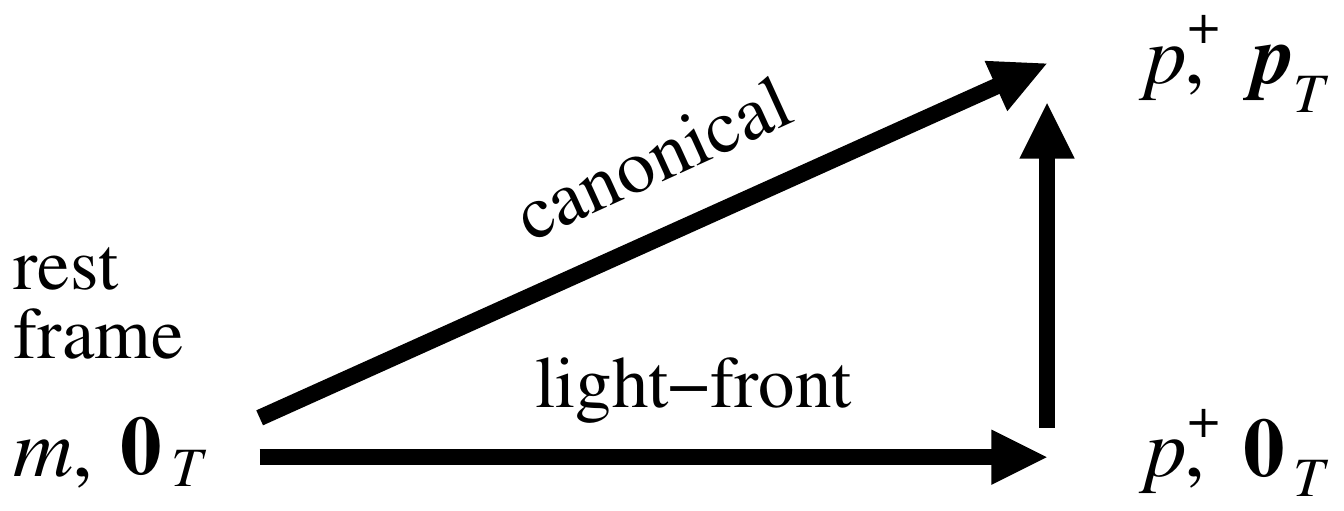}
\caption[]{LF helicity and canonical spin states. The diagram shows the sequence of boosts
leading from the rest-frame to the particle state with LF momentum $p^+$ and $\bm{p}_T$.}
\label{fig:lf_helicity}
\end{figure}

In LF quantization the spin degrees of freedom of particles and composite systems are described by 
light-front helicity states (see Fig.~\ref{fig:lf_helicity}). 
They are prepared by starting from the spin states in the rest frame, 
$p^+ = m$ and $\bm{p}_T = 0$, quantized along the $z$-direction, performing first a longitudinal boost 
to the desired plus momentum $p^+ \neq m$, and then a transverse boost to the desired transverse 
momentum $\bm{p}_T \neq 0$. The states thus defined are invariant under longitudinal boosts
and transform kinematically under transverse boosts. They differ from the so-called canonical 
spin states, which are prepared by performing a standard boost along the particle's 3-momentum
direction as in equal-time quantization, because boosts along different directions do not commute.
The difference between the two states is a spin rotation, the so-called Melosh rotation.
The explicit form of the nucleon bispinors for LF helicity states and canonical spin states,
and the Melosh rotation connecting them, is given in Appendix~\ref{app:spinors}.
In the following we use a representation of the deuteron LF wave function in which the LF helicity
character of the nucleon spin states is encoded in the explicit form of the bispinors and
thus no explicit Melosh rotations appear; the rotations are needed only in proving the
equivalence of this representation to the 3-dimensional canonical spin structure in the
CM frame in Sec.~\ref{subsec:cmframe} and Appendix~\ref{app:wave_function}.
%
%
\subsection{Light-front wave function in 4-dimensional form}
\label{subsec:lf_wave_function}
\begin{figure}[t]
\includegraphics[width=.22\textwidth]{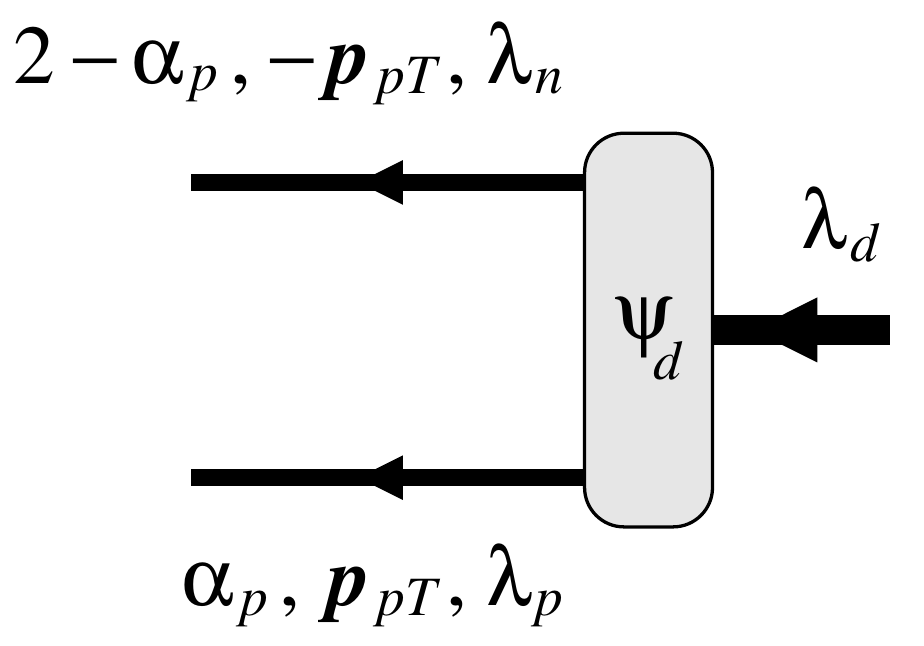}
\caption[]{The deuteron LF wave function, Eq.~(\ref{wf_general}), describing the expansion
of the deuteron state in $pn$ states.}
\label{fig:wf}
\end{figure}

The LF quantization is performed in the collinear frames, Sec.~\ref{subsec:collinear_frames}, where the
deuteron has LF plus momentum $p_d^+$ (arbitrary) and transverse momentum $\bm{p}_{dT} = 0$;
the deuteron 4-momentum components are [cf.\ Eq.~(\ref{collinear_frame})]
\be
p_d \; &=& \; \left[ p_d^+, \frac{M_d^2}{p_d^+}, \bm{0}_T \right]
\ee
The deuteron spin is described by the LF helicity, $\lambda_d$, which coincides with the rest-frame
spin projection because the states have zero transverse momentum. The expansion of the deuteron state 
in $pn$ states is described by LF wave function (see Fig.~\ref{fig:wf})
\beq
\Psi_d (\alpha_p , \bm{p}_{pT}; \lambda_p, \lambda_n | \lambda_d);
\label{wf_general}
\eeq
the definition of the matrix element and normalization of the states are given in Ref.~\cite{Strikman:2017koc}.
The wave function is normalized such that
\begin{align}
& \sum_{\lambda_p, \lambda_n}
\int \frac{d\alpha_p \; d^2 p_{pT}}{\alpha_p (2 - \alpha_p)} \;
\Psi_d^\ast (\alpha_p, \bm{p}_{pT}; \lambda_p, \lambda_n | \lambdadp)
\nonumber \\
& \times \Psi_d (\alpha_p, \bm{p}_{pT}; \lambda_p, \lambda_n | \lambda_d)
\;\; = \;\; \delta (\lambdadp, \lambda_d) .
\label{wf_general_normalization}
\end{align}
It is a function of the proton LF momentum fraction $\alpha_p$ and transverse momentum $\bm{p}_{pT}$;
the corresponding values for the neutron are
\beq
\alpha_n \; = \; 2 - \alpha_p, \hspace{2em} \bm{p}_{nT} \; = \; -\bm{p}_{pT} .
\eeq
The wave function is symmetric under the interchange of proton and neutron variables 
(momentum and spin) and satisfies the relation
\begin{align}
& \; \Psi_d (\alpha_p, \bm{p}_{pT}; \lambda_p, \lambda_n | \lambda_d)
\nonumber \\
=& \; 
\Psi_d (2 - \alpha_p, -\bm{p}_{pT}; \lambda_n, \lambda_p | \lambda_d) .
\label{lfwf_symmetry}
\end{align}
%
%

The spin structure of the deuteron LF wave function can be established on general grounds. 
It describes the coupling of the proton and neutron LF helicities,
and the LF orbital angular momentum, to the overall LF helicity of the deuteron. 
In the following we use a representation in which this coupling is expressed in 4-dimensional form, 
through invariants constructed out of the nucleon LF bispinors and a deuteron polarization 4-vector
\cite{Kondratyuk:1983kq}. The LF wave function is written in the form
\begin{align}
& \Psi_d (\alpha_p , \bm{p}_{pT}; \lambda_p, \lambda_n | \lambda_d)
\nonumber \\[1ex]
& = \; \bar u_{\rm LF}(p_n, \lambda_n) \, \Gamma_\alpha (p_p, p_n) \, v_{\rm LF} (p_p, \lambda_p)
\; \epsilon_{pn}^\alpha (p_{pn}, \lambda_d) .
\label{wf_bilinear}
\end{align}
Here $p_p$ and $p_n$ are the 4-momenta of the proton and neutron, whose LF components are
\begin{subequations}
\begin{align}
p_p \; &= \; \left[ \frac{\alpha_p p_d^+}{2}, \;
\frac{2(|\bm{p}_{pT}^2| + m^2)}{\alpha_p p_d^+}, \; \bm{p}_{pT} \right] ,
\\[2ex]
p_n \; &= \; \left[ \frac{(2 - \alpha_p) p_d^+}{2}, \;
\frac{2(|\bm{p}_{pT}^2| + m^2)}{(2 - \alpha_p) p_d^+}, \; -\bm{p}_{pT} \right] ,
\\[2ex]
p_p^2 \; &= \; p_n^2 \; = \; m^2 .
\label{p_p_p_n_mass_shell}
\end{align}
\end{subequations}
The LF plus and transverse components are determined by the variables $\alpha_p$ and $\bm{p}_{pT}$;
the minus components are fixed by the mass-shell conditions Eq.~(\ref{p_p_p_n_mass_shell}). 
Furthermore,
\begin{subequations}
\label{nucleon_spinors}
\begin{align}
& u_{\rm LF}(p_n, \lambda_n), \hspace{1em} v_{\rm LF}(p_p, \lambda_p),
\\[1ex]
& (p_n \gamma - m ) u_{\rm LF} = 0, \hspace{1em} (p_p\gamma + m ) v_{\rm LF} = 0, 
\end{align}
\end{subequations}
are the LF bispinor wave functions of the nucleon states with 4-momenta $p_p$ and $p_n$ and
LF helicities $\lambda_p$ and $\lambda_n$,
whose explicit form is given in Appendix~\ref{app:spinors}. In Eq.~(\ref{wf_bilinear}), $p_{pn}$ is
the total 4-momentum of the $pn$ pair,
\begin{subequations}
\begin{align}
p_{pn} \; &\equiv \; p_p + p_n \; = \; 
\displaystyle \left[ p_d^+, \; \frac{M_{pn}^2}{p_d^+}, \; \bm{0}_T \right],
\label{P_pn_def}
\\[2ex]
p_{pn}^2 &= \; M^2_{pn} \; \equiv \; \frac{4(|\bm{p}_{pT}^2| + m^2)}{\alpha_p (2 - \alpha_p)}.
\label{invariant_mass_def}
\end{align}
\end{subequations}
$M_{pn}$ is known as the invariant mass of the $pn$ pair. Note that the plus and transverse 
4-momentum components (LF momenta) of the $pn$ pair are the same as those of the external 
deuteron state, but the minus component (LF energy) is different,
\begin{subequations}
\begin{align}
p_{pn}^+ \; &= \; p_d^+,  \hspace{2em} \bm{P}_{pn, T} \; = \; \bm{p}_{dT} \; (= \; 0),
\\[1ex]
p_{pn}^- \; &\neq \; p_d^- ,
\end{align}
\end{subequations}
and the invariant mass of the $pn$ pair is different from the deuteron mass
\begin{align}
M_{pn}^2 \; &\neq \; M_d^2 .
\end{align}
These relations reflect the choice of momentum and energy variables specific to LF quantization.
Finally, in Eq.~(\ref{wf_bilinear}), $\epsilon_{pn}$ is the 4-vector spin wave function of the 
$pn$ system with 4-momentum $p_{pn}$ and mass $M_{pn}$,
\begin{subequations}
\label{epsilon_lf_components}
\begin{align}
&\epsilon_{pn} (p_{pn}, \lambda_d) \; = \; 
\left[ \frac{p_d^+}{M_{pn}} \epsilon_d^z, -\frac{M_{pn}}{p_d^+} \epsilon_d^z, \bm{\epsilon}_{dT} \right],
\\[2ex]
& \epsilon_{pn} p_{pn} \; = \; 0,
\hspace{2em}
\epsilon_{pn}^2 \; = \; -(\epsilon_d^z)^2 -\bm{\epsilon}_{dT}^2 \; = \; -1,
\end{align}
\end{subequations}
in which $\epsilon_d^z$ and $\bm{\epsilon}_{dT} \equiv (\epsilon_d^x, \epsilon_d^y)$ 
are the components of the deuteron 3-vector spin wave function in the deuteron rest frame,
\beq
\bm{\epsilon}_d \; \equiv \; \bm{\epsilon}_d(\lambda_d ) 
\; = \; (\bm{\epsilon}_{dT}, \epsilon_d^z), \hspace{2em} \bm{\epsilon}_d^2 \; = \; 1.
\eeq
Note that the $pn$ 4-vector Eq.~(\ref{epsilon_lf_components}) is different from the
deuteron 4-vector,
\begin{subequations}
\label{epsilon_d}
\begin{align}
\epsilon_d (p_d, \lambda_d) &=  
\left[ \frac{p_d^+}{M_d} \epsilon_d^z, -\frac{M_d}{p_d^+} \epsilon_d^z, \bm{\epsilon}_{dT} \right],
\\[2ex]
\epsilon_d p_d \; &= 0, \hspace{2em} \epsilon_d^2 \; = \; -(\epsilon_d^z)^2 -\bm{\epsilon}_{dT}^2 \; = -1.
\end{align}
\end{subequations}
The particular form of Eq.~(\ref{epsilon_lf_components}) is necessary to ensure the equivalence
of Eq.~(\ref{wf_bilinear}) with the 3-dimensional spin structure of the $pn$ pair in the CM frame,
as explained in Sec.~\ref{subsec:cmframe} and Appendix~\ref{app:wave_function}.

The function $\Gamma_\alpha$ in Eq.~(\ref{wf_bilinear}), a matrix in bispinor space and a 4-vector, 
connects the nucleon bispinors and the deuteron 4-vector to an invariant form.
It may be regarded as a function of the 4-momenta $p_p$ and $p_n$, and its form is constrained 
by standard 4-dimensional relativistic covariance. Taking into account the equations for the
nucleon spinors, Eq.~(\ref{nucleon_spinors}), it can be decomposed 
in independent covariant structures as\footnote{The decomposition of the nucleon-deuteron
coupling Eq.~(\ref{Gamma_decomposition}) is analogous to that of the nucleon coupling to the
electromagnetic current and involves the same number of independent structures.}
\be
\Gamma^\alpha \; \equiv \; \Gamma^\alpha  (p_p, p_n)
\; &=& \; \gamma^\alpha G_1 \; + \; \Delta^\alpha G_2, 
\label{Gamma_decomposition}
\ee
where $\Delta$ is the difference of the nucleon 4-momenta,
\begin{align}
& \Delta \equiv  p_p - p_n,
\hspace{1em} \Delta p_{pn} = 0,  \hspace{1em} \Delta^2 = -M_{pn}^2 + 4 m^2 ,
\end{align}
and $G_{1,2}$ are scalar functions of the invariant mass of the $pn$ pair,
\be
G_{1, 2} &\equiv& G_{1, 2}(M_{pn}) .
\label{G_12_def}
\ee
The functions $G_{1, 2}$ contain the dynamical information about deuteron structure in LF quantization. 
They can be matched with the 3-dimensional radial wave functions in equal-time quantization,
and their explicit form is given in Sec.~\ref{subsec:cmframe}.

Together, Eqs.~(\ref{wf_bilinear}) and (\ref{Gamma_decomposition}) provide a representation of
the LF spin structure of the deuteron in 4-dimensional form. Its advantages are: 
(a)~The representation of Eqs.~(\ref{wf_bilinear}) and (\ref{Gamma_decomposition}) avoids 
the use of explicit Melosh rotations, which appear in the standard construction
of the LF wave function starting from a 3-dimensional wave function with canonical spinors.
The rotations are contained in the explicit form of the LF bispinors.
(b)~The representation of Eqs.~(\ref{wf_bilinear}) and (\ref{Gamma_decomposition}) permits 
efficient evaluation of the sums over nucleon spin degrees of freedom in observables, 
given by overlap integrals of the LF wave functions. Sums over the nucleon LF helicities 
can be converted to traces over spin density matrices in bispinor representation, which 
can be evaluated using standard techniques. (c)~Overall, the representation enables a 
4-dimensional treatment of spin structure within the essentially 3-dimensional approach
of LF quantization.
\subsection{Center-of-mass frame variables}
\label{subsec:cmframe}
In the LF wave function Eq.~(\ref{wf_bilinear}) the $pn$ configuration is specified by the
LF momentum variables $\alpha_p$ and $\bm{p}_{pT}$. An alternative representation of the LF 
wave function is obtained by using as variables the proton 3-momentum in the CM frame of the $pn$ pair.
This representation offers a simple way of realizing rotational invariance in LF quantization,
permits matching of the invariant functions Eq.~(\ref{G_12_def}) with the equal-time wave functions, 
and enables the construction of a nonrelativistic approximation to the LF wave functions.
In the following calculations we deal with the LF components and the ordinary components of 
4-vectors at the same time and use the notation [cf.\ Eq.~(\ref{notation_lf})]
\beq
[v^+, v^-, \bm{v}_T], \hspace{2em} (v^0, \bm{v}),
\eeq
to distinguish both sets of components in a given frame.

%
%
\begin{figure}[t]
\includegraphics[width=.48\textwidth]{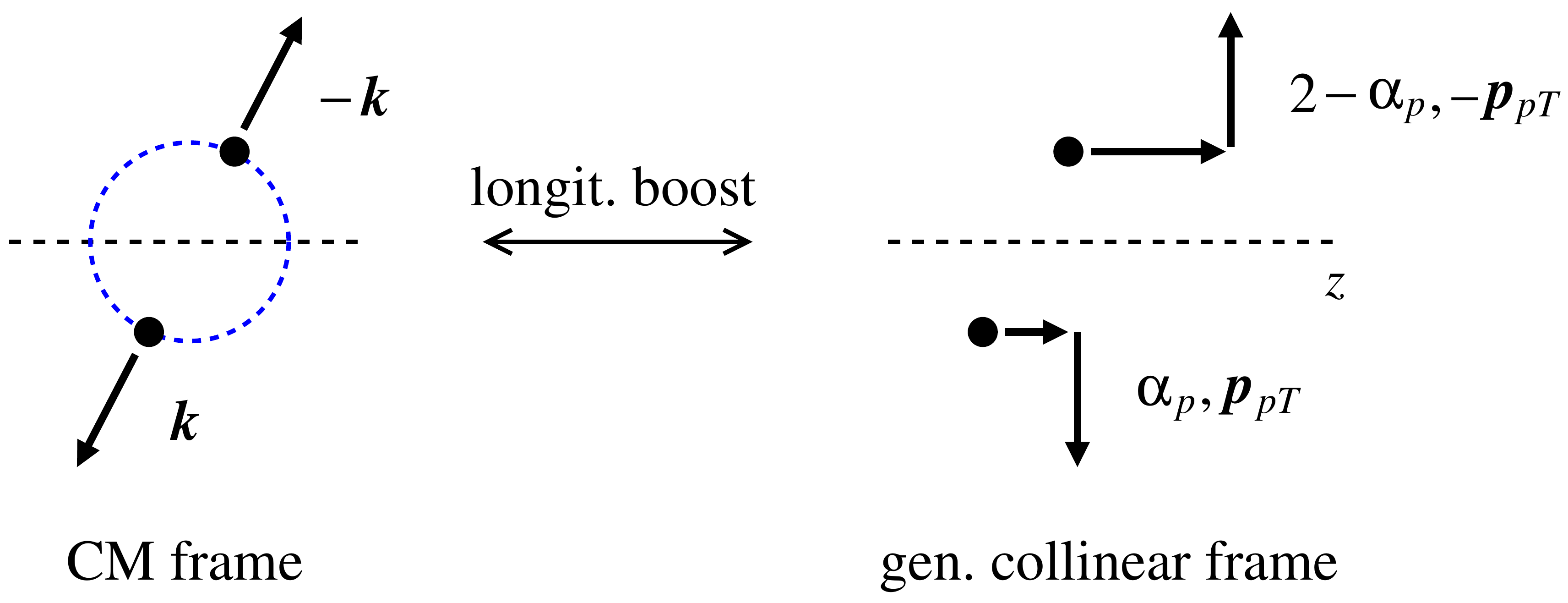}
\caption[]{The CM frame variables for the deuteron LF wave function.
In a general collinear frame the wave function depends on the longitudinal 
and transverse momenta, $\alpha_p$ and $\bm{p}_{pT}$ (right). 
By a longitudinal boost any such frame is connected with the CM frame of the $pn$ pair, 
in which the wave function depends on the 3-momentum $\bm{k}$ and exhibits
rotational symmetry (left).}
\label{fig:cmframe}
\end{figure}
The CM frame of a given $pn$ configuration is defined as the frame in which the proton and neutron have opposite 
3-momenta in the sense of ordinary vector components. This frame is a member of the class of collinear 
frames and can be reached from any other collinear frame by a longitudinal boost 
(see Fig.~\ref{fig:cmframe}). To show this, we use the fact that a collinear frame in the class
is specified by the value of $p_d^+$ in that frame (see Sec.~\ref{subsec:collinear_frames}).
The CM frame of the $pn$ configuration is the special collinear frame with
\beq
p_d^+ [\textrm{CM}] \; = \; M_{pn}.
\eeq
In this frame the total 4-momentum of the $pn$ configuration, Eq.~(\ref{P_pn_def}), has LF components
\beq
p_{pn}[\textrm{CM}] \; = \; [M_{pn}, M_{pn}, \bm{0}_T],
\eeq
and therefore ordinary components
\beq
p_{pn}[\textrm{CM}] \; = \; (M_{pn}, \bm{0}).
\eeq
The individual proton and neutron 4-momenta have LF components
\begin{subequations}
\begin{align}
p_p[\textrm{CM}] &= \left[\frac{\alpha_p M_{pn}}{2}, \frac{(2 - \alpha_p) M_{pn}}{2},  \bm{p}_{pT}\right],
\\[2ex]
p_n[\textrm{CM}] &= \left[\frac{(2 - \alpha_p) M_{pn}}{2}, \frac{\alpha_p M_{pn}}{2}, -\bm{p}_{pT}\right],
\end{align}
\end{subequations}
such that the ordinary components satisfy (we suppress the label [CM])
\begin{subequations}
\begin{align}
p_n^0 &= {\textstyle\frac{1}{2}} (p_n^+ + p_n^-)  \; = \; \phantom{-}{\textstyle\frac{1}{2}} (p_p^+ + p_p^-) 
\; = \; \phantom{-}p_p^0,
\\[2ex]
p_n^z &= {\textstyle\frac{1}{2}} (p_n^+ - p_n^-) \; = \; -{\textstyle\frac{1}{2}} (p_p^+ - p_p^-) \; = \; -p_p^z,
\\[2ex]
\bm{p}_{nT} &= -\bm{p}_{pT},
\end{align}
\end{subequations}
i.e., they have the same energy and opposite 3-momenta. In this frame the proton and neutron
4-momenta can therefore be expressed in terms of the common 3-momentum vector as
\begin{subequations}
\label{p_p_p_n_CM}
\begin{align}
p_p[\textrm{CM}] \; &= \; (E, \bm{k}), 
\\[2ex]
p_n[\textrm{CM}] \; &= \; (E, -\bm{k}),
\\[2ex]
E \; \equiv \; E(\bm{k}) \; &\equiv \; \sqrt{|\bm{k}|^2 + m^2}
\end{align}
\end{subequations}
The relation of the CM momentum $\bm{k}$ to the LF variables $\alpha_p$ and $\bm{p}_{pT}$ is
\begin{subequations}
\label{CM_LF_variables}
\begin{align}
k^z \; &= \; \frac{M_{pn}}{2} (\alpha_p - 1), 
\hspace{2em}
\bm{k}_T \; = \; \bm{p}_{pT} ,
\\[1ex]
\alpha_p \; &= \; 1 + \frac{k^z}{E},
\hspace{2em}
M_{pn} \; = \; 2 E.
\end{align}
\end{subequations}
The CM momentum can therefore serve as a kinematic variable alternative to the LF variables,
\begin{align}
\bm{k} \; \leftrightarrow \; \{ \alpha_p, \bm{p}_{pT} \} .
\end{align}
The relation between the integration measures is
\be
\int\frac{d\alpha_p \; d^2 p_{pT}}{\alpha_p (2 - \alpha_p)}  
\; &=& \; \int \frac{d^3 k}{E(k)} .
\label{integration_k}
\ee

In the CM frame the polarization vector of the $pn$ system, Eq.~(\ref{epsilon_lf_components}),
has 4-vector components (LF and ordinary)
\be
\epsilon_{pn}[\textrm{CM}] \; &=& \; [\epsilon_d^z, -\epsilon_d^z, \bm{\epsilon}_{dT}] 
\; = \;
(0, \bm{\epsilon}_d) .
\label{epsilon_pn_cm}
\ee
This is the same form that the deuteron polarization vector has in the deuteron rest 
frame, cf.\ Eq.~(\ref{epsilon_d}) with $p_d^+ = M_d$. The CM frame thus permits a
particularly simple representation of the deuteron spin structure. We use this
representation extensively in the calculations in Secs.~\ref{sec:nucleon_operators}
and \ref{sec:ia}.

In the CM frame the $pn$ LF wave function can be can be formulated as a 3-dimensional 
relativistic wave function in the 3-momentum variable $\bm{k}$. It is constructed using
angular momentum wave functions (S and D waves), canonical nucleon spinors, and the
Melosh rotations mediating the transition from canonical spin to LF helicity
(see Appendix~\ref{app:wave_function}). The dynamical information is 
contained in the radial wave functions of the S- and D-waves,
\beq
f_0(k), \; f_2(k) \hspace{2em} [k \equiv |\bm{k}|],
\label{radial}
\eeq
which are normalized such that
\beq
4\pi \int \frac{dk \, k^2}{E(k)} \; [f_0^2 (k) + f_2^2(k)] \;\; = \;\; 1 .
\label{normalization_radial}
\eeq
Using the fact that the CM frame is a special collinear frame, one can then establish the 
correspondence between the general LF wave function Eq.~(\ref{wf_bilinear}) and the 
3-dimensional wave function in the CM frame.
The proof involves expressing the LF bispinors in Eq.~(\ref{wf_bilinear}) in terms of 
canonical bispinors in the CM frame, reducing the bilinear form in the canonical bispinors 
to two-component spinors, and comparing with the 3-dimensional wave function
(see Appendix~\ref{app:wave_function}). As a result, one obtains the relation
between the invariant functions $G_{1, 2} (M_{pn}^2)$, Eq.~(\ref{G_12_def}) and the 
3-dimensional radial wave functions in the CM frame, $f_0(k)$ and $f_2(k)$:
\begin{subequations}
\label{vertex_radial}
\begin{align}
G_1 &= \frac{1}{4 E} \left( \sqrt{2}\, f_0 - f_2 \right),
\label{vertex_radial_1}
\\[2ex]
G_2 &= \frac{1}{8 E k^2} \left[ \sqrt{2} (E - m) \, f_0 + (2 E + m) \, f_2 \right]
\label{vertex_radial_2}
\\[3ex]
\nonumber
& \left[ G_{1, 2} \; \equiv \; G_{1, 2}(M_{pn}^2), \hspace{1em} f_0, f_2 \; \equiv \; f_0(k), f_2(k) \right] ,
\end{align}
\end{subequations}
where the LF and CM variables are related by Eq.~(\ref{CM_LF_variables}). In particular, the correct
normalization of the LF wave function, Eq.~(\ref{wf_general_normalization}), is obtained from the 
normalization condition for the radial wave functions Eq.~(\ref{normalization_radial}).
Equation~(\ref{vertex_radial}) allows one to express the dynamical elements in the 4-dimensional 
representation of the deuteron LF wave function in terms of 3-dimensional wave functions with 
well-known properties and represents an essential tool in the LF structure calculations.
\subsection{Nonrelativistic approximation}
\label{subsec:nonrel}
The dynamical elements in the deuteron LF wave function can be determined by solving the dynamical equation 
for the two-nucleon bound state (in its differential or integral form) with an effective $NN$ interaction 
formulated at fixed LF time. The specific form of the dynamical equation, the physical conditions
for the truncation to the two-nucleon sector, and the technical issues relating to rotational invariance,
are discussed in Refs.~\cite{Frankfurt:1981mk,Frankfurt:1992ny}. Alternatively, one may construct an
approximation to the deuteron LF wave function from the nonrelativistic wave function obtained with
an effective nonrelativistic $NN$ interaction (potential). This approach allows one to incorporate
the extensive knowledge of $NN$ interactions in nonrelativistic nuclear theory into the LF nuclear
structure calculations. The nonrelativistic approximation turns out to be fully adequate for nucleon 
rest-frame momenta $|\bm{p}_p| \lesssim$ 300 MeV and is used in the present study.

In the nonrelativistic limit $k^2 \ll m^2$, the relativistic radial wave functions in the CM frame, 
Eq.~(\ref{radial}), approach the nonrelativistic radial wave functions,
\beq
f_L(k) \;\; \rightarrow \;\; \sqrt{m} \, f_{L, {\rm nr}} (k) \hspace{2em} (L = 0, 2).
\eeq
The factor $\sqrt{m}$ results from the normalization convention for the nonrelativistic
radial functions, which differs from Eq.~(\ref{normalization_radial}),
\beq
4 \pi \int dk \, k^2 \, [f_{0, {\rm nr}}^2(k) + f_{2, {\rm nr}}^2(k)] \;\; = \;\; 1.
\eeq
A nonrelativistic approximation to the relativistic radial functions is provided by
\beq
f_L(k) \;\; = \;\; \sqrt{E(k)} \, f_{L, {\rm nr}}(k) \hspace{2em} (L = 0, 2).
\label{nonrel_approx}
\eeq
The approximation becomes exact at small momenta $k^2 \ll m^2$; it satisfies the relativistic
normalization condition Eq.~(\ref{normalization_radial}) and is therefore correct ``on average'' 
also at large momenta; altogether the approximation thus has an interpolating 
quality.\footnote{In Eq.~(\ref{nonrel_approx}) the nonrelativistic wave function on the right-hand side
is evaluated at the LF CM momentum $k \equiv |\bm{k}|$ defined in Eq.~(\ref{CM_LF_variables}), 
which is not identical to the proton 3-momentum in the deuteron rest frame, $|\bm{p}_p|$, but differs
from it by corrections of the order $\bm{p}_p/m$ in the nonrelativistic limit. By expanding the
variable $|\bm{k}|$ in $\bm{p}_p/m$ one can derive a simplified version of the nonrelativistic approximation, 
in which the nonrelativistic wave function is evaluated directly at the rest-frame momentum $|\bm{p}_p|$,
and certain factors $(1 - p_p^z/m)$ account for the anisotropy of the LF description;
see Ref.~\cite{Strikman:2017koc} for details. This approximation no longer has the interpolating quality
of Eq.~(\ref{nonrel_approx}).} In the numerical studies in this work we use Eq.~(\ref{nonrel_approx}) 
with the non-relativistic deuteron wave functions obtained from AV18 $NN$ potential \cite{Wiringa:1994wb}. 
\section{Nucleon operators}
\label{sec:nucleon_operators}
\subsection{Matrix elements of nucleon operators}
\label{subsec:matrix_elements}
We now present the methods for evaluating matrix elements of nucleon operators in the 
polarized deuteron state described by the LF wave function of Sec.~\ref{sec:deuteron}.
In the following applications we consider nucleon one-body operators (vector and axial current,
tagged DIS cross section in IA); the methods can easily be extended to two-body operators 
(correlations, FSI). The matrix elements of nucleon one-body operators in the polarized
deuteron state can be computed in a simple manner using the 4-dimensional form of the 
deuteron LF wave function, Eq.~(\ref{wf_bilinear}). We limit ourselves to matrix elements 
at zero momentum transfer (charges, nucleon LF momentum densities); the formulas can easily 
be generalized to include momentum transfer.

\begin{figure}[t]
\includegraphics[width=.35\textwidth]{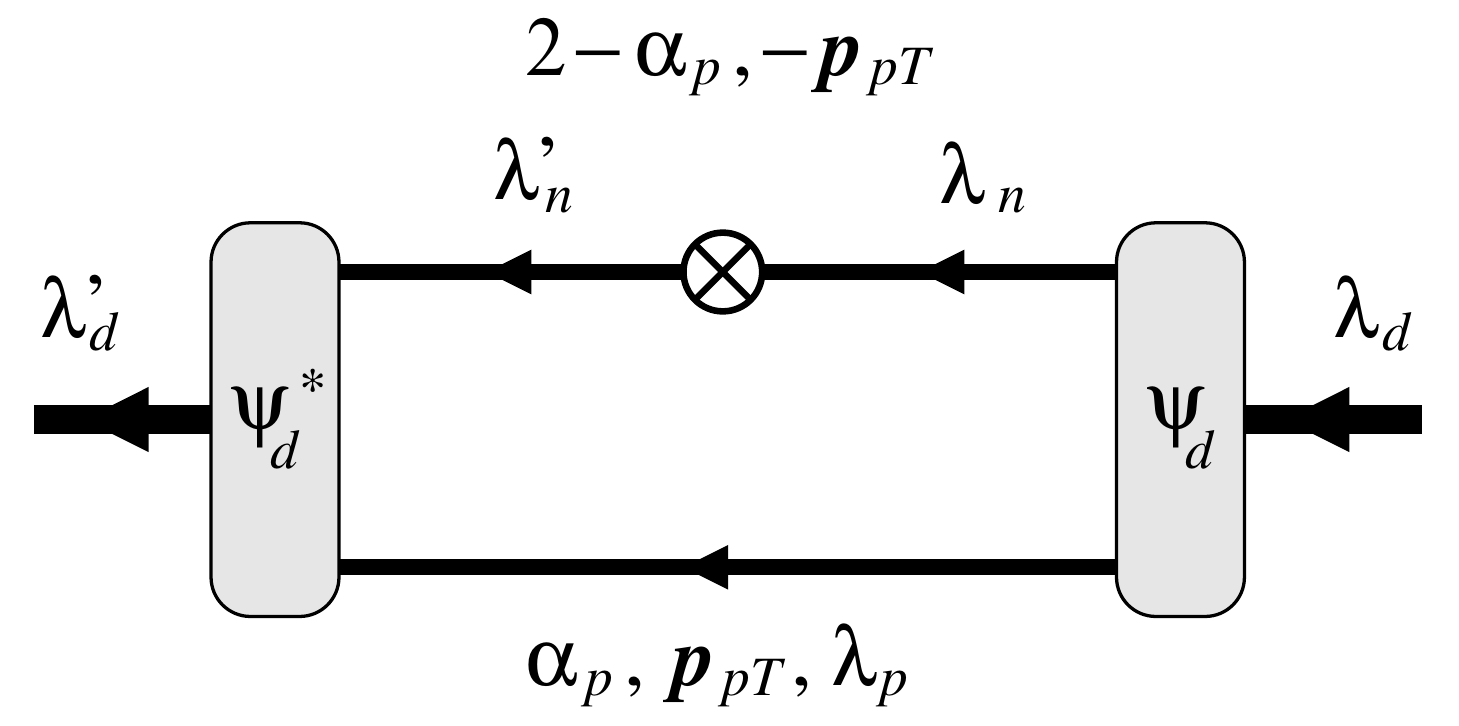}
\caption[]{The deuteron matrix element of a one-body neutron operator (zero momentum transfer).
The intermediate proton state has LF momentum $\alpha_p$ and $\bm{p}_{pT}$ and LF helicity
$\lambda_p$; the neutron states have LF momentum $2 - \alpha_p$ and $-\bm{p}_{pT}$,
and helicities $\lambda_n$ and $\lambdanp$. The variables describing the intermediate 
state are integrated/summed over.}
\label{fig:onebody}
\end{figure}

Let $O_n$ denote a generic nucleon one-body operator coupling to the neutron. 
Its expectation value in a polarized deuteron ensemble is given by
\begin{align}
\langle O_n \rangle \, \equiv \, 
\sum_{\lambdadp, \lambda_d} \rho_d (\lambda_d, \lambdadp ) \,
\langle d, p_d, \lambdadp | O_n | d, p_d, \lambda_d \rangle ,
\label{onebody_deuteron_average}
\end{align}
where $\rho_d$ is the deuteron spin density matrix, and $\lambdadp$ and $\lambda_d$ 
are the LF helicities of the deuteron states, which coincide with the rest-frame spin
projections ($\bm{p}_{dT} = 0$). The matrix element of the nucleon operator between deuteron 
states with LF helicities $\lambda_d$ and $\lambdadp$ in Eq.~(\ref{onebody_deuteron_average})
can be calculated with standard methods, by inserting a complete set of proton-neutron 
intermediate states, and is obtained as \cite{Strikman:2017koc}  (see Fig.~\ref{fig:onebody})
\begin{align}
& \langle d, p_d, \lambdadp | O_n | d, p_d, \lambda_d \rangle
\nonumber \\[1ex]
&= \; \int\frac{d\alpha_p}{\alpha_p} \, d^2 p_{pT} \;
\sum_{\lambda_p, \lambdanp, \lambda_n} \;
\frac{2}{(2 - \alpha_p)^2} 
\nonumber \\[1ex]
& \times \; \Psi_d^\ast (\alpha_p , \bm{p}_{pT}; \lambda_p, \lambdanp | \lambdadp) \;
\Psi_d (\alpha_p , \bm{p}_{pT}; \lambda_p, \lambda_n | \lambda_d) \;
\nonumber \\[2ex]
&\times \; \langle n, p_n, \lambdanp | O_n | n, p_n, \lambda_n \rangle .
\label{onebody_deuteron_me}
\end{align}
The integration is over the LF momentum variables of the proton; the LF momentum of the 
neutron is given by
\beq
\alpha_n \; = \; 2 - \alpha_p, \hspace{2em} \bm{p}_{nT} \; = \; -\bm{p}_{pT} .
\label{proton_neutron_momentum}
\eeq
The factor $2/(2 - \alpha_p)^2$ results from the normalization of the deuteron and
nucleon states in LF quantization \cite{Strikman:2017koc}. In Eq.~(\ref{onebody_deuteron_me})
the summation is over the LF helicities of the proton and neutron intermediate states,
$\lambda_p, \lambda_n$ and $\lambdanp$. $\Psi_d$ and $\Psi_d^\ast$
are the deuteron LF wave function Eq.~(\ref{wf_general}) and its complex conjugate.
\beq
\langle n, p_n, \lambdanp | O_n | n, p_n, \lambda_n \rangle
\eeq
denotes the matrix element of the operator between neutron states with 4-momentum $p_n$ 
[with LF components given by Eq.~(\ref{proton_neutron_momentum})] and LF helicities 
$\lambda_n$ and $\lambdanp$.

%
%
To evaluate the spin sums in Eq.~(\ref{onebody_deuteron_me}), we substitute
the deuteron LF wave functions by their representation as bilinear forms
in LF bispinors, Eq.~(\ref{wf_bilinear}). The matrix element of the nucleon 
operator between neutron states can likewise be represented as a bilinear form
\beq
\langle n, p_n, \lambdanp | O_n | n, p_n, \lambda_n \rangle
\;\; = \;\; \bar u_{\rm LF}(p_n, \lambdanp) \, \Gamma_n \, u_{\rm LF} (p_n, \lambda_n),
\eeq
where the bispinor matrix $\Gamma_n$ is specific to the operator and governs the dependence 
of the matrix element on the LF helicities $\lambdanp$ and $\lambda_n$. The summations 
over $\lambda_p$, $\lambdanp$, and $\lambda_n$ in Eq.~(\ref{onebody_deuteron_me}) can
then be carried out and give rise to the bispinor projectors [cf.~Appendix~\ref{app:spinors}]
\begin{subequations}
\label{projector_uv}
\begin{align}
\sum_{\lambda_p} v_{\rm LF} (p_p, \lambda_p) \bar v_{\rm LF} (p_p, \lambda_n)
\; &= \;
(p_p\gamma - m),
\label{projector_v}
\\[1ex]
\sum_{\lambda_n} u_{\rm LF} (p_n, \lambda_n) \bar u_{\rm LF} (p_n, \lambda_n)
\; &= \; 
(p_n\gamma + m)
\nonumber
\\[0ex]
\textrm{(same for $\lambdanp$).}
\label{projector_u}
\end{align}
\end{subequations}
The deuteron expectation value Eq.~(\ref{onebody_deuteron_average}) is then obtained as a
bispinor trace,
\begin{subequations}
\begin{align}
\langle O_n \rangle 
\; &= \; 
\int\frac{d\alpha_p}{\alpha_p} \, d^2 p_{pT} \;
\frac{2 \, \textrm{tr}[\Pi_n \Gamma_n]}{(2 - \alpha_p)^2} ,
\label{deuteron_onebody_trace}
\\[3ex]
\Pi_n 
\; &\equiv \;
(\rho_{pn})^{\alpha\beta} (p_n\gamma + m) \Gamma_\alpha (p_p\gamma - m)
\Gamma_\beta (p_n\gamma + m) ,
\label{neutron_density}
\\[3ex]
(\rho_{pn})^{\alpha\beta} 
\; &\equiv \; \langle \epsilon^\alpha_{pn} \epsilon^{\beta\ast}_{pn} \rangle
\; = \;
\sum_{\lambdadp, \lambda_d}
\rho_d (\lambda_d, \lambdadp) \epsilon^\alpha_{pn}(\lambda_d) \epsilon^{\beta\ast}_{pn}(\lambdadp) .
\label{rho_pn}
\end{align}
\end{subequations}
$\Pi_n$ in Eq.~(\ref{neutron_density}) is a matrix in bispinor indices and represents the effective neutron 
spin density matrix in the deuteron for a given proton LF momentum $\alpha_p$ and $\bm{p}_{pT}$ 
[or equivalently, for a given neutron momentum Eq.~(\ref{proton_neutron_momentum})]. 
It depends covariantly on the neutron and proton 4-momenta, $p_p$ and $p_n$, and the parameters
characterizing the deuteron polarization. It satisfies the conditions
\beq
(p_n\gamma - m) \Pi_n \; = \; 0, \hspace{2em}
\Pi_n (p_n\gamma - m)  \; = \; 0 .
\eeq
which follow from the Dirac equation of the neutron bispinors. $\rho_{pn}$ in Eq.~(\ref{rho_pn})
is the ensemble average of the $pn$ polarization vectors, Eq.~(\ref{epsilon_lf_components}), with the deuteron 
spin density matrix and represents the spin density matrix of the $pn$ configuration 
in the polarized deuteron in LF quantization. It depends covariantly on the total 
4-momentum of the $pn$ system, $p_{pn}$, as well as on the axial 4-vector and 4-tensor
characterizing the deuteron polarization, $s_d$ and $t_d$. It satisfies the conditions
\beq
p_{pn, \alpha} (\rho_{pn})^{\alpha\beta} \; = \; 0,
\hspace{2em}
(\rho_{pn})^{\alpha\beta} p_{pn, \beta} \; = \; 0 ,
\label{rho_pn_conditions}
\eeq
which follow from Eq.~(\ref{epsilon_lf_components}); they are analogous to those satisfied by 
the deuteron density matrix but involve the 4-momentum of the $pn$ system rather than that of the deuteron. 
Explicit expressions for $\Pi_n$ and $\rho_{pn}$ will be derived in the following.

The representation of the expectation value of Eq.~(\ref{deuteron_onebody_trace}) has
several important advantages: (a)~The projectors resulting from the summation over the 
nucleon LF helicities, Eq.~(\ref{projector_uv}), do not ``remember'' the specific form
of the LF helicity spinors; they are the same as for canonical spinors summed over 
the canonical spin. (In the 3-dimensional formulation of spin structure, this comes about
because the Melosh rotations cancel when summing over the nucleon LF helicities; see Appendix~\ref{app:spinors})
This implies that the ``knowledge'' of the specific choice of the LF spin states
resides only in the function $\Gamma^\alpha$, Eq.~(\ref{Gamma_decomposition}), and
the spin density matrix of the $pn$ configuration, $\rho_{pn}$, Eq.~(\ref{rho_pn}).
(b)~The covariant definition of $\rho_{pn}$ and $\Pi_n$ constrains their dependence on
the proton and neutron 4-momenta and the deuteron polarization vector and tensor
and greatly simplifies the calculation. (c)~The neutron spin density matrix Eq.~(\ref{neutron_density}) 
can be computed in closed form and has a simple structure. It can be compared with the density matrix 
of a free neutron, Eq.~(\ref{density_matrix_spin12}), and interpreted accordingly. 
(d)~Products of gamma matrices and bispinor traces are easily evaluated.
Equation~(\ref{deuteron_onebody_trace}) will be our main tool in the subsequent deuteron
structure calculations.
\subsection{Neutron spin density matrix}
We now evaluate the effective neutron spin density matrix in the polarized deuteron,
Eq.~(\ref{neutron_density}). We first derive explicit expressions for the spin density matrix 
of the $pn$ system, Eq.~(\ref{rho_pn}), taking into account the properties of the LF
spin wave functions Eq.~(\ref{epsilon_lf_components}) and the conditions Eq.~(\ref{rho_pn_conditions}).
We then calculate the contraction in Eq.~(\ref{neutron_density}) with these expressions
and convert the result to a form that can be compared with the free neutron density matrix.
We consider separately the contributions of the unpolarized, vector-polarized, and tensor-polarized 
parts of the deuteron spin density matrix, Eq.~(\ref{density_matrix_spin1_decomposition}), 
to the neutron density matrix, Eq.~(\ref{neutron_density}), and write
\begin{align}
\Pi_n \; &= \; \Pi_n[\textrm{unpol}] \, + \, \Pi_n[\textrm{vector}] \, + \, \Pi_n[\textrm{tensor}] .
\label{neutron_density_decomposition}
\end{align}

The unpolarized part of the $pn$ spin density matrix is given by
\beq
(\rho_{pn})^{\alpha\beta}[\textrm{unpol}] \;\; = \;\;
\frac{1}{3} \left( -g^{\alpha\beta} + \frac{p_{pn}^\alpha p_{pn}^\beta}{M_{pn}^2} \right) .
\label{rho_pn_unpol}
\eeq
It corresponds to the general form of Eq.~(\ref{rho_deuteron_unpol}) but involves the 4-momentum 
of the $pn$ configuration instead of that of the deuteron, cf.\ Eq.~(\ref{epsilon_lf_components}). 
Substituting Eq.~(\ref{rho_pn_unpol}) in Eq.~(\ref{neutron_density}), we obtain the effective 
neutron density matrix in the unpolarized as
\begin{align}
\Pi_n[\textrm{unpol}] \; =& \; \frac{1}{3} (p_n\gamma + m) 
\; \left[ (12 m^2 - 2 \Delta^2) G_1^2 \right.
\nonumber \\[2ex]
& \left. - 4 m \Delta^2 G_1 G_2 + (\Delta^2)^2 G_2^2 \right]
\nonumber
\\[2ex]
& [G_{1, 2} \equiv G_{1, 2}(M_{pn})].
\label{neutron_density_unpol}
\end{align}
The result has been simplified by making use of gamma matrix identities and the properties 
of the projectors Eq.~(\ref{projector_uv}). The effective neutron density matrix has the form of the 
density matrix of a free neutron with 4-momentum $p_n$, multiplied by a certain function of the invariant 
mass of the $pn$ pair, which is quadratic in the scalar functions $G_{1, 2}$, Eq.~(\ref{G_12_def}).

The vector-polarized part of the $pn$ spin density matrix is given by
\beq
(\rho_{pn})^{\alpha\beta} [\textrm{vector}] \;\; = \;\;
\frac{i}{2 M_{pn}} \epsilon^{\alpha\beta\gamma\delta} p_{pn, \gamma} s_{pn, \delta} .
\label{rho_pn_pol}
\eeq
Here $s_{pn}$ is the polarization vector of the $pn$ configuration in the collinear frame,
\begin{subequations}
\label{A_def}
\begin{align}
s_{pn} \; &\equiv \; \left[ \frac{p_d^+}{M_{pn}} S_d^z, \; -\frac{M_{pn}}{p_d^+} S_d^z, 
\; \bm{S}_{dT} \right], 
\\[2ex]
(s_{pn} p_{pn}) \; &= \; 0, 
\\[2ex]
s_{pn}^2 \; &= \; -(S_d^z)^2 - |\bm{S}_{dT}|^2 \; = \; -|\bm{S}_d|^2 ,
\end{align}
\end{subequations}
where $S_d^z$ and $\bm{S}_{dT}$ are the components of the deuteron polarization vector in the rest frame,
\begin{align}
\bm{S}_d \; &= \; (\bm{S}_{dT}, S_d^z), 
\hspace{2em}
|\bm{S}_d|^2 \; \leq \; 1.
\label{vector_restframe}
\end{align}
Again the form of Eqs.~(\ref{rho_pn_pol}) and (\ref{A_def}) corresponds to that of the deuteron spin
density matrix and polarization vector, cf.\ Eq.~(\ref{deuteron_polarization_vector_collinear}),
only the quantities are constructed with the $pn$ 4-momentum rather than the deuteron 4-momentum.
Substituting Eq.~(\ref{rho_pn_pol}) in Eq.~(\ref{neutron_density}), we obtain the effective neutron 
density matrix induced by deuteron vector polarization,
\begin{subequations}
\label{neutron_density_vector}
\begin{align}
\Pi_n [\textrm{vector}] \; &= \; {\textstyle\frac{1}{2}} (p_n\gamma + m) (s_n \gamma) \gamma^5 ,
\label{neutron_density_vector_1}
\\[3ex]
s_n^\alpha \; &= \; 2 M_{pn} \left( 2 m G_1 a_1^\alpha - \Delta^2 G_2 a_2^\alpha \right) G_1 ,
\label{neutron_spin_vector}
\\[2ex]
a_1^\alpha \; &= \; \left( g_{\alpha\beta}  - \frac{p_{n, \alpha} p_{n, \beta}}{m^2} \right) s_{pn}^\beta ,
\label{a_1_def}
\\[2ex]
a_2^\alpha \; &= \; \left( g_{\alpha\beta}  - \frac{\Delta_\alpha \Delta_\beta}{\Delta^2} \right) s_{pn}^\beta ,
\\[2ex]
\label{a_2_def}
s_n p_n \; & = \; 0, \hspace{2em} a_1 p_n = 0, \hspace{2em} a_2 p_n = 0 .
\end{align}
\end{subequations}
The specific form of Eq.~(\ref{neutron_density_vector}) is obtained by making extensive 
use of gamma matrix identities and the properties of the projectors Eqs.~(\ref{projector_uv}).
The effective neutron density matrix Eq.~(\ref{neutron_density_vector_1})
depends on the axial vector $s_n$, Eq.~(\ref{neutron_spin_vector}), which may be regarded 
as the effective polarization vector of the neutron in the deuteron. It is constructed from the
the auxiliary 4-vectors $a_1$ and $a_2$, which are projections of the $pn$ polarization 4-vector $s_{pn}$
on the subspaces orthogonal to $p_n$ and $\Delta$. Note that the two structures in 
Eq.~(\ref{neutron_spin_vector}) individually satisfy the condition $p_n s_n = 0$
for any value of the scalar functions $G_1$ and $G_2$.

The tensor-polarized part of the $pn$ spin density matrix is parameterized as
\beq
(\rho_{pn})^{\alpha\beta}[\textrm{tensor}] \; = \;  -(t_{pn})^{\alpha\beta} ,
\label{rho_pn_tensor}
\eeq
where $t_{pn}$ is the polarization 4-tensor of the $pn$ configuration and satisfies
\begin{subequations}
\begin{align}
(t_{pn})^{\alpha\beta} \; &= \; (t_{pn})^{\beta\alpha}, \hspace{2em} (t_{pn})^\alpha_{\;\; \alpha} \; = \; 0,
\\[2ex]
p_{pn, \alpha} (t_{pn})^{\alpha\beta} \; &= \; (t_{pn})^{\alpha\beta} p_{pn, \beta} \; = \; 0.
\end{align}
\end{subequations}
The tensor can be constructed in analogy to the polarization 4-vector 
of Eq.~(\ref{A_def}). Its LF components in the collinear frame are 
[in the notation of Eq.~(\ref{notation_lf_tensor})]
\be
t_{pn}
\; &=& \; 
\left[\begin{array}{rrr} 
\displaystyle \frac{(p_d^+)^2}{M_{pn}^2} T_d^{zz} & -T_d^{zz} & \displaystyle \frac{p_d^+}{M_{pn}} T_d^{zj} \\[2ex]
-T_d^{zz} & \displaystyle \frac{M_{pn}^2}{(p_d^+)^2} T_d^{zz} & \displaystyle -\frac{M_{pn}}{p_d^+} T_d^{zj} \\[2ex]
\displaystyle\frac{p_d^+}{M_{pn}} T_d^{zi} & \displaystyle -\frac{M_{pn}}{p_d^+} T_d^{zi} & T_d^{ij}
\end{array} \right] ,
\label{T_pn_collinear}
\ee
where $T_d^{zz}, T_d^{zi}$ and $T_d^{ij}$ ($i, j = x, y$) 
are the components of is the deuteron polarization 3-tensor in the rest frame.
The $pn$ tensor in Eq.~(\ref{T_pn_collinear}) has the same form as the 
deuteron tensor, Eq.~(\ref{deuteron_polarization_tensor_collinear}), only with the
deuteron mass in the boost parameter replaced by the $pn$ invariant mass.
Substituting Eq.~(\ref{rho_pn_tensor}) in Eq.~(\ref{neutron_density}),
the effective neutron density matrix induced by the deuteron tensor polarization is obtained as
\begin{align}
\Pi_n[\textrm{tensor}] \; &= \; \frac{1}{2} (p_n\gamma + m) \;(p_n t_{pn} p_n) \; 
\nonumber \\[1ex]
&\times \; 8 \left( G_1^2 + \Delta^2 G_2^2 - 4 m G_1 G_2 \right) .
\label{neutron_density_tensor}
\end{align}
One observes that the neutron density matrix induced by deuteron tensor polarization has
has the same form as the unpolarized density matrix Eq.~(\ref{neutron_density_unpol}). 
The deuteron tensor polarization enters in the neutron density matrix only through
the contraction $(p_n t_{pn} p_n)$. This encodes the constraints from rotational and
relativistic covariance. It implies that the tensor polarization of the deuteron cannot
induce an effective spin polarization of the neutron. 
Altogether, Eq.~(\ref{neutron_density_decomposition}), with the parts given by 
Eqs.~(\ref{neutron_density_unpol}), (\ref{neutron_density_vector}) and 
(\ref{neutron_density_tensor}), describes the effective neutron spin density matrix 
in a polarized deuteron ensemble with general vector and tensor polarization, 

The neutron spin density matrix becomes particularly simple when expressed in terms of the CM frame variables
[see Sec.~\ref{subsec:cmframe}, Eq.~(\ref{vertex_radial})]. This representation allows one to identify the contributions of the 
S- and D-waves and relate the effective neutron polarization to 3-dimensional deuteron structure. The unpolarized 
part of the neutron density matrix, Eq.~(\ref{neutron_density_unpol}), becomes
\be
\Pi_n[\textrm{unpol}] &=& \frac{1}{2} (p_n\gamma + m) (f_0^2 + f_2^2) 
\nonumber
\\[2ex]
&& [f_{0, 2} \equiv f_{0, 2}(k)].
\label{neutron_density_unpol_cm}
\ee
In the vector-polarized part, the $pn$ polarization 4-vector Eq.~(\ref{A_def}) is in the CM frame 
given by (we give both the LF and ordinary components, cf.~Sec.~\ref{subsec:cmframe})
\be
s_{pn}[\textrm{CM}] \; &=& \; [S_d^z, -S_d^z, \bm{S}_{dT}] \; = \; (0, \bm{S}_d) ,
\ee
i.e., it has the same components as the deuteron polarization 4-vector in the deuteron rest frame.
It is straightforward to compute the neutron polarization vector from the formulas 
in Eq.~(\ref{neutron_density_vector}). The 4-vector products in Eq.~(\ref{a_1_def}) 
and (\ref{a_2_def}) can be evaluated directly in the CM frame and become
\begin{subequations}
\begin{align}
s_{pn} p_n \; &= \; s_{pn}[\textrm{CM}]\; p_n[\textrm{CM}] \; = \; \bm{S}_d\bm{k} ,
\\[2ex]
s_{pn} \Delta \; &= \; s_{pn}[\textrm{CM}]\; \Delta[\textrm{CM}] \; = \; -2\bm{S}_d\bm{k} .
\end{align}
\end{subequations}
Altogether we obtain
the vector-polarized part of the neutron spin density matrix in the CM frame as
\begin{subequations}
\label{neutron_density_vector_cm}
\begin{align}
\Pi_n [\textrm{vector}] \; &= \; {\textstyle\frac{1}{2}} (p_n\gamma + m) (s_n \gamma) \gamma^5 ,
\\[3ex]
s_n[\textrm{CM}] \; &= \; (s_n^0, \bm{s}_n) ,
\\[2ex]
s_n^0 \; &= \; -\frac{\bm{S}_d\bm{k}}{m} \left( f_0 - \frac{f_2}{\sqrt{2}} \right)^2 ,
\\[2ex]
\bm{s}_n \; &= \; \left[ 
\left( \bm{S}_d + \frac{E - m}{m} \frac{(\bm{S}_d\bm{k}) \bm{k} }{|\bm{k}|^2} \right) f_0 \right.
\nonumber \\
&+ \; \left.
\left( 2 \bm{S}_d - \frac{E + 2m}{m} \frac{(\bm{S}_d\bm{k}) \bm{k} }{|\bm{k}|^2} \right) \frac{f_2}{\sqrt{2}}
\right]
\nonumber \\
& \times \; 
\left( f_0 - \frac{f_2}{\sqrt{2}} \right) .
\end{align}
\end{subequations}
The components of $s_n$ in an arbitrary collinear frame can be obtained from 
Eq.~(\ref{neutron_density_vector_cm}) by forming the 
LF plus and minus components in the CM frame, and performing a longitudinal boost
to the desired value of $p_d^+$ (cf.\ Secs.~\ref{subsec:collinear_frames} and \ref{subsec:cmframe})
\begin{subequations}
\begin{align}
s_n^\pm [\textrm{CM}] \; &= \; (s_n^0 \, \pm \, s_n^z) [\textrm{CM}] ,
\\[2ex]
s_n^+ [\textrm{arb.~coll.}] \; &= \; \frac{p_d^+}{M_{pn}} \, s_n^+ [\textrm{CM}] ,
\\[2ex]
s_n^- [\textrm{arb.~coll.}] \; &= \; \frac{M_{pn}}{p_d^+} \, s_n^- [\textrm{CM}] ,
\\[2ex]
\bm{s}_{nT} [\textrm{arb.~coll.}] \; &= \; \bm{s}_{nT} [\textrm{CM}] .
\end{align}
\end{subequations}
In the tensor-polarized part, the $pn$ polarization 4-tensor Eq.~(\ref{rho_pn_tensor})
is in the CM frame given by (we give the ordinary components)
\begin{subequations}
\label{tensor_cm}
\begin{align}
(t_{pn})^{\alpha 0}[\textrm{CM}] \; &= \;
(t_{pn})^{0\beta}[\textrm{CM}] \; = \; 0,
\\[1ex]
(t_{pn})^{ij}[\textrm{CM}] \; &= \; (T_d)^{ij} \; \neq \; 0 ,
\end{align}
\end{subequations}
where $(T_d)^{ij}$ are the components of the 3-dimensional deuteron polarization tensor 
in the deuteron rest frame. The contraction in Eq.~(\ref{neutron_density_tensor}) becomes
\be
(p_n t_{pn} p_n) \; &=& \; (\bm{k} T_d \bm{k}) .
\ee
We obtain the tensor-polarized part of the neutron spin density matrix in the CM frame as
\begin{align}
\Pi_n[\textrm{tensor}] \; &= \; -\frac{1}{2} (p_n\gamma + m) \; 
\frac{3 (\bm{k} T_d \bm{k})}{|\bm{k}|^2}
\nonumber
\\[1ex]
& \times \; \left( 2 f_0 + \frac{f_2}{\sqrt{2}} \right) \frac{f_2}{\sqrt{2}} .
\label{neutron_density_tensor_cm}
\end{align}
Together, Eqs.~(\ref{neutron_density_unpol_cm}), (\ref{neutron_density_vector_cm})
and (\ref{neutron_density_tensor_cm}) specify the neutron spin density matrix 
in terms of the CM frame variables. 

The structure of the neutron spin density matrix in terms of the 
CM frame variables shows several interesting features. 
(a)~The unpolarized part, Eq.~(\ref{neutron_density_unpol_cm}),
is the sum of S- and D-wave probabilities of the CM wave function, as expected. (b)~The vector-polarized 
part of the neutron spin density matrix, Eq.~(\ref{neutron_density_vector_cm}), involves mixing 
of the S- and D-waves. This is a consequence of the relativistic spin rotations involved in the 
transition from canonical spin states to LF helicity states (Melosh rotations; in our formulation
they are contained in the explicit form of the LF helicity bispinors). (c)~The tensor-polarized 
part of the neutron spin density matrix, Eq.~(\ref{neutron_density_tensor_cm}), is proportional 
to the D-wave. This is natural, as tensor polarization would be absent for a pure S-wave bound 
state in the CM frame and cannot be induced by relativistic spin rotations ($|\Delta L| \leq 1$).

The neutron density matrix Eq.~(\ref{neutron_density_decomposition}), with the parts
given by Eqs.~(\ref{neutron_density_unpol}), (\ref{neutron_density_vector})
and (\ref{neutron_density_tensor}), or equivalently by 
Eqs.~(\ref{neutron_density_unpol_cm}), (\ref{neutron_density_vector_cm})
and (\ref{neutron_density_tensor_cm}), contains the full information on the effective
neutron polarization (longitudinal and transverse spin, spin-orbit correlations) 
in a general polarized deuteron ensemble. We use this density matrix to calculate
the neutron LF helicity distributions in Sec.~\ref{subsec:neutron_lf_momentum_distributions}
and the tagged DIS structure functions in Sec.~\ref{sec:ia}

In the following applications we need the effective neutron spin density matrix in a pure deuteron
spin state polarized along a given direction. It can be obtained by evaluating the general expressions
for the neutron density matrix derived in this section, Eqs.~(\ref{neutron_density_unpol_cm}), 
(\ref{neutron_density_vector_cm}) and (\ref{neutron_density_tensor_cm}), with the specific values 
of the rest-frame polarization 3-vector and 3-tensor corresponding to the pure spin state given 
in Sec.~\ref{subsec:spin_density_spin_1}. For a state with spin projection $\lambda_d$ along a rest-frame
direction $\bm{N}$, with $|\bm{N}|^2 = 1$, the neutron density is obtained as
\begin{subequations}
\label{neutron_density_pure_cm}
\begin{align}
\Pi_n[\textrm{pure}](\bm{N}, \lambda_d)
\; &= \; 
\Pi_n [\textrm{unpol}] \; + \; \Pi_n [\textrm{vector}] (\bm{S}_d)
\nonumber \\[1ex]
&+ \; \Pi_n [\textrm{tensor}] (T_d)
\label{neutron_density_pure_cm_terms}
\end{align}
\begin{align}
\bm{S}_d \; &= \; \lambda_d \bm{N} ,
\\[2ex]
T_d \; &= \; -\frac{1}{6} \left( \delta^{ij} - 3 N^i N^j \right)
\nonumber
\\[1ex]
&\times \; \left\{ 
\begin{array}{rl} 1, & \lambda_d = \pm 1 \\
(-2), & \lambda_d = 0
\end{array}
\right\} ,
\label{tensor_special_cm}
\\[2ex]
(\bm{k} T_d \bm{k}) \; &= \; -\frac{1}{6} \left[ |\bm{k}|^2 - 3 (\bm{N}\bm{k})^2 \right] 
\nonumber \\[1ex]
& \times \; 
\left\{ 
\begin{array}{rl} 1, & \lambda_d = \pm 1 \\
(-2), & \lambda_d = 0
\end{array}
\right\} ,
\label{tensor_contraction_cm}
\end{align}
\end{subequations}
where the general expressions of the terms in Eq.~(\ref{neutron_density_pure_cm_terms}) are given by 
Eqs.~(\ref{neutron_density_unpol_cm}), (\ref{neutron_density_vector_cm})
and (\ref{neutron_density_tensor_cm}). The 3-tensor of Eq.~(\ref{tensor_special_cm}) represents
the rest-frame components of the special tensor Eq.~(\ref{tensor_special}).
\subsection{Neutron light-front momentum distributions}
\label{subsec:neutron_lf_momentum_distributions}
We now want to calculate the LF momentum distributions of nucleons with given LF helicity in the deuteron.
The simplest way is to calculate the expectation value of the plus component of the vector and axial vector 
current in the deuteron (i.e., the deuteron's vector and axial charge) in the IA, and represent it as 
an integral over the LF momentum of the nucleons. A formal definition of the LF momentum distributions 
can be given using the second-quantized nucleon number operators or light-ray operators.

The expectation value of the isoscalar vector current in a deuteron ensemble is of the general form
\be
\langle J_V^+ \rangle \; &=& \; 2 p_d^+ \, g^{\phantom +}_{Vd} ,
\ee
where $g_{Vd} = 2$ is the vector charge (baryon number) of the deuteron. In the IA the current is the 
sum of the proton and neutron currents. The deuteron expectation value can be computed 
as the sum of the nucleon contributions, using the general formulas of Sec.~\ref{subsec:matrix_elements}.
The matrix element of the vector current between neutron states with LF helicity $\lambda_n = \pm 1/2$ is
\begin{subequations}
\label{vector_current_nucleon}
\begin{align}
& \langle n, p_n, \lambdanp | J_V^+ | n, p_n, \lambda_n \rangle
\nonumber
\\[2ex]
&= \; \bar u_{\rm LF} (p_n, \lambdanp) \gamma^+ u_{\rm LF} (p_n, \lambda_n) \; g_V
\\[2ex]
&= \; 2 p_n^+ \delta (\lambda_n, \lambdanp) \; g_V ,
\end{align}
\end{subequations}
where $g_V = 1$ is the isoscalar vector charge of the nucleon. The matrix element
is diagonal in the LF helicities and independent of their value. The vector current thus 
``counts'' the helicity-independent (or averaged) number of neutrons. 
According to Eq.~(\ref{deuteron_onebody_trace}) the neutron contribution 
to the deuteron expectation value is given by
\begin{align}
\langle J_V^+ \rangle 
\; &= \; g_V
\int\frac{d\alpha_p}{\alpha_p} \, d^2 p_{pT} \;
\frac{2 \; \textrm{tr}[\Pi_n \gamma^+]}{(2 - \alpha_p)^2} 
\nonumber
\\[2ex]
&= \; p_d^+ \, g_{Vd} .
\label{vector_current_integral}
\end{align}
We identify the function
\be
\mathcal{S}_{d}(\alpha_p, \bm{p}_{pT}) \; &\equiv& \; 
\frac{\textrm{tr}[\Pi_n \gamma^+]}{(2 - \alpha_p)^2 \; p_d^+} 
\label{distribution_unpol_trace}
\ee
as the helicity-independent LF momentum distribution of neutrons in the deuteron ensemble.
In accordance with nuclear physics terminology we refer to it as the deuteron LF spectral function
(see explanation below). It is a function of the proton LF momentum variables and satisfies
\be
\int\frac{d\alpha_p}{\alpha_p} \, d^2 p_{pT} \; 
\mathcal{S}_{d}(\alpha_p, \bm{p}_{pT}) 
\; &=& \; 
\frac{g_{Vd}}{2 g_V} \; = \; 1 .
\label{neutron_density_unpol_normalization}
\ee
The trace in Eq.~(\ref{distribution_unpol_trace})
receives contributions from the unpolarized and tensor-polarized parts of the neutron density matrix.
Substituting the explicit form of the neutron density matrix in the CM frame variable,
Eqs.~(\ref{neutron_density_unpol_cm}) and (\ref{neutron_density_tensor_cm}), and using 
\be
\frac{1}{2} {\rm tr}[(p_n\gamma + m) \gamma^+] \; &=& \; 
2 p_n^+ \; = \; (2 - \alpha_p) p_d^+
\ee
we obtain
\begin{subequations}
\label{f_UT}
\begin{align}
\mathcal{S}_{d} \; &= \; \mathcal{S}_{d}[\textrm{unpol}] \; + \; \mathcal{S}_{d}[\textrm{tensor}] ,
\end{align}
\begin{align}
\mathcal{S}_{d}(\alpha_p, \bm{p}_{pT})[\textrm{unpol}] \; &= \; \frac{f_0^2 + f_2^2}{2 - \alpha_p} ,
\label{f_U}
\\[1ex]
\mathcal{S}_{d}(\alpha_p, \bm{p}_{pT})[\textrm{tensor}]
\; &= \; -\frac{3}{2 - \alpha_p} \; \frac{(\bm{k} T_d \bm{k})}{|\bm{k}|^2} 
\nonumber
\\[2ex]
& \times \; \left( 2 f_0 + \frac{f_2}{\sqrt{2}} \right) \frac{f_2}{\sqrt{2}} .
\label{f_T}
\end{align}
\end{subequations}
The LF momentum-dependent prefactor could equivalently be expressed in terms of the CM momentum 
variable as
\beq
\frac{1}{2 - \alpha_p} \; = \; \frac{E}{E - k^z} ;
\eeq
we leave it in its original form, as in this form it can be combined with similar factors
appearing in integrals over the LF momentum.
Using the relation Eq.~(\ref{integration_k}) one easily verifies that the unpolarized part
Eq.~(\ref{f_U}) satisfies the normalization condition Eq.~(\ref{neutron_density_unpol_normalization}),
and that the tensor-polarized part Eq.~(\ref{f_T}) averages to zero,
\begin{subequations}
\begin{align}
& \int\frac{d\alpha_p}{\alpha_p} \, d^2 p_{pT} \; 
\mathcal{S}_{d}(\alpha_p, \bm{p}_{pT}) [\textrm{unpol}]
\nonumber
\\[1ex]
&= \; \int \frac{d^3 k}{E} \; (f_0^2 + f_2^2) \; = \; 1 ,
\label{f_U_integral}
\\[1ex]
& \int\frac{d\alpha_p}{\alpha_p} \, d^2 p_{pT} \; 
\mathcal{S}_{d}(\alpha_p, \bm{p}_{pT}) [\textrm{tensor}]
\nonumber
\\[1ex]
&\propto \; \int d\Omega_k \frac{(\bm{k} T_d \bm{k})}{|\bm{k}|^2}
\; = \; 0 .
\label{f_T_integral}
\end{align}
\end{subequations}
Eq.~(\ref{f_T_integral}) holds because the polarization tensor in the CM frame is traceless,
cf.\ Eq.~(\ref{tensor_cm}). These relations ensure the conservation of baryon number in the 
deuteron LF structure.

The helicity-independent neutron distributions also satisfy a LF momentum sum rule.
Because the functions 
\beq
\frac{\mathcal{S}_{d}(\alpha_p, \bm{p}_{pT})}{\alpha_p} [\textrm{unpol}]
\hspace{1em} \textrm{and} \hspace{1em}
\frac{\mathcal{S}_{d}(\alpha_p, \bm{p}_{pT})}{\alpha_p} [\textrm{tensor}]
\eeq
are symmetric under $\alpha_p \rightarrow 2 - \alpha_p$, Eqs.~(\ref{f_U_integral}) and
(\ref{f_T_integral}) imply
\begin{subequations}
\begin{align}
\int\frac{d\alpha_p}{\alpha_p} \, d^2 p_{pT} \; (2 - \alpha_p) \; 
\mathcal{S}_{d}(\alpha_p, \bm{p}_{pT}) [\textrm{unpol}]
\; &= \; 1,
\label{f_U_momentum_sr}
\\[1ex]
\int\frac{d\alpha_p}{\alpha_p} \, d^2 p_{pT} \; (2 - \alpha_p) \;  
\mathcal{S}_{d}(\alpha_p, \bm{p}_{pT}) [\textrm{tensor}]
\; &= \; 0.
\label{f_T_momentum_sr}
\end{align}
\end{subequations}
This relation ensures LF momentum conservation in tagged DIS on the deuteron (see Sec.~\ref{sec:ia}).

In a similar way we can obtain the spin-dependent neutron LF momentum distribution in a 
vector-polarized deuteron. The expectation value of the isoscalar axial vector current in 
the deuteron ensemble is 
\be
\langle J_A^+ \rangle \; &=& \; 2 M_d s_d^+ \, g_{Ad} \;\; = \;\; 2 S_d^z \, p_d^+ \, g_{Ad} ,
\ee
where we used the explicit form of polarization vector in collinear frame,
Eq.~(\ref{deuteron_polarization_vector_collinear}). Here $g_{Ad}$ is the axial charge 
of the deuteron, which is a property of the deuteron bound state and cannot be determined
from first principles. The axial current is the sum of proton and neutron currents.
The matrix element of the neutron axial current between nucleon states with LF helicity 
$\lambda_n = \pm 1/2$ is 
\begin{subequations}
\label{axial_current_nucleon}
\begin{align}
&\langle n, p_n, \lambdanp | J_A^+ | n, p_n, \lambda_n \rangle
\nonumber
\\[2ex]
&= \; \bar u_{\rm LF} (p_n, \lambdanp) (-\gamma^+ \gamma^5) u_{\rm LF} (p_n, \lambda_n) \, g_A
\\[2ex]
&= \; 2 p_n^+ \;  (2 \lambda_n ) \, \delta (\lambda_n, \lambdanp) \, g_A ,
\end{align}
\end{subequations}
where $g_A$ is the isoscalar axial coupling of the nucleon.
The matrix element is again diagonal in the LF helicities, but the diagonal value 
is now proportional to the LF helicity. The axial vector current with matrix $-\gamma^+\gamma^5$
therefore counts the difference between the number of neutrons with LF helicities $+1/2$ and $-1/2$. 
The calculation of
the deuteron expectation value proceeds analogously to Eq.~(\ref{vector_current_integral}),
and we identify 
\be
\Delta \mathcal{S}_{d}(\alpha_p, \bm{p}_{pT}) \; &\equiv& \; 
\frac{\textrm{tr}[\Pi_n (-\gamma^+ \gamma^5)]}{(2 - \alpha_p)^2 \; p_d^+} 
\label{neutron_density_pol_trace}
\ee
as the helicity-dependent LF momentum distribution of neutrons in the deuteron ensemble. 
It satisfies
\be
\int\frac{d\alpha_p}{\alpha_p} \, d^2 p_{pT} \;
\Delta \mathcal{S}_{d}(\alpha_p, \bm{p}_{pT}) \; &=& \; 
S_d^z \, \left(\frac{g_{Ad}}{2 g_A}\right) ,
\label{neutron_density_pol_normalization}
\ee
which may be regarded as the nucleon spin sum rule. Only the vector-polarized part of 
the neutron density matrix Eq.~(\ref{neutron_density_decomposition})
contributes to the trace in Eq.~(\ref{neutron_density_pol_trace}),
\be
\Delta \mathcal{S}_{d} \; &\equiv& \; \Delta \mathcal{S}_{d} [\textrm{vector}] .
\ee
With Eq.~(\ref{neutron_density_vector_cm}) we obtain
\begin{align}
\textrm{tr}[\Pi_n (-\gamma^+ \gamma^5)] 
\; &= \frac{1}{2} {\rm tr}[(p_n\gamma + m) (s_n \gamma) \gamma^5 (-\gamma^+ \gamma^5) ] 
\nonumber
\\[1ex]
&= \; 2 m s_n^+ .
\end{align}
Using the explicit expressions for the neutron polarization vector in the CM frame, 
Eq.~(\ref{neutron_density_vector_cm}), we obtain
\begin{align}
\Delta \mathcal{S}_{d}(\alpha_p, \bm{p}_{pT}) \; &= \; \frac{1}{2 - \alpha_p} \; 
\left( f_0 - \frac{f_2}{\sqrt{2}} \right)
\nonumber \\
& \times \; \left( C_0 f_0 - \frac{C_2 f_2}{\sqrt{2}} \right) ,
\label{Delta_f_d_k}
\end{align}
where
\begin{subequations}
\begin{align}
C_0 &\equiv C_0(\bm{k})
\nonumber \\[1ex]
&\equiv 
\frac{m}{(2 - \alpha_p) E} 
\left[ S_d^z \; - \; \frac{\bm{S}_d\bm{k}}{m} \; + \; \frac{(\bm{S}_d\bm{k}) k^z}{m (E + m)} \right] ,
\label{C_0_orig}
\\[2ex]
C_2 &\equiv C_2(\bm{k})
\nonumber \\[1ex]
&\equiv \frac{m}{(2 - \alpha_p) E} 
\left[ 
-2 S_d^z - \frac{\bm{S}_d\bm{k}}{m} + \frac{(E + 2m) \, (\bm{S}_d\bm{k}) k^z}{m |\bm{k}|^2} \right] .
\label{C_2_orig}
\end{align}
\end{subequations}
These factors can also be written in the form
\begin{subequations}
\begin{align}
C_0(\bm{k}) 
&= \; S_d^z 
- \frac{(E + k^z) |\bm{k}_T|^2}{(E + m)(m^2 + |\bm{k}_T|^2)} \, S_d^z
\nonumber \\[1ex]
&- \frac{(E + k^z) (E - k^z + m)}{(E + m)(m^2 + |\bm{k}_T|^2)}  \, (\bm{S}_{dT} \bm{k}_T) ,
\label{C_0}
\\[2ex]
C_2(\bm{k})
&= \; S_d^z - \frac{(E + 2 m) (E + k^z) |\bm{k}_T|^2}{(m^2 + |\bm{k}_T|^2) |\bm{k}|^2} \,  S_d^z
\nonumber \\[1ex]
&+ \; \frac{(E + k^z) [-|\bm{k}|^2 + (E + 2 m) k^z]}{(m^2 + |\bm{k}_T|^2) |\bm{k}|^2} \, (\bm{S}_{dT} \bm{k}_T) ,
\label{C_2}
\end{align}
\end{subequations}
where we have used that
\begin{subequations}
\begin{align}
(E + k^z) (E - k^z) \; &= \; E^2 - (k^z)^2
\nonumber
\\[2ex]
&= \; m^2 + |\bm{k}_T|^2 ,
\\[2ex]
2 - \alpha_p \; &= \; 1 - \frac{k^z}{E} ,
\\[1ex]
\frac{1}{2 - \alpha_p} \; &= \; \frac{E}{E - k^z}
\; = \; \frac{E (E + k^z)}{m^2 + |\bm{k}_T|^2} .
\end{align}
\end{subequations}
The factors $C_0$ and $C_2$ describe the effect of the nucleons' orbital motion on the 
neutron LF helicity. In the 3-dimensional representation of the deuteron LF wave function
they arise from the Melosh rotations connecting the canonical nucleon spinors with the
LF helicity spinors (see Appendix~\ref{app:wave_function}). In the 4-dimensional representation employed here
this information is encoded in the specific form of the effective neutron polarization 
vector $s_n$, Eq.~(\ref{neutron_spin_vector}). Note that the factors are equal to unity at
zero transverse momentum, cf.~Eqs.~(\ref{C_0}) and (\ref{C_2}),
\beq
C_0(\bm{k}), C_2(\bm{k}) \; = \; 1 \hspace{1em} \textrm{at} \hspace{1em} 
\bm{k}_T = \bm{p}_{pT} = 0, \; k^z \; \textrm{arbitrary},
\label{C_0_W_zero_transverse_momentum}
\eeq 
as is expected of the Melosh rotation.

In the CM representation Eq.~(\ref{Delta_f_d_k}) one can easily evaluate the normalization
integral of the polarized nucleon distribution, Eq.~(\ref{neutron_density_pol_normalization}).
Using the relation Eq.~(\ref{integration_k}) one obtains
\begin{align}
& \left( \frac{g_{Ad}}{2 g_A} \right) \, S_d^z 
\nonumber \\[1ex]
&= \; 
\int\frac{d\alpha_p}{\alpha_p} \, d^2 p_{pT} \Delta \mathcal{S}_{d}(\alpha_p, \bm{p}_{pT}) 
\nonumber \\[1ex]
&= \; 
\int \frac{d^3 k}{E} \; \left( f_0 - \frac{f_2}{\sqrt{2}} \right)
\left( C_0 f_0 - \frac{C_2 f_2}{\sqrt{2}} \right)
\nonumber \\[1ex]
&= \; 
4\pi \int \frac{dk \; k^2}{E} \; \left( f_0 - \frac{f_2}{\sqrt{2}} \right)
\left( \bar C_0 f_0  - \frac{\bar C_2 f_2}{\sqrt{2}} \right) ,
\label{g_A_integral_relativistic}
\end{align}
where $\bar C_0, \bar C_2$ are the angular averages of the factors $C_0, C_2$,
\begin{subequations}
\begin{align}
\bar C_0 (k) \; &\equiv \; \int \frac{d\Omega_k}{4\pi} \; C_0 (\bm{k})
\nonumber \\[1ex]
& = \; \frac{m \, S_d^z}{E + m} \left( \frac{m}{k} \log \frac{E + k}{m} + 1 \right) ,
\\[1ex]
\bar C_2 (k) \; &\equiv \; \int \frac{d\Omega_k}{4\pi} \; C_2 (\bm{k})
\nonumber \\[1ex]
& = \; \frac{m \, S_d^z}{k^2} \left[ (E + 2 m) \frac{m}{k} \log \frac{E + k}{m} - 2E - m \right] .
\end{align}
\end{subequations}
Notice that structures proportional to $\bm{S}_{dT}\bm{k}_T$ average to zero, so that the averages
are proportional to $S_d^z$, as it should be. The expansion in $k/m$ of the averages is
\begin{subequations}
\begin{align}
\bar C_0 (k) \; &= \; S_d^z \,
\left[
\phantom{-}1 \; - \; \frac{1}{3} \frac{k^2}{m^2} \; + \; \mathcal{O}\left(\frac{k^4}{m^4}\right) 
\right] ,
\\[1ex]
\bar C_2 (k) 
&= \; S_d^z \,
\left[ -1 \; + \; \frac{4}{15} \frac{k^2}{m^2} \; + \; \mathcal{O}\left(\frac{k^4}{m^4}\right) \right] .
\end{align}
\end{subequations}
Neglecting terms $\mathcal{O}(k^2/m^2)$, the
integral Eq.~(\ref{g_A_integral_relativistic}) becomes
\begin{align}
\frac{g_{Ad}}{2 g_A} \; &= \; 
4\pi \int \frac{dk \; k^2}{E} \; \left( f_0 - \frac{f_2}{\sqrt{2}} \right)
\left( f_0 + \frac{f_2}{\sqrt{2}} \right)
\nonumber
\\[1ex]
\; &= \; 
4\pi \int \frac{dk \; k^2}{E} \; \left( f_0^2 - \frac{f_2^2}{2} \right)
\nonumber
\\[1ex]
\; &= \; 
4\pi \int \frac{dk \; k^2}{E} \; \left( f_0^2 + f_2^2 - \frac{3 f_2^2}{2} \right)
\nonumber
\\[1ex]
\; &= \; 
1 - \frac{3}{2} \omega_2 ,
\label{g_A_integral_nonrelativistic}
\end{align}
where we have canceled the factor $S_d^z$ on both sides. Here
\beq
\omega_2 \; \equiv \;
4\pi \int \frac{dk \, k^2}{E(k)} \; f_2^2(k) 
\eeq
is the D-state probability of the deuteron wave function in the CM frame. Equation~(\ref{g_A_integral_nonrelativistic}) 
has the same form as the nonrelativistic result for the deuteron axial charge including the D-state correction.

Some explanations are in order regarding the definition of the LF spectral function and the
correspondence with nuclear physics terminology. (a)~The deuteron spectral function $\mathcal{S}_d$
describes the LF momentum distribution of neutrons in the deuteron as a function of the proton LF
momentum and is normalized as Eq.~(\ref{neutron_density_unpol_normalization}); this definition is
appropriate for tagged DIS experiments where the proton momentum is measured and will be used
in the following. The conventional neutron LF momentum density is defined as a function of the neutron
LF momentum and related to the spectral function as
\begin{align}
\mathcal{N}_d (\alpha_n, \bm{p}_{nT}) \; =& \; \frac{\alpha_n}{\alpha_p} \mathcal{S}_d (\alpha_p, \bm{p}_{pT}) \,
\nonumber \\
& \; [\alpha_p = 2 - \alpha_n, \; \bm{p}_{pT} = - \bm{p}_{nT}] ,
\end{align}
such that its normalization is
\begin{align}
\int \frac{d\alpha_n}{\alpha_n} \int d^2 p_{nT} \; \mathcal{N}_d (\alpha_n, \bm{p}_{nT}) 
\; &= \; 1 .
\label{momentum_density_normalization}
\end{align}
(b)~The spectral function represents the nuclear structure information entering in the IA.
In the general case of a nucleus with $A > 2$, it describes the probability density for
removing a nucleon with a given momentum, while leaving the $A - 1$ remnant system with
a given excitation energy (the momentum of the $A - 1$ system is fixed by the removed
nucleon momentum) \cite{Benhar:2005dj}. The nucleon momentum density is then obtained
by integrating over the excitation energy of the $A - 1$ system. In the particular case of
the deuteron with $A = 2$, the remnant system is the single spectator nucleon, whose energy
is fixed by the momentum, so that the spectral function depends on the momentum variable only
and coincides with the momentum density up to a factor accounting for the normalization,
cf.\ Eq.~(\ref{momentum_density_normalization}).
\subsection{Distribution in unpolarized deuteron}
The expressions Eqs.~(\ref{f_UT}) and (\ref{Delta_f_d_k}) describe the LF momentum distributions
of neutrons in a deuteron ensemble with arbitrary polarization (vector, tensor). It is instructive
to consider the distributions in some special cases. In an unpolarized ensemble ($\bm{S} = 0, T = 0$)
only the helicity-independent neutron distribution Eqs.~(\ref{f_U}) is present,
\begin{subequations}
\begin{align}
\mathcal{S}_{d}(\alpha_p, \bm{p}_{pT}) \; &= \; \frac{f_0^2 + f_2^2}{2 - \alpha_p} ,
\label{neutron_density_deut_unpol}
\\[1ex]
\Delta \mathcal{S}_{d}(\alpha_p, \bm{p}_{pT}) \; &= \; 0 .
\end{align}
\end{subequations}
One notes that (a)~in the helicity-independent distribution the S- and D-waves of the CM-frame wave function 
do not mix; (b)~the helicity-dependent distribution is zero for arbitrary neutron LF momentum, i.e., 
no LF helicity polarization is induced by the orbital motion.

%
%
\begin{figure}
\includegraphics[width=.48\textwidth]{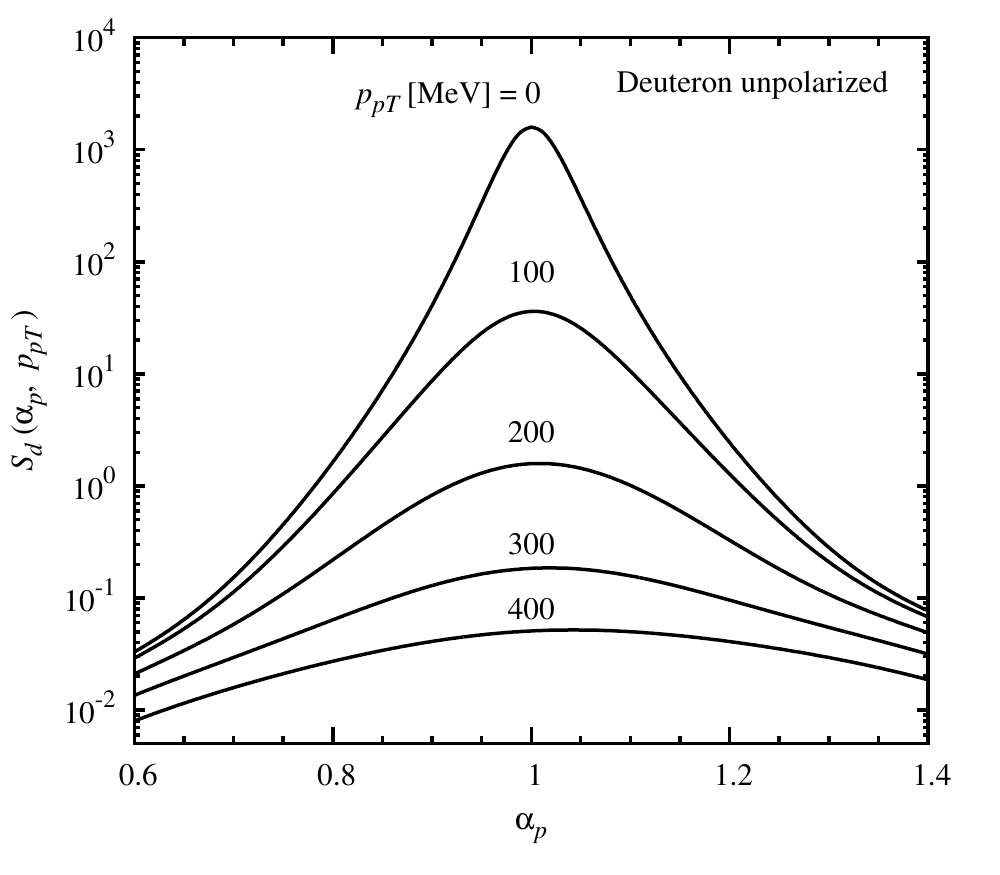}
\caption[]{The deuteron spectral function $\mathcal{S}_{d}$ in the unpolarized deuteron, Eq.~(\ref{neutron_density_deut_unpol}).
It describes the helicity-independent LF momentum distribution of neutrons in the unpolarized deuteron
as a function of the proton LF momentum. The plot shows the dependence on $\alpha_p$ for several
values of $|\bm{p}_{pT}|$.}
\label{fig:spectral_unpol}
\end{figure}
\begin{figure*}
\includegraphics[width=.7\textwidth]{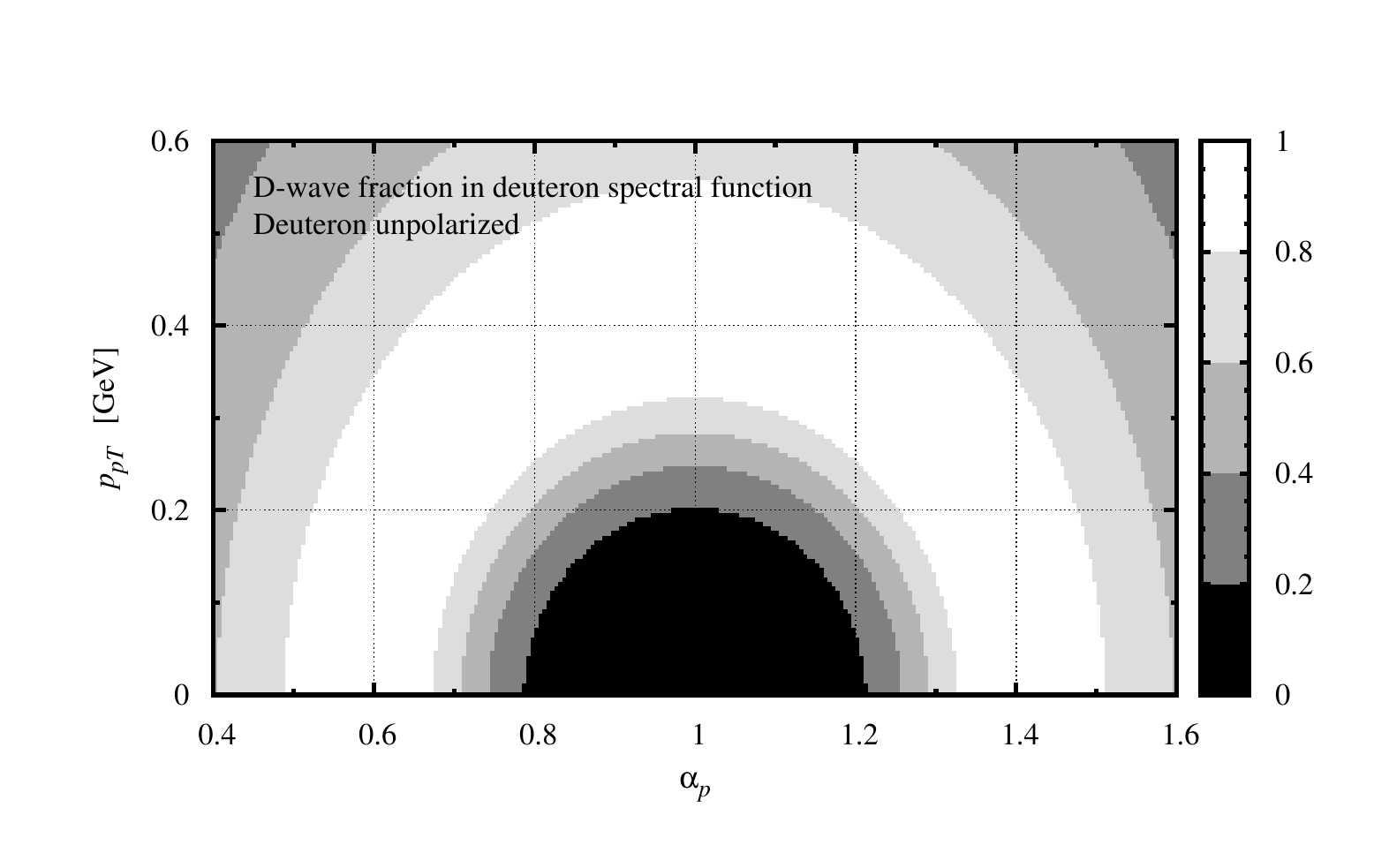}
\caption[]{The D-wave fraction in the deuteron spectral function,
$\mathcal{S}_{d}[\textrm{D-wave}]/\mathcal{S}_{d}[\textrm{S+D-waves}]$,
in the unpolarized deuteron,
Eq.~(\ref{neutron_density_dwave_ratio}). The 2-dimensional plot shows the ratio as a 
function of $\alpha_p$ and $|\bm{p}_{pT}|$.}
\label{fig:spectral_unpol_dwave}
\end{figure*}

Figure~\ref{fig:spectral_unpol} shows the helicity-independent neutron distribution in the unpolarized
deuteron as a function of $\alpha_p$ for several values of $|\bm{p}_{pT}|$. One observes:
(a) The distribution is maximal at $\alpha_p = 1$ and $|\bm{p}_{pT}| = 0$ and drops steeply
when increasing $|\alpha_p - 1|$ or $|\bm{p}_{pT}|$, as implied by the nucleon momentum 
distribution in the CM frame. (b) An asymmetry between the distributions at $\alpha_p > 1$ and 
$< 1$ is caused by the flux factor $1/(2 - \alpha_p)$ in Eq.~(\ref{neutron_density_deut_unpol}).
Figure~\ref{fig:spectral_unpol_dwave} shows the ratio of the neutron distribution resulting from the D-wave
only to that resulting from S- and D-waves (i.e., the total distribution),
\beq
\frac{\mathcal{S}_{d} (\alpha_p, \bm{p}_{pT}) \, [\textrm{D-wave}]}{\mathcal{S}_{d} (\alpha_p, \bm{p}_{pT}) \, [\textrm{S+D-waves}]} .
\label{neutron_density_dwave_ratio}
\eeq
This ratio can be regarded the probability for sampling the D-wave component of the CM motion
when observing a proton (or neutron) with given LF momentum. One observes: (a) The D-wave probability
vanishes at $\alpha_p = 1$ and $|\bm{p}_{pT}| = 0$, corresponding to CM momentum $|\bm{k}| = 0$.
(b) The D-wave probability becomes unity at LF momenta corresponding to $|\bm{k}| \approx$ 400 MeV,
where the S-wave function in the CM frame goes through zero \cite{Wiringa:1994wb}; (c) 
The D-wave probability decreases again at large CM momenta.
%
%
\subsection{Distribution in deuteron helicity states}
In a pure deuteron state with LF helicity $+1$ 
the helicity-independent neutron distribution is
\begin{align}
\mathcal{S}_{d}(\alpha_p, \bm{p}_{pT})[\textrm{pure $+1$}] \; &= \; 
\mathcal{S}_{d}(\alpha_p, \bm{p}_{pT})[\textrm{unpol}]
\nonumber \\[1ex]
&+ \; \mathcal{S}_{d}(\alpha_p, \bm{p}_{pT})[\textrm{tensor}].
\end{align}
Here the tensor-polarized part is given by Eq.~(\ref{f_T}), evaluated with the special tensor 
Eq.~(\ref{tensor_special_cm}), and with the contraction given by Eq.~(\ref{tensor_contraction_cm}),
for polarization along the $z$-direction, $\bm{N} = \bm{e}_z$,
\begin{subequations}
\begin{align}
& \mathcal{S}_{d}(\alpha_p, \bm{p}_{pT})[\textrm{tensor}]
\nonumber \\[1ex]
&= \; - \frac{1}{2-\alpha_p} C_T \left( 2 f_0 + \frac{f_2}{\sqrt{2}} \right) \frac{f_2}{\sqrt{2}} ,
\label{f_T_pure}
\end{align}
\begin{align}
C_T \; &\equiv \; C_T(\bm{k}) \; = \; 3\; \frac{(\bm{k} T \bm{k})}{|\bm{k}|^2}
\; = \; \frac{1}{2} \left[ -1 + \frac{3 (k^z)^2}{|\bm{k}|^2} \right]
\nonumber \\[1ex]
& = \; 1 - \frac{3 |\bm{k}_T|^2}{2 |\bm{k}|^2} .
\label{C_T_pure}
\end{align}
\end{subequations}
Combining this with the unpolarized part given by Eq.~(\ref{f_U}), we obtain
\begin{align}
& \mathcal{S}_{d}(\alpha_p, \bm{p}_{pT})[\textrm{pure $+1$}] \;
\nonumber \\[1ex]
&= \; \frac{1}{2 - \alpha_p} 
\left[f_0^2 + f_2^2 - C_T \left( 2 f_0 + \frac{f_2}{\sqrt{2}} \right) \frac{f_2}{\sqrt{2}}
\right] .
\label{neutron_densities_pure}
\end{align}
The helicity-dependent neutron distribution is
\begin{align}
& \Delta \mathcal{S}_{d}(\alpha_p, \bm{p}_{pT})[\textrm{pure $+1$}]
\nonumber \\[1ex]
&= \; 
\frac{1}{2 - \alpha_p} \; 
\left( f_0 - \frac{f_2}{\sqrt{2}} \right)
\left(  C_0 f_0 - \frac{C_2 f_2}{\sqrt{2}} \right) ,
\label{Delta_f_pure}
\end{align}
where $C_0, C_2$ are the factors of Eqs.~(\ref{C_0_orig}) and (\ref{C_2_orig}) with
$S_d^z = 1$ and $\bm{S}_{dT} = 0$
\begin{subequations}
\begin{align}
C_0 &\equiv C_0(\bm{k})
\nonumber \\[1ex]
&\equiv \;
\frac{m}{(2 - \alpha_p) E} 
\left[ 1 \; - \; \frac{k^z}{m} \; + \; \frac{(k^z)^2}{m (E + m)} \right] ,
\label{C_0_helicity}
\\[2ex]
C_2 &\equiv C_2(\bm{k})
\nonumber \\[1ex]
 &\equiv \;
\frac{m}{(2 - \alpha_p) E} 
\left[ 
-2 \; - \; \frac{k^z}{m} \; + \; \frac{(E + 2m) \, (k^z)^2}{m |\bm{k}|^2} \right] ,
\label{C_2_helicity}
\end{align}
\end{subequations}
which can also be written in the form
\begin{subequations}
\begin{align}
C_0(\bm{k}) 
\; &= \; 1 - \frac{(E + k^z) |\bm{k}_T|^2}{(E + m)(m^2 + |\bm{k}_T|^2)} ,
\label{C_0_helicity_kt}
\\[2ex]
C_2(\bm{k}) 
\; &= \; 1 - \frac{(E + 2 m)(E + k^z) |\bm{k}_T|^2}{(m^2 + |\bm{k}_T|^2) |\bm{k}|^2} .
\label{C_2_helicity_kt}
\end{align}
\end{subequations}
One may also consider the distributions of neutrons with LF helicity $+1/2$
and $-1/2$ in the deuteron state with LF helicity $+1$,
\begin{subequations}
\begin{align}
&\mathcal{S}_{d\pm}[\textrm{pure $+1$}] \; \equiv \; \frac{1}{2} \left( \mathcal{S}_{d} \pm
\Delta \mathcal{S}_{d} \right) [\textrm{pure $+1$}] ,
\\[1ex]
& \{ \mathcal{S}_{d} , \, \Delta \mathcal{S}_{d} \}[\textrm{pure $+1$}] \; = \; \left( \mathcal{S}_{d+} 
\pm \mathcal{S}_{d-} \right)[\textrm{pure $+1$}] .
\end{align}
\end{subequations}
The functions $S_{d\pm}$ are positive for arbitrary LF momenta
\begin{align}
\mathcal{S}_{d\pm} \; \geq \; 0,
\hspace{2em}
|\Delta \mathcal{S}_{d}| \; \leq \; \mathcal{S}_{d} .
\end{align}
The positivity can be proved using the explicit expressions given above and 
will be demonstrated by the numerical results below.
The functions $S_{d\pm}$ have a probabilistic interpretation as the distributions of neutrons 
with helicities ``along'' or ``against'' the deuteron helicity. 

%
%
\begin{figure*}[t]
\includegraphics[width=.7\textwidth]{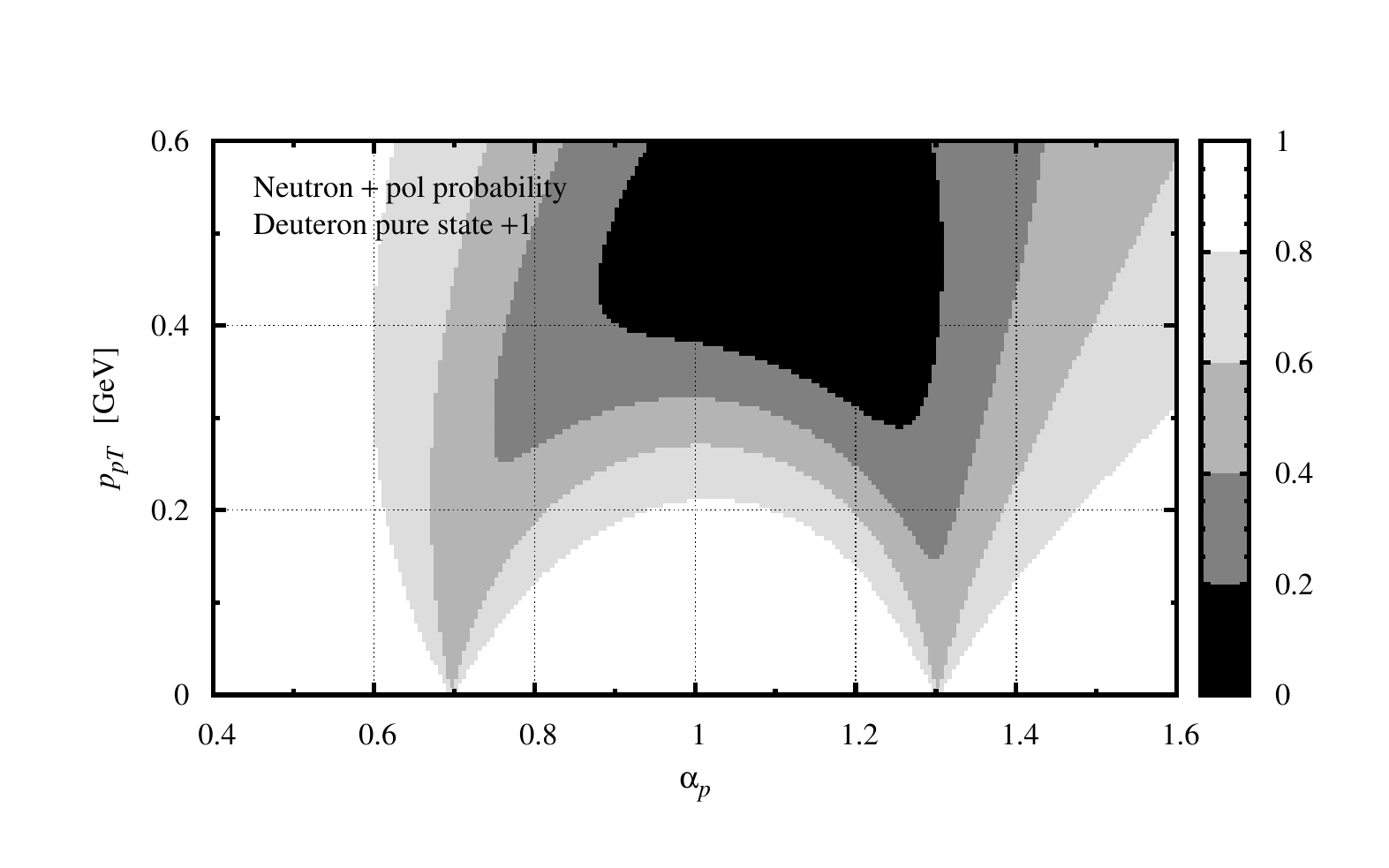}
\caption[]{The deuteron spectral function ratio $\mathcal{S}_{d+}/(\mathcal{S}_{d+} + \mathcal{S}_{d-})$, 
Eq.~(\ref{neutron_helicity_ratio}), in the pure deuteron spin state with LF helicity $+1$.
It describes the probability that a neutron observed at the given LF momentum has LF helicity $+1/2$,
i.e., is polarized along the deuteron spin direction. The 2-dimensional plot shows the ratio as a 
function of $\alpha_p$ and $|\bm{p}_{pT}|$.}
\label{fig:spectral_pol_probability}
\end{figure*}
The effective polarization of the neutron in the deuteron depends on the proton LF momentum.
An interesting effect is observed at zero transverse momentum, $\bm{p}_{pT} = \bm{k}_T = 0$,
and arbitrary longitudinal momentum $\alpha_p \neq 1, k^z \neq 0$.
In this kinematic limit the factors $C_0$ and $C_2$ in $\Delta \mathcal{S}_{d}$ are unity, 
Eqs.~(\ref{C_0_helicity_kt}) and (\ref{C_2_helicity_kt}), and the factor $C_T$ in $\mathcal{S}_{d}$ is also unity,
Eq.~(\ref{C_T_pure}). The distributions Eq.~(\ref{neutron_densities_pure})
thus become
\begin{subequations}
\label{neutron_densities_zero_transverse_momentum}
\begin{align}
& \mathcal{S}_{d} (\alpha_p, \bm{p}_{pT} = 0)[\textrm{pure $+1$}]
\nonumber \\[3ex]
&= \;
\Delta \mathcal{S}_{d} (\alpha_p, \bm{p}_{pT} = 0)[\textrm{pure $+1$}] 
\nonumber
\\[3ex]
\; &= \; 
\mathcal{S}_{d+} (\alpha_p, \bm{p}_{pT} = 0)[\textrm{pure $+1$}]
\nonumber
\\[1ex]
\; &= \; 
\frac{1}{2 - \alpha_p} \left( f_0 - \frac{f_2}{\sqrt{2}} \right)^2 ,
\\[1ex]
&\mathcal{S}_{d-} (\alpha_p, \bm{p}_{pT} = 0)[\textrm{pure $+1$}]
\; = \; 0 ,
\end{align}
\end{subequations}
and the neutron is completely polarized along the direction of the deuteron LF helicity. 
This happens because configurations with neutron LF helicity opposite to the deuteron
involve LF orbital angular momentum $L_z \neq 0$, which requires non-zero transverse momentum.
It is remarkable that the 4-dimensional representation of the deuteron LF wave function
reproduces this effect without explicit reference to orbital angular momentum states.
Notice that the distribution at $\bm{p}_{pT}$ includes contributions from  the S- and D-wave 
of the CM wave function.

Figure~\ref{fig:spectral_pol_probability} shows the ratio
\beq
\frac{\mathcal{S}_{d+}}{\mathcal{S}_{d+} + \mathcal{S}_{d-}} 
\; = \; \frac{1}{2}\left( 1 + \frac{\Delta \mathcal{S}_{d}}{\mathcal{S}_{d}} \right),
\hspace{1em}
0 \leq (\textrm{ratio}) \leq 1,
\label{neutron_helicity_ratio}
\eeq
in a pure deuteron spin state with LF helicity $+1$. It describes the probability that a neutron
observed at a given LF momentum has LF helicity $+1/2$ (along the deuteron spin direction);
the probability that it has LF helicity $-1/2$ (opposite to the deuteron spin direction)
is given by 1 minus the ratio. The dependence on the LF momentum shows
several interesting features: (a) The ratio is close to unity in the region
$0.8 < \alpha_p < 1.2$ and $|\bm{p}_{pT}| <$ 200 MeV, which corresponds to
CM momenta $|\bm{k}| \lesssim$ 200 MeV, where the S-wave dominates in the
CM wave function. (b) The ratio is equal to unity at $|\bm{p}_{pT}| = 0$ for 
arbitrary $\alpha_p$, as implied by Eq.~(\ref{neutron_densities_zero_transverse_momentum}).
Exceptions are the points $\alpha_p = 0.7$ and 1.3, which correspond to 
values of $k$ at which the combination of radial wave functions
$(f_0 - f_2/\sqrt{2})$ in Eq.~(\ref{neutron_densities_zero_transverse_momentum})
vanishes (singular points). (c) The ratio becomes small at $\alpha_p \sim 1$
and $|\bm{p}_{pT}| \geq$ 400 MeV. In this region the D-wave dominates
in the CM wave function, and the effects of relativistic spin rotations are large.

In sum, one sees that the neutron polarization in the polarized deuteron can 
effectively be controlled by selecting certain regions of the LF momentum.
This feature can be exploited in tagged DIS experiments (see below) and represents
an important advantage of spectator tagging. 

\begin{figure}[t]
\includegraphics[width=.48\textwidth]{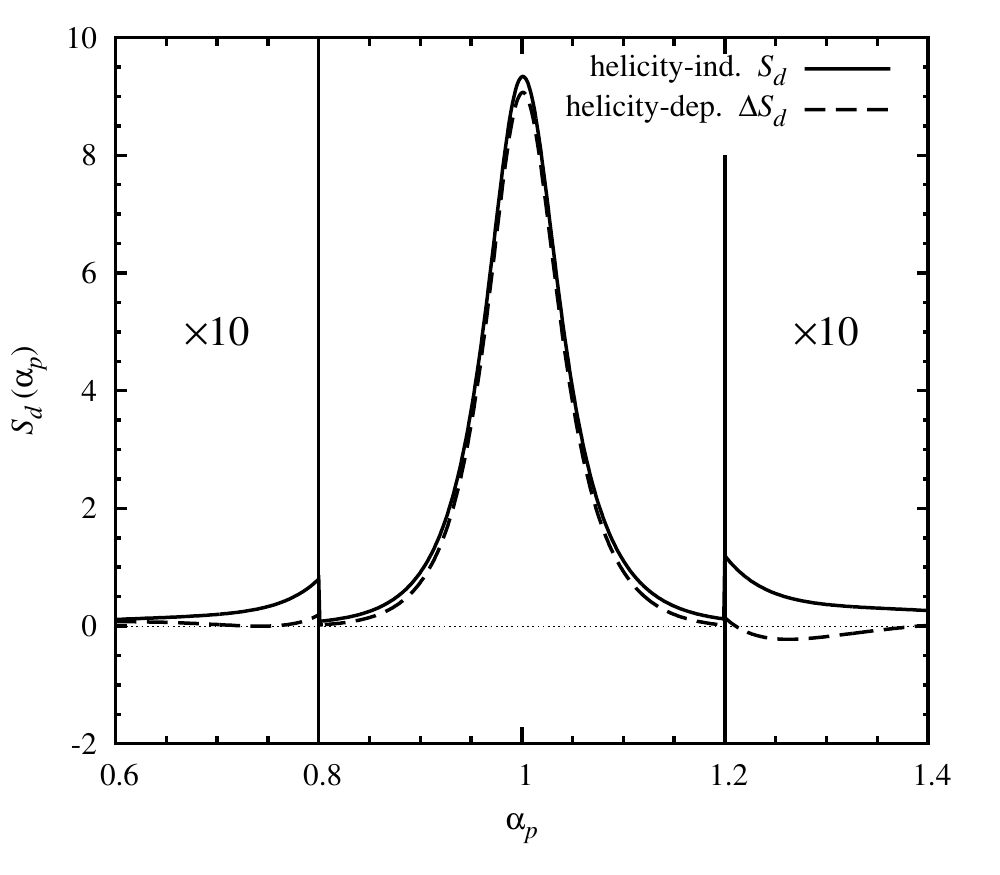}
\caption[]{The deuteron spectral functions $\mathcal{S}_{d}$ and $\Delta \mathcal{S}_{d}$,
integrated over the proton transverse momentum, in the pure deuteron spin state with LF helicity $+1$.
These functions describe the helicity-independent and helicity-dependent LF momentum distributions
of neutrons in the deuteron with LF helicity $+1$.}
\label{fig:spectral_int}
\end{figure}

\begin{figure}
\includegraphics[width=.48\textwidth]{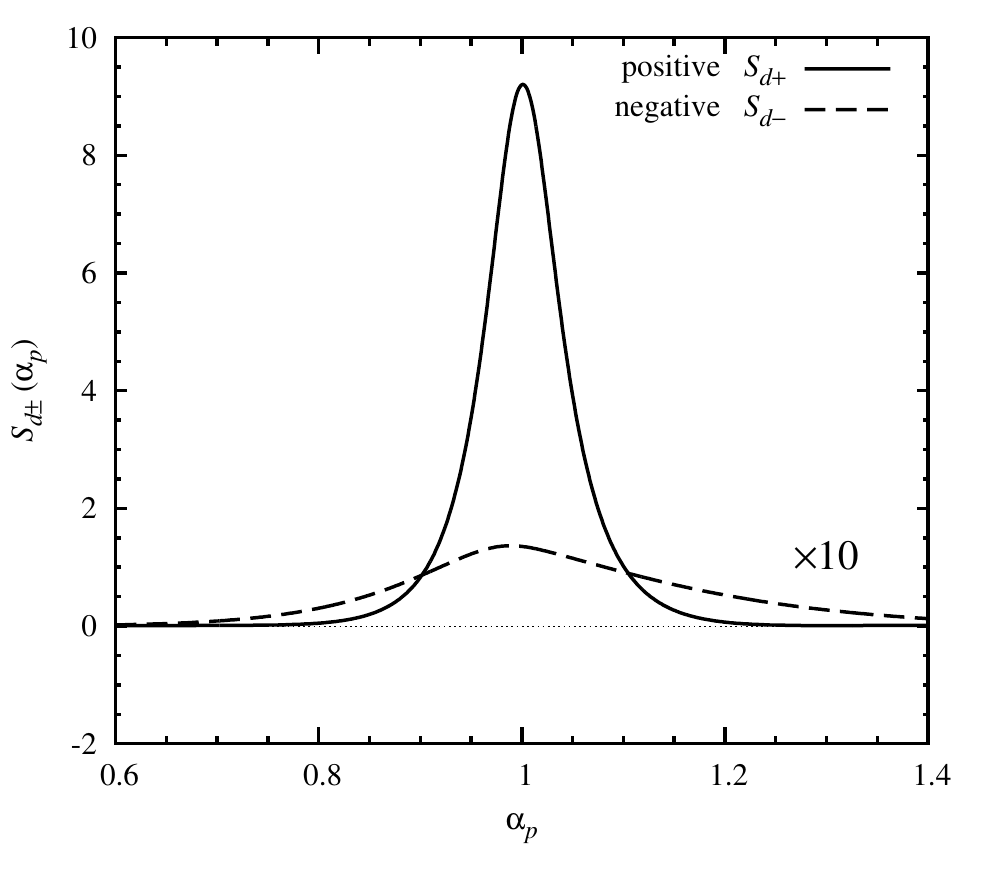}
\caption[]{The deuteron spectral functions $\mathcal{S}_{d+}$ and $\Delta \mathcal{S}_{d-}$,
integrated over the proton transverse momentum, in the pure deuteron spin state with LF helicity $+1$
(cf.\ Fig.~\ref{fig:spectral_int}).
These functions describe the LF momentum distributions of neutrons with LF helicity $+1/2$ and
$-1/2$ in the deuteron with LF helicity $+1$. The plot shows $\mathcal{S}_{d-}$ multiplied by 10.}
\label{fig:spectral_int_pm}
\end{figure}

The integrals of the deuteron spectral functions over the proton transverse momentum $\bm{p}_{pT}$
describe the distributions of neutrons with respect to the longitudinal LF momentum fraction
$\alpha_p$ only,
\begin{subequations}
\begin{align}
\mathcal{S}_{d} (\alpha_p) \; &\equiv \;
\int d^2 p_{pT} \; \mathcal{S}_{d} (\alpha_p, \bm{p}_{pT}) ,
\\[1ex]
\Delta \mathcal{S}_{d} (\alpha_p) \; &\equiv \;
\int d^2 p_{pT} \; \Delta \mathcal{S}_{d} (\alpha_p, \bm{p}_{pT}) ,
\end{align}
\end{subequations}
which are normalized such that
\be
\int_0^2 \frac{d\alpha_p}{\alpha_p} \; \left\{ \mathcal{S}_{d} (\alpha_p), \,
\Delta \mathcal{S}_{d} (\alpha_p) 
\right\} \; &=& \; \left\{ 1, \, \frac{g_{Ad}}{2 g_A} \right\} .
\ee
These functions are the analog of the parton distributions in the parton model of hadron structure
and appear in the description of $\bm{p}_{pT}$-integrated tagged DIS measurements on the deuteron.
Figures~\ref{fig:spectral_int} and \ref{fig:spectral_int_pm}
show the integrated LF momentum densities obtained with
our deuteron LF wave function. One observes: (a) The integrated distributions are concentrated in the
region $\alpha_p \approx (0.9, 1.1)$, corresponding to nucleon CM momenta $k \lesssim$ 100~MeV.
(b)~The neutrons are overwhelmingly polarized along the direction of the deuteron LF helicity.
The integrated distribution $\mathcal{S}_{d-}$ is at the level of a few percent of $\mathcal{S}_{d+}$. 
The magnitude of the distribution is consistent with their normalization integrals: in the nonrelativistic limit,
using Eq.~(\ref{g_A_integral_nonrelativistic}), we find
\beq
\left. \left[ \int \frac{d\alpha_p}{\alpha_p} \mathcal{S}_{d-} (\alpha_p) \right] \right/
\left[ \int \frac{d\alpha_p}{\alpha_p} \mathcal{S}_{d+} (\alpha_p) \right]
\;\; \approx \;\; \frac{3}{4} \omega_2 .
\eeq 
%
%
%
%
\section{Tagged DIS in impulse approximation}
\label{sec:ia}
\subsection{Impulse approximation}
\label{subsec:ia}
We now calculate the cross section of polarized tagged DIS on the deuteron, using the LF methods 
developed in Secs.~\ref{sec:deuteron} and \ref{sec:nucleon_operators}. We restrict ourselves to
the IA, where it is assumed that (i) the current operator couples only to a single nucleon;
(ii) the DIS final state produced from the active nucleon evolves independently from spectator.
The IA result contains the nucleon pole and dominates the tagged DIS cross section at small proton momenta.
The justification of the IA and steps of calculation are described in Ref.~\cite{Strikman:2017koc}.
Here we focus on the aspects specific to electron and deuteron polarization.

\begin{figure}[t]
\includegraphics[width=.36\textwidth]{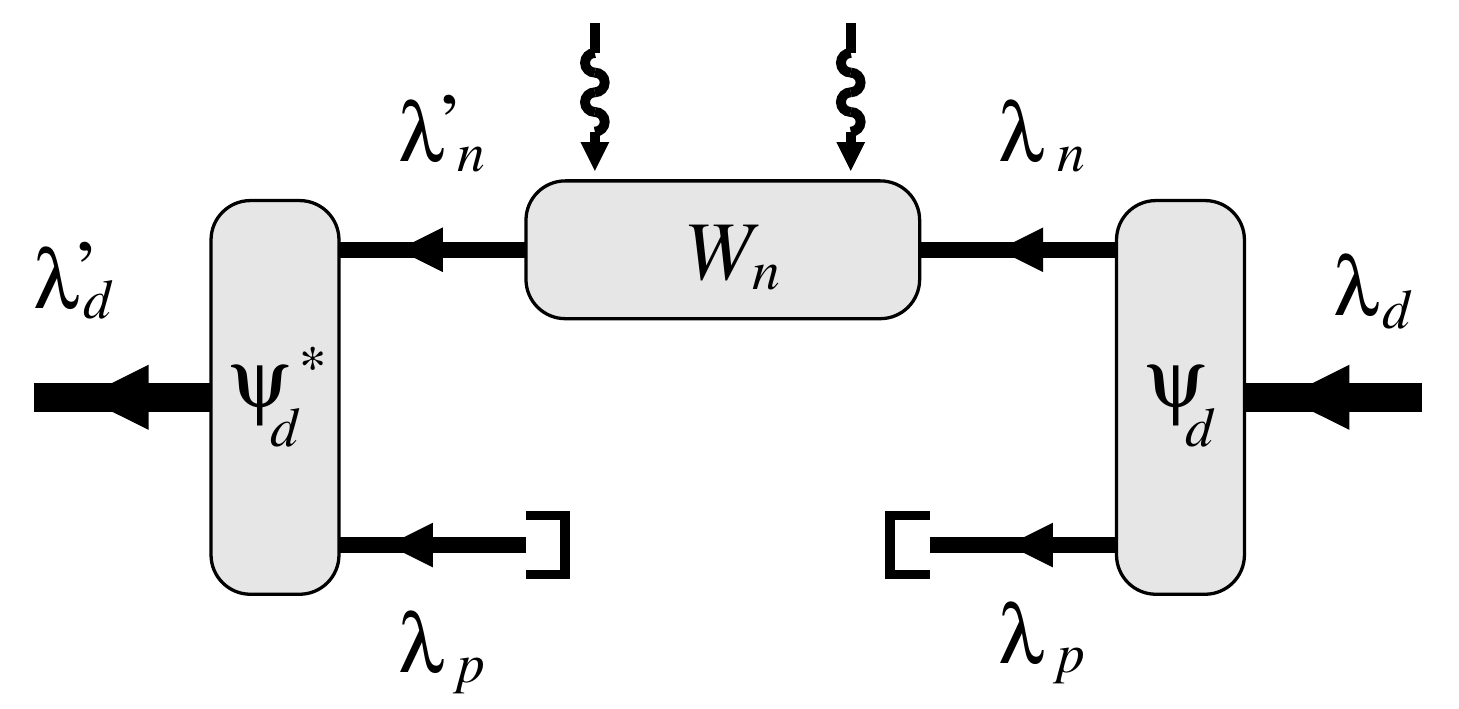}
\caption[]{The deuteron tensor in tagged DIS in the IA. Only the spin quantum numbers of
the deuteron and nucleon states are shown.}
\label{fig:ia_spin}
\end{figure}

In tagged DIS on the deuteron in a general ensemble of spin states, described by the density matrix 
$\rho_d (\lambda_d, \lambdadp)$, the deuteron tensor given by
\begin{align}
& W_d^{\mu\nu} (p_d, q, p_p) \; = \; 
\langle W_d^{\mu\nu} \rangle
\nonumber
\\
& = \; 
\sum_{\lambdadp, \lambda_d} \rho_d (\lambda_d, \lambdadp) \;
W_d^{\mu\nu} (p_d, q, p_p| \lambdadp, \lambda_d),
\label{deuteron_tensor_ensemble_alt}
\end{align}
where the tensor on the right-hand side is the nondiagonal deuteron tensor between LF helicity states with 
$\lambda_d$ and $\lambdadp$, cf.\ Eqs.~(\ref{deuteron_tensor_ensemble}) and (\ref{w_mu_nu_current}).
In the IA in a collinear frame, the latter is obtained as (see Fig.~\ref{fig:ia_spin}) \cite{Strikman:2017koc}
\begin{align}
& W_d^{\mu\nu} (p_d, q, p_p| \lambdadp, \lambda_d)
\nonumber \\[2ex]
&= \;
\sum_{\lambda_p, \lambdanp, \lambda_n} 
[2(2\pi)^3] \; \frac{2}{(2 - \alpha_p)^2} \;
\nonumber \\[2ex]
&\times 
\Psi_d^\ast (\alpha_p , \bm{p}_{pT}; \lambda_p, \lambdanp | \lambdadp) \;
\Psi_d (\alpha_p , \bm{p}_{pT}; \lambda_p, \lambda_n | \lambda_d)
\nonumber \\[3ex]
&\times 
W_n^{\mu\nu} (p_n, \widetilde{q}; \lambdanp, \lambda_n) .
\label{deuteron_tensor_nondiagonal_ia}
\end{align}
The $\alpha_p$ dependent factor and the quadratic expression in $\Psi_d^\ast$ and $\Psi_d$ 
represent a particular momentum density of the deuteron LF wave function. The tensor 
\beq
W_n^{\mu\nu} (p_n, \widetilde{q}; \lambdanp, \lambda_n)
\eeq
is the tensor for inclusive DIS on the neutron. It is evaluated at the on-shell neutron 4-momentum 
with LF components Eq.~(\ref{proton_neutron_momentum}), and at the effective 4-momentum transfer
\begin{subequations}
\begin{align}
\widetilde{q} \; &\equiv \; q + p_d - p_n - p_p, 
\\[1ex]
p_n + \widetilde{q} \; &= \; q + p_d - p_p ,
\end{align}
\end{subequations}
which accounts for the effects of LF energy nonconservation in the intermediate state \cite{Strikman:2017koc}.
The neutron spin states are described by their LF helicities $\lambdanp$ and $\lambda_n$, and the tensor 
is generally non-diagonal in these variables.
%
%

The summation over the nucleon LF helicities in Eq.~(\ref{deuteron_tensor_nondiagonal_ia})
and deuteron spin in Eq.~(\ref{deuteron_tensor_ensemble_alt}) can be performed using the 
neutron spin density matrix formalism described in Sec.~\ref{subsec:matrix_elements}.
The neutron tensor is represented as a bilinear form in LF bispinors,
\begin{align}
W_n^{\mu\nu} (p_n, \widetilde{q}; \lambdanp, \lambda_n)
\; &= \; \bar u_{\rm LF}(p_n, \lambdanp) \; 
\Gamma_n^{\mu\nu} (p_n, \widetilde{q}) \; 
\nonumber \\[1ex]
& \times u_{\rm LF}(p_n, \lambda_n),
\label{neutron_tensor_bilinear}
\end{align}
where $\Gamma_n^{\mu\nu}$ is matrix in bispinor indices, whose specific form is given below.
The deuteron tensor of the ensemble, Eq.~(\ref{deuteron_tensor_ensemble_alt}), can then 
be expressed as a bispinor trace,
\begin{align}
W_d^{\mu\nu}(p_d, q, p_p) \; &= \; 
[2(2\pi)^3]\; \frac{2}{(2 - \alpha_p)^2}
\nonumber \\[1ex]
&\times \; \textrm{tr}[ \, \Pi_n (p_d, p_p; s_n) \;
\Gamma_n^{\mu\nu} (p_n, \widetilde{q}) \, ],
\label{deuteron_tensor_trace}
\end{align}
where $\Pi_n$ is the effective neutron spin density matrix Eq.~(\ref{neutron_density}).
[In contrast to Eq.~(\ref{deuteron_onebody_trace}), there is no integration over
the proton LF momentum variables $\alpha_p$ and $\alpha_p$, as those are fixed 
to the tagged proton values.] Equation~(\ref{neutron_tensor_bilinear}) represents
the IA result for a general deuteron ensemble.

To determine the tagged DIS structure functions, 
one evaluates the tensor equation Eq.~(\ref{deuteron_tensor_trace}) with the specific form 
of the neutron tensor (unpolarized, polarized) and the neutron spin density matrix
(unpolarized, vector, tensor), and derives equations for the structure functions by taking
suitable components of the tensor equation. To separate the various structure functions one
needs to use both $+$ and $T$ LF components of the deuteron and nucleon tensors.
It is known that in LF quantization of interacting spin-1/2 systems the $+$ and $T$ components
of the current operator have different status: the $+$ current components are formed from
the independent canonical degrees of freedom and are free of interactions (``good components'');
the $T$ components involve dependent degrees of freedom and thus depend on the interactions
(``bad components''). The use of bad components in the IA therefore generally must be regarded
as an approximation. However, studies have shown that the interaction effects in the $T$ components
of the current are suppressed in the DIS limit $Q^2 \gg m^2, x$ fixed, i.e., that they do not affect
the result for the leading structure functions in the DIS limit
(leading-twist approximation) \cite{Frankfurt:1988nt,Lev:1998qz}.
In the following calculations we consider only the leading structure functions in the DIS limit
(unpolarized, polarized), for which the use of $T$ components is justified.
This is supported by the fact that the IA structure functions in the DIS limit satisfy
the sum rules for baryon number, LF momentum, and spin (see below).

We note that the calculations could in principle be extended to power-suppressed structures in the DIS limit,
such as the spin structure functions $F_{[LS_T]d}$ or $g_{2d}$
(see Sec.~\ref{subsec:spin_asymmetries}). In this case the interaction effects in the
$T$ components of the current could no longer be neglected, and new considerations would be needed --- the
so-called angular conditions on the current matrix elements \cite{Frankfurt:1988nt,Lev:1998qz}. 
\subsection{Unpolarized electron scattering}
Unpolarized electron scattering involves the symmetric parts of the deuteron and neutron tensors.
The symmetric part of the neutron tensor is diagonal in the neutron LF helicities
and independent of the value of the helicity,\footnote{We neglect the transverse spin 
dependence of the unpolarized nucleon tensor due to two-photon exchange effects
\cite{Christ:1966zz,Afanasev:2007ii}; cf.\ Sec.~\ref{subsec:cross_section_vector}.} 
\begin{align}
& W^{\mu\nu}_n(p_n, \widetilde{q}; \lambdanp, \lambda_n)[\textrm{symm}]
\nonumber \\[1ex]
& = \; \delta(\lambdanp, \lambda_n) \; W^{\mu\nu}_n(p_n, \widetilde{q})[\textrm{symm}] .
\end{align}
In the bilinear form Eq.~(\ref{neutron_tensor_bilinear}) the tensor is therefore represented 
by the unit matrix in bispinor indices
\beq
\Gamma^{\mu\nu}_n \;\; = \;\; \frac{1}{2 m} \; W^{\mu\nu}_n [\textrm{symm}] .
\eeq
The unpolarized symmetric deuteron tensor is then obtained from Eq.~(\ref{deuteron_tensor_trace}) as
\begin{align}
W^{\mu\nu}_d[\textrm{unpol}] \, = \, [2 (2\pi)^3] \; \frac{2 \mathcal{S}_{d} [\textrm{unpol}]}{2 - \alpha_p}
\, W^{\mu\nu}_n [\textrm{symm}] .
\label{tensor_ia_unpol}
\end{align}
The relation between the tensors is, up to a factor, given by the helicity-independent neutron 
distribution in the unpolarized deuteron, Eq.~(\ref{f_U}).

The decomposition of the symmetric neutron tensor is of the same form as that of the unpolarized 
deuteron tensor Eq.~(\ref{deuteron_tensor_unpol}),
\be
W_n^{\mu\nu}[\textrm{symm}] \; &= \; 
\frac{1}{2} (e_L^\mu e_L^\nu - e_q^\mu e_q^\nu - g^{\mu\nu}) \, F_{[UU,T] n}
\nonumber
\\[1ex]
&+ \; \frac{1}{2} e_L^\mu e_L^\nu \, F_{[UU,L] n} ,
\ee
where now the basis vectors $e_L$ and $e_q$ are constructed with the 4-momenta $p_n$ and $\widetilde{q}$,
and the structure functions depend on invariants formed with the latter.
Substituting this form in Eq.~(\ref{tensor_ia_unpol}), taking suitable components of the
tensor equation ($\mu\nu = ++$ and $TT$), one obtains expressions for the
tagged deuteron structure functions \cite{Strikman:2017koc}
\begin{subequations}
\begin{align}
& F_{[UU, T]d}(x, Q^2; \alpha_p, \bm{p}_{pT}) 
\nonumber \\[1ex]
&= \;  [2(2\pi)^3] \; \frac{2 \mathcal{S}_{d} (\alpha_p, \bm{p}_{pT}) [\textrm{unpol}]}{2 - \alpha_p}
\; F_{[UU, T]n} (\widetilde x, Q^2),
\label{FT_ia}
\\[2ex]
& F_{[UU, L]d}(x, Q^2; \alpha_p, \bm{p}_{pT}) 
\nonumber \\[1ex]
&= \;  [2(2\pi)^3] \; \frac{2 \mathcal{S}_{d} (\alpha_p, \bm{p}_{pT}) [\textrm{unpol}]}{2 - \alpha_p}
\; F_{[UU, L]n} (\widetilde x, Q^2),
\label{FL_ia}
\\[2ex]
& \widetilde x \; \equiv \; \frac{x}{2 - \alpha_p} .
\label{xtilde}
\end{align}
\end{subequations}
$\tilde x$ is the effective scaling variable for scattering on the neutron, which takes
into account its longitudinal LF momentum in the deuteron. The expressions Eqs.~(\ref{FT_ia})--(\ref{xtilde})
are valid in the DIS limit $Q^2 \gg m^2$, neglecting kinematic corrections ${\mathcal O}(m^2/Q^2)$.
The corresponding expressions for the tagged deuteron structure functions $F_{1d}$ and $F_{2d}$, 
Eq.~(\ref{F1_F2}), are
\begin{subequations}
\begin{align}
& F_{1d}(x, Q^2; \alpha_p, \bm{p}_{pT}) 
\nonumber \\[1ex]
&= \;  [2(2\pi)^3] \; \frac{2 \mathcal{S}_{d} (\alpha_p, \bm{p}_{pT}) [\textrm{unpol}]}{2 - \alpha_p}
\; F_{1n} (\widetilde x, Q^2) ,
\\[1ex]
& F_{2d}(x, Q^2; \alpha_p, \bm{p}_{pT}) 
\nonumber \\[1ex]
&= \; [2(2\pi)^3] \; \; \mathcal{S}_{d}(\alpha_p, \bm{p}_{pT}) [\textrm{unpol}] \; F_{2n} (\widetilde x, Q^2) .
\end{align}
\end{subequations}
The tagged deuteron $F_{2d}$ and the neutron $F_{2n}$ are related directly by the 
neutron momentum distribution in the deuteron. 

The neutron momentum distribution satisfies the LF momentum sum rule 
Eq.~(\ref{f_U_momentum_sr}). It implies that the integrated tagged structure function 
[see Eq.~(\ref{phase_space_alpha_pt}); note the kinematic limit $\alpha_p < 2 - x$]
\begin{align}
F_{2d}^{\rm int} (x, Q^2) \; &\equiv \; 
[2(2\pi)^3]^{-1} 
\int_0^{2 - x} \frac{d\alpha_p}{\alpha_p} \int d^2 p_{pT}
\nonumber \\[1ex]
&\times \; F_{2d} (x, Q^2; \alpha_p, p_{pT}) 
\end{align}
satisfies the sum rule
\beq
\int_0^2 dx \; F_{2d}^{\rm int} (x, Q^2) \;\; = \;\; 
\int_0^1 d\widetilde{x} \; F_{2n} (\widetilde{x}, Q^2) .
\label{sumrule_f2}
\eeq
It shows that spectator tagging only fixes the neutron LF momentum in the deuteron but does not 
change the integral of the distribution.

The tensor-polarized part of the symmetric deuteron tensor is obtained from Eq.~(\ref{deuteron_tensor_trace}) as
\begin{align}
W^{\mu\nu}_d [\textrm{tensor}] \; &= \; [2 (2\pi)^3] \; \frac{2 \mathcal{S}_{d} [\textrm{tensor}]}{2 - \alpha_p}
\nonumber \\[1ex]
& \times \; W^{\mu\nu}_n [\textrm{symm}] ,
\label{tensor_ia_tensor}
\end{align}
where the tensor-polarized neutron distribution is given in Eq.~(\ref{f_T}), and the symmetric neutron
tensor is the same as above. Equation~(\ref{tensor_ia_tensor}) holds for an ensemble with general 
tensor polarization. The calculation of the tensor-polarized structure functions proceeds in the same 
way as for the unpolarized deuteron. Here we quote only the expressions for the tensor 
structure functions $F_{[UT_{LL}, T]d}$ and $F_{[UT_{LL}, L]d}$, which appear in the calculation
of spin asymmetries in the scaling limit (see Sec.~\ref{subsec:spin_asymmetries} and below). 
They can be obtained with the special polarization tensor associated with pure states, 
Eq.~(\ref{tensor_special_cm}), choosing the polarization along the $z$-direction, 
$\bm{N} = \bm{e}_z$. We obtain
\begin{subequations}
\begin{align}
& T_{LL} F_{[UT_{LL}, T]d}(x, Q^2; \alpha_p, \bm{p}_{pT}) 
\nonumber \\[1ex]
&= \; [2(2\pi)^3]\; \frac{2 \mathcal{S}_{d} (\alpha_p, \bm{p}_{pT})[\textrm{tensor}]}{2 - \alpha_p}
\; F_{[UU, T]n} (\widetilde x, Q^2),
\label{F_UTLL_T_ia}
\\[2ex]
& T_{LL} F_{[UT_{LL}, L]d}(x, Q^2; \alpha_p, \bm{p}_{pT}) 
\nonumber \\[1ex]
&= \; [2(2\pi)^3] \; \frac{2 \mathcal{S}_{d} (\alpha_p, \bm{p}_{pT}) [\textrm{tensor}]}{2 - \alpha_p}
\; F_{[UU, L]n} (\widetilde x, Q^2),
\label{F_UTLL_L_ia}
\end{align}
\end{subequations}
where $\mathcal{S}_{d} (\alpha_p, \bm{p}_{pT}) [\textrm{tensor}]$ is the neutron distribution 
given in Eq.~(\ref{f_T_pure}).
\subsection{Polarized electron scattering}
Polarized electron scattering involves the antisymmetric parts of the deuteron and neutron tensors.
The antisymmetric part of the neutron tensor depends on the neutron LF helicities. This dependence
can be expressed in the form
\begin{subequations}
\begin{align}
& W^{\mu\nu}_n(p_n, \widetilde{q}; \lambdanp, \lambda_n)[\textrm{antisymm}]
\nonumber \\[1ex]
&= \; A^{\mu\nu}_{\phantom{\mu\nu}\rho} (p_n, \widetilde{q}) \;
s_{n, \textrm{gen}}^\rho (p_n, \lambdanp, \lambda_n) ,
\label{neutron_tensor_general}
\\[2ex]
& s_{n, \textrm{gen}}^\rho (p_n, \lambdanp, \lambda_n)
\nonumber \\[1ex]
& \equiv \; 
\frac{1}{2m} \, \bar u_{\rm LF} (p_n, \lambdanp) \,
(-\gamma^\rho \gamma^5) \, u_{\rm LF} (p_n, \lambda_n) .
\end{align}
\end{subequations}
The axial 4-vector $s_{n, \textrm{gen}}$ is defined as a function of the LF helicities, 
$\lambda_n$ and $\lambdanp$, and represents the general polarization 4-vector of the 
free neutron. The tensor $A^{\mu\nu}_{\phantom{\mu\nu}\rho}$ is given by 
[cf.\ Eq.~(\ref{deuteron_tensor_polarized})]
\be
A^{\mu\nu}_{\phantom{\mu\nu}\rho}
\; &= \; \frac{i}{2} \epsilon^{\mu\nu\sigma\tau} e_{q, \sigma}
\left\{ e_{L\ast\tau} e_{L\ast\rho} \, \gamma F_{[LS_L]n}
\right.
\nonumber \\
& + \; \left. \left( e_{L\ast\tau} e_{L\ast\rho} + g_{\tau\rho} \right) F_{[LS_T]n} \right\}
\nonumber
\label{neutron_tensor_polarized}
\ee
where the basis vectors $e_{L\ast}$ and $e_q$ are constructed with the 4-momenta $p_n$ and $\widetilde{q}$,
and the structure functions depend on invariants formed with the latter.
In the bilinear form Eq.~(\ref{neutron_tensor_bilinear}) the antisymmetric 
neutron tensor is therefore represented by the matrix
\be
\Gamma^{\mu\nu}_n \; &=& \; \frac{1}{2m} \, A^{\mu\nu}_{\phantom{\mu\nu}\rho} 
\; (-\gamma^\rho \gamma^5) .
\ee
The trace in Eq.~(\ref{deuteron_tensor_trace}) now involves the vector-polarized part 
of the neutron density matrix, Eq.~(\ref{neutron_density_vector}),
\begin{subequations}
\begin{align}
\textrm{tr} \left[ \Pi_n \Gamma_n^{\mu\nu} \right] \; &= \;
\frac{1}{2m} \, A^{\mu\nu}_{\phantom{\mu\nu}\rho} \, 
\textrm{tr}[{\textstyle\frac{1}{2}} (p_n\gamma + m)
(s_n \gamma) \gamma^5 \; (-\gamma^\rho \gamma^5)]
\nonumber
\\[1ex]
&= \; A^{\mu\nu}_{\phantom{\mu\nu}\rho} \; s_n^\rho ,
\end{align}
\end{subequations}
where $s_n$ is now the effective neutron polarization vector in the deuteron, Eq.~(\ref{neutron_spin_vector}).
Altogether, the vector-polarized deuteron tensor is obtained from Eq.~(\ref{deuteron_tensor_trace}) as
\begin{align}
W^{\mu\nu}_d(p_d, q, p_p) [\textrm{vector}] \; &= \; [2 (2\pi)^3] \;
\frac{2}{(2 - \alpha_p)^2} \; A^{\mu\nu}_{\phantom{\mu\nu}\rho} \; s_n^\rho .
\label{tensor_ia_pol}
\end{align}
In the last expressions 
\beq
A^{\mu\nu}_{\phantom{\mu\nu}\rho} \; s_n^\rho
\;\; \equiv \;\; W^{\mu\nu}_n(p_n, \widetilde q, s_n)[\textrm{antisymm}]
\eeq
is the antisymmetric neutron tensor, evaluated with the effective polarization 4-vector $s_n$, 
Eq.~(\ref{neutron_spin_vector}). All the information pertaining to the deuteron 
wave function (radial functions, polarization, etc.) is contained in $s_n$.
The entire effect of averaging with the effective neutron density matrix has been to 
replace the general neutron polarization 4-vector of the free neutron in Eq.~(\ref{neutron_tensor_general})
with the specific polarization 4-vector of the neutron in the deuteron of Eq.~(\ref{neutron_spin_vector}).
This simple result is achieved thanks to the covariant representation of the deuteron spin structure
cf.\ Secs.~\ref{sec:deuteron} and \ref{sec:nucleon_operators}.

We can now derive the expression for the tagged deuteron spin structure functions. 
As before we consider the DIS limit, in which the structure function $F_{[L S_L]d}$ is leading
and $F_{[L S_T]d}$ is suppressed, and neglect terms ${\mathcal O}(m/Q)$. The derivation
proceeds as follows: (a) Take the tensor equation Eq.~(\ref{tensor_ia_pol}) in a general collinear 
frame with $p_d^+ \neq M_d$; (b) take the deuteron longitudinally polarized, corresponding to a rest-frame 
polarization vector $\bm{S} = (\bm{0}_T, S^z)$, cf.\ Eq.~(\ref{vector_restframe}).
(c) take the non-zero transverse component of the tensor equation, $\mu\nu = 12$.
In this way we obtain, after a brief calculation,
\begin{align}
& F_{[L S_L]d}(x, Q^2; \alpha_p, \bm{p}_{pT})
\nonumber \\[1ex]
&= \; 
[2 (2\pi)^3] \; \frac{4 m s_n^+}{(2 - \alpha_p)^3 \, p_d^+ \, S^z} \; F_{[L S_L]n}(\widetilde x, Q^2) .
\end{align}
$\widetilde x$ is defined in Eq.~(\ref{xtilde}). 
The ratio on the R.H.S.\ is, up to a factor, equal to the helicity-dependent LF momentum distribution
of neutrons in the deuteron state with LF helicity $+1$, Eq.~(\ref{Delta_f_pure}), and we can write
\begin{align}
& F_{[L S_L]d}(x, Q^2; \alpha_p, \bm{p}_{pT}) 
\nonumber \\[1ex]
&= \; [2 (2\pi)^3] \; \frac{2 \Delta \mathcal{S}_{d}(\alpha_p, \bm{p}_{pT})[\textrm{pure $+1$}]}{2 - \alpha_p}  
\; F_{[L S_L]n}(\widetilde x, Q^2) .
\end{align}
The corresponding expression for the tagged spin structure function $g_{1d}$ is
\begin{align}
& g_{1d}(x, Q^2; \alpha_p, \bm{p}_{pT})  
\nonumber \\[1ex]
&= \; [2 (2\pi)^3] \; \frac{2 \Delta \mathcal{S}_{d}(\alpha_p, \bm{p}_{pT}) 
[\textrm{pure $+1$}]}{2 - \alpha_p} \; g_{1n}(\widetilde x, Q^2) .
\end{align}
The formula applies in the DIS limit, where $\gamma \rightarrow 0$ and the contribution of the $g_2$ 
structure function is suppressed; the formulas including power corrections can be derived easily.

The helicity-dependent neutron distribution in the deuteron satisfies the spin sum rule 
Eq.~(\ref{neutron_density_pol_normalization}). It implies that the integrated 
tagged spin structure function
\begin{align}
g_{1d}^{\rm int} (x, Q^2) \; &\equiv \; [2(2\pi)^3]^{-1} \int_0^{2 - x} \frac{d\alpha_p}{\alpha_p}
\int d^2 p_{pT}
\nonumber \\[1ex]
&\times \; g_{1d} (x, Q^2; \alpha_p, p_{pT}) 
\end{align}
satisfies the sum rule
\begin{subequations}
\begin{align}
\int_0^2 dx \; g_{1d}^{\rm int} (x, Q^2) \; &= \; \frac{g_{Ad}}{g_A} \;
\int_0^1 d\widetilde{x} \; g_{1n} (\widetilde{x}, Q^2) 
\label{sumrule_g1}
\\[1ex]
\; &= \; 2 \left( 1 - \frac{3}{2} \omega_2 \right) \;
\int_0^1 d\widetilde{x} \; g_{1n} (\widetilde{x}, Q^2) .
\end{align}
\end{subequations}
Eq.~(\ref{sumrule_g1}) is obtained by changing the order of integrations over $x$ and $\alpha_p$ 
and substituting $x$ by $\widetilde x$, Eq.~(\ref{xtilde}). The last expression applies
when the axial charge is evaluated in the non-relativistic limit, Eq.~(\ref{g_A_integral_nonrelativistic}).
We see that our result for the integrated tagged spin structure function reproduces the nonrelativistic 
formula for the D-state correction to the deuteron spin structure function.
\subsection{Spin asymmetries}
\label{subsec:spin_asymmetries_ia}
Using the results for the tagged spin structure functions in the IA, we now calculate the spin 
asymmetries in tagged DIS. In the DIS limit $\gamma \rightarrow 0$ only the parallel spin asymmetries 
$A_\parallel$ are present. The scaling variable $y$ and the virtual photon polarization parameter $\epsilon$,
and therefore the depolarization factors $D_{\parallel [S_L]}$, are the same for DIS on the deuteron and the 
(bound) neutron up to power corrections,
\begin{subequations}
\begin{align}
\epsilon[\textrm{deuteron}] \; &= \; \epsilon[\textrm{neutron}]
\; + \; {\mathcal O}(m^2/Q^2),
\\[1ex]
D_{\parallel [S_L]}[\textrm{deuteron}] \; &= \; D_{\parallel [S_L]}[\textrm{neutron}]
\; + \; {\mathcal O}(m^2/Q^2);
\end{align}
\end{subequations}
we do not distinguish between the deuteron and neutron in the notation of these quantities. 
Following Sec.~\ref{subsec:spin_asymmetries} we consider both the three-state and
the two-state asymmetries, Eqs.~(\ref{A_parallel_3}) and (\ref{A_parallel_2}).
Substituting the expressions for the tagged structure functions in the IA, 
the deuteron three-state asymmetry Eq.~(\ref{A_parallel_3}) becomes
\begin{subequations}
\begin{align}
& A_{\parallel (3) d} (x, Q^2; \alpha_p, \bm{p}_{pT}) 
\nonumber \\[1ex]
&= \;
\frac{D_{\parallel [S_L]} \; F_{[LS_L]d}(x, Q^2; \alpha_p, \bm{p}_{pT}) }
{[F_{[UU,T]d} \; + \; \epsilon F_{[UU,L]d}](x, Q^2; \alpha_p, \bm{p}_{pT}) }
\\[2ex]
&= \;
\frac{\Delta \mathcal{S}_{d} (\alpha_p, \bm{p}_{pT})[\textrm{pure $+1$}]  \; D_{\parallel [S_L]} \; F_{[LS_L]n}(\widetilde{x}, Q^2)}
{\mathcal{S}_{d}(\alpha_p, \bm{p}_{pT})[\textrm{unpol}] \; [F_{[UU,T]n} \; + \; \epsilon F_{[UU,L]n}](\widetilde{x}, Q^2)}
\\[2ex]
&= \; 
\mathcal{D}_{\parallel (3) d} (\alpha_p, \bm{p}_{pT}) \; A_{\parallel n} (y, \widetilde{x}, Q^2) ,
\label{A_3_factorized}
\end{align}
\begin{align}
\mathcal{D}_{\parallel (3) d} (\alpha_p, \bm{p}_{pT})
\; &\equiv \;
\frac{\Delta \mathcal{S}_{d} (\alpha_p, \bm{p}_{pT})[\textrm{pure $+1$}]}
{\mathcal{S}_{d} (\alpha_p, \bm{p}_{pT})[\textrm{unpol}]} ,
\label{D_3_def}
\\[2ex]
A_{\parallel n} (y, \widetilde{x}, Q^2) 
\; &\equiv \;
\frac{D_{\parallel [S_L]} \; F_{[LS_L]n}(\widetilde{x}, Q^2) }
{[F_{[UU,T]n} \; + \; \epsilon F_{[UU,L]n}](\widetilde{x}, Q^2)} .
\end{align}
\end{subequations}
The tagged deuteron asymmetry can be factorized into the neutron spin asymmetry and a function 
describing deuteron structure. $A_{\parallel n}$ is the ratio of the polarized and unpolarized
neutron structure functions and the depolarization factor; it represents the parallel spin asymmetry 
for scattering on the neutron and depends on the nucleon subprocess variables, 
$\widetilde x$ and $Q^2$, as well as on $y$. 
$\mathcal{D}_{\parallel (3) d}$ is the ratio of helicity-dependent and -independent neutron distributions in the deuteron 
and depends on the proton momentum variables $\alpha_p$ and $\bm{p}_{pT}$. We refer to this
function as the deuteron depolarization factor. Note that the unpolarized neutron distribution in 
Eq.~(\ref{D_3_def}) represents the average of the three pure deuteron states with LF helicities $\pm 1$ and $0$, 
so that
\begin{align}
\mathcal{D}_{\parallel (3) d} 
&=
\frac{\Delta \mathcal{S}_{d} [\textrm{pure $+1$}]}{\frac{1}{3} \left( \mathcal{S}_{d}[\textrm{pure $+1$}] + \mathcal{S}_{d}[\textrm{pure $-1$}]
+ \mathcal{S}_{d}[\textrm{pure $0$}] \right)} .
\end{align}
Its explicit form as a function of the CM momentum variable is
\begin{align}
\mathcal{D}_{\parallel (3) d} =& \frac{\displaystyle \left( f_0 - \frac{f_2}{\sqrt{2}}\right) \left( C_0 f_0
- \frac{C_2 f_2}{\sqrt{2}} \right)}{f_0^2 + f_2^2}
\nonumber
\\[1ex]
& [f_{0, 2} \equiv f_{0, 2}(k); \, C_{0, 2} \equiv C_{0, 2}(\bm{k})] .
\label{D_3_cm}
\end{align}

The two-state asymmetry Eq.~(\ref{A_parallel_2}) in the DIS limit is given by
\begin{widetext}
\begin{align}
A_{\parallel (2) d} (x, Q^2; \alpha_p, \bm{p}_{pT})  \; &= \;
\frac{D_{\parallel [S_L]} \; F_{[LS_L]}(x, Q^2; \alpha_p, \bm{p}_{pT})}
{\left[ F_{[UU,T]d} \; + \; \epsilon F_{[UU,L]d} 
\; + \; D_{\parallel [T_{LL}]} \, (F_{[UT_{LL},T]d} 
+ \epsilon F_{[UT_{LL},L]d}) \right](x, Q^2; \alpha_p, \bm{p}_{pT})} .
\label{A_parallel_2_ia}
\end{align}
\end{widetext}
Now, in addition to the unpolarized structure functions, the tensor-polarized structure functions 
$F_{[UT_{LL},T]d}$ and $F_{[UT_{LL},L]d}$ appear in denominator. 
In the IA they are given by Eqs.~(\ref{F_UTLL_T_ia}) and (\ref{F_UTLL_L_ia}) in terms of the
neutron distribution in the tensor-polarized deuteron and the unpolarized neutron structure functions.
In Eq.~(\ref{A_parallel_2_ia}) $D_{\parallel [T_{LL}]}$ is the depolarization factor given 
by Eq.~(\ref{depolarization_TLL_parallel}). In the DIS limit it becomes
\beq
D_{\parallel [T_{LL}]} \; = \; \frac{1}{3} \; + \; {\mathcal O}(m^2/Q^2).
\eeq
Substituting the expressions for the tagged structure functions in the IA, the asymmetry can
be factorized in a similar way as Eq.~(\ref{A_3_factorized}),
\begin{subequations}
\begin{align}
& A_{\parallel (2) d} (x, Q^2; \alpha_p, \bm{p}_{pT})
\nonumber \\[1ex]
&= \mathcal{D}_{\parallel (2) d}(\alpha_p, \bm{p}_{pT}) \; A_{\parallel n} (\widetilde{x}, Q^2) ,
\label{A_2_factorized}
\\[2ex]
& \mathcal{D}_{\parallel (2) d}(\alpha_p, \bm{p}_{pT})
\nonumber \\[1ex]
&\equiv 
\frac{\Delta \mathcal{S}_{d} (\alpha_p, \bm{p}_{pT})[\textrm{pure $+1$}]}
{\mathcal{S}_{d} (\alpha_p, \bm{p}_{pT})[\textrm{unpol}] + \mathcal{S}_{d} (\alpha_p, \bm{p}_{pT})[\textrm{tensor}]} .
\label{D_2_def}
\end{align}
\end{subequations}
The deuteron depolarization factor now includes the tensor-polarized neutron distribution
in the denominator. Note that the sum of unpolarized and tensor neutron distributions in 
Eq.~(\ref{D_2_def}) represents the sum of the two pure deuteron states with LF helicities $\pm 1$, 
so that
\begin{align}
\mathcal{D}_{\parallel (2) d}
&= \frac{\Delta \mathcal{S}_{d} [\textrm{pure $+1$}]}
{\frac{1}{2} \left( \mathcal{S}_{d}[\textrm{pure $+1$}] + \mathcal{S}_{d}[\textrm{pure $-1$}] \right)} .
\end{align}
This function is related to the effective neutron polarization in the pure $+1$ state, 
Eq.~(\ref{neutron_helicity_ratio}), and is bounded as
\beq
-1 \; \leq \; \mathcal{D}_{\parallel (2) d} \; \leq \; 1 .
\label{D_2_bound}
\eeq
Its explicit form as a function of the CM momentum variable is
\be
\mathcal{D}_{\parallel (2) d}
=& \frac{\displaystyle \left( f_0 - \frac{f_2}{\sqrt{2}}\right) \left( C_0 f_0 - \frac{C_2 f_2}{\sqrt{2}} \right)}
{\displaystyle f_0^2 + f_2^2 - C_T \left( 2 f_0 + \frac{f_2}{\sqrt{2}} \right) \frac{f_2}{\sqrt{2}}}
\nonumber
\\[1ex]
& [f_{0, 2} \equiv f_{0, 2}(k), \, C_{0, 2, T} \equiv C_{0, 2, T}(\bm{k})] .
\label{D_2_cm}
\ee

%
%
\begin{figure}[t]
\includegraphics[width=.48\textwidth]{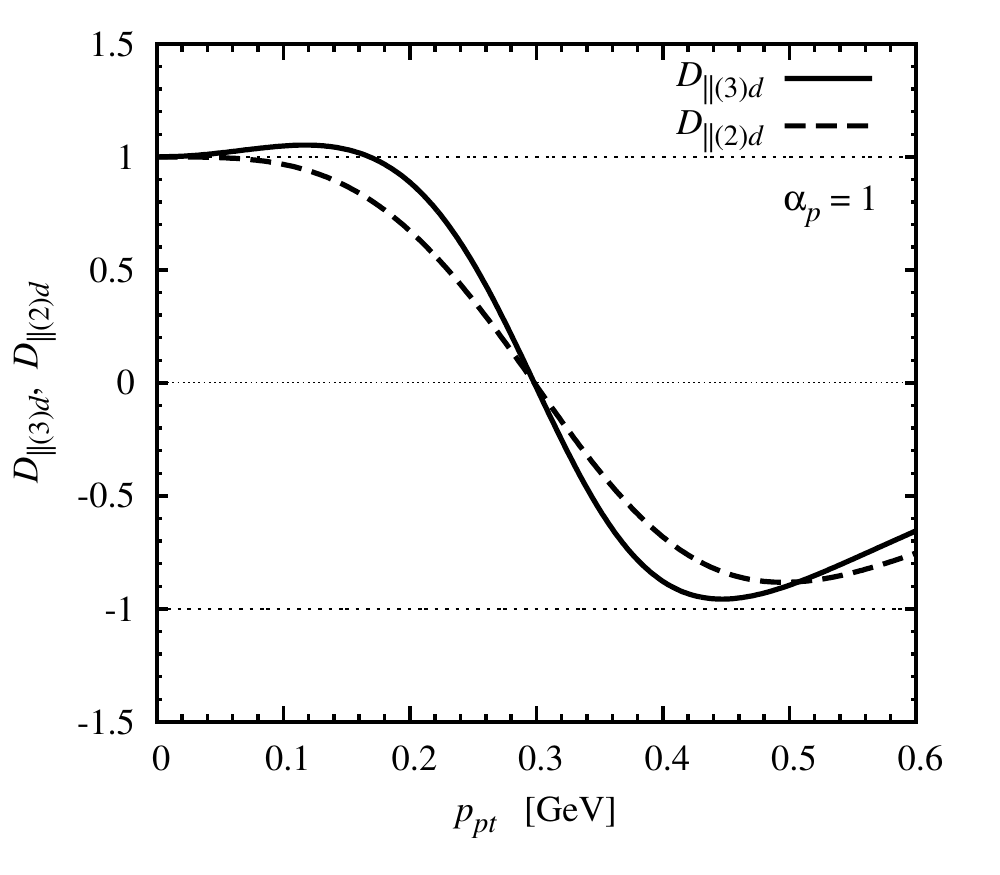}
\caption[]{The deuteron depolarization factors $\mathcal{D}_{\parallel (3) d}$ (three-state asymmetry)
and $\mathcal{D}_{\parallel (2) d}$ (two-state asymmetry) in polarized tagged DIS.
The graph shows the depolarization factors at $\alpha_p = 1$ as functions of $|\bm{p}_{pT}|$.}
\label{fig:depol_pt}
\end{figure}
The deuteron depolarization factors $\mathcal{D}_{\parallel (3) d}$ and $\mathcal{D}_{\parallel (2) d}$ describe the effective neutron polarization 
``selected'' by the tagged proton momentum and summarize the deuteron structure effects in spin asymmetry 
measurements in tagged DIS. Figure~\ref{fig:depol_pt} shows $\mathcal{D}_{\parallel (3) d}$ and $\mathcal{D}_{\parallel (2) d}$ at $\alpha_p = 1$
as functions of $|\bm{p}_{pT}|$. One observes: (a)~Both $\mathcal{D}_{\parallel (3) d}$ and $\mathcal{D}_{\parallel (2) d}$ are unity  
at $\alpha_p = 1$ and $|\bm{p}_{pT}| = 0$, where $|\bm{k}| = 0$ and only the S-wave is present. 
(b)~$\mathcal{D}_{\parallel (3) d}$ and $\mathcal{D}_{\parallel (2) d}$ 
remain close to unity for $|\bm{p}_{pT}| \lesssim$ 150 MeV, where the S-wave dominates. The D-wave 
contributions raise $\mathcal{D}_{\parallel (3) d}$ above unity but lower $\mathcal{D}_{\parallel (2) d}$, in accordance with the 
bound Eq.~(\ref{D_2_bound}), showing the effect of the tensor polarized structure in $\mathcal{D}_{\parallel (2) d}$.
(c)~Both $\mathcal{D}_{\parallel (3) d}$ and $\mathcal{D}_{\parallel (2) d}$ decrease significantly at $|\bm{p}_{pT}| \gtrsim$ 150 MeV,
pass through zero at $|\bm{p}_{pT}| \approx$ 300 MeV, where the combination $(f_0 - f_2/\sqrt{2})$ vanishes,
and become negative at larger momenta, where the D-wave dominates.

%
%
\begin{figure}
\includegraphics[width=.48\textwidth]{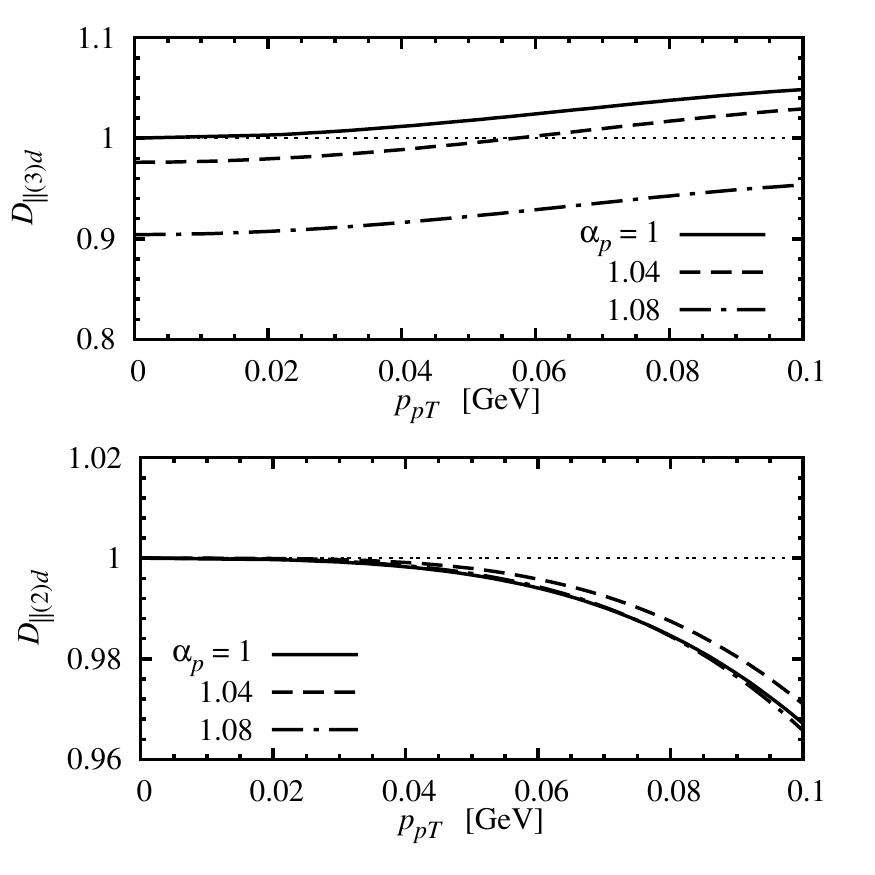}
\caption[]{The deuteron depolarization factors $\mathcal{D}_{\parallel (3) d}$ (three-state asymmetry, upper graph)
and $\mathcal{D}_{\parallel (2) d}$ (two-state asymmetry, lower graph), as functions of $|\bm{p}_{pT}|$, 
for several values of $\alpha_p$. Note the different scales on the vertical axis in the 
upper and lower plots.}
\label{fig:depol_pt_aplha}
\end{figure}
Important differences between the three-state and two-state depolarization factors appear in the behavior at
small proton momenta. Because the factors $C_0, C_2$ and $C_T$ become unity at $\bm{k}_T = 0$,
\be
\mathcal{D}_{\parallel (2) d} \; &=& \; 1 \hspace{1em}
\textrm{at} \hspace{1em} \textrm{$|\bm{p}_{pT}| = 0$, \, $\alpha_p$ arbitrary.}
\ee
In contrast, $\mathcal{D}_{\parallel (3) d} \neq 1$ for $|\bm{p}|_{pT} = 0$ and $\alpha_p$ arbitrary; the value 
at $|\bm{p}_{pT}| = 0$ is unity only if $\alpha_p = 1$. Figure~\ref{fig:depol_pt_aplha} shows $\mathcal{D}_{\parallel (2) d}$
and $\mathcal{D}_{\parallel (3) d}$ as functions of $|\bm{p}_{pT}|$ for several values of $\alpha_p$. One observes 
that at small proton momenta the values of $\mathcal{D}_{\parallel (2) d}$ are much closer to unity than those of 
$\mathcal{D}_{\parallel (3) d}$. This difference can also be seen in the expressions Eqs.~(\ref{D_3_cm}) and (\ref{D_2_cm}),
when one substitutes the factors $C_0, C_2$ and $C_T$ by their approximate forms for small CM momenta 
$\bm{k} \ll m$,
\begin{subequations}
\begin{align}
C_0 \; &= \; 1 \; + \; {\mathcal O}(\bm{k}/m) ,
\\[2ex]
C_2 \; &= \; 1 - \frac{3|\bm{k}_T|^2}{|\bm{k}|^2} \; + \; {\mathcal O}(\bm{k}/m) ,
\\[1ex]
C_T \; &= \; 1 - \frac{3|\bm{k}_T|^2}{2 |\bm{k}|^2} \; + \; {\mathcal O}(\bm{k}/m) .
\end{align}
\end{subequations}
In this approximation
\begin{subequations}
\begin{align}
\lefteqn{\left( f_0 - \frac{f_2}{\sqrt{2}}\right) \left( C_0 f_0 - \frac{C_2 f_2}{\sqrt{2}} \right) } &&
\nonumber
\\[1ex]
&= \; f_0^2 - \left( 2 - \frac{3|\bm{k}_T|^2}{|\bm{k}|^2} \right) \frac{f_0 f_2}{\sqrt{2}}
+ \left( 1 - \frac{3|\bm{k}_T|^2}{|\bm{k}|^2} \right) \frac{f_2^2}{2} ,
\label{D_2_numerator_approx}
\\[1ex]
\lefteqn{f_0^2 + f_2^2 - C_T \left( 2 f_0 + \frac{f_2}{\sqrt{2}} \right) \frac{f_2}{\sqrt{2}}} &&
\nonumber
\\[1ex]
&= \; f_0^2 - \left( 2 - \frac{3|\bm{k}_T|^2}{|\bm{k}|^2} \right) \frac{f_0 f_2}{\sqrt{2}}
+ \left( 1 + \frac{3|\bm{k}_T|^2}{2|\bm{k}|^2} \right) \frac{f_2^2}{2} .
\label{D_2_denominator_approx}
\end{align}
\end{subequations}
One sees that in $\mathcal{D}_{\parallel (2) d}$ the numerator and denominator coincide in terms linear in the 
D-wave and differ only in terms quadratic in the D-wave; in $\mathcal{D}_{\parallel (3) d}$ the numerator and
denominator differ already in terms linear in the D-wave. Consequently,
\begin{align}
\left.
\begin{array}{lcl}
\mathcal{D}_{\parallel (2) d} \; &=& \; 1 \; + \; \textrm{terms} \; f_2^2/f_0^2
\\[1ex]
\mathcal{D}_{\parallel (3) d} \; &=& \; 1 \; + \; \textrm{terms} \; f_2/f_0
\end{array}
\right\} \hspace{1em} (|\bm{k}| \ll m).
\end{align}
D-wave corrections affect $\mathcal{D}_{\parallel (2) d}$ at only quadratic order at low momenta, but 
$\mathcal{D}_{\parallel (3) d}$ already at linear order.

The deviation of $\mathcal{D}_{\parallel (2) d}$ from unity can be computed using a simple approximate formula.
At CM momenta $k \lesssim 0.1$ GeV the ratio of D- and S-wave functions is approximately given by
\beq
\frac{f_2(k)}{f_0(k)} \; \approx \; \frac{k^2}{\lambda^2} ,
\eeq
where $\lambda^2 \approx 0.087$ GeV$^2$ for the AV18 wave functions \cite{Wiringa:1994wb}. 
Expanding the ratio of Eqs.~(\ref{D_2_numerator_approx}) and (\ref{D_2_denominator_approx})
to first order in $f_2^2/f_0^2$, we then obtain
\begin{align}
\mathcal{D}_{\parallel (2) d}
\; &\approx \;
1 - \frac{9 |\bm{k}_T|^2}{4 |\bm{k}|^2} \left[ \frac{f_2(k)}{f_0(k)} \right]^2
\nonumber
\\[1ex]
\; &\approx \;
1 - \frac{9 \, |\bm{k}_T|^2 \, |\bm{k}|^2}{4 (\lambda^2)^2} .
\end{align}
The expression of the CM momentum variable in terms of the original LF variables $\alpha_p$ and 
$\bm{p}_{pT}$ is given in Eq.~(\ref{CM_LF_variables}).
\section{Neutron spin structure extraction}
\label{sec:neutron_spin_structure}
\subsection{Analytic properties of light-front wave function}
We are now ready to study the extraction of neutron structure functions from tagged deuteron DIS measurements. 
In general, for arbitrary nonzero values of the proton momentum, the tagged cross section is modified 
by initial-state nuclear binding effects beyond those included in the $pn$ LF wave function 
(non-nucleonic degrees of freedom, modified nucleon structure), and by final-state interactions 
beyond the IA (scattering of slow hadrons in the DIS final state from the spectator nucleon).
These effects can be eliminated by pole extrapolation in the proton
momentum \cite{Sargsian:2005rm,Strikman:2017koc}. The method uses the analytic properties of the 
deuteron wave function to select $pn$ configurations in the deuteron with infinite transverse separation, 
in which both nuclear binding effects and final-state interactions are suppressed.
It permits model-independent extraction of free neutron structure.
Here we review its use in the unpolarized case, and then apply it to the polarized case.

%
%
\begin{figure}[t]
\includegraphics[width=.45\textwidth]{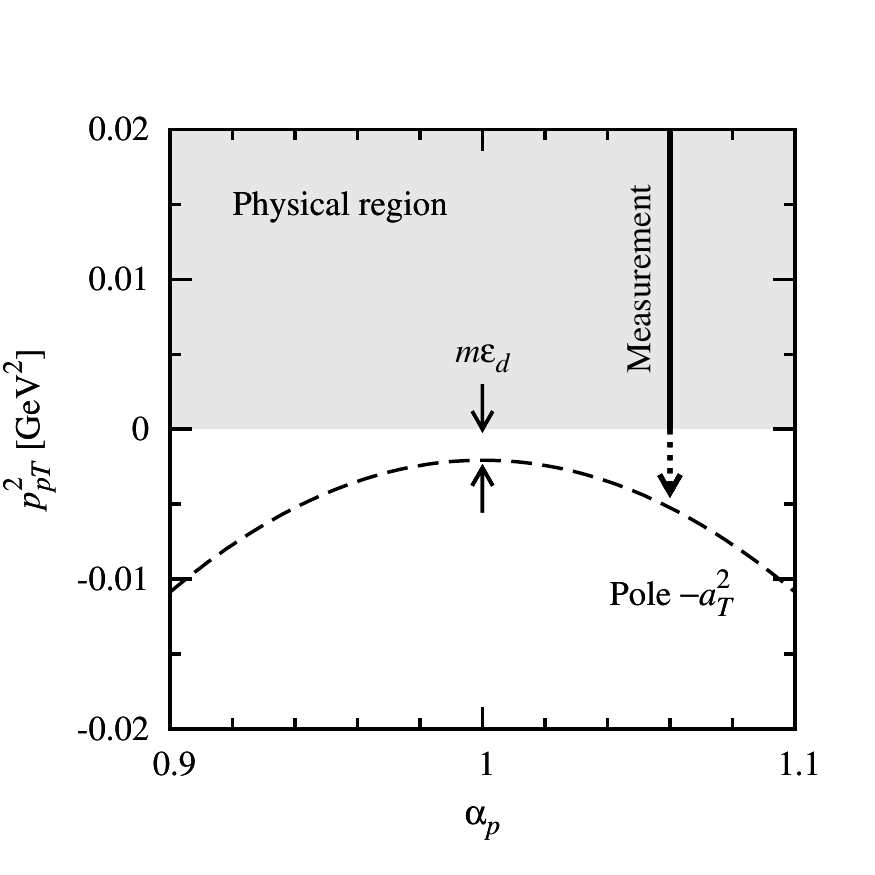}
\caption[]{The kinematic variables $\alpha_p$ and $p_{pT}^2 \equiv |\bm{p}_{pT}|^2$ in neutron structure 
measurements with pole extrapolation. The physical region $p_{pT}^2 > 0$ is shown by the shaded area.
The pole position at $p_{pT}^2 = - a_T^2$ is shown by the dashed line. The tagged structure 
functions are measured at fixed $\alpha_p$ as functions of $p_{pT}^2$ in the physical region
as indicated by the solid line. The measurements are then extrapolated to the pole as indicated 
by the dotted line.}
\label{fig:alpha_pt_pole}
\end{figure}
To discuss the analytic properties in the proton transverse momentum, we regard the deuteron wave function 
as a function of the invariant mass of the $pn$ pair, Eq.~(\ref{invariant_mass_def})
\begin{align}
M_{pn}^2 \; &= \; \frac{4 (|\bm{p}_{pT}|^2 + m^2)}{\alpha_p (2 - \alpha_p)},
\label{inv_mass}
\end{align}
which takes values $M_{pn}^2 \geq 4 m^2$ for physical proton momenta. In this variable the wave function 
has a pole singularity of the form (we suppress the spin structure for the moment)
\begin{align}
\Psi_d (\alpha_p, \bm{p}_{pT}) \; &= \; \frac{\textrm{Residue}}{M_{pn}^2 - M_d^2} \; + \; \textrm{(less singular)} ,
\label{pole_inv_mass}
\end{align}
which follows from the general properties of the LF bound-state equation. It lies outside 
the physical region of proton momenta because $M_{pn}^2 \geq 4 m^2 > M_d^2$. The pole in 
the invariant mass implies a pole in the transverse momentum of the form
\begin{align}
\Psi_d (\alpha_p, \bm{p}_{pT}) \; &= \; \frac{\textrm{Residue}}{|\bm{p}_{pT}|^2 + a_T^2} \; + \; \textrm{(less singular)} ,
\label{pole_transverse}
\\[1ex]
a_T^2 \; &= \; a_T^2(\alpha_p) \; \equiv \; m^2 - \alpha_p (2 - \alpha_p) \frac{M_d^2}{4} .
\label{a_T_def}
\end{align}
It lies in the unphysical region $|\bm{p}_{pT}|^2 < 0$
(see Fig.~\ref{fig:alpha_pt_pole}). The pole position $a_T^2$ depends on $\alpha_p$.
The minimum value of $a_T^2$ occurs at $\alpha_p = 1$ and is 
\begin{align}
a_T^2[\textrm{min}] = m^2 - \frac{M_d^2}{4} = m \epsilon_d + \mathcal{O}(\epsilon_d^2) = a^2 ,
\label{a_T_min}
\end{align}
where $\epsilon_d \equiv 2 m - M_d$ is the deuteron binding energy and $a^2 \equiv m \epsilon_d$ 
is the inverse squared Bethe-Peierls radius of the deuteron. 

The singularity in the transverse momentum Eq.~(\ref{pole_transverse}) has a simple physical interpretation.
It is related to the asymptotic behavior of the transverse coordinate-space wave function at large transverse separations 
of the $pn$ pair. When the transverse momentum-space wave function is represented as a Fourier transform of the 
transverse-coordinate wave function, the singularity arises from the contribution of large transverse
separations $r_T \rightarrow \infty$ to the Fourier integral. In this sense the singularity describes the
presence of $pn$ configurations with large (infinite) transverse separation in the deuteron wave function 
at unphysical momenta. (The interpretation of the pole in the invariant formulation in terms of
Feynman diagrams is discussed below.)

In the representation of the deuteron LF wave function in terms of the CM momentum,
the invariant mass of the $pn$ configuration is
\begin{align}
M_{pn}^2 \; &= \; 4 (|\bm{k}|^2 + m^2) .
\end{align}
The pole Eq.~(\ref{pole_inv_mass}) appears through the pole of the S-wave radial function
\begin{align}
f_0(k) \; &= \; \frac{\sqrt{m} \, \Gamma}{|\bm{k}|^2 + a^2} \; + \; \textrm{(less singular)} ,
\label{pole_radial}
\end{align}
which is obtained from the 3-dimensional bound state equation and represents a general feature 
of the weakly bound system. The residue $\Gamma$ can be inferred from the non-relativistic
approximation (see Sec.~\ref{subsec:nonrel}).\footnote{In the nonrelativistic approximation
of Eq.~(\ref{nonrel_approx}), the residue of the relativistic wave function would involve the
CM energy evaluated at the pole, $E(k^2 = - a^2) = \sqrt{m^2 - a^2}$. We neglect the small effect 
of the binding energy on the residue and set $E(k^2 = - a^2) \approx m$. We note that corrections 
to this approximation should not be inferred from Eq.~(\ref{nonrel_approx}), since the factor 
$\sqrt{E}$ is motivated by the normalization condition involving large CM momenta.}
This again shows the close correspondence between the LF and the nonrelativistic description
of the two-body bound state.

%
%
\begin{figure}
\includegraphics[width=.43\textwidth]{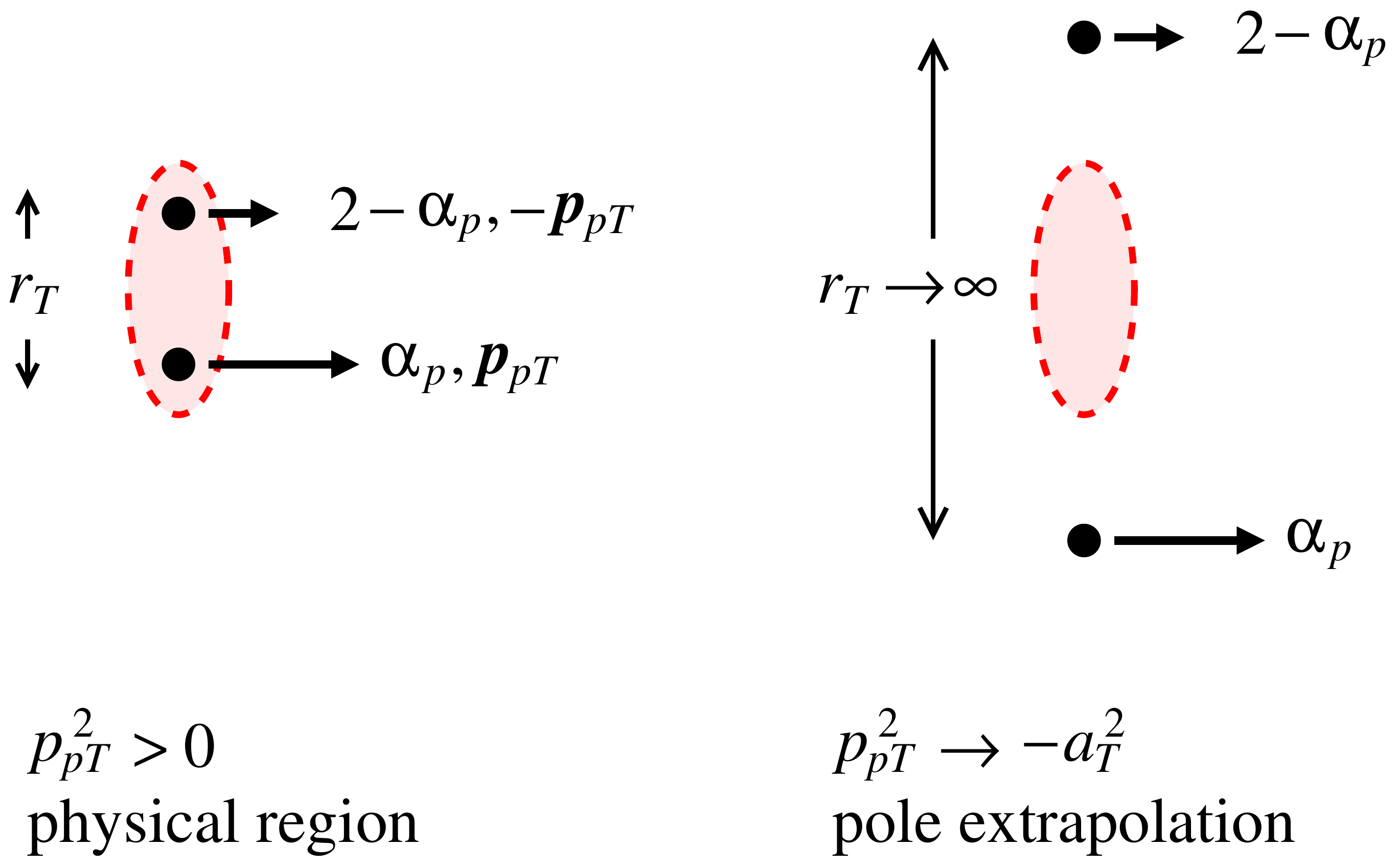}
\caption[]{Coordinate-space interpretation of the pole extrapolation.
Left: Physical transverse momenta $p_{pT}^2 > 0$ correspond to $pn$ configurations of
finite transverse separations of the order of the order of the deuteron size.
Right: The pole at $p_{pT}^2 = - a_T^2$ corresponds to $pn$ configurations with
infinite transverse separatio. The extrapolation $p_{pT}^2 \rightarrow - a_T^2$ effectively
selects such large-size configurations.}
\label{fig:pole_coordinate}
\end{figure}
In the large-separation $pn$ configurations described by the singularity Eq.~(\ref{pole_transverse}),
nuclear binding effects are absent, as such configurations are outside of the range of the 
nucleon-nucleon interactions. If tagged DIS could be performed in such configurations, the IA
cross section could therefore be computed exactly in terms of the \textrm{free neutron} structure function.
Likewise, final-state interactions would be suppressed, since they require the active nucleon and
the spectator to be aligned along the reaction axis. In this way one could effectively realize DIS 
on the free neutron and use it to extract free neutron structure from tagged DIS.
Obviously this is not possible in measurements at physical proton momenta, as the singularity
representing the large-separation $pn$ configurations lies in the unphysical region.
However, the singularity can be reached by analytic continuation in $|\bm{p}_{pT}|^2$ 
to unphysical negative values,
\begin{align}
|\bm{p}_{pT}|^2 \; &\rightarrow \; -a_T^2 .
\label{extrapolation_pt}
\end{align}
Because the values of $a_T^2$ for $\alpha \approx 1$ are small, Eq.~(\ref{a_T_def}) and (\ref{a_T_min}),
the singularity is very close to the physical region and can be reached by extrapolation in $|\bm{p}_{pT}|^2$. 
This opens a practical way of accessing non-interacting large-size $pn$ configurations in tagged
DIS on the deuteron and extracting the free neutron structure functions
(see Fig.~\ref{fig:pole_coordinate}).\footnote{It is important
to understand that the tagged DIS cross section is not ``diagonal'' in the transverse coordinate
representation, i.e., it cannot be viewed as taking place at a fixed impact parameter in a transversely
localized deuteron state. The cross section and the deuteron spectral function are given
by the product of the deuteron LF wave functions in the transverse momentum representation,
Eq.~(\ref{deuteron_tensor_nondiagonal_ia}),
which becomes a convolution integral in the transverse coordinate representation. The production
of the spectator nucleon thus happens at different transverse positions in the amplitude and
the complex-conjugate amplitude, and the observed transverse momentum $\bm{p}_{pT}$
is Fourier-conjugate to the vector difference between the positions. The pole extrapolation
of the cross section forces both deuteron wave functions in the convolution integral
to be in large-size configurations. Our arguments are to be understood in this sense.}
\subsection{Pole extrapolation of unpolarized cross section}
In unpolarized tagged DIS the free neutron structure functions are extracted as follows.
The unpolarized deuteron spectral function Eq.~(\ref{neutron_density_deut_unpol}) has a pole in $|\bm{p}_{pT}|^2$
\begin{align}
& \frac{2 \mathcal{S}_{d} (\alpha_p, \bm{p}_{pT}) [\textrm{unpol}]}{2 - \alpha_p}
\nonumber \\[1ex]
&\sim \;
\frac{R}{(|\bm{p}_{pT}|^2 + a_T^2)^2} 
\nonumber \\[2ex]
& + \; \textrm{(less singular for $|\bm{p}_{pT}|^2 \rightarrow -a_T^2$)},
\end{align}
with residue
\begin{align}
R \; &\equiv \; R(\alpha_p) \; = \; 2 \alpha_p^2 m \Gamma^2 ,
\end{align}
where $\Gamma$ is the residue of the 3-dimensional wave function in Eq.~(\ref{pole_radial}).
The unpolarized tagged structure functions therefore have a pole
\begin{align}
& F_{[UU, T]d}(x, Q^2; \alpha_p, \bm{p}_{pT})
\nonumber \\[1ex]
&\sim \;
[2 (2\pi)^3] \;
\frac{R}{(|\bm{p}_T|^2 + a_T^2)^2} \; F_{[UU, T]n} (\widetilde x, Q^2)
\nonumber
\\[2ex]
&+ \; \textrm{(less singular)},
\label{analytic_general}
\end{align}
and similarly for $F_{[UU, L]d}$. The arguments presented above imply that 
this pole is a general feature of the tagged structure function; the pole is contained 
completely in the IA cross section; the residue is given by the free neutron 
structure function. To extract the free neutron structure functions, one follows these
steps (see Fig.~\ref{fig:alpha_pt_pole}): (i) Measure the unpolarized tagged DIS cross section for fixed $\alpha_p$ as 
a function of $|\bm{p}_{pT}|^2$ in the physical region $|\bm{p}_{pT}|^2 > 0$.
The ranges of these variables are chosen such that they 
correspond to nucleon CM momenta of the order of the typical nucleon momenta in 
the deuteron, $|\bm{k}| \lesssim$ 100 MeV (ideally, much smaller than that), which implies
\begin{align}
0.9 \lesssim \; \alpha_p \; \lesssim 1.1,
\hspace{2em}
0 < \; |\bm{p}_{pT}| \; \lesssim \textrm{0.1 GeV} .
\label{range_onshell}
\end{align}
(ii) Tabulate the measured tagged deuteron structure functions in $|\bm{p}_{pT}|^2$, and
multiply by 
\begin{align}
\frac{(|\bm{p}_{pT}|^2 + a_T^2)^2}{R}
\end{align}
to remove the pole factor.
(iii) Extrapolate to $|\bm{p}_{pT}|^2 \rightarrow -a_T^2$:
\begin{align}
F_{[UU, T]n} (\widetilde x, Q^2)
\; &= \; \lim_{|\bm{p}_{pT}|^2 \rightarrow -a_T^2} \left[ \frac{(|\bm{p}_{pT}|^2 + a_T^2)^2}{R}
\right.
\nonumber \\[1ex]
& \left. \times  \; F_{[UU, T]d}(x, Q^2; \alpha_p, \bm{p}_{pT}) \phantom{\frac{0}{0}} \hspace{-.6em}\right] .
\label{extrapolation}
\end{align}
In the region of $|\bm{p}_{pT}|$ sufficiently close to zero, where there are no other singularities 
besides the nucleon pole, the extrapolation can be performed through a polynomial fit
and is model-independent. Notice that the neutron structure function on the left-hand 
side of Eq.~(\ref{extrapolation}) depends only on the effective scaling variable 
$\tilde x = x / (2 - \alpha_p)$, not on $x$ and $\alpha_p$ individually. This makes it 
possible to extract the structure function at the same value of $\tilde x$ using measurements
at different $x$ and $\alpha_p$, thus testing the extraction procedure. 

\begin{figure}
\includegraphics[width=.48\textwidth]{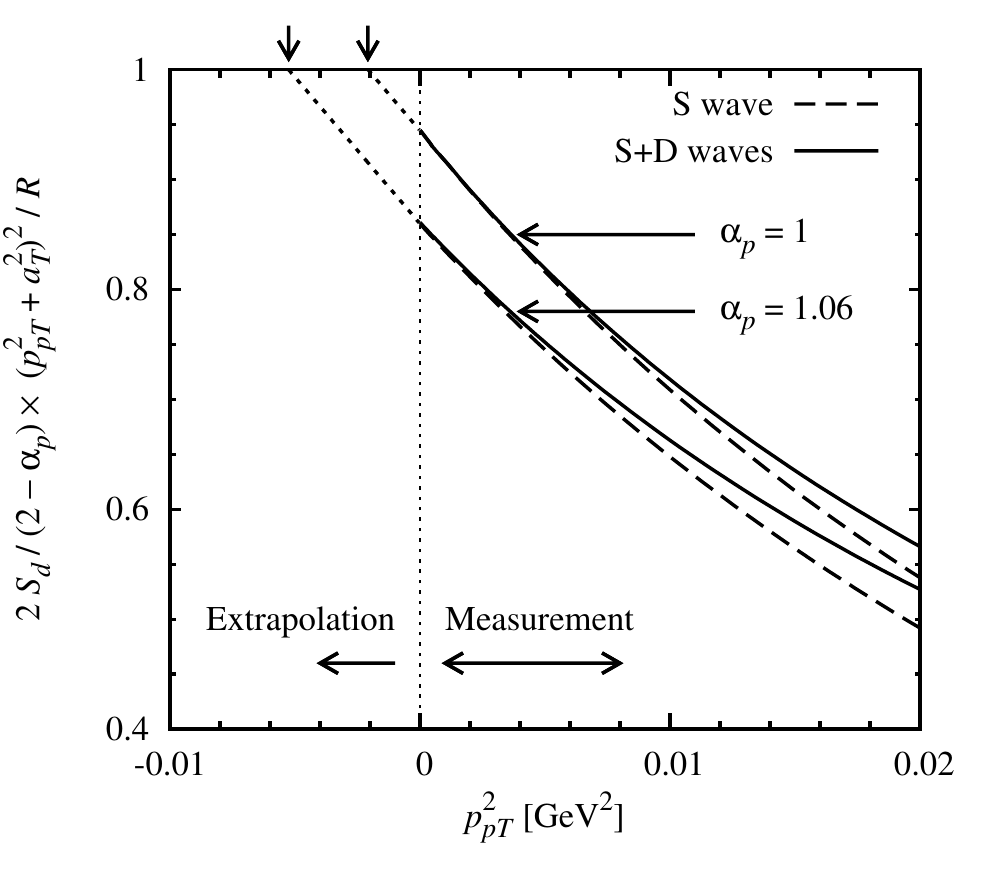}
\caption[]{Pole extrapolation of the unpolarized tagged deuteron structure function.
The plot shows the unpolarized deuteron spectral functions with the pole factor extracted,
Eq.~(\ref{extrapolation_normalized}), at fixed values of $\alpha_p$,
as functions of $p_{pT}^2$ in the physical region $p_{pT}^2 > 0$.
Dashed line: S wave contribution only.
Solid line: S+D waves (total). Dotted line: Extrapolation to $p_{pT}^2 = -a_T^2$.
The pole positions $-a_T^2$ are indicated by the vertical arrows above the plot.}
\label{fig:extrapolation_unpol}
\end{figure}

Figure~\ref{fig:extrapolation_unpol} shows the $|\bm{p}_{pT}|^2$ dependence of the 
IA spectral function with the pole factor removed,
\begin{align}
&\frac{2 \mathcal{S}_{d}}{2 - \alpha_p} \times \frac{(|\bm{p}_{pT}|^2 + a_T^2)^2}{R} ,
\nonumber
\\[1ex]
&\lim_{|\bm{p}_{pT}|^2 \rightarrow -a_T^2} \; [...] \; = \; 1 ,
\label{extrapolation_normalized}
\end{align}
which represents the normalized $|\bm{p}_{pT}|^2$ dependence of the expression in brackets in Eq.~(\ref{extrapolation}) 
in the IA. One observes that: (a) The expected variation of the expression Eq.~(\ref{extrapolation_normalized})
over the range used for extrapolation, Eq.~(\ref{range_onshell}), is $<50\%$. (b) The D-wave contribution is small 
throughout this range and vanishes at $|\bm{p}_{pT}| = 0$. (c) The extrapolation distance increases as
$\alpha_p$ moves away from 1, because the pole position $-a_T^2$ moves away from the physical region;
see Fig.~\ref{fig:alpha_pt_pole} and Eq.~(\ref{a_T_def}). Note that further modifications of the 
$|\bm{p}_{pT}|$ dependence of Eq.~(\ref{extrapolation_normalized}) arise from FSI; 
however, they vanish at the pole $|\bm{p}_{pT}|^2 \rightarrow -a_T^2$ and do not affect the 
result of the extrapolation \cite{Sargsian:2005rm,Strikman:2017koc}.
%
%

We have presented here the pole extrapolation in the context of LF quantization, where it
emerges from the analytic properties of the deuteron LF wave function. An alternative view 
is obtained in the context of the invariant formulation of electron-deuteron scattering
using Feynman diagrams with 4-momentum conservation and off-shell nucleons.
There one regards the tagged DIS cross section as a function of the invariant 4-momentum 
transfer between the deuteron and the proton, 
\begin{align}
t' \; &\equiv \; (p_d - p_p)^2 - m^2, 
\end{align}
which is a function of $\alpha_p$ and $|\bm{p}_{pT}|$. The kinematic variable $t'$ 
can be interpreted as the ``off-shellness'' of the active neutron in the Feynman diagram sense.
The $e + d \rightarrow e' + X + p$ cross section has a pole at $t' = 0$, 
corresponding to ``neutron exchange'' between the deuteron breakup and the DIS process.
The residue at the pole is given by the product of the $d \rightarrow p + n$ vertex function
squared and the free on-shell neutron structure function; this can be derived formally from 
general principles of S-matrix theory (residue factorization). The pole extrapolation of the
cross section is therefore referred to as ``on-shell extrapolation.'' 
The connection with the LF formulation is established by noting that 
\begin{align}
t' \; &= \; - \frac{2 - \alpha_p}{2} (M_{pn}^2 - M_d^2) 
\nonumber \\[1ex]
&= \; - \frac{2}{\alpha_p} (|\bm{p}_{pT}|^2 + a_T^2) .
\end{align}
The off-shellness in the Feynman formulation is proportional to the invariant mass
difference in the LF formulation. The approach to the on-shell point in the Feynman 
formulation corresponds to the selection of large-distance configurations and the
vanishing of binding in the LF formulation. Further aspects of the Feynman formulation of the 
pole extrapolation (absence of FSI at the pole, magnitude of FSI effects at $t' \neq 0$)
are discussed in Refs.~\cite{Sargsian:2005rm,Strikman:2017koc}.
\subsection{Pole extrapolation of spin asymmetries}
The neutron spin structure functions could be extracted from the polarized tagged deuteron
structure functions in the same manner as Eq.~(\ref{extrapolation}), by performing the
pole extrapolation at the level of the structure functions. However, it is more convenient
to perform the extrapolation at the level of the spin asymmetries. Advantages of this
approach are: (a)~the nucleon pole factors in the structure functions cancel between the 
numerator and denominator of the asymmetry, so that one does not have to remove them externally; 
(b)~the asymmetries overall depend only weakly on $|\bm{p}_{pT}|^2$;
(c)~certain systematic experimental uncertainties cancel between the numerator and denominator.

For the neutron spin structure extraction we use the two-state asymmetry, in which the deuteron 
structure effects at small proton momenta cancel to a higher degree than in the three-state asymmetry
(see Sec.~\ref{subsec:spin_asymmetries_ia}). In the IA, which contains the pole terms in the
cross sections and becomes exact at the pole, the asymmetry is given by Eq.~(\ref{A_2_factorized}), 
\begin{align}
& A_{\parallel (2) d} (x, Q^2; \alpha_p, \bm{p}_{pT})
\nonumber \\[1ex]
&= \;
\mathcal{D}_{\parallel (2) d}(\alpha_p, \bm{p}_{pT}) \; A_{\parallel n} (\widetilde{x}, Q^2) .
\end{align}
The $|\bm{p}_{pT}|$ dependence of the deuteron depolarization factor $\mathcal{D}_{\parallel (2) d}$
near the pole can be determined from the explicit expression in terms of the CM momentum,
Eq.~(\ref{D_2_cm}), by expanding in $|\bm{p}_{pT}|/m$, counting the binding energy as
$\epsilon_d/m \sim |\bm{p}_{pT}|/m$, and retaining only the S-wave radial wave function,
\begin{align}
& \mathcal{D}_{\parallel (2) d}(\alpha_p, |\bm{p}_{pT}|^2)
\nonumber \\[1ex]
&= \; 1 \; + \; \frac{(\alpha_p - 1)^2}{2 (2 - \alpha_p)}
\nonumber 
\\[1ex]
&+ \; 
\frac{1}{2 (2 - \alpha_p)} \left[1 - \frac{(\alpha_p - 1)^2}{4} \right]
\left[ \frac{\epsilon_d}{m} - \frac{|\bm{p}_{pT}|^2 + a_T^2}{\alpha_p (2 - \alpha_p) m^2} \right]
\nonumber 
\\[1ex]
&+ \; \textrm{terms} \, \left\{ \frac{(|\bm{p}_{pT}|^2 + a_T^2)^2}{m^4},
\frac{(m\epsilon_d)^2}{m^4}, \frac{m\epsilon_d (|\bm{p}_{pT}|^2 + a_T^2)}{m^4} \right\} .
\label{D_2_pt2_unphysical}
\end{align}
The approximate expression is valid in the physical and unphysical regions and can be used in the extrapolation.

%
%
\begin{figure}
\includegraphics[width=.48\textwidth]{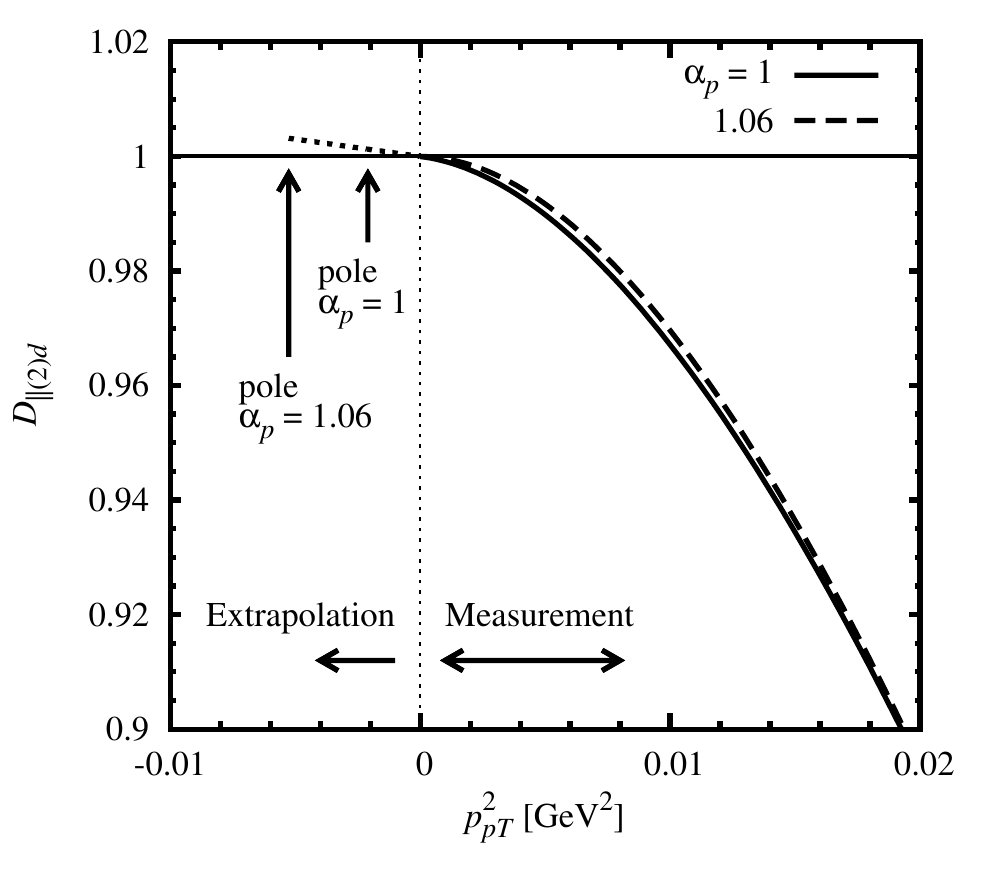}
\caption[]{Pole extrapolation of the deuteron depolarization factor $\mathcal{D}_{\parallel (2) d}$.
Solid line: $\mathcal{D}_{\parallel (2) d}$ at $\alpha_p = 1$ as a function of $p_{pT}^2$
in the physical region $p_{pT}^2 > 0$, evaluated using Eq.~(\ref{D_2_cm}).
Dashed line: Same function for $\alpha_p = 1.06$.
Dotted line: $\mathcal{D}_{\parallel (2) d}$ in the unphysical region $p_{pT}^2 < 0$,
evaluated using Eq.~(\ref{D_2_pt2_unphysical}). The functions for the two values
of $\alpha_p$ are practically the same and show up as a the single dotted line in the plot.
The positions of the pole at $p_{pT}^2 = -a_T^2$ for the two values of $\alpha_p$ are
indicated by the two vertical arrows.}
\label{extrapolation_pt2_depol}
\end{figure}
Figure~\ref{extrapolation_pt2_depol} shows $\mathcal{D}_{\parallel (2) d}$ at fixed $\alpha_p$ as a function of
$|\bm{p}_{pT}|^2$, including the physical region $|\bm{p}_{pT}|^2 > 0$,
where it is evaluated using Eq.~(\ref{D_2_cm}),
and the unphysical region $|\bm{p}_{pT}|^2 < 0$, where it is approximated by Eq.~(\ref{D_2_pt2_unphysical}).
One observes: (a)~The deviations of $\mathcal{D}_{\parallel (2) d}$ from unity
are $\lesssim$5\% for $0 < |\bm{p}_{pT}|^2 < 0.01$ GeV$^2$,
the range typically used for pole extrapolation.
(b)~The approximate expression in the unphysical region matches well with the 
exact result in the physical region at the boundary $|\bm{p}_{pT}|^2 = 0$.
(c) The values of $\mathcal{D}_{\parallel (2) d}$ at the pole are very close to unity
$|\mathcal{D}_{\parallel (2) d}(|\bm{p}_{pT}|^2 = - a_T^2) - 1| \lesssim 0.01$ for $|\alpha_p - 1| < 0.1$.

To extract the neutron spin asymmetry, one should measure the deuteron two-state asymmetry
$A_{\parallel (2) d}(x, Q^2; \alpha_p, \bm{p}_{pT})$ at fixed $\alpha_p$ over a range 
of $|\bm{p}_{pT}|$ down to $|\bm{p}_{pT}| = 0$, tabulate the result in $|\bm{p}_{pT}|^2$,
and extrapolate to $|\bm{p}_{pT}|^2 \rightarrow - a_T^2$.
Because of the very weak expected $|\bm{p}_{pT}|^2$ dependence, the extrapolation should be performed
by fitting it with a constant (zeroth-order polynomial); deviations from the constant would
be difficult to reproduce through a higher-order polynomial because of their complex 
$|\bm{p}_{pT}|^2$ dependence (instabilities) and are likely beyond the accuracy of any measurement. 
The neutron asymmetry is then obtained as
\begin{align}
&A_{\parallel n} (\widetilde{x}, Q^2)
\nonumber \\[1ex]
&= \frac{\lim_{|\bm{p}_{pT}|^2 \rightarrow -a_T^2} \; 
A_{\parallel (2) d}(x, Q^2; \alpha_p, \bm{p}_{pT})}
{\mathcal{D}_{\parallel (2) d} (\alpha_p, |\bm{p}_{pT}|^2 = - a_T^2)} ,
\end{align}
where the denominator is given by Eq.~(\ref{D_2_pt2_unphysical}) at the pole
\begin{align}
& \mathcal{D}_{\parallel (2) d} (\alpha_p, |\bm{p}_{pT}|^2 = -a_T^2)
\nonumber \\[1ex]
&= \; 1 \; + \; \frac{1}{2 (2 - \alpha_p)} \left\{ (\alpha_p - 1)^2
\; + \; \left[1 - \frac{(\alpha_p - 1)^2}{4} \right]  \frac{\epsilon_d}{m} \right\} 
\nonumber \\[1ex]
&\approx \; 1 \; + \; \frac{1}{2 (2 - \alpha_p)} \left[ (\alpha_p - 1)^2
\; + \; \frac{\epsilon_d}{m} \right] .
\end{align}
A similar extrapolation procedure could be formulated for the three-state asymmetry
$A_{\parallel (3) d}$, Eq.~(\ref{A_3_factorized}). 
\section{Summary and perspective}
\label{sec:summary}
We have presented a theoretical framework for polarized electron-deuteron DIS with spectator nucleon 
tagging, suitable for measurements with a future EIC. It includes the general form of the observables 
(cross sections, spin asymmetries), the description of nuclear structure in the high-energy process 
(LF quantization, deuteron LF wave function, spin effects, nucleon momentum distributions, IA), 
and the prospects for neutron structure extraction (proton momentum dependence, pole extrapolation).
We now summarize the main theoretical and practical conclusions of our work and suggest directions 
for further study. 

The main theoretical conclusions, related to the description of polarized deuteron structure in LF 
quantization, are:

\textit{(a) Advantages of 4-dimensional formulation of LF spin structure.}
The 4-dimensional formulation of the deuteron LF spin structure offers several advantages
compared to the 3-dimensional one. It avoids the use of explicit Melosh rotations, 
permits efficient evaluation of the spin sums as bispinor traces, and deals with expressions 
whose form is familiar from covariant field theory (neutron density matrix, polarization 4-vector). 
Rotational invariance is ensured by the constraints resulting from the covariant dependence of 
the nucleon polarization 4-vector on the deuteron polarization 4-vector and tensor.

\textit{(b) Simplicity of CM momentum variable.}
The use of the CM momentum variable in the two-body LF wave function further simplifies the 
calculation and interpretation of the results. It results in expressions that have a form similar 
to the nonrelativistic wave function and densities and exhibit 3-dimensional rotational symmetry
(angular momentum content, S and D waves). The fact that the $pn$ CM frame is related to the
deuteron rest frame by a collinear boost permits an extremely simple representation of the
spin structure, in which the deuteron polarization 3-vector and 3-tensor in the rest frame
directly determine the 3-dimensional vector and tensor polarization of the $pn$ configuration
in the CM frame.

\textit{(c) Coordinate-space picture of LF wave function pole.}
The pole in the deuteron LF wave function describes the behavior of the two-body system 
at asymptotically large transverse distances and fixed LF momenta. This interpretation explains
both the absence of nuclear binding at the pole and the absence of FSI in scattering processes
after extrapolation to the unphysical pole kinematics. It appears naturally in LF quantization
and is alternative to the interpretation as a mass-shell singularity (nucleon pole)
in the Feynman formulation.

The main practical conclusions, related to the results for tagged DIS observables and the 
the prospects for neutron structure extraction, are:

\textit{(d) Control of D-wave through tagged proton momentum.} The D-wave contribution to the polarized
cross section is controlled by the proton momentum and is absent at zero transverse momentum,
$|\bm{p}_{pT}| = 0$, for any longitudinal momentum $\alpha_p$. This remarkable feature is due to the 
LF description of the deuteron's spin structure, which permits relativistic spin rotations of the
3-dimensional D-wave only at nonzero transverse momentum. It offers a practical way to eliminate
the D-wave depolarization and reduce the quantum system to a completely polarized configuration.

\textit{(e) Neutron spin structure extraction from two-state spin asymmetry.}
The neutron spin structure function $g_{1n}$ can be extracted efficiently from the tagged deuteron 
spin asymmetry between the two longitudinal spin states with projection $\pm 1$ only.
The fact that the spin state with projection 0 is not needed represents an important simplification,
as the systematic uncertainties in the preparation of the $\pm 1$ states are likely correlated 
and cancel in the asymmetry. While the two-state asymmetry appears theoretically more complex because 
it involves tensor polarization in the denominator, it is in fact a simpler observable than the 
three-state asymmetry.

\textit{(f) Simple pole extrapolation.} The two-state spin asymmetry
depends only weakly on the proton transverse momentum; the variation is
$\lesssim$5\% for $|\bm{p}_{pT}| <$ 100 MeV. The pole extrapolation can therefore
safely be performed by fitting with a constant, with very small theoretical corrections.
Altogether, our results provide a practical procedure for extracting neutron spin structure 
from future polarized tagged measurements with minimal theoretical and experimental uncertainties.

The methods developed in this work can be applied to other types of tagged DIS measurements 
on the deuteron. This includes: (a)~Measurements of tensor-polarized tagged cross sections 
or spin asymmetries, with the aim to extract the leading-twist tensor-polarized structure 
functions $F_{[UT_{LL}, T]d}$ and $F_{[UT_{LL}, L]d}$. The IA formulas for these observables
can easily be obtained by combining the expressions given in Secs.~\ref{sec:scattering} and \ref{sec:ia}.
(b)~Measurements of the vector-polarized transverse spin asymmetries, with the aim to extract the 
power-suppressed tagged spin structure functions $F_{[LS_T]d}$ or $g_{2d}$. The IA formulas for 
these observables can be derived from the general expressions given in 
Secs.~\ref{sec:scattering} and \ref{sec:ia}. The calculation of these power-suppressed
structure requires an assessment of the interaction effects in the $T$ components of the
LF current operator (see comments in Sec.~\ref{subsec:ia}).
(c)~Measurements of the azimuthal angle ($\phi_p$-) 
dependence of the polarized tagged cross section. Such measurements can offer new ways to probe the 
$\phi_p$-integrated structure functions considered in the present study. They can also access the 
numerous $\phi_p$-dependent responses sensitive to spin-orbit effects in the deuteron breakup, 
including time-reversal-odd (T-odd) structures that are zero in the IA and require FSI.
The general form of the polarized tagged DIS cross section including the azimuthal angle dependence,
and the LF calculation of the corresponding tagged structure functions, will be described in a
forthcoming article \cite{Cosyn:inprep}. (d)~Measurements of exclusive processes in tagged
high-energy scattering on the deuteron, such as deeply-virtual Compton scattering (DVCS) or
hard exclusive meson production, with the aim of extracting the generalized parton distributions
of the neutron; see Refs.~\cite{Goeke:2001tz,Belitsky:2001ns,Diehl:2003ny} for a review.
(e)~Studies of spin effects in deuteron breakup in diffractive high-energy scattering with
tagged nucleons \cite{Tu:2020ymk}.

The methods and results described here can also be applied to study the nuclear modifications of 
partonic spin structure (polarized EMC effect, antishadowing, shadowing) and its dynamical origin.
Spectator tagging effectively controls the size of the $pn$ configuration during the DIS processes
and can therefore provide insight into the dynamical origin of the nuclear modifications 
(dependence on the distance between the interacting nucleons, connection with short-range $NN$ correlations).
Such studies require tagged measurements at larger proton momenta $|\bm{p}_{pT}| \sim$ 300--600 MeV, 
where FSI effects are of the same order as the IA cross sections. FSI effects in unpolarized tagged DIS 
at $x \gtrsim 0.1$ were studied in Ref.~\cite{Strikman:2017koc}; the results could be extended to
polarized DIS with appropriate dynamical input (spin-dependent nucleon fragmentation). Investigations
of the EMC effect with spectator tagging might be possible in kinematic regions where the FSI effects
are approximately independent of the proton momentum, so that they cancel in ratio observables 
(backward region, spectator momentum opposite to the $\bm{q}$ vector in the deuteron rest frame). 
How such studies could be extended to polarization observables and the spin-dependent EMC effect 
is an interesting question for further study.

The techniques used in the present study could in principle be extended to describe high-energy 
scattering on more complex light nuclei ($A > 2$) with detection of the nuclear breakup state,
e.g.\ DIS on $A = 3$ nuclei with quasi-two-body breakup, $e + {}^3{\rm He} ({}^3{\rm H}) \rightarrow e' + X + d$.
Such measurements could be used to test the universality of the extracted neutron structure, 
or to investigate the isospin dependence of nuclear modifications. The theoretical description of 
breakup processes with $A > 2$ nuclei is much more challenging than spectator nucleon tagging with the 
the deuteron because (a)~the initial state is much more complex (3-body system, $NN$ pairs with isospin 0 and 1);
(b)~the nuclear remnant in the final state is a multinucleon system with internal degrees of freedom
on the nuclear scale (bound states, scattering states); (c) the amplitude for the nuclear breakup in
the high-energy process generally involves multiple trajectories with different possibilities of FSI.
The matching of the high-energy scattering process and low-energy nuclear structure in these reactions
is therefore much more complex and requires a genuine interdisciplinary effort.
\section*{Acknowledgments}
We are indebted to M.~Sargsian and M.~Strikman for helpful advice during the course of this work. 
We have also benefited from the collaboration with V.~Guzey, D.~Higinbotham, Ch.~Hyde, Kijun~Park 
and P.~Nadel-Turonski on various aspects of spectator tagging and EIC simulations. We thank
Stijn van Geyte for carefully reading the manuscript and pointing out numerous errors.

This material is based upon work supported by the U.S.~Department of Energy, Office of Science, 
Office of Nuclear Physics, under contract DE-AC05-06OR23177. 
The work was supported in part by Jefferson Lab's Laboratory-Directed Research and
Development (LDRD) Project ``Physics potential of polarized light ions with EIC@JLab'' (2014/15).
\appendix
\section{Light-front helicity spinors}
\label{app:spinors}
In this appendix we present the explicit expressions of the bispinors describing nucleon LF helicity states,
which are used in the covariant formulation of the deuteron LF wave function (cf.\ Sec.~\ref{sec:deuteron}).
We describe the bispinors for both canonical and LF helicity states, and the Melosh rotation relating them.
We derive the expressions for the positive-energy ($u$-) spinors; the negative-energy ($v$-) spinors are then
obtained by applying the charge conjugation transformation; this ensures that they are defined consistently 
for both types of spin states. 

A spin-1/2 particle with mass $m$ and 4-momentum $p$ is described by a bispinor wave function $u(p, \lambda)$
satisfying
\begin{subequations}
\label{u_spinor_relations}
\begin{align}
(p\gamma - m) u \; &= \; 0,  
\\[2ex]
\bar u(p, \lambda') \, u(p, \lambda) \; &= \; 2 m \, \delta(\lambda', \lambda), 
\\[1ex]
\sum_{\lambda} u(p, \lambda) \, \bar u(p, \lambda)  \; &= \; p\gamma + m .
\end{align}
\end{subequations}
Here $\lambda = \pm 1/2$ denotes a generic spin quantum number characterizing the states. For canonical spin states, 
obtained by ordinary boosts of the bispinor at rest to the desired 3-momentum $\bm{p}$, 
the bispinor is of the form (in the standard representation of the gamma matrices)
\begin{align}
u_{\rm can}(p, \lambda) \; &= \; 
\left(
\begin{array}{l}
\sqrt{E + m} 
\\[1ex]
\sqrt{E - m} \; \bm{p}\bm{\sigma}/|\bm{p}|
\end{array} 
\right) \; \chi(\lambda) ,
\label{u_spinor_canonical}
\end{align}
where $\bm{\sigma} = (\sigma^1, \sigma^2, \sigma^3)$ are the Pauli matrices and 
$\chi (\lambda)$ is the two-component spinor describing the spin wave function of the particle at rest,
\begin{align}
\chi^\dagger (\lambda') \chi (\lambda) \; &= \; \delta(\lambda', \lambda) .
\end{align}
For LF helicity states, obtained by a LF boosts of the bispinor at rest to the desired LF momentum 
$p^+$ and $\bm{p}_T$, the bispinor is of the form 
\begin{subequations}
\label{u_spinor_LF}
\begin{align}
u_{\rm LF}(p, \lambda) &= \frac{1}{\sqrt{2 p^+}}
\left[ p^+ \gamma^- + (m - \bm{p}_T \bm{\gamma}_T) \gamma^+ \right]
\left( 
\begin{array}{l}
\chi (\lambda ) \\[1ex] 0 
\end{array} 
\right)
\\[1ex]
&=  
\frac{1}{\sqrt{2 p^+}}
\left( 
\begin{array}{r}
p^+ + m + \bm{p}_T \bm{\sigma}_T \sigma^3 \;
\\[1ex]
\sigma^3 (p^+ - m - \bm{p}_T \bm{\sigma}_T \sigma^3)
\end{array} 
\right) \, \chi (\lambda) ,
\end{align}
\end{subequations}
where again $\chi (\lambda)$ describes the spin wave function of the particle at rest. The bispinors 
Eqs.~(\ref{u_spinor_canonical}) and (\ref{u_spinor_LF}) are related by a rotation (Melosh rotation).
This rotation can be implemented as a transformation acting on the rest-frame 2-spinor,
\begin{subequations}
\label{melosh_spinor}
\begin{align}
u_{\rm LF}(p, \lambda) \; &= \; \left( 
\begin{array}{l} \sqrt{E + m}
\\[1ex]
\displaystyle \sqrt{E - m} \; \bm{p}\bm{\sigma}/|\bm{p}|
\end{array} 
\right) \; \mathcal{U}(p) \, \chi(\lambda) 
\\[1ex]
&= \; u_{\rm can}(p, \lambda)[\chi(\lambda) \rightarrow \mathcal{U}(p) \chi(\lambda)] ,
\label{u_spinor_rotation_phi}
\end{align}
\end{subequations}
where $\mathcal{U}$ is a unitary matrix in 2-spinor indices, 
\be
\mathcal{U}(p) &\equiv& \frac{p^+ + m + \bm{p}_T \bm{\sigma}_T \sigma^3}{\sqrt{2 p^+} \sqrt{E + m}} ,
\hspace{2em} \mathcal{U}^\dagger \mathcal{U} \; = \; 1.
\label{U_matrix_def}
\ee
In this interpretation, the LF bispinor is equal to the canonical bispinor with the same momentum, in which the 
rest-frame 2-spinor has been rotated by the matrix $\mathcal{U}$. Alternatively, the relation between
the bispinors Eqs.~(\ref{u_spinor_canonical}) and (\ref{u_spinor_LF}) can be expressed as a transformation 
acting on the spin quantum numbers. Using the completeness of the two-spinor basis one can write
\begin{subequations}
\label{melosh_quantum_numbers}
\begin{align}
\mathcal{U}_{ij}(p) \chi^j (\lambda) 
\; &= \; \sum_{\lambda'} \chi^i (\lambda') \chi^{k\ast}(\lambda') \mathcal{U}_{kj}(p) \chi^j (\lambda)
\nonumber \\[1ex]
&\equiv \; \sum_{\lambda'} \chi^i (\lambda') U(p, \lambda', \lambda),
\\[1ex]
U(p, \lambda', \lambda) \; &\equiv \; \chi^{k\ast}(\lambda') \mathcal{U}_{kj} (p) \chi^j (\lambda)
\nonumber \\[2ex]
&= \; \chi^\dagger (\lambda') \mathcal{U} (p) \chi(\lambda) ,
\label{U_function_def}
\\[2ex]
u_{\rm LF}(p, \lambda) \; &= \; \sum_{\lambda'} u_{\rm can}(p, \lambda') \, U(p, \lambda', \lambda) .
\end{align}
\end{subequations}
The function $U (p, \lambda', \lambda)$ represents the ``matrix element'' of the ``operator'' $\mathcal{U}$
between the spin states with quantum numbers $\lambda'$ and $\lambda$ (note that the order of the arguments
$\lambda'$ and $\lambda$ is important in this and the following expressions). In this interpretation, the
LF bispinor is equal to a certain combination of the canonical spinors corresponding to the same set
of rest-frame spin quantum numbers. 

The unitary matrix Eq.~(\ref{U_matrix_def}) is a function of the momentum of the state 
(expressed either in terms of its ordinary or the LF vector components) and satisfies
\begin{subequations}
\begin{align}
\mathcal{U}^\dagger (p^+, \bm{p}_T) \; &= \;
\mathcal{U}(p^+, \bm{p}_T \rightarrow -\bm{p}_T) ,
\\[1ex]
\mathcal{U}(p^+, \bm{p}_T = 0) \; &= 1 .
\end{align}
\end{subequations}
For zero transverse momentum the LF spinors coincide with the canonical ones, because in both cases
only longitudinal boosts (along the $z$-axis) are involved in the preparation of the states 
(see Fig.~\ref{fig:lf_helicity}).

The conjugate bispinors are defined as
\beq
\bar u \; \equiv \; u^\dagger \gamma^0
\eeq
for both canonical and LF spin states; their explicit expressions can be obtained by taking the 
hermitean conjugate of Eqs.~(\ref{u_spinor_canonical}) and (\ref{u_spinor_LF}) and mutiplying 
with $\gamma^0$ in the standard representation. The relation between the canonical and LF
conjugate spinors is
\begin{subequations}
\begin{align}
\bar u_{\rm LF}(p, \lambda) &= \bar u_{\rm can}(p, \lambda)[\chi^\dagger(\lambda) \rightarrow \chi^\dagger (\lambda) 
\mathcal{U}^\dagger(p)]
\\[2ex]
&= \sum_{\lambda'}
\bar u_{\rm can}(p, \lambda') U^\ast (p, \lambda', \lambda) .
\label{ubar_spinor_rotation_quantum_numbers}
\end{align}
\end{subequations}

Antiparticle states are described by the bispinor wave function $v(p, \lambda)$ satisfying 
[cf.\ Eq.~(\ref{u_spinor_relations})]
\begin{subequations}
\begin{align}
(p\gamma + m) v \; &= \; 0,  
\\[2ex]
\bar v(p, \lambda') \, v(p, \lambda) \; &= \; -2 m \, \delta(\lambda', \lambda), 
\\[1ex]
\sum_{\lambda} v(p, \lambda) \, \bar v(p, \lambda)  \; &= \; p\gamma - m .
\end{align}
\end{subequations}
It is convenient to define the specific form of these spinors such that they correspond to 
the result of a charge conjugation transformation (with a unique phase factor) applied 
to the particle spinors. The transformation is of the form
\be
v(p, \lambda) &=& C \bar u^T(p, \lambda) \; = \; C (\gamma^0)^T u^\ast(p, \lambda) ,
\label{v_spinor_from_charge_conjugation}
\ee
where the bispinor matrix $C$ satisfies the conditions
\begin{align}
C (\gamma^\mu)^T = -\gamma^\mu C,
\hspace{2em}
C^T = -C,
\hspace{2em}
C^\dagger C = 1.
\end{align}
These conditions specify the form of the matrix $C$ up to a phase factor.
We choose the phase factor such that (in the standard representation of the gamma matrices)
\begin{align}
C \equiv i \gamma^2 \gamma^0 \; = \; 
\left(
\begin{array}{rr} & -i \sigma^2
\\[1ex]
-i \sigma^2 & 
\end{array} 
\right) ,
\hspace{1em}
C^\dagger = -C.
\end{align}
With this choice the $v$-spinors for canonical spin states are obtained as
\begin{align}
v_{\rm can}(p, \lambda) =  
\left(
\begin{array}{l}
\sqrt{E - m} \; \bm{p}\bm{\sigma}/|\bm{p}|
\\[1ex]
\sqrt{E + m} 
\end{array} 
\right) \, (-i \sigma^2) \chi^\ast (\lambda),
\label{v_spinor_canonical}
\end{align}
where we have used the identity
\beq
\sigma^2 \bm{\sigma}^\ast \sigma^2 \; = \; -\bm{\sigma} .
\label{sigma_complex_identity}
\eeq
The two-component spinor $\chi^\ast (\lambda)$ describes the spin wave function of the antiparticle
in the rest frame. The $v$-spinors for the LF spin states are obtained as
\begin{align}
v_{\rm LF}(p, \lambda) 
\; &= \; 
\frac{1}{\sqrt{2 p^+}}
\left( 
\begin{array}{r}
\sigma^3 (p^+ - m - \bm{p}_T \bm{\sigma}_T \sigma^3)
\\[1ex]
p^+ + m + \bm{p}_T \bm{\sigma}_T \sigma^3 \;
\end{array} 
\right)
\nonumber \\[1ex]
&\times \; (-i \sigma^2) \chi^\ast (\lambda) .
\end{align}
The relation between the canonical and LF $v$-spinors is then
\begin{subequations}
\begin{align}
v_{\rm LF}(p, \lambda) \; &= \; v_{\rm can}(p, \lambda)[\chi^\ast (\lambda) \rightarrow \mathcal{U}^\ast (p)
\chi^\ast (\lambda)]
\label{v_spinor_rotation_phi}
\\[2ex]
&= \; \sum_{\lambda'}
v_{\rm can}(p, \lambda') U^\ast (p, \lambda', \lambda) .
\label{v_spinor_rotation_quantum_numbers}
\end{align}
\end{subequations}
Notice that the unitary matrix Eq.~(\ref{U_matrix_def}) satisfies
\begin{align}
\sigma^2 \mathcal{U}^\ast \sigma^2 \; &= \; \mathcal{U} ,
\end{align}
which follows from Eq.~(\ref{sigma_complex_identity}) and relates the formulas for the 
$u$- and $v$-spinors in Eqs.~(\ref{u_spinor_rotation_phi}) and (\ref{v_spinor_rotation_phi}).
Equation~(\ref{v_spinor_rotation_quantum_numbers}) follows directly from the definition 
Eq.~(\ref{v_spinor_from_charge_conjugation}).

We have presented the formulas for canonical and LF bispinors for a general spin wave function
of the particle at rest, described by the 2-spinors $\chi (\lambda)$. This has helped us distinguish
the two implementations of the Melosh rotation, as acting on the bispinor wave function or the spin 
quantum numbers, Eq.~(\ref{melosh_spinor}) and (\ref{melosh_quantum_numbers}), especially when 
complex-conjugate spinors are involved. The set of spinors $\chi (\lambda = \pm \frac{1}{2})$ can be 
quantized along any fixed direction. In the LF helicity states in the proper sense 
[cf.\ Sec.~\ref{subsec:light_front_structure}] the spinors are quantized along the 
$z$-axis, i.e., chosen as eigenspinors of $\sigma^3$,
\beq
\chi(\lambda = {\textstyle\frac{1}{2}}) \; = \; \left( \begin{array}{c} 1 \\ 0 \end{array}\right),
\hspace{2em}
\chi(\lambda = -{\textstyle\frac{1}{2}}) \; = \; \left( \begin{array}{c} 0 \\ 1 \end{array}\right) .
\eeq
With this choice of spinors the function $U(p, \lambda', \lambda)$ in Eq.~(\ref{U_function_def})
formally coincides with the element of the matrix $\mathcal{U}(p)$, Eq.~(\ref{U_matrix_def}), 
indexed by $\lambda'$ and $\lambda$, with $(\frac{1}{2}, -\frac{1}{2})$ identified with the 
matrix indices (1, 2),
\beq
U(p, \lambda', \lambda) \; = \; \mathcal{U}_{\lambda'\lambda}(p) .
\eeq
When using this compact notation one needs to be careful regarding the order of the matrix indices.
In the calculations in Appendix~\ref{app:wave_function} we use the form with the general 2-spinors, 
which is more transparent.
\section{Deuteron light-front wave function}
\label{app:wave_function}
In this appendix we present the deuteron LF wave function in the 3-dimensional representation
in the CM frame and demonstrate its equivalence to the 4-dimensional representation of 
Sec.~\ref{subsec:lf_wave_function}. In particular, we derive the relation between the 
radial wave functions in the 3-dimensional representation invariant functions and the 
4-dimensional representation, Eq.~(\ref{vertex_radial}).

In the CM frame of the $pn$ configuration, the proton and neutron 3-momenta (ordinary vector components)
are equal and opposite, Eq.~(\ref{p_p_p_n_CM}),
\begin{align}
p_p[\textrm{CM}] \; &= \; (E, \bm{k}), 
\hspace{2em}
p_n[\textrm{CM}] \; = \; (E, -\bm{k}),
\label{p_p_p_n_CM_app}
\end{align}
and the configuration is specified by the 3-momentum $\bm{k}$. In this frame the LF wave function
can be constructed as a 3-dimensional wave function in the variable $\bm{k}$. One first constructs a 
3-dimensional relativistic wave function in canonical spin states, which couples the orbital motion in
$\bm{k}$ with the canonical spin degrees of freedom, and then applies the Melosh rotation to convert
from canonical to LF helicity states. The result is \cite{Kondratyuk:1983kq}
\begin{widetext}
\begin{subequations}
\label{wf_3d}
\begin{align}
\Psi_d (\alpha_p , \bm{p}_{pT}; \lambda_p, \lambda_n | \lambda_d)
=& \sum_{\lambdapp, \lambdanp}
\widetilde \Psi_d (\bm{k}, \lambdanp, \lambdapp | \lambda_d )  \;
U^\ast (\bm{k}, \lambdapp, \lambda_p ) \; U^\ast(-\bm{k}, \lambdanp, \lambda_n ) ,
\label{wf_k_3d_lf}
\\[2ex]
\widetilde \Psi_d (\bm{k}, \lambdanp, \lambdapp | \lambda_d )
=& \frac{1}{\sqrt{2}} \epsilon_d^i (\lambda_d) \; 
\left[ 
\delta^{ij} f_0(k) \; + \; \frac{1}{\sqrt{2}} \left( \frac{3 k^i k^j}{k^2} - \delta^{ij} 
\right) f_2(k)
\right]
\chi^\dagger (\lambdanp) \left[ \sigma^j (i \sigma^2) \right] \, \chi^\ast (\lambdapp) ,
\label{wf_k_3d}
\\[2ex]
U (\bm{k}, \lambdapp, \lambda_p ) \; \equiv& \; U (p_p, \lambdapp, \lambda_p )[\textrm{CM}]
\; = \; \chi^\dagger (\lambdapp) \left[
\frac{E + k^3 + m + \bm{k}_T \bm{\sigma}_T \sigma^3}{\sqrt{2 (E + k^3) (E + m)}} \right]
\chi(\lambda_p) ,
\label{U_proton}
\\[2ex]
U (-\bm{k}, \lambdanp, \lambda_n )
\; \equiv& \; U (p_n, \lambdanp, \lambda_n )[\textrm{CM}]
\; = \; \chi^\dagger (\lambdanp) \left[
\frac{E - k^3 + m - \bm{k}_T \bm{\sigma}_T \sigma^3}{\sqrt{2 (E - k^3) (E + m)}} 
\right] \chi(\lambda_n) .
\label{U_neutron}
\end{align}
\end{subequations}
\end{widetext}
$\widetilde \Psi_d$ in Eq.~(\ref{wf_k_3d}) is the 3-dimensional relativistic wave function in the
CM momentum $\bm{k}$ and the canonical spin variables $\lambdapp$ and $\lambdanp$. The 3-vector
$\bm{\epsilon}_d$ is the polarization 3-vector of the $pn$ configuration in the CM frame 
[which is identical to the deuteron polarization 3-vector in the deuteron rest frame, cf.\ 
Eq.~(\ref{epsilon_pn_cm})], and the 2-spinors $\chi$ describe the spin wave 
function of the proton and neutron at rest. $\widetilde\Psi_d$ includes the 
S- and D-waves of the orbital motion. It has the same form as the nonrelativistic
deuteron wave function; the only difference is in the normalization of the radial
wave functions \cite{Strikman:2017koc,Kondratyuk:1983kq},
\begin{align}
& 4\pi \int \frac{dk \, k^2}{E(k)} [f_0^2(k) + f_2^2(k)] \;\; = \;\; 1 .
\end{align}
The functions $U$ in Eqs.~(\ref{U_proton}) and (\ref{U_neutron})
are the Melosh rotations Eq.~(\ref{U_function_def}) associated with the proton and neutron 4-momenta 
in the CM frame, Eq.~(\ref{p_p_p_n_CM_app}) (we regard them here as functions of the 3-vector $\bm{k}$).
They connect the canonical spin variables $\lambdapp$ and $\lambdanp$ with the LF helicities $\lambda_p$ 
and $\lambda_n$. The functions $U$ enter in Eq.~(\ref{wf_k_3d_lf}) as their complex conjugates, because the proton and 
neutron states are described by complex-conjugate spinors (the wave function corresponds to the 
$\langle p n| d\rangle$ matrix element).

The 3-dimensional representation of the LF wave function, Eq.~(\ref{wf_3d}), is equivalent to the
4-dimensional representation of Eqs.~(\ref{wf_bilinear}) and (\ref{Gamma_decomposition}),
\begin{align}
& \Psi_d (\alpha_p , \bm{p}_{pT}; \lambda_p, \lambda_n | \lambda_d)
\nonumber \\[1ex]
&= \; \bar u_{\rm LF}(p_n, \lambda_n) \, 
[\gamma_\alpha G_1 \; + \; (p_p - p_n)_\alpha G_2] \, v_{\rm LF} (p_p, \lambda_p)
\nonumber \\[1ex]
&\times \; \epsilon_{pn}^\alpha (p_{pn}, \lambda_d) 
\nonumber
\\[1ex]
& \hspace{1em} [G_{1, 2} \equiv G_{1, 2}(M_{pn}^2)].
\label{wf_bilinear_app}
\end{align}
This can be demonstrated using the invariance properties of the 4-dimensional expressions.
The bilinear forms in Eq.~(\ref{wf_bilinear_app}) are Lorentz invariants and can therefore
be evaluated with the 4-vectors and bispinors in any collinear frame. Specifically, we may 
evaluate them with the 4-vectors and bispinors in the CM frame,
\begin{subequations}
\label{bilinear_CM}
\begin{align}
& \bar u_{\rm LF}(p_n, \lambda_n) \gamma_\alpha v_{\rm LF}(p_p, \lambda_p) 
\epsilon_{pn}^\alpha (p_{pn}, \lambda_d) [\textrm{any coll. frame}] 
\nonumber
\\[1ex]
&= 
\bar u_{\rm LF}(p_n, \lambda_n) \gamma_\alpha v_{\rm LF}(p_p, \lambda_p) 
\epsilon_{pn}^\alpha (p_{pn}, \lambda_d) [\textrm{CM}] 
\\[3ex]
& \bar u_{\rm LF}(p_n, \lambda_n) v_{\rm LF}(p_p, \lambda_p) (p_p - p_n)_\alpha
\epsilon_{pn}^\alpha (p_{pn}, \lambda_d) [\textrm{any coll.}] 
\nonumber
\\[1ex]
&=
\bar u_{\rm LF}(p_n, \lambda_n) v_{\rm LF}(p_p, \lambda_p) (p_p - p_n)^\alpha 
\epsilon_{pn}^\alpha (p_{pn}, \lambda_d) [\textrm{CM}] 
\end{align}
\end{subequations}
The LF bispinors in the CM frame can be expressed in terms of the canonical bispinors in the
CM frame and the Melosh rotations corresponding to the CM momentum, 
cf.\ Eqs.~(\ref{ubar_spinor_rotation_quantum_numbers})
and (\ref{v_spinor_rotation_quantum_numbers}),
\begin{subequations}
\label{Melosh_CM}
\begin{align}
& \bar u_{\rm LF}(p_n, \lambda_n)[\textrm{CM}]  
\nonumber \\[1ex]
&= \;  \sum_{\lambdanp}
\bar u_{\rm can}(p_n, \lambdanp) \; U^\ast (p_n, \lambdanp, \lambda_n) [\textrm{CM}]
\nonumber
\\[1ex]
\; &= \; \sum_{\lambdanp} 
\bar u_{\rm can}(p_n, \lambdanp)[\textrm{CM}] \; U^\ast (-\bm{k}, \lambdanp, \lambda_n) ,
\\[3ex]
& v_{\rm LF}(p_p, \lambda_p)[\textrm{CM}]  
\nonumber \\[1ex]
&= \;  \sum_{\lambdapp}
v_{\rm can}(p_p, \lambdapp) \; U^\ast (p_p, \lambdanp, \lambda_n) [\textrm{CM}]
\nonumber
\\[1ex]
&= \;  \sum_{\lambdapp}
v_{\rm can}(p_p, \lambdapp)[\textrm{CM}] \; U^\ast (\bm{k}, \lambdanp, \lambda_n) ,
\end{align}
\end{subequations}
where the $\bm{k}$-dependent functions in the last expressions are the ones given in
Eqs.~(\ref{U_neutron}) and (\ref{U_proton}). Substituting the specific expressions of
Eqs.~(\ref{bilinear_CM}) and (\ref{Melosh_CM}) in Eq.~(\ref{wf_bilinear_app}),
and comparing with Eq.~(\ref{wf_3d}), one sees that the 4-dimensional wave function 
coincides with the 3-dimensional one, provided that the 3-dimensional canonical wave function 
$\widetilde\Psi_d$ in Eq.~(\ref{wf_k_3d_lf}) can be identified as
\begin{align}
& \widetilde\Psi_d (\bm{k}, \lambdanp, \lambdapp | \lambda_d )
\nonumber \\[1ex]
& \stackrel{!}{=} \bar u_{\rm can}(p_n, \lambdanp) \, 
[\gamma_\alpha G_1 \, + \, (p_p - p_n)_\alpha G_2 ]
\, v_{\rm can} (p_p, \lambdapp)
\nonumber \\[1ex]
& \times \; \epsilon_{pn}^\alpha (p_{pn}, \lambda_d) [\textrm{CM}] ,
\label{identification}
\end{align}
which is the same 4-dimensional bilinear form as in Eq.~(\ref{wf_bilinear_app}) but evaluated with
the canonical bispinors in the CM frame. To show that the identification Eq.~(\ref{identification}) 
is possible we need to express the bilinear form in terms of the two-component spinors and
match it with the explicit form of Eq.~(\ref{wf_k_3d}). The canonical bispinors in the CM frame 
are given by the general expressions Eqs.~(\ref{u_spinor_canonical}) and (\ref{v_spinor_canonical}),
evaluated with the CM 4-momenta of Eq.~(\ref{p_p_p_n_CM_app}). The components of the $pn$ 
spin 4-vector in the CM frame are given by Eq.~(\ref{epsilon_pn_cm}),
\begin{align}
\epsilon_{pn}^\alpha (p_{pn}, \lambda_d) [\textrm{CM}] 
\; &= \; (0, \bm{\epsilon}_d) .
\end{align}
Reducing the bilinear forms in Eq.~(\ref{identification}) to two-component form, we get
\begin{widetext}
\begin{subequations}
\begin{align}
u_{\rm can}(p_n, \lambdanp) v_{\rm can} (p_p, \lambdapp) \; [\textrm{CM}]  
\; &= \;
\chi^\dagger (\lambdanp) \left[ -2 \bm{k}\cdot\bm{\sigma}
\right] \, (i\sigma^2) \, \chi^\ast (\lambdapp) ,
\\[1ex]
u_{\rm can}(p_n, \lambdanp) \gamma^i v_{\rm can} (p_p, \lambdapp) \; [\textrm{CM}]  
\; &= \; \chi^\dagger (\lambdanp) \left[ - 2 E \sigma^i + 2(E - m) \frac{k^i \bm{k}\cdot\bm{\sigma}}{k^2} \right] \, 
(i\sigma^2) \, \chi^\ast (\lambdapp) ,
\\[1ex]
u_{\rm can}(p_n, \lambdanp) \gamma^0 v_{\rm can} (p_p, \lambdapp) \; [\textrm{CM}]  
\; &= \; 0.
\end{align}
\end{subequations}
Altogether, Eq.~(\ref{identification}) becomes
\be
\widetilde\Psi_d (\bm{k}, \lambdanp, \lambdapp | \lambda_d )
\; & \stackrel{!}{=}& \;
\epsilon_d^i (\lambda_d) \, \chi^\dagger (\lambdanp) \,
\left\{
\left[ 2 E \sigma^i - 2(E - m) \frac{k^i \bm{k}\cdot\bm{\sigma}}{k^2} \right] G_1 
+ 4 k^i \bm{k}\cdot\bm{\sigma} G_2 \right\} (i\sigma^2) \, \chi^\ast (\lambdapp)
\ee
\end{widetext}
The expression on the R.H.S.\ has the same structure as Eq.~(\ref{wf_k_3d}).
Matching the coefficients of the S- and D-wave tensor structures, we obtain the 
relation of Eq.~(\ref{vertex_radial}),
\begin{subequations}
\begin{align}
G_1 \; =& \; \frac{1}{4 E} \left( \sqrt{2}\, f_0 - f_2 \right),
\\[1ex]
G_2 \; =& \; \frac{1}{8 E k^2} \left[ \sqrt{2} (E - m) \, f_0 + (2 E + m) \, f_2 \right]
\\[3ex]
\nonumber
& \left[ G_{1, 2} \equiv G_{1, 2}(M_{pn}^2), \; f_{0, 2} \equiv f_{0, 2}(k) \right] ,
\end{align}
\end{subequations}
where the arguments are related by [cf.\ Eq.~(\ref{CM_LF_variables})]
\beq
M_{pn}^2 \; = \; 4 E^2 \; = \; 4 (|\bm{k}|^2 + m^2) .
\eeq
This completes the proof of equivalence of the 3-dimensional and 4-dimensional representations
of the deuteron LF wve function and determines the 4-dimensional invariant functions in terms
of the 3-dimensional radial wave functions.
\bibliography{poltag.bib}

\begin{thebibliography}{96}%
\makeatletter
\providecommand \@ifxundefined [1]{%
 \@ifx{#1\undefined}
}%
\providecommand \@ifnum [1]{%
 \ifnum #1\expandafter \@firstoftwo
 \else \expandafter \@secondoftwo
 \fi
}%
\providecommand \@ifx [1]{%
 \ifx #1\expandafter \@firstoftwo
 \else \expandafter \@secondoftwo
 \fi
}%
\providecommand \natexlab [1]{#1}%
\providecommand \enquote  [1]{``#1''}%
\providecommand \bibnamefont  [1]{#1}%
\providecommand \bibfnamefont [1]{#1}%
\providecommand \citenamefont [1]{#1}%
\providecommand \href@noop [0]{\@secondoftwo}%
\providecommand \href [0]{\begingroup \@sanitize@url \@href}%
\providecommand \@href[1]{\@@startlink{#1}\@@href}%
\providecommand \@@href[1]{\endgroup#1\@@endlink}%
\providecommand \@sanitize@url [0]{\catcode `\\12\catcode `\$12\catcode
  `\&12\catcode `\#12\catcode `\^12\catcode `\_12\catcode `\%12\relax}%
\providecommand \@@startlink[1]{}%
\providecommand \@@endlink[0]{}%
\providecommand \url  [0]{\begingroup\@sanitize@url \@url }%
\providecommand \@url [1]{\endgroup\@href {#1}{\urlprefix }}%
\providecommand \urlprefix  [0]{URL }%
\providecommand \Eprint [0]{\href }%
\providecommand \doibase [0]{http://dx.doi.org/}%
\providecommand \selectlanguage [0]{\@gobble}%
\providecommand \bibinfo  [0]{\@secondoftwo}%
\providecommand \bibfield  [0]{\@secondoftwo}%
\providecommand \translation [1]{[#1]}%
\providecommand \BibitemOpen [0]{}%
\providecommand \bibitemStop [0]{}%
\providecommand \bibitemNoStop [0]{.\EOS\space}%
\providecommand \EOS [0]{\spacefactor3000\relax}%
\providecommand \BibitemShut  [1]{\csname bibitem#1\endcsname}%
\let\auto@bib@innerbib\@empty
\bibitem [{\citenamefont {Anselmino}\ \emph {et~al.}(1995)\citenamefont
  {Anselmino}, \citenamefont {Efremov},\ and\ \citenamefont
  {Leader}}]{Anselmino:1994gn}%
  \BibitemOpen
  \bibfield  {author} {\bibinfo {author} {\bibfnamefont {M.}~\bibnamefont
  {Anselmino}}, \bibinfo {author} {\bibfnamefont {A.}~\bibnamefont {Efremov}},
  \ and\ \bibinfo {author} {\bibfnamefont {E.}~\bibnamefont {Leader}},\ }\href
  {\doibase 10.1016/0370-1573(95)00011-5} {\bibfield  {journal} {\bibinfo
  {journal} {Phys. Rept.}\ }\textbf {\bibinfo {volume} {261}},\ \bibinfo
  {pages} {1} (\bibinfo {year} {1995})},\ \bibinfo {note} {[Erratum: Phys.
  Rept.281,399(1997)]},\ \Eprint {http://arxiv.org/abs/hep-ph/9501369}
  {arXiv:hep-ph/9501369 [hep-ph]} \BibitemShut {NoStop}%
\bibitem [{\citenamefont {Burkardt}\ \emph {et~al.}(2010)\citenamefont
  {Burkardt}, \citenamefont {Miller},\ and\ \citenamefont
  {Nowak}}]{Burkardt:2008jw}%
  \BibitemOpen
  \bibfield  {author} {\bibinfo {author} {\bibfnamefont {M.}~\bibnamefont
  {Burkardt}}, \bibinfo {author} {\bibfnamefont {C.~A.}\ \bibnamefont
  {Miller}}, \ and\ \bibinfo {author} {\bibfnamefont {W.~D.}\ \bibnamefont
  {Nowak}},\ }\href {\doibase 10.1088/0034-4885/73/1/016201} {\bibfield
  {journal} {\bibinfo  {journal} {Rept. Prog. Phys.}\ }\textbf {\bibinfo
  {volume} {73}},\ \bibinfo {pages} {016201} (\bibinfo {year}
  {2010})}\BibitemShut {NoStop}%
\bibitem [{\citenamefont {Kuhn}\ \emph {et~al.}(2009)\citenamefont {Kuhn},
  \citenamefont {Chen},\ and\ \citenamefont {Leader}}]{Kuhn:2008sy}%
  \BibitemOpen
  \bibfield  {author} {\bibinfo {author} {\bibfnamefont {S.~E.}\ \bibnamefont
  {Kuhn}}, \bibinfo {author} {\bibfnamefont {J.~P.}\ \bibnamefont {Chen}}, \
  and\ \bibinfo {author} {\bibfnamefont {E.}~\bibnamefont {Leader}},\ }\href
  {\doibase 10.1016/j.ppnp.2009.02.001} {\bibfield  {journal} {\bibinfo
  {journal} {Prog. Part. Nucl. Phys.}\ }\textbf {\bibinfo {volume} {63}},\
  \bibinfo {pages} {1} (\bibinfo {year} {2009})},\ \Eprint
  {http://arxiv.org/abs/0812.3535} {arXiv:0812.3535 [hep-ph]} \BibitemShut
  {NoStop}%
\bibitem [{\citenamefont {Aidala}\ \emph {et~al.}(2013)\citenamefont {Aidala},
  \citenamefont {Bass}, \citenamefont {Hasch},\ and\ \citenamefont
  {Mallot}}]{Aidala:2012mv}%
  \BibitemOpen
  \bibfield  {author} {\bibinfo {author} {\bibfnamefont {C.~A.}\ \bibnamefont
  {Aidala}}, \bibinfo {author} {\bibfnamefont {S.~D.}\ \bibnamefont {Bass}},
  \bibinfo {author} {\bibfnamefont {D.}~\bibnamefont {Hasch}}, \ and\ \bibinfo
  {author} {\bibfnamefont {G.~K.}\ \bibnamefont {Mallot}},\ }\href {\doibase
  10.1103/RevModPhys.85.655} {\bibfield  {journal} {\bibinfo  {journal} {Rev.
  Mod. Phys.}\ }\textbf {\bibinfo {volume} {85}},\ \bibinfo {pages} {655}
  (\bibinfo {year} {2013})},\ \Eprint {http://arxiv.org/abs/1209.2803}
  {arXiv:1209.2803 [hep-ph]} \BibitemShut {NoStop}%
\bibitem [{\citenamefont {de~Florian}\ \emph {et~al.}(2009)\citenamefont
  {de~Florian}, \citenamefont {Sassot}, \citenamefont {Stratmann},\ and\
  \citenamefont {Vogelsang}}]{deFlorian:2009vb}%
  \BibitemOpen
  \bibfield  {author} {\bibinfo {author} {\bibfnamefont {D.}~\bibnamefont
  {de~Florian}}, \bibinfo {author} {\bibfnamefont {R.}~\bibnamefont {Sassot}},
  \bibinfo {author} {\bibfnamefont {M.}~\bibnamefont {Stratmann}}, \ and\
  \bibinfo {author} {\bibfnamefont {W.}~\bibnamefont {Vogelsang}},\ }\href
  {\doibase 10.1103/PhysRevD.80.034030} {\bibfield  {journal} {\bibinfo
  {journal} {Phys. Rev. D}\ }\textbf {\bibinfo {volume} {80}},\ \bibinfo
  {pages} {034030} (\bibinfo {year} {2009})},\ \Eprint
  {http://arxiv.org/abs/0904.3821} {arXiv:0904.3821 [hep-ph]} \BibitemShut
  {NoStop}%
\bibitem [{\citenamefont {Leader}\ \emph {et~al.}(2010)\citenamefont {Leader},
  \citenamefont {Sidorov},\ and\ \citenamefont {Stamenov}}]{Leader:2010rb}%
  \BibitemOpen
  \bibfield  {author} {\bibinfo {author} {\bibfnamefont {E.}~\bibnamefont
  {Leader}}, \bibinfo {author} {\bibfnamefont {A.~V.}\ \bibnamefont {Sidorov}},
  \ and\ \bibinfo {author} {\bibfnamefont {D.~B.}\ \bibnamefont {Stamenov}},\
  }\href {\doibase 10.1103/PhysRevD.82.114018} {\bibfield  {journal} {\bibinfo
  {journal} {Phys. Rev. D}\ }\textbf {\bibinfo {volume} {82}},\ \bibinfo
  {pages} {114018} (\bibinfo {year} {2010})},\ \Eprint
  {http://arxiv.org/abs/1010.0574} {arXiv:1010.0574 [hep-ph]} \BibitemShut
  {NoStop}%
\bibitem [{\citenamefont {Blumlein}\ and\ \citenamefont
  {Bottcher}(2010)}]{Blumlein:2010rn}%
  \BibitemOpen
  \bibfield  {author} {\bibinfo {author} {\bibfnamefont {J.}~\bibnamefont
  {Blumlein}}\ and\ \bibinfo {author} {\bibfnamefont {H.}~\bibnamefont
  {Bottcher}},\ }\href {\doibase 10.1016/j.nuclphysb.2010.08.005} {\bibfield
  {journal} {\bibinfo  {journal} {Nucl. Phys. B}\ }\textbf {\bibinfo {volume}
  {841}},\ \bibinfo {pages} {205} (\bibinfo {year} {2010})},\ \Eprint
  {http://arxiv.org/abs/1005.3113} {arXiv:1005.3113 [hep-ph]} \BibitemShut
  {NoStop}%
\bibitem [{\citenamefont {Nocera}\ \emph {et~al.}(2014)\citenamefont {Nocera},
  \citenamefont {Ball}, \citenamefont {Forte}, \citenamefont {Ridolfi},\ and\
  \citenamefont {Rojo}}]{Nocera:2014gqa}%
  \BibitemOpen
  \bibfield  {author} {\bibinfo {author} {\bibfnamefont {E.~R.}\ \bibnamefont
  {Nocera}}, \bibinfo {author} {\bibfnamefont {R.~D.}\ \bibnamefont {Ball}},
  \bibinfo {author} {\bibfnamefont {S.}~\bibnamefont {Forte}}, \bibinfo
  {author} {\bibfnamefont {G.}~\bibnamefont {Ridolfi}}, \ and\ \bibinfo
  {author} {\bibfnamefont {J.}~\bibnamefont {Rojo}} (\bibinfo {collaboration}
  {NNPDF}),\ }\href {\doibase 10.1016/j.nuclphysb.2014.08.008} {\bibfield
  {journal} {\bibinfo  {journal} {Nucl. Phys. B}\ }\textbf {\bibinfo {volume}
  {887}},\ \bibinfo {pages} {276} (\bibinfo {year} {2014})},\ \Eprint
  {http://arxiv.org/abs/1406.5539} {arXiv:1406.5539 [hep-ph]} \BibitemShut
  {NoStop}%
\bibitem [{\citenamefont {Sato}\ \emph {et~al.}(2016)\citenamefont {Sato},
  \citenamefont {Melnitchouk}, \citenamefont {Kuhn}, \citenamefont {Ethier},\
  and\ \citenamefont {Accardi}}]{Sato:2016tuz}%
  \BibitemOpen
  \bibfield  {author} {\bibinfo {author} {\bibfnamefont {N.}~\bibnamefont
  {Sato}}, \bibinfo {author} {\bibfnamefont {W.}~\bibnamefont {Melnitchouk}},
  \bibinfo {author} {\bibfnamefont {S.}~\bibnamefont {Kuhn}}, \bibinfo {author}
  {\bibfnamefont {J.}~\bibnamefont {Ethier}}, \ and\ \bibinfo {author}
  {\bibfnamefont {A.}~\bibnamefont {Accardi}} (\bibinfo {collaboration}
  {Jefferson Lab Angular Momentum}),\ }\href {\doibase
  10.1103/PhysRevD.93.074005} {\bibfield  {journal} {\bibinfo  {journal} {Phys.
  Rev. D}\ }\textbf {\bibinfo {volume} {93}},\ \bibinfo {pages} {074005}
  (\bibinfo {year} {2016})},\ \Eprint {http://arxiv.org/abs/1601.07782}
  {arXiv:1601.07782 [hep-ph]} \BibitemShut {NoStop}%
\bibitem [{\citenamefont {Altarelli}\ \emph {et~al.}(1997)\citenamefont
  {Altarelli}, \citenamefont {Ball}, \citenamefont {Forte},\ and\ \citenamefont
  {Ridolfi}}]{Altarelli:1996nm}%
  \BibitemOpen
  \bibfield  {author} {\bibinfo {author} {\bibfnamefont {G.}~\bibnamefont
  {Altarelli}}, \bibinfo {author} {\bibfnamefont {R.~D.}\ \bibnamefont {Ball}},
  \bibinfo {author} {\bibfnamefont {S.}~\bibnamefont {Forte}}, \ and\ \bibinfo
  {author} {\bibfnamefont {G.}~\bibnamefont {Ridolfi}},\ }\href {\doibase
  10.1016/S0550-3213(97)00201-0} {\bibfield  {journal} {\bibinfo  {journal}
  {Nucl. Phys.}\ }\textbf {\bibinfo {volume} {B496}},\ \bibinfo {pages} {337}
  (\bibinfo {year} {1997})}\BibitemShut {NoStop}%
\bibitem [{\citenamefont {Pasechnik}\ \emph {et~al.}(2010)\citenamefont
  {Pasechnik}, \citenamefont {Shirkov}, \citenamefont {Teryaev}, \citenamefont
  {Solovtsova},\ and\ \citenamefont {Khandramai}}]{Pasechnik:2009yc}%
  \BibitemOpen
  \bibfield  {author} {\bibinfo {author} {\bibfnamefont {R.~S.}\ \bibnamefont
  {Pasechnik}}, \bibinfo {author} {\bibfnamefont {D.~V.}\ \bibnamefont
  {Shirkov}}, \bibinfo {author} {\bibfnamefont {O.~V.}\ \bibnamefont
  {Teryaev}}, \bibinfo {author} {\bibfnamefont {O.~P.}\ \bibnamefont
  {Solovtsova}}, \ and\ \bibinfo {author} {\bibfnamefont {V.~L.}\ \bibnamefont
  {Khandramai}},\ }\href {\doibase 10.1103/PhysRevD.81.016010} {\bibfield
  {journal} {\bibinfo  {journal} {Phys. Rev.}\ }\textbf {\bibinfo {volume}
  {D81}},\ \bibinfo {pages} {016010} (\bibinfo {year} {2010})}\BibitemShut
  {NoStop}%
\bibitem [{\citenamefont {Baikov}\ \emph {et~al.}(2010)\citenamefont {Baikov},
  \citenamefont {Chetyrkin},\ and\ \citenamefont {Kuhn}}]{Baikov:2010je}%
  \BibitemOpen
  \bibfield  {author} {\bibinfo {author} {\bibfnamefont {P.}~\bibnamefont
  {Baikov}}, \bibinfo {author} {\bibfnamefont {K.}~\bibnamefont {Chetyrkin}}, \
  and\ \bibinfo {author} {\bibfnamefont {J.}~\bibnamefont {Kuhn}},\ }\href
  {\doibase 10.1103/PhysRevLett.104.132004} {\bibfield  {journal} {\bibinfo
  {journal} {Phys. Rev. Lett.}\ }\textbf {\bibinfo {volume} {104}},\ \bibinfo
  {pages} {132004} (\bibinfo {year} {2010})},\ \Eprint
  {http://arxiv.org/abs/1001.3606} {arXiv:1001.3606 [hep-ph]} \BibitemShut
  {NoStop}%
\bibitem [{\citenamefont {Deur}\ \emph {et~al.}(2014)\citenamefont {Deur},
  \citenamefont {Prok}, \citenamefont {Burkert}, \citenamefont {Crabb},
  \citenamefont {Girod}, \citenamefont {Griffioen}, \citenamefont {Guler},
  \citenamefont {Kuhn},\ and\ \citenamefont {Kvaltine}}]{Deur:2014vea}%
  \BibitemOpen
  \bibfield  {author} {\bibinfo {author} {\bibfnamefont {A.}~\bibnamefont
  {Deur}}, \bibinfo {author} {\bibfnamefont {Y.}~\bibnamefont {Prok}}, \bibinfo
  {author} {\bibfnamefont {V.}~\bibnamefont {Burkert}}, \bibinfo {author}
  {\bibfnamefont {D.}~\bibnamefont {Crabb}}, \bibinfo {author} {\bibfnamefont
  {F.~X.}\ \bibnamefont {Girod}}, \bibinfo {author} {\bibfnamefont {K.~A.}\
  \bibnamefont {Griffioen}}, \bibinfo {author} {\bibfnamefont {N.}~\bibnamefont
  {Guler}}, \bibinfo {author} {\bibfnamefont {S.~E.}\ \bibnamefont {Kuhn}}, \
  and\ \bibinfo {author} {\bibfnamefont {N.}~\bibnamefont {Kvaltine}},\ }\href
  {\doibase 10.1103/PhysRevD.90.012009} {\bibfield  {journal} {\bibinfo
  {journal} {Phys. Rev.}\ }\textbf {\bibinfo {volume} {D90}},\ \bibinfo {pages}
  {012009} (\bibinfo {year} {2014})}\BibitemShut {NoStop}%
\bibitem [{\citenamefont {Cveti\v{c}}\ and\ \citenamefont
  {Kataev}(2016)}]{Cvetic:2016rot}%
  \BibitemOpen
  \bibfield  {author} {\bibinfo {author} {\bibfnamefont {G.}~\bibnamefont
  {Cveti\v{c}}}\ and\ \bibinfo {author} {\bibfnamefont {A.}~\bibnamefont
  {Kataev}},\ }\href {\doibase 10.1103/PhysRevD.94.014006} {\bibfield
  {journal} {\bibinfo  {journal} {Phys. Rev. D}\ }\textbf {\bibinfo {volume}
  {94}},\ \bibinfo {pages} {014006} (\bibinfo {year} {2016})},\ \Eprint
  {http://arxiv.org/abs/1604.00509} {arXiv:1604.00509 [hep-ph]} \BibitemShut
  {NoStop}%
\bibitem [{\citenamefont {Kotlorz}\ and\ \citenamefont
  {Mikhailov}(2019)}]{Kotlorz:2018bxp}%
  \BibitemOpen
  \bibfield  {author} {\bibinfo {author} {\bibfnamefont {D.}~\bibnamefont
  {Kotlorz}}\ and\ \bibinfo {author} {\bibfnamefont {S.}~\bibnamefont
  {Mikhailov}},\ }\href {\doibase 10.1103/PhysRevD.100.056007} {\bibfield
  {journal} {\bibinfo  {journal} {Phys. Rev. D}\ }\textbf {\bibinfo {volume}
  {100}},\ \bibinfo {pages} {056007} (\bibinfo {year} {2019})},\ \Eprint
  {http://arxiv.org/abs/1810.02973} {arXiv:1810.02973 [hep-ph]} \BibitemShut
  {NoStop}%
\bibitem [{\citenamefont {Ayala}\ \emph {et~al.}(2018)\citenamefont {Ayala},
  \citenamefont {Cveti\v{c}}, \citenamefont {Kotikov},\ and\ \citenamefont
  {Shaikhatdenov}}]{Ayala:2018ulm}%
  \BibitemOpen
  \bibfield  {author} {\bibinfo {author} {\bibfnamefont {C.}~\bibnamefont
  {Ayala}}, \bibinfo {author} {\bibfnamefont {G.}~\bibnamefont {Cveti\v{c}}},
  \bibinfo {author} {\bibfnamefont {A.~V.}\ \bibnamefont {Kotikov}}, \ and\
  \bibinfo {author} {\bibfnamefont {B.~G.}\ \bibnamefont {Shaikhatdenov}},\
  }\href {\doibase 10.1140/epjc/s10052-018-6490-9} {\bibfield  {journal}
  {\bibinfo  {journal} {Eur. Phys. J. C}\ }\textbf {\bibinfo {volume} {78}},\
  \bibinfo {pages} {1002} (\bibinfo {year} {2018})},\ \Eprint
  {http://arxiv.org/abs/1812.01030} {arXiv:1812.01030 [hep-ph]} \BibitemShut
  {NoStop}%
\bibitem [{\citenamefont {Deur}\ \emph {et~al.}(2016)\citenamefont {Deur},
  \citenamefont {Brodsky},\ and\ \citenamefont {de~Teramond}}]{Deur:2016tte}%
  \BibitemOpen
  \bibfield  {author} {\bibinfo {author} {\bibfnamefont {A.}~\bibnamefont
  {Deur}}, \bibinfo {author} {\bibfnamefont {S.~J.}\ \bibnamefont {Brodsky}}, \
  and\ \bibinfo {author} {\bibfnamefont {G.~F.}\ \bibnamefont {de~Teramond}},\
  }\href {\doibase 10.1016/j.ppnp.2016.04.003} {\bibfield  {journal} {\bibinfo
  {journal} {Prog. Part. Nucl. Phys.}\ }\textbf {\bibinfo {volume} {90}},\
  \bibinfo {pages} {1} (\bibinfo {year} {2016})}\BibitemShut {NoStop}%
\bibitem [{\citenamefont {Bjorken}(1970)}]{Bjorken:1969mm}%
  \BibitemOpen
  \bibfield  {author} {\bibinfo {author} {\bibfnamefont {J.~D.}\ \bibnamefont
  {Bjorken}},\ }\href {\doibase 10.1103/PhysRevD.1.1376} {\bibfield  {journal}
  {\bibinfo  {journal} {Phys. Rev.}\ }\textbf {\bibinfo {volume} {D1}},\
  \bibinfo {pages} {1376} (\bibinfo {year} {1970})}\BibitemShut {NoStop}%
\bibitem [{\citenamefont {Braun}\ and\ \citenamefont
  {Kolesnichenko}(1987)}]{Braun:1986ty}%
  \BibitemOpen
  \bibfield  {author} {\bibinfo {author} {\bibfnamefont {V.~M.}\ \bibnamefont
  {Braun}}\ and\ \bibinfo {author} {\bibfnamefont {A.}~\bibnamefont
  {Kolesnichenko}},\ }\href {\doibase 10.1016/0550-3213(87)90295-1} {\bibfield
  {journal} {\bibinfo  {journal} {Nucl. Phys. B}\ }\textbf {\bibinfo {volume}
  {283}},\ \bibinfo {pages} {723} (\bibinfo {year} {1987})}\BibitemShut
  {NoStop}%
\bibitem [{\citenamefont {Balitsky}\ \emph {et~al.}(1990)\citenamefont
  {Balitsky}, \citenamefont {Braun},\ and\ \citenamefont
  {Kolesnichenko}}]{Balitsky:1989jb}%
  \BibitemOpen
  \bibfield  {author} {\bibinfo {author} {\bibfnamefont {I.}~\bibnamefont
  {Balitsky}}, \bibinfo {author} {\bibfnamefont {V.~M.}\ \bibnamefont {Braun}},
  \ and\ \bibinfo {author} {\bibfnamefont {A.}~\bibnamefont {Kolesnichenko}},\
  }\href {\doibase 10.1016/0370-2693(90)91465-N} {\bibfield  {journal}
  {\bibinfo  {journal} {Phys. Lett. B}\ }\textbf {\bibinfo {volume} {242}},\
  \bibinfo {pages} {245} (\bibinfo {year} {1990})},\ \bibinfo {note} {[Erratum:
  Phys.Lett.B 318, 648 (1993)]},\ \Eprint {http://arxiv.org/abs/hep-ph/9310316}
  {arXiv:hep-ph/9310316} \BibitemShut {NoStop}%
\bibitem [{\citenamefont {Stein}\ \emph {et~al.}(1995)\citenamefont {Stein},
  \citenamefont {Gornicki}, \citenamefont {Mankiewicz},\ and\ \citenamefont
  {Schafer}}]{Stein:1995si}%
  \BibitemOpen
  \bibfield  {author} {\bibinfo {author} {\bibfnamefont {E.}~\bibnamefont
  {Stein}}, \bibinfo {author} {\bibfnamefont {P.}~\bibnamefont {Gornicki}},
  \bibinfo {author} {\bibfnamefont {L.}~\bibnamefont {Mankiewicz}}, \ and\
  \bibinfo {author} {\bibfnamefont {A.}~\bibnamefont {Schafer}},\ }\href
  {\doibase 10.1016/0370-2693(95)00544-U} {\bibfield  {journal} {\bibinfo
  {journal} {Phys. Lett.}\ }\textbf {\bibinfo {volume} {B353}},\ \bibinfo
  {pages} {107} (\bibinfo {year} {1995})}\BibitemShut {NoStop}%
\bibitem [{\citenamefont {Balla}\ \emph {et~al.}(1998)\citenamefont {Balla},
  \citenamefont {Polyakov},\ and\ \citenamefont {Weiss}}]{Balla:1997hf}%
  \BibitemOpen
  \bibfield  {author} {\bibinfo {author} {\bibfnamefont {J.}~\bibnamefont
  {Balla}}, \bibinfo {author} {\bibfnamefont {M.~V.}\ \bibnamefont {Polyakov}},
  \ and\ \bibinfo {author} {\bibfnamefont {C.}~\bibnamefont {Weiss}},\ }\href
  {\doibase 10.1016/S0550-3213(98)00638-5} {\bibfield  {journal} {\bibinfo
  {journal} {Nucl. Phys. B}\ }\textbf {\bibinfo {volume} {510}},\ \bibinfo
  {pages} {327} (\bibinfo {year} {1998})},\ \Eprint
  {http://arxiv.org/abs/hep-ph/9707515} {arXiv:hep-ph/9707515} \BibitemShut
  {NoStop}%
\bibitem [{\citenamefont {Meziani}\ \emph {et~al.}(2005)\citenamefont {Meziani}
  \emph {et~al.}}]{Meziani:2004ne}%
  \BibitemOpen
  \bibfield  {author} {\bibinfo {author} {\bibfnamefont {Z.~E.}\ \bibnamefont
  {Meziani}} \emph {et~al.},\ }\href {\doibase
  10.1016/j.j.physletb.2005.03.046, 10.1016/j.physletb.2005.03.046} {\bibfield
  {journal} {\bibinfo  {journal} {Phys. Lett.}\ }\textbf {\bibinfo {volume}
  {B613}},\ \bibinfo {pages} {148} (\bibinfo {year} {2005})}\BibitemShut
  {NoStop}%
\bibitem [{\citenamefont {Sidorov}\ and\ \citenamefont
  {Weiss}(2006)}]{Sidorov:2006vu}%
  \BibitemOpen
  \bibfield  {author} {\bibinfo {author} {\bibfnamefont {A.~V.}\ \bibnamefont
  {Sidorov}}\ and\ \bibinfo {author} {\bibfnamefont {C.}~\bibnamefont
  {Weiss}},\ }\href {\doibase 10.1103/PhysRevD.73.074016} {\bibfield  {journal}
  {\bibinfo  {journal} {Phys. Rev.}\ }\textbf {\bibinfo {volume} {D73}},\
  \bibinfo {pages} {074016} (\bibinfo {year} {2006})}\BibitemShut {NoStop}%
\bibitem [{\citenamefont {Anselmino}\ \emph {et~al.}(2010)\citenamefont
  {Anselmino}, \citenamefont {Boglione}, \citenamefont {D'Alesio},
  \citenamefont {Melis}, \citenamefont {Murgia},\ and\ \citenamefont
  {Prokudin}}]{Anselmino:2009pn}%
  \BibitemOpen
  \bibfield  {author} {\bibinfo {author} {\bibfnamefont {M.}~\bibnamefont
  {Anselmino}}, \bibinfo {author} {\bibfnamefont {M.}~\bibnamefont {Boglione}},
  \bibinfo {author} {\bibfnamefont {U.}~\bibnamefont {D'Alesio}}, \bibinfo
  {author} {\bibfnamefont {S.}~\bibnamefont {Melis}}, \bibinfo {author}
  {\bibfnamefont {F.}~\bibnamefont {Murgia}}, \ and\ \bibinfo {author}
  {\bibfnamefont {A.}~\bibnamefont {Prokudin}},\ }\href {\doibase
  10.1103/PhysRevD.81.034007} {\bibfield  {journal} {\bibinfo  {journal} {Phys.
  Rev.}\ }\textbf {\bibinfo {volume} {D81}},\ \bibinfo {pages} {034007}
  (\bibinfo {year} {2010})}\BibitemShut {NoStop}%
\bibitem [{\citenamefont {Anselmino}\ \emph {et~al.}(2014)\citenamefont
  {Anselmino}, \citenamefont {Boglione}, \citenamefont {D'Alesio},
  \citenamefont {Melis}, \citenamefont {Murgia},\ and\ \citenamefont
  {Prokudin}}]{Anselmino:2014eza}%
  \BibitemOpen
  \bibfield  {author} {\bibinfo {author} {\bibfnamefont {M.}~\bibnamefont
  {Anselmino}}, \bibinfo {author} {\bibfnamefont {M.}~\bibnamefont {Boglione}},
  \bibinfo {author} {\bibfnamefont {U.}~\bibnamefont {D'Alesio}}, \bibinfo
  {author} {\bibfnamefont {S.}~\bibnamefont {Melis}}, \bibinfo {author}
  {\bibfnamefont {F.}~\bibnamefont {Murgia}}, \ and\ \bibinfo {author}
  {\bibfnamefont {A.}~\bibnamefont {Prokudin}},\ }\href {\doibase
  10.1103/PhysRevD.89.114026} {\bibfield  {journal} {\bibinfo  {journal} {Phys.
  Rev.}\ }\textbf {\bibinfo {volume} {D89}},\ \bibinfo {pages} {114026}
  (\bibinfo {year} {2014})}\BibitemShut {NoStop}%
\bibitem [{\citenamefont {Anthony}\ \emph {et~al.}(1996)\citenamefont {Anthony}
  \emph {et~al.}}]{Anthony:1996mw}%
  \BibitemOpen
  \bibfield  {author} {\bibinfo {author} {\bibfnamefont {P.~L.}\ \bibnamefont
  {Anthony}} \emph {et~al.},\ }\href {\doibase 10.1103/PhysRevD.54.6620}
  {\bibfield  {journal} {\bibinfo  {journal} {Phys. Rev.}\ }\textbf {\bibinfo
  {volume} {D54}},\ \bibinfo {pages} {6620} (\bibinfo {year}
  {1996})}\BibitemShut {NoStop}%
\bibitem [{\citenamefont {Abe}\ \emph {et~al.}(1998)\citenamefont {Abe} \emph
  {et~al.}}]{Abe:1998wq}%
  \BibitemOpen
  \bibfield  {author} {\bibinfo {author} {\bibfnamefont {K.}~\bibnamefont
  {Abe}} \emph {et~al.},\ }\href {\doibase 10.1103/PhysRevD.58.112003}
  {\bibfield  {journal} {\bibinfo  {journal} {Phys. Rev.}\ }\textbf {\bibinfo
  {volume} {D58}},\ \bibinfo {pages} {112003} (\bibinfo {year}
  {1998})}\BibitemShut {NoStop}%
\bibitem [{\citenamefont {Abe}\ \emph {et~al.}(1997{\natexlab{a}})\citenamefont
  {Abe} \emph {et~al.}}]{Abe:1997cx}%
  \BibitemOpen
  \bibfield  {author} {\bibinfo {author} {\bibfnamefont {K.}~\bibnamefont
  {Abe}} \emph {et~al.},\ }\bibfield  {booktitle} {\emph {\bibinfo {booktitle}
  {{The spin structure of the nucleon. Proceedings, International School of
  Nucleon Structure, 1st Course, Erice, Italy, August 3-10, 1995}}},\ }\href
  {\doibase 10.1103/PhysRevLett.79.26} {\bibfield  {journal} {\bibinfo
  {journal} {Phys. Rev. Lett.}\ }\textbf {\bibinfo {volume} {79}},\ \bibinfo
  {pages} {26} (\bibinfo {year} {1997}{\natexlab{a}})}\BibitemShut {NoStop}%
\bibitem [{\citenamefont {Abe}\ \emph {et~al.}(1997{\natexlab{b}})\citenamefont
  {Abe} \emph {et~al.}}]{Abe:1997qk}%
  \BibitemOpen
  \bibfield  {author} {\bibinfo {author} {\bibfnamefont {K.}~\bibnamefont
  {Abe}} \emph {et~al.},\ }\href {\doibase 10.1016/S0370-2693(97)00613-8}
  {\bibfield  {journal} {\bibinfo  {journal} {Phys. Lett.}\ }\textbf {\bibinfo
  {volume} {B404}},\ \bibinfo {pages} {377} (\bibinfo {year}
  {1997}{\natexlab{b}})}\BibitemShut {NoStop}%
\bibitem [{\citenamefont {Anthony}\ \emph
  {et~al.}(1999{\natexlab{a}})\citenamefont {Anthony} \emph
  {et~al.}}]{Anthony:1999py}%
  \BibitemOpen
  \bibfield  {author} {\bibinfo {author} {\bibfnamefont {P.~L.}\ \bibnamefont
  {Anthony}} \emph {et~al.},\ }\href {\doibase 10.1016/S0370-2693(99)00590-0}
  {\bibfield  {journal} {\bibinfo  {journal} {Phys. Lett.}\ }\textbf {\bibinfo
  {volume} {B458}},\ \bibinfo {pages} {529} (\bibinfo {year}
  {1999}{\natexlab{a}})}\BibitemShut {NoStop}%
\bibitem [{\citenamefont {Anthony}\ \emph
  {et~al.}(1999{\natexlab{b}})\citenamefont {Anthony} \emph
  {et~al.}}]{Anthony:1999rm}%
  \BibitemOpen
  \bibfield  {author} {\bibinfo {author} {\bibfnamefont {P.~L.}\ \bibnamefont
  {Anthony}} \emph {et~al.},\ }\href {\doibase 10.1016/S0370-2693(99)00940-5}
  {\bibfield  {journal} {\bibinfo  {journal} {Phys. Lett.}\ }\textbf {\bibinfo
  {volume} {B463}},\ \bibinfo {pages} {339} (\bibinfo {year}
  {1999}{\natexlab{b}})}\BibitemShut {NoStop}%
\bibitem [{\citenamefont {Anthony}\ \emph {et~al.}(2000)\citenamefont {Anthony}
  \emph {et~al.}}]{Anthony:2000fn}%
  \BibitemOpen
  \bibfield  {author} {\bibinfo {author} {\bibfnamefont {P.~L.}\ \bibnamefont
  {Anthony}} \emph {et~al.},\ }\href {\doibase 10.1016/S0370-2693(00)01014-5}
  {\bibfield  {journal} {\bibinfo  {journal} {Phys. Lett.}\ }\textbf {\bibinfo
  {volume} {B493}},\ \bibinfo {pages} {19} (\bibinfo {year}
  {2000})}\BibitemShut {NoStop}%
\bibitem [{\citenamefont {Anthony}\ \emph {et~al.}(2003)\citenamefont {Anthony}
  \emph {et~al.}}]{Anthony:2002hy}%
  \BibitemOpen
  \bibfield  {author} {\bibinfo {author} {\bibfnamefont {P.~L.}\ \bibnamefont
  {Anthony}} \emph {et~al.},\ }\href {\doibase 10.1016/S0370-2693(02)03015-0}
  {\bibfield  {journal} {\bibinfo  {journal} {Phys. Lett.}\ }\textbf {\bibinfo
  {volume} {B553}},\ \bibinfo {pages} {18} (\bibinfo {year}
  {2003})}\BibitemShut {NoStop}%
\bibitem [{\citenamefont {Ackerstaff}\ \emph {et~al.}(1997)\citenamefont
  {Ackerstaff} \emph {et~al.}}]{Ackerstaff:1997ws}%
  \BibitemOpen
  \bibfield  {author} {\bibinfo {author} {\bibfnamefont {K.}~\bibnamefont
  {Ackerstaff}} \emph {et~al.},\ }\bibfield  {booktitle} {\emph {\bibinfo
  {booktitle} {{The spin structure of the nucleon. Proceedings, International
  School of Nucleon Structure, 1st Course, Erice, Italy, August 3-10, 1995}}},\
  }\href {\doibase 10.1016/S0370-2693(97)00611-4} {\bibfield  {journal}
  {\bibinfo  {journal} {Phys. Lett.}\ }\textbf {\bibinfo {volume} {B404}},\
  \bibinfo {pages} {383} (\bibinfo {year} {1997})}\BibitemShut {NoStop}%
\bibitem [{\citenamefont {Airapetian}\ \emph {et~al.}(2007)\citenamefont
  {Airapetian} \emph {et~al.}}]{Airapetian:2006vy}%
  \BibitemOpen
  \bibfield  {author} {\bibinfo {author} {\bibfnamefont {A.}~\bibnamefont
  {Airapetian}} \emph {et~al.},\ }\href {\doibase 10.1103/PhysRevD.75.012007}
  {\bibfield  {journal} {\bibinfo  {journal} {Phys. Rev.}\ }\textbf {\bibinfo
  {volume} {D75}},\ \bibinfo {pages} {012007} (\bibinfo {year}
  {2007})}\BibitemShut {NoStop}%
\bibitem [{\citenamefont {Adeva}\ \emph {et~al.}(1998)\citenamefont {Adeva}
  \emph {et~al.}}]{Adeva:1998vv}%
  \BibitemOpen
  \bibfield  {author} {\bibinfo {author} {\bibfnamefont {B.}~\bibnamefont
  {Adeva}} \emph {et~al.},\ }\href {\doibase 10.1103/PhysRevD.58.112001}
  {\bibfield  {journal} {\bibinfo  {journal} {Phys. Rev.}\ }\textbf {\bibinfo
  {volume} {D58}},\ \bibinfo {pages} {112001} (\bibinfo {year}
  {1998})}\BibitemShut {NoStop}%
\bibitem [{\citenamefont {Alexakhin}\ \emph {et~al.}(2007)\citenamefont
  {Alexakhin} \emph {et~al.}}]{Alexakhin:2006oza}%
  \BibitemOpen
  \bibfield  {author} {\bibinfo {author} {\bibfnamefont {V.~{\relax Yu}.}\
  \bibnamefont {Alexakhin}} \emph {et~al.},\ }\href {\doibase
  10.1016/j.physletb.2006.12.076} {\bibfield  {journal} {\bibinfo  {journal}
  {Phys. Lett.}\ }\textbf {\bibinfo {volume} {B647}},\ \bibinfo {pages} {8}
  (\bibinfo {year} {2007})}\BibitemShut {NoStop}%
\bibitem [{\citenamefont {Alekseev}\ \emph {et~al.}(2009)\citenamefont
  {Alekseev} \emph {et~al.}}]{Alekseev:2009ac}%
  \BibitemOpen
  \bibfield  {author} {\bibinfo {author} {\bibfnamefont {M.}~\bibnamefont
  {Alekseev}} \emph {et~al.},\ }\href {\doibase 10.1016/j.physletb.2009.08.065}
  {\bibfield  {journal} {\bibinfo  {journal} {Phys. Lett.}\ }\textbf {\bibinfo
  {volume} {B680}},\ \bibinfo {pages} {217} (\bibinfo {year}
  {2009})}\BibitemShut {NoStop}%
\bibitem [{\citenamefont {Adolph}\ \emph {et~al.}(2017)\citenamefont {Adolph}
  \emph {et~al.}}]{Adolph:2016myg}%
  \BibitemOpen
  \bibfield  {author} {\bibinfo {author} {\bibfnamefont {C.}~\bibnamefont
  {Adolph}} \emph {et~al.} (\bibinfo {collaboration} {COMPASS}),\ }\href
  {\doibase 10.1016/j.physletb.2017.03.018} {\bibfield  {journal} {\bibinfo
  {journal} {Phys. Lett. B}\ }\textbf {\bibinfo {volume} {769}},\ \bibinfo
  {pages} {34} (\bibinfo {year} {2017})},\ \Eprint
  {http://arxiv.org/abs/1612.00620} {arXiv:1612.00620 [hep-ex]} \BibitemShut
  {NoStop}%
\bibitem [{\citenamefont {Zheng}\ \emph {et~al.}(2004)\citenamefont {Zheng}
  \emph {et~al.}}]{Zheng:2004ce}%
  \BibitemOpen
  \bibfield  {author} {\bibinfo {author} {\bibfnamefont {X.}~\bibnamefont
  {Zheng}} \emph {et~al.},\ }\href {\doibase 10.1103/PhysRevC.70.065207}
  {\bibfield  {journal} {\bibinfo  {journal} {Phys. Rev.}\ }\textbf {\bibinfo
  {volume} {C70}},\ \bibinfo {pages} {065207} (\bibinfo {year}
  {2004})}\BibitemShut {NoStop}%
\bibitem [{\citenamefont {Posik}\ \emph {et~al.}(2014)\citenamefont {Posik}
  \emph {et~al.}}]{Posik:2014usi}%
  \BibitemOpen
  \bibfield  {author} {\bibinfo {author} {\bibfnamefont {M.}~\bibnamefont
  {Posik}} \emph {et~al.},\ }\href {\doibase 10.1103/PhysRevLett.113.022002}
  {\bibfield  {journal} {\bibinfo  {journal} {Phys. Rev. Lett.}\ }\textbf
  {\bibinfo {volume} {113}},\ \bibinfo {pages} {022002} (\bibinfo {year}
  {2014})}\BibitemShut {NoStop}%
\bibitem [{\citenamefont {Prok}\ \emph {et~al.}(2014)\citenamefont {Prok} \emph
  {et~al.}}]{Prok:2014ltt}%
  \BibitemOpen
  \bibfield  {author} {\bibinfo {author} {\bibfnamefont {Y.}~\bibnamefont
  {Prok}} \emph {et~al.},\ }\href {\doibase 10.1103/PhysRevC.90.025212}
  {\bibfield  {journal} {\bibinfo  {journal} {Phys. Rev.}\ }\textbf {\bibinfo
  {volume} {C90}},\ \bibinfo {pages} {025212} (\bibinfo {year}
  {2014})}\BibitemShut {NoStop}%
\bibitem [{\citenamefont {Chen}\ \emph {et~al.}(2011)\citenamefont {Chen},
  \citenamefont {Deur}, \citenamefont {Kuhn},\ and\ \citenamefont
  {Meziani}}]{Chen:2011zzp}%
  \BibitemOpen
  \bibfield  {author} {\bibinfo {author} {\bibfnamefont {J.~P.}\ \bibnamefont
  {Chen}}, \bibinfo {author} {\bibfnamefont {A.}~\bibnamefont {Deur}}, \bibinfo
  {author} {\bibfnamefont {S.}~\bibnamefont {Kuhn}}, \ and\ \bibinfo {author}
  {\bibfnamefont {Z.~E.}\ \bibnamefont {Meziani}},\ }\href {\doibase
  10.1088/1742-6596/299/1/012005} {\bibfield  {journal} {\bibinfo  {journal}
  {J. Phys. Conf. Ser.}\ }\textbf {\bibinfo {volume} {299}},\ \bibinfo {pages}
  {012005} (\bibinfo {year} {2011})}\BibitemShut {NoStop}%
\bibitem [{\citenamefont {Malace}\ \emph {et~al.}(2014)\citenamefont {Malace},
  \citenamefont {Gaskell}, \citenamefont {Higinbotham},\ and\ \citenamefont
  {Cloet}}]{Malace:2014uea}%
  \BibitemOpen
  \bibfield  {author} {\bibinfo {author} {\bibfnamefont {S.}~\bibnamefont
  {Malace}}, \bibinfo {author} {\bibfnamefont {D.}~\bibnamefont {Gaskell}},
  \bibinfo {author} {\bibfnamefont {D.~W.}\ \bibnamefont {Higinbotham}}, \ and\
  \bibinfo {author} {\bibfnamefont {I.}~\bibnamefont {Cloet}},\ }\href
  {\doibase 10.1142/S0218301314300136} {\bibfield  {journal} {\bibinfo
  {journal} {Int. J. Mod. Phys.}\ }\textbf {\bibinfo {volume} {E23}},\ \bibinfo
  {pages} {1430013} (\bibinfo {year} {2014})}\BibitemShut {NoStop}%
\bibitem [{\citenamefont {Frankfurt}\ and\ \citenamefont
  {Strikman}(1988)}]{Frankfurt:1988nt}%
  \BibitemOpen
  \bibfield  {author} {\bibinfo {author} {\bibfnamefont {L.~L.}\ \bibnamefont
  {Frankfurt}}\ and\ \bibinfo {author} {\bibfnamefont {M.~I.}\ \bibnamefont
  {Strikman}},\ }\href {\doibase 10.1016/0370-1573(88)90179-2} {\bibfield
  {journal} {\bibinfo  {journal} {Phys. Rept.}\ }\textbf {\bibinfo {volume}
  {160}},\ \bibinfo {pages} {235} (\bibinfo {year} {1988})}\BibitemShut
  {NoStop}%
\bibitem [{\citenamefont {Arneodo}(1994)}]{Arneodo:1992wf}%
  \BibitemOpen
  \bibfield  {author} {\bibinfo {author} {\bibfnamefont {M.}~\bibnamefont
  {Arneodo}},\ }\href {\doibase 10.1016/0370-1573(94)90048-5} {\bibfield
  {journal} {\bibinfo  {journal} {Phys. Rept.}\ }\textbf {\bibinfo {volume}
  {240}},\ \bibinfo {pages} {301} (\bibinfo {year} {1994})}\BibitemShut
  {NoStop}%
\bibitem [{\citenamefont {Ciofi~degli Atti}\ \emph {et~al.}(1993)\citenamefont
  {Ciofi~degli Atti}, \citenamefont {Scopetta}, \citenamefont {Pace},\ and\
  \citenamefont {Salme}}]{CiofidegliAtti:1993zs}%
  \BibitemOpen
  \bibfield  {author} {\bibinfo {author} {\bibfnamefont {C.}~\bibnamefont
  {Ciofi~degli Atti}}, \bibinfo {author} {\bibfnamefont {S.}~\bibnamefont
  {Scopetta}}, \bibinfo {author} {\bibfnamefont {E.}~\bibnamefont {Pace}}, \
  and\ \bibinfo {author} {\bibfnamefont {G.}~\bibnamefont {Salme}},\ }\href
  {\doibase 10.1103/PhysRevC.48.R968} {\bibfield  {journal} {\bibinfo
  {journal} {Phys. Rev.}\ }\textbf {\bibinfo {volume} {C48}},\ \bibinfo {pages}
  {R968} (\bibinfo {year} {1993})},\ \Eprint
  {http://arxiv.org/abs/nucl-th/9303016} {arXiv:nucl-th/9303016 [nucl-th]}
  \BibitemShut {NoStop}%
\bibitem [{\citenamefont {Melnitchouk}\ \emph {et~al.}(1995)\citenamefont
  {Melnitchouk}, \citenamefont {Piller},\ and\ \citenamefont
  {Thomas}}]{Melnitchouk:1994tx}%
  \BibitemOpen
  \bibfield  {author} {\bibinfo {author} {\bibfnamefont {W.}~\bibnamefont
  {Melnitchouk}}, \bibinfo {author} {\bibfnamefont {G.}~\bibnamefont {Piller}},
  \ and\ \bibinfo {author} {\bibfnamefont {A.~W.}\ \bibnamefont {Thomas}},\
  }\href {\doibase 10.1016/0370-2693(94)01690-E} {\bibfield  {journal}
  {\bibinfo  {journal} {Phys. Lett.}\ }\textbf {\bibinfo {volume} {B346}},\
  \bibinfo {pages} {165} (\bibinfo {year} {1995})},\ \Eprint
  {http://arxiv.org/abs/hep-ph/9501282} {arXiv:hep-ph/9501282 [hep-ph]}
  \BibitemShut {NoStop}%
\bibitem [{\citenamefont {Kulagin}\ \emph {et~al.}(1995)\citenamefont
  {Kulagin}, \citenamefont {Melnitchouk}, \citenamefont {Piller},\ and\
  \citenamefont {Weise}}]{Kulagin:1994cj}%
  \BibitemOpen
  \bibfield  {author} {\bibinfo {author} {\bibfnamefont {S.~A.}\ \bibnamefont
  {Kulagin}}, \bibinfo {author} {\bibfnamefont {W.}~\bibnamefont
  {Melnitchouk}}, \bibinfo {author} {\bibfnamefont {G.}~\bibnamefont {Piller}},
  \ and\ \bibinfo {author} {\bibfnamefont {W.}~\bibnamefont {Weise}},\ }\href
  {\doibase 10.1103/PhysRevC.52.932} {\bibfield  {journal} {\bibinfo  {journal}
  {Phys. Rev.}\ }\textbf {\bibinfo {volume} {C52}},\ \bibinfo {pages} {932}
  (\bibinfo {year} {1995})},\ \Eprint {http://arxiv.org/abs/hep-ph/9504377}
  {arXiv:hep-ph/9504377 [hep-ph]} \BibitemShut {NoStop}%
\bibitem [{\citenamefont {Piller}\ \emph {et~al.}(1996)\citenamefont {Piller},
  \citenamefont {Melnitchouk},\ and\ \citenamefont {Thomas}}]{Piller:1995mf}%
  \BibitemOpen
  \bibfield  {author} {\bibinfo {author} {\bibfnamefont {G.}~\bibnamefont
  {Piller}}, \bibinfo {author} {\bibfnamefont {W.}~\bibnamefont {Melnitchouk}},
  \ and\ \bibinfo {author} {\bibfnamefont {A.~W.}\ \bibnamefont {Thomas}},\
  }\href {\doibase 10.1103/PhysRevC.54.894} {\bibfield  {journal} {\bibinfo
  {journal} {Phys. Rev.}\ }\textbf {\bibinfo {volume} {C54}},\ \bibinfo {pages}
  {894} (\bibinfo {year} {1996})},\ \Eprint
  {http://arxiv.org/abs/nucl-th/9605045} {arXiv:nucl-th/9605045 [nucl-th]}
  \BibitemShut {NoStop}%
\bibitem [{\citenamefont {Frankfurt}\ \emph {et~al.}(1996)\citenamefont
  {Frankfurt}, \citenamefont {Guzey},\ and\ \citenamefont
  {Strikman}}]{Frankfurt:1996nf}%
  \BibitemOpen
  \bibfield  {author} {\bibinfo {author} {\bibfnamefont {L.}~\bibnamefont
  {Frankfurt}}, \bibinfo {author} {\bibfnamefont {V.}~\bibnamefont {Guzey}}, \
  and\ \bibinfo {author} {\bibfnamefont {M.}~\bibnamefont {Strikman}},\ }\href
  {\doibase 10.1016/0370-2693(96)00625-9} {\bibfield  {journal} {\bibinfo
  {journal} {Phys. Lett.}\ }\textbf {\bibinfo {volume} {B381}},\ \bibinfo
  {pages} {379} (\bibinfo {year} {1996})},\ \Eprint
  {http://arxiv.org/abs/hep-ph/9602301} {arXiv:hep-ph/9602301 [hep-ph]}
  \BibitemShut {NoStop}%
\bibitem [{\citenamefont {Bissey}\ \emph {et~al.}(2002)\citenamefont {Bissey},
  \citenamefont {Guzey}, \citenamefont {Strikman},\ and\ \citenamefont
  {Thomas}}]{Bissey:2001cw}%
  \BibitemOpen
  \bibfield  {author} {\bibinfo {author} {\bibfnamefont {F.~R.~P.}\
  \bibnamefont {Bissey}}, \bibinfo {author} {\bibfnamefont {V.~A.}\
  \bibnamefont {Guzey}}, \bibinfo {author} {\bibfnamefont {M.}~\bibnamefont
  {Strikman}}, \ and\ \bibinfo {author} {\bibfnamefont {A.~W.}\ \bibnamefont
  {Thomas}},\ }\href {\doibase 10.1103/PhysRevC.65.064317} {\bibfield
  {journal} {\bibinfo  {journal} {Phys. Rev.}\ }\textbf {\bibinfo {volume}
  {C65}},\ \bibinfo {pages} {064317} (\bibinfo {year} {2002})},\ \Eprint
  {http://arxiv.org/abs/hep-ph/0109069} {arXiv:hep-ph/0109069 [hep-ph]}
  \BibitemShut {NoStop}%
\bibitem [{\citenamefont {Ethier}\ and\ \citenamefont
  {Melnitchouk}(2013)}]{Ethier:2013hna}%
  \BibitemOpen
  \bibfield  {author} {\bibinfo {author} {\bibfnamefont {J.~J.}\ \bibnamefont
  {Ethier}}\ and\ \bibinfo {author} {\bibfnamefont {W.}~\bibnamefont
  {Melnitchouk}},\ }\href {\doibase 10.1103/PhysRevC.88.054001} {\bibfield
  {journal} {\bibinfo  {journal} {Phys. Rev.}\ }\textbf {\bibinfo {volume}
  {C88}},\ \bibinfo {pages} {054001} (\bibinfo {year} {2013})},\ \Eprint
  {http://arxiv.org/abs/1308.3723} {arXiv:1308.3723 [nucl-th]} \BibitemShut
  {NoStop}%
\bibitem [{\citenamefont {Friar}\ \emph {et~al.}(1990)\citenamefont {Friar},
  \citenamefont {Gibson}, \citenamefont {Payne}, \citenamefont {Bernstein},\
  and\ \citenamefont {Chupp}}]{Friar:1990vx}%
  \BibitemOpen
  \bibfield  {author} {\bibinfo {author} {\bibfnamefont {J.~L.}\ \bibnamefont
  {Friar}}, \bibinfo {author} {\bibfnamefont {B.~F.}\ \bibnamefont {Gibson}},
  \bibinfo {author} {\bibfnamefont {G.~L.}\ \bibnamefont {Payne}}, \bibinfo
  {author} {\bibfnamefont {A.~M.}\ \bibnamefont {Bernstein}}, \ and\ \bibinfo
  {author} {\bibfnamefont {T.~E.}\ \bibnamefont {Chupp}},\ }\href {\doibase
  10.1103/PhysRevC.42.2310} {\bibfield  {journal} {\bibinfo  {journal} {Phys.
  Rev.}\ }\textbf {\bibinfo {volume} {C42}},\ \bibinfo {pages} {2310} (\bibinfo
  {year} {1990})}\BibitemShut {NoStop}%
\bibitem [{\citenamefont {Frankfurt}\ and\ \citenamefont
  {Strikman}(1981)}]{Frankfurt:1981mk}%
  \BibitemOpen
  \bibfield  {author} {\bibinfo {author} {\bibfnamefont {L.~L.}\ \bibnamefont
  {Frankfurt}}\ and\ \bibinfo {author} {\bibfnamefont {M.~I.}\ \bibnamefont
  {Strikman}},\ }\href {\doibase 10.1016/0370-1573(81)90129-0} {\bibfield
  {journal} {\bibinfo  {journal} {Phys. Rept.}\ }\textbf {\bibinfo {volume}
  {76}},\ \bibinfo {pages} {215} (\bibinfo {year} {1981})}\BibitemShut
  {NoStop}%
\bibitem [{\citenamefont {Sargsian}\ and\ \citenamefont
  {Strikman}(2006)}]{Sargsian:2005rm}%
  \BibitemOpen
  \bibfield  {author} {\bibinfo {author} {\bibfnamefont {M.}~\bibnamefont
  {Sargsian}}\ and\ \bibinfo {author} {\bibfnamefont {M.}~\bibnamefont
  {Strikman}},\ }\href {\doibase 10.1016/j.physletb.2006.05.091} {\bibfield
  {journal} {\bibinfo  {journal} {Phys. Lett.}\ }\textbf {\bibinfo {volume}
  {B639}},\ \bibinfo {pages} {223} (\bibinfo {year} {2006})},\ \Eprint
  {http://arxiv.org/abs/hep-ph/0511054} {arXiv:hep-ph/0511054 [hep-ph]}
  \BibitemShut {NoStop}%
\bibitem [{\citenamefont {Baillie}\ \emph {et~al.}(2012)\citenamefont {Baillie}
  \emph {et~al.}}]{Baillie:2011za}%
  \BibitemOpen
  \bibfield  {author} {\bibinfo {author} {\bibfnamefont {N.}~\bibnamefont
  {Baillie}} \emph {et~al.} (\bibinfo {collaboration} {CLAS}),\ }\href
  {\doibase 10.1103/PhysRevLett.108.142001} {\bibfield  {journal} {\bibinfo
  {journal} {Phys. Rev. Lett.}\ }\textbf {\bibinfo {volume} {108}},\ \bibinfo
  {pages} {142001} (\bibinfo {year} {2012})},\ \bibinfo {note} {[Erratum:
  Phys.Rev.Lett. 108, 199902 (2012)]},\ \Eprint
  {http://arxiv.org/abs/1110.2770} {arXiv:1110.2770 [nucl-ex]} \BibitemShut
  {NoStop}%
\bibitem [{\citenamefont {Tkachenko}\ \emph {et~al.}(2014)\citenamefont
  {Tkachenko} \emph {et~al.}}]{Tkachenko:2014byy}%
  \BibitemOpen
  \bibfield  {author} {\bibinfo {author} {\bibfnamefont {S.}~\bibnamefont
  {Tkachenko}} \emph {et~al.} (\bibinfo {collaboration} {CLAS}),\ }\href
  {\doibase 10.1103/PhysRevC.89.045206} {\bibfield  {journal} {\bibinfo
  {journal} {Phys. Rev. C}\ }\textbf {\bibinfo {volume} {89}},\ \bibinfo
  {pages} {045206} (\bibinfo {year} {2014})},\ \bibinfo {note} {[Addendum:
  Phys.Rev.C 90, 059901 (2014)]},\ \Eprint {http://arxiv.org/abs/1402.2477}
  {arXiv:1402.2477 [nucl-ex]} \BibitemShut {NoStop}%
\bibitem [{\citenamefont {Bueltmann}\ \emph {et~al.}()\citenamefont {Bueltmann}
  \emph {et~al.}}]{Bonus12}%
  \BibitemOpen
  \bibfield  {author} {\bibinfo {author} {\bibfnamefont {S.}~\bibnamefont
  {Bueltmann}} \emph {et~al.},\ }\href@noop {} {\bibinfo  {journal} {Report
  JLAB-PR12-06-113 (2006),
  \url{https://www.jlab.org/exp_prog/proposals/10/PR12-06-113-pac36.pdf}}\
  }\BibitemShut {NoStop}%
\bibitem [{\citenamefont {Armstrong}\ \emph {et~al.}(2017)\citenamefont
  {Armstrong} \emph {et~al.}}]{Armstrong:2017zqr}%
  \BibitemOpen
\bibfield  {journal} {  }\bibfield  {author} {\bibinfo {author} {\bibfnamefont
  {W.}~\bibnamefont {Armstrong}} \emph {et~al.},\ }\href@noop {} {\  (\bibinfo
  {year} {2017})},\ \Eprint {http://arxiv.org/abs/1708.00891} {arXiv:1708.00891
  [nucl-ex]} \BibitemShut {NoStop}%
\bibitem [{\citenamefont {Klimenko}\ \emph {et~al.}(2006)\citenamefont
  {Klimenko} \emph {et~al.}}]{Klimenko:2005zz}%
  \BibitemOpen
  \bibfield  {author} {\bibinfo {author} {\bibfnamefont {A.~V.}\ \bibnamefont
  {Klimenko}} \emph {et~al.} (\bibinfo {collaboration} {CLAS}),\ }\href
  {\doibase 10.1103/PhysRevC.73.035212} {\bibfield  {journal} {\bibinfo
  {journal} {Phys. Rev.}\ }\textbf {\bibinfo {volume} {C73}},\ \bibinfo {pages}
  {035212} (\bibinfo {year} {2006})},\ \Eprint
  {http://arxiv.org/abs/nucl-ex/0510032} {arXiv:nucl-ex/0510032 [nucl-ex]}
  \BibitemShut {NoStop}%
\bibitem [{\citenamefont {Hen}\ \emph {et~al.}(2015)\citenamefont {Hen} \emph
  {et~al.}}]{Hen:2011}%
  \BibitemOpen
  \bibfield  {author} {\bibinfo {author} {\bibfnamefont {O.}~\bibnamefont
  {Hen}} \emph {et~al.},\ }\href@noop {} {\bibfield  {journal} {\bibinfo
  {journal} {{Jefferson Lab Experiment E12-11-003A,
  \url{https://www.jlab.org/exp_prog/proposals/15/E12-11-003A.pdf}}}\ }
  (\bibinfo {year} {2015})}\BibitemShut {NoStop}%
\bibitem [{\citenamefont {Hen}\ \emph {et~al.}(2014)\citenamefont {Hen},
  \citenamefont {Weinstein}, \citenamefont {Gilad},\ and\ \citenamefont
  {Wood}}]{Hen:2014vua}%
  \BibitemOpen
  \bibfield  {author} {\bibinfo {author} {\bibfnamefont {O.}~\bibnamefont
  {Hen}}, \bibinfo {author} {\bibfnamefont {L.}~\bibnamefont {Weinstein}},
  \bibinfo {author} {\bibfnamefont {S.}~\bibnamefont {Gilad}}, \ and\ \bibinfo
  {author} {\bibfnamefont {S.}~\bibnamefont {Wood}},\ }\href@noop {} {\
  (\bibinfo {year} {2014})},\ \Eprint {http://arxiv.org/abs/1409.1717}
  {arXiv:1409.1717 [nucl-ex]} \BibitemShut {NoStop}%
\bibitem [{\citenamefont {Boer}\ \emph {et~al.}(2011)\citenamefont {Boer} \emph
  {et~al.}}]{Boer:2011fh}%
  \BibitemOpen
  \bibfield  {author} {\bibinfo {author} {\bibfnamefont {D.}~\bibnamefont
  {Boer}} \emph {et~al.},\ }\href@noop {} {\  (\bibinfo {year} {2011})},\
  \Eprint {http://arxiv.org/abs/1108.1713} {arXiv:1108.1713 [nucl-th]}
  \BibitemShut {NoStop}%
\bibitem [{\citenamefont {Accardi}\ \emph {et~al.}(2016)\citenamefont {Accardi}
  \emph {et~al.}}]{Accardi:2012qut}%
  \BibitemOpen
  \bibfield  {author} {\bibinfo {author} {\bibfnamefont {A.}~\bibnamefont
  {Accardi}} \emph {et~al.},\ }\href {\doibase 10.1140/epja/i2016-16268-9}
  {\bibfield  {journal} {\bibinfo  {journal} {Eur. Phys. J.}\ }\textbf
  {\bibinfo {volume} {A52}},\ \bibinfo {pages} {268} (\bibinfo {year}
  {2016})},\ \Eprint {http://arxiv.org/abs/1212.1701} {arXiv:1212.1701
  [nucl-ex]} \BibitemShut {NoStop}%
\bibitem [{\citenamefont {Aschenauer}\ \emph {et~al.}(2014)\citenamefont
  {Aschenauer} \emph {et~al.}}]{Aschenauer:2014cki}%
  \BibitemOpen
  \bibfield  {author} {\bibinfo {author} {\bibfnamefont {E.~C.}\ \bibnamefont
  {Aschenauer}} \emph {et~al.},\ }\href@noop {} {\  (\bibinfo {year} {2014})},\
  \Eprint {http://arxiv.org/abs/1409.1633} {arXiv:1409.1633 [physics.acc-ph]}
  \BibitemShut {NoStop}%
\bibitem [{\citenamefont {Beebe-Wang}\ \emph {et~al.}()\citenamefont
  {Beebe-Wang} \emph {et~al.}}]{EICpCDR}%
  \BibitemOpen
  \bibfield  {author} {\bibinfo {author} {\bibfnamefont {J.}~\bibnamefont
  {Beebe-Wang}} \emph {et~al.},\ }\bibfield  {booktitle} {\emph {\bibinfo
  {booktitle} {An Electron-Ion Collider Study}},\ }\href@noop {} {\bibinfo
  {journal} {Pre-Conceptual Design Report, Brookhaven National Laboratory,
  public version for scientists available at
  \url{https://wiki.bnl.gov/eic/upload/EIC.Design.Study.pdf}}\ }\BibitemShut
  {NoStop}%
\bibitem [{\citenamefont {Jentsch}()}]{Jentsch:2020}%
  \BibitemOpen
\bibfield  {journal} {  }\bibfield  {author} {\bibinfo {author} {\bibfnamefont
  {A.}~\bibnamefont {Jentsch}},\ }\bibfield  {booktitle} {\emph {\bibinfo
  {booktitle} {DVCS and e+D spectator tagging in the FF region}},\ }\href@noop
  {} {\bibinfo  {journal} {Presentation at 2nd EIC Yellow Report Workshop at
  Pavia University, 20-22 May 2020,
  \url{https://indico.bnl.gov/event/8231/contributions/37699/}}\ }\BibitemShut
  {NoStop}%
\bibitem [{\citenamefont {Higinbotham}()}]{Higinbotham:2020}%
  \BibitemOpen
\bibfield  {journal} {  }\bibfield  {author} {\bibinfo {author} {\bibfnamefont
  {D.}~\bibnamefont {Higinbotham}},\ }\bibfield  {booktitle} {\emph {\bibinfo
  {booktitle} {Magic beam energies for polarized deuteron}},\ }\href@noop {}
  {\bibinfo  {journal} {Presentation at 2nd EIC Yellow Report Workshop at Pavia
  University, 20-22 May 2020,
  \url{https://indico.bnl.gov/event/8231/contributions/37701/}}\ }\BibitemShut
  {NoStop}%
\bibitem [{\citenamefont {Abeyratne}\ \emph {et~al.}(2012)\citenamefont
  {Abeyratne} \emph {et~al.}}]{Abeyratne:2012ah}%
  \BibitemOpen
\bibfield  {journal} {  }\bibfield  {author} {\bibinfo {author} {\bibfnamefont
  {S.}~\bibnamefont {Abeyratne}} \emph {et~al.},\ }\href@noop {} {\  (\bibinfo
  {year} {2012})},\ \Eprint {http://arxiv.org/abs/1209.0757} {arXiv:1209.0757
  [physics.acc-ph]} \BibitemShut {NoStop}%
\bibitem [{\citenamefont {Weiss}\ \emph {et~al.}()\citenamefont {Weiss} \emph
  {et~al.}}]{LD1506}%
  \BibitemOpen
  \bibfield  {author} {\bibinfo {author} {\bibfnamefont {C.}~\bibnamefont
  {Weiss}} \emph {et~al.},\ }\href@noop {} {\bibinfo  {journal} {Jefferson Lab
  2014/2015 Laboratory-directed Research and Development Project, Webpage:
  \url{https://www.jlab.org/theory/tag/}}\ }\BibitemShut {NoStop}%
\bibitem [{\citenamefont {Guzey}\ \emph {et~al.}(2014)\citenamefont {Guzey},
  \citenamefont {Higinbotham}, \citenamefont {Hyde}, \citenamefont
  {Nadel-Turonski}, \citenamefont {Park}, \citenamefont {Sargsian},
  \citenamefont {Strikman},\ and\ \citenamefont {Weiss}}]{Guzey:2014jva}%
  \BibitemOpen
\bibfield  {journal} {  }\bibfield  {author} {\bibinfo {author} {\bibfnamefont
  {V.}~\bibnamefont {Guzey}}, \bibinfo {author} {\bibfnamefont
  {D.}~\bibnamefont {Higinbotham}}, \bibinfo {author} {\bibfnamefont
  {C.}~\bibnamefont {Hyde}}, \bibinfo {author} {\bibfnamefont {P.}~\bibnamefont
  {Nadel-Turonski}}, \bibinfo {author} {\bibfnamefont {K.}~\bibnamefont
  {Park}}, \bibinfo {author} {\bibfnamefont {M.}~\bibnamefont {Sargsian}},
  \bibinfo {author} {\bibfnamefont {M.}~\bibnamefont {Strikman}}, \ and\
  \bibinfo {author} {\bibfnamefont {C.}~\bibnamefont {Weiss}},\ }\bibfield
  {booktitle} {\emph {\bibinfo {booktitle} {{Proceedings, 22nd International
  Workshop on Deep-Inelastic Scattering and Related Subjects (DIS 2014):
  Warsaw, Poland, April 28-May 2, 2014}}},\ }\href@noop {} {\bibfield
  {journal} {\bibinfo  {journal} {PoS}\ }\textbf {\bibinfo {volume}
  {DIS2014}},\ \bibinfo {pages} {234} (\bibinfo {year} {2014})},\ \Eprint
  {http://arxiv.org/abs/1407.3236} {arXiv:1407.3236 [hep-ph]} \BibitemShut
  {NoStop}%
\bibitem [{\citenamefont {Cosyn}\ \emph {et~al.}(2014)\citenamefont {Cosyn},
  \citenamefont {Guzey}, \citenamefont {Higinbotham}, \citenamefont {Hyde},
  \citenamefont {Kuhn}, \citenamefont {Nadel-Turonski}, \citenamefont {Park},
  \citenamefont {Sargsian}, \citenamefont {Strikman},\ and\ \citenamefont
  {Weiss}}]{Cosyn:2014zfa}%
  \BibitemOpen
  \bibfield  {author} {\bibinfo {author} {\bibfnamefont {W.}~\bibnamefont
  {Cosyn}}, \bibinfo {author} {\bibfnamefont {V.}~\bibnamefont {Guzey}},
  \bibinfo {author} {\bibfnamefont {D.~W.}\ \bibnamefont {Higinbotham}},
  \bibinfo {author} {\bibfnamefont {C.}~\bibnamefont {Hyde}}, \bibinfo {author}
  {\bibfnamefont {S.}~\bibnamefont {Kuhn}}, \bibinfo {author} {\bibfnamefont
  {P.}~\bibnamefont {Nadel-Turonski}}, \bibinfo {author} {\bibfnamefont
  {K.}~\bibnamefont {Park}}, \bibinfo {author} {\bibfnamefont {M.}~\bibnamefont
  {Sargsian}}, \bibinfo {author} {\bibfnamefont {M.}~\bibnamefont {Strikman}},
  \ and\ \bibinfo {author} {\bibfnamefont {C.}~\bibnamefont {Weiss}},\
  }\bibfield  {booktitle} {\emph {\bibinfo {booktitle} {{Proceedings, Tensor
  Polarized Solid Target Workshop: Newport News, USA, March 10-12, 2014}}},\
  }\href {\doibase 10.1088/1742-6596/543/1/012007} {\bibfield  {journal}
  {\bibinfo  {journal} {J. Phys. Conf. Ser.}\ }\textbf {\bibinfo {volume}
  {543}},\ \bibinfo {pages} {012007} (\bibinfo {year} {2014})},\ \Eprint
  {http://arxiv.org/abs/1409.5768} {arXiv:1409.5768 [hep-ph]} \BibitemShut
  {NoStop}%
\bibitem [{\citenamefont {Frankfurt}\ and\ \citenamefont
  {Strikman}(1983)}]{Frankfurt:1983qs}%
  \BibitemOpen
  \bibfield  {author} {\bibinfo {author} {\bibfnamefont {L.~L.}\ \bibnamefont
  {Frankfurt}}\ and\ \bibinfo {author} {\bibfnamefont {M.~I.}\ \bibnamefont
  {Strikman}},\ }\href {\doibase 10.1016/0375-9474(83)90518-3} {\bibfield
  {journal} {\bibinfo  {journal} {Nucl. Phys.}\ }\textbf {\bibinfo {volume}
  {A405}},\ \bibinfo {pages} {557} (\bibinfo {year} {1983})}\BibitemShut
  {NoStop}%
\bibitem [{\citenamefont {Cosyn}\ and\ \citenamefont
  {Weiss}(2019)}]{Cosyn:2019hem}%
  \BibitemOpen
  \bibfield  {author} {\bibinfo {author} {\bibfnamefont {W.}~\bibnamefont
  {Cosyn}}\ and\ \bibinfo {author} {\bibfnamefont {C.}~\bibnamefont {Weiss}},\
  }\href {\doibase 10.1016/j.physletb.2019.135035} {\bibfield  {journal}
  {\bibinfo  {journal} {Phys. Lett. B}\ }\textbf {\bibinfo {volume} {799}},\
  \bibinfo {pages} {135035} (\bibinfo {year} {2019})},\ \Eprint
  {http://arxiv.org/abs/1906.11119} {arXiv:1906.11119 [hep-ph]} \BibitemShut
  {NoStop}%
\bibitem [{\citenamefont {Strikman}\ and\ \citenamefont
  {Weiss}(2018)}]{Strikman:2017koc}%
  \BibitemOpen
  \bibfield  {author} {\bibinfo {author} {\bibfnamefont {M.}~\bibnamefont
  {Strikman}}\ and\ \bibinfo {author} {\bibfnamefont {C.}~\bibnamefont
  {Weiss}},\ }\href {\doibase 10.1103/PhysRevC.97.035209} {\bibfield  {journal}
  {\bibinfo  {journal} {Phys. Rev.}\ }\textbf {\bibinfo {volume} {C97}},\
  \bibinfo {pages} {035209} (\bibinfo {year} {2018})},\ \Eprint
  {http://arxiv.org/abs/1706.02244} {arXiv:1706.02244 [hep-ph]} \BibitemShut
  {NoStop}%
\bibitem [{\citenamefont {Coester}(1992)}]{Coester:1992cg}%
  \BibitemOpen
  \bibfield  {author} {\bibinfo {author} {\bibfnamefont {F.}~\bibnamefont
  {Coester}},\ }\href {\doibase 10.1016/0146-6410(92)90002-J} {\bibfield
  {journal} {\bibinfo  {journal} {Prog. Part. Nucl. Phys.}\ }\textbf {\bibinfo
  {volume} {29}},\ \bibinfo {pages} {1} (\bibinfo {year} {1992})}\BibitemShut
  {NoStop}%
\bibitem [{\citenamefont {Brodsky}\ \emph {et~al.}(1998)\citenamefont
  {Brodsky}, \citenamefont {Pauli},\ and\ \citenamefont
  {Pinsky}}]{Brodsky:1997de}%
  \BibitemOpen
  \bibfield  {author} {\bibinfo {author} {\bibfnamefont {S.~J.}\ \bibnamefont
  {Brodsky}}, \bibinfo {author} {\bibfnamefont {H.-C.}\ \bibnamefont {Pauli}},
  \ and\ \bibinfo {author} {\bibfnamefont {S.~S.}\ \bibnamefont {Pinsky}},\
  }\href {\doibase 10.1016/S0370-1573(97)00089-6} {\bibfield  {journal}
  {\bibinfo  {journal} {Phys. Rept.}\ }\textbf {\bibinfo {volume} {301}},\
  \bibinfo {pages} {299} (\bibinfo {year} {1998})},\ \Eprint
  {http://arxiv.org/abs/hep-ph/9705477} {arXiv:hep-ph/9705477 [hep-ph]}
  \BibitemShut {NoStop}%
\bibitem [{\citenamefont {Heinzl}(1998)}]{Heinzl:1998kz}%
  \BibitemOpen
  \bibfield  {author} {\bibinfo {author} {\bibfnamefont {T.}~\bibnamefont
  {Heinzl}},\ }\emph {\bibinfo {title} {{Light cone dynamics of particles and
  fields}}},\ \href@noop {} {\bibinfo {type} {Habilitation thesis}} (\bibinfo
  {year} {1998}),\ \Eprint {http://arxiv.org/abs/hep-th/9812190}
  {arXiv:hep-th/9812190} \BibitemShut {NoStop}%
\bibitem [{\citenamefont {Kondratyuk}\ and\ \citenamefont
  {Strikman}(1984)}]{Kondratyuk:1983kq}%
  \BibitemOpen
  \bibfield  {author} {\bibinfo {author} {\bibfnamefont {L.~A.}\ \bibnamefont
  {Kondratyuk}}\ and\ \bibinfo {author} {\bibfnamefont {M.~I.}\ \bibnamefont
  {Strikman}},\ }\href {\doibase 10.1016/0375-9474(84)90165-9} {\bibfield
  {journal} {\bibinfo  {journal} {Nucl. Phys.}\ }\textbf {\bibinfo {volume}
  {A426}},\ \bibinfo {pages} {575} (\bibinfo {year} {1984})}\BibitemShut
  {NoStop}%
\bibitem [{\citenamefont {Cosyn}\ and\ \citenamefont
  {Weiss}(2020)}]{Cosyn:inprep}%
  \BibitemOpen
  \bibfield  {author} {\bibinfo {author} {\bibfnamefont {W.}~\bibnamefont
  {Cosyn}}\ and\ \bibinfo {author} {\bibfnamefont {C.}~\bibnamefont {Weiss}},\
  }\href@noop {} {\bibfield  {journal} {\bibinfo  {journal} {in preparation}\ }
  (\bibinfo {year} {2020})}\BibitemShut {NoStop}%
\bibitem [{\citenamefont {Berestetskii}\ \emph {et~al.}(1973)\citenamefont
  {Berestetskii}, \citenamefont {Lifshitz},\ and\ \citenamefont
  {Pitayevskii}}]{LLIV}%
  \BibitemOpen
  \bibfield  {author} {\bibinfo {author} {\bibfnamefont {V.~B.}\ \bibnamefont
  {Berestetskii}}, \bibinfo {author} {\bibfnamefont {E.~M.}\ \bibnamefont
  {Lifshitz}}, \ and\ \bibinfo {author} {\bibfnamefont {L.~P.}\ \bibnamefont
  {Pitayevskii}},\ }\href@noop {} {\emph {\bibinfo {title} {{Course of
  Theoretical Physics, Vol.~IV: Relativistic Quantum Theory}}}}\ (\bibinfo
  {publisher} {Pergamon Press, Oxford},\ \bibinfo {year} {1973})\BibitemShut
  {NoStop}%
\bibitem [{\citenamefont {Hoodbhoy}\ \emph {et~al.}(1989)\citenamefont
  {Hoodbhoy}, \citenamefont {Jaffe},\ and\ \citenamefont
  {Manohar}}]{Hoodbhoy:1988am}%
  \BibitemOpen
  \bibfield  {author} {\bibinfo {author} {\bibfnamefont {P.}~\bibnamefont
  {Hoodbhoy}}, \bibinfo {author} {\bibfnamefont {R.}~\bibnamefont {Jaffe}}, \
  and\ \bibinfo {author} {\bibfnamefont {A.}~\bibnamefont {Manohar}},\ }\href
  {\doibase 10.1016/0550-3213(89)90572-5} {\bibfield  {journal} {\bibinfo
  {journal} {Nucl.Phys.}\ }\textbf {\bibinfo {volume} {B312}},\ \bibinfo
  {pages} {571} (\bibinfo {year} {1989})}\BibitemShut {NoStop}%
\bibitem [{\citenamefont {Leader}(2005)}]{Leader:2001gr}%
  \BibitemOpen
  \bibfield  {author} {\bibinfo {author} {\bibfnamefont {E.}~\bibnamefont
  {Leader}},\ }\href@noop {} {\emph {\bibinfo {title} {{Spin in Particle
  Physics (Cambridge Monographs on Particle Physics, Nuclear Physics and
  Cosmology)}}}}\ (\bibinfo  {publisher} {Cambridge University Press},\
  \bibinfo {year} {2005})\BibitemShut {NoStop}%
\bibitem [{\citenamefont {Bacchetta}\ \emph {et~al.}(2007)\citenamefont
  {Bacchetta}, \citenamefont {Diehl}, \citenamefont {Goeke}, \citenamefont
  {Metz}, \citenamefont {Mulders},\ and\ \citenamefont
  {Schlegel}}]{Bacchetta:2006tn}%
  \BibitemOpen
  \bibfield  {author} {\bibinfo {author} {\bibfnamefont {A.}~\bibnamefont
  {Bacchetta}}, \bibinfo {author} {\bibfnamefont {M.}~\bibnamefont {Diehl}},
  \bibinfo {author} {\bibfnamefont {K.}~\bibnamefont {Goeke}}, \bibinfo
  {author} {\bibfnamefont {A.}~\bibnamefont {Metz}}, \bibinfo {author}
  {\bibfnamefont {P.~J.}\ \bibnamefont {Mulders}}, \ and\ \bibinfo {author}
  {\bibfnamefont {M.}~\bibnamefont {Schlegel}},\ }\href {\doibase
  10.1088/1126-6708/2007/02/093} {\bibfield  {journal} {\bibinfo  {journal}
  {JHEP}\ }\textbf {\bibinfo {volume} {02}},\ \bibinfo {pages} {093} (\bibinfo
  {year} {2007})}\BibitemShut {NoStop}%
\bibitem [{\citenamefont {Christ}\ and\ \citenamefont
  {Lee}(1966)}]{Christ:1966zz}%
  \BibitemOpen
  \bibfield  {author} {\bibinfo {author} {\bibfnamefont {N.}~\bibnamefont
  {Christ}}\ and\ \bibinfo {author} {\bibfnamefont {T.~D.}\ \bibnamefont
  {Lee}},\ }\href {\doibase 10.1103/PhysRev.143.1310} {\bibfield  {journal}
  {\bibinfo  {journal} {Phys. Rev.}\ }\textbf {\bibinfo {volume} {143}},\
  \bibinfo {pages} {1310} (\bibinfo {year} {1966})}\BibitemShut {NoStop}%
\bibitem [{\citenamefont {Afanasev}\ \emph {et~al.}(2008)\citenamefont
  {Afanasev}, \citenamefont {Strikman},\ and\ \citenamefont
  {Weiss}}]{Afanasev:2007ii}%
  \BibitemOpen
  \bibfield  {author} {\bibinfo {author} {\bibfnamefont {A.}~\bibnamefont
  {Afanasev}}, \bibinfo {author} {\bibfnamefont {M.}~\bibnamefont {Strikman}},
  \ and\ \bibinfo {author} {\bibfnamefont {C.}~\bibnamefont {Weiss}},\ }\href
  {\doibase 10.1103/PhysRevD.77.014028} {\bibfield  {journal} {\bibinfo
  {journal} {Phys. Rev.}\ }\textbf {\bibinfo {volume} {D77}},\ \bibinfo {pages}
  {014028} (\bibinfo {year} {2008})},\ \Eprint {http://arxiv.org/abs/0709.0901}
  {arXiv:0709.0901 [hep-ph]} \BibitemShut {NoStop}%
\bibitem [{\citenamefont {Frankfurt}\ and\ \citenamefont
  {Strikman}(1992)}]{Frankfurt:1992ny}%
  \BibitemOpen
  \bibfield  {author} {\bibinfo {author} {\bibfnamefont {L.}~\bibnamefont
  {Frankfurt}}\ and\ \bibinfo {author} {\bibfnamefont {M.}~\bibnamefont
  {Strikman}},\ }\href@noop {} {\emph {\bibinfo {title} {{Modern topics in
  electron scattering (eds.\ B.~Frois, I.~Sick)}}}}\ (\bibinfo  {publisher}
  {{World Scientific, Singapore}},\ \bibinfo {year} {1992})\ pp.\ \bibinfo
  {pages} {645--694}\BibitemShut {NoStop}%
\bibitem [{\citenamefont {Wiringa}\ \emph {et~al.}(1995)\citenamefont
  {Wiringa}, \citenamefont {Stoks},\ and\ \citenamefont
  {Schiavilla}}]{Wiringa:1994wb}%
  \BibitemOpen
  \bibfield  {author} {\bibinfo {author} {\bibfnamefont {R.~B.}\ \bibnamefont
  {Wiringa}}, \bibinfo {author} {\bibfnamefont {V.~G.~J.}\ \bibnamefont
  {Stoks}}, \ and\ \bibinfo {author} {\bibfnamefont {R.}~\bibnamefont
  {Schiavilla}},\ }\href {\doibase 10.1103/PhysRevC.51.38} {\bibfield
  {journal} {\bibinfo  {journal} {Phys. Rev.}\ }\textbf {\bibinfo {volume}
  {C51}},\ \bibinfo {pages} {38} (\bibinfo {year} {1995})},\ \Eprint
  {http://arxiv.org/abs/nucl-th/9408016} {arXiv:nucl-th/9408016 [nucl-th]}
  \BibitemShut {NoStop}%
\bibitem [{\citenamefont {Benhar}\ \emph {et~al.}(2005)\citenamefont {Benhar},
  \citenamefont {Farina}, \citenamefont {Nakamura}, \citenamefont {Sakuda},\
  and\ \citenamefont {Seki}}]{Benhar:2005dj}%
  \BibitemOpen
  \bibfield  {author} {\bibinfo {author} {\bibfnamefont {O.}~\bibnamefont
  {Benhar}}, \bibinfo {author} {\bibfnamefont {N.}~\bibnamefont {Farina}},
  \bibinfo {author} {\bibfnamefont {H.}~\bibnamefont {Nakamura}}, \bibinfo
  {author} {\bibfnamefont {M.}~\bibnamefont {Sakuda}}, \ and\ \bibinfo {author}
  {\bibfnamefont {R.}~\bibnamefont {Seki}},\ }\href {\doibase
  10.1103/PhysRevD.72.053005} {\bibfield  {journal} {\bibinfo  {journal} {Phys.
  Rev. D}\ }\textbf {\bibinfo {volume} {72}},\ \bibinfo {pages} {053005}
  (\bibinfo {year} {2005})},\ \Eprint {http://arxiv.org/abs/hep-ph/0506116}
  {arXiv:hep-ph/0506116} \BibitemShut {NoStop}%
\bibitem [{\citenamefont {Lev}\ \emph {et~al.}(1998)\citenamefont {Lev},
  \citenamefont {Pace},\ and\ \citenamefont {Salme}}]{Lev:1998qz}%
  \BibitemOpen
  \bibfield  {author} {\bibinfo {author} {\bibfnamefont {F.~M.}\ \bibnamefont
  {Lev}}, \bibinfo {author} {\bibfnamefont {E.}~\bibnamefont {Pace}}, \ and\
  \bibinfo {author} {\bibfnamefont {G.}~\bibnamefont {Salme}},\ }\href
  {\doibase 10.1016/S0375-9474(98)00469-2} {\bibfield  {journal} {\bibinfo
  {journal} {Nucl. Phys.}\ }\textbf {\bibinfo {volume} {A641}},\ \bibinfo
  {pages} {229} (\bibinfo {year} {1998})},\ \Eprint
  {http://arxiv.org/abs/hep-ph/9807255} {arXiv:hep-ph/9807255 [hep-ph]}
  \BibitemShut {NoStop}%
\bibitem [{\citenamefont {Goeke}\ \emph {et~al.}(2001)\citenamefont {Goeke},
  \citenamefont {Polyakov},\ and\ \citenamefont
  {Vanderhaeghen}}]{Goeke:2001tz}%
  \BibitemOpen
  \bibfield  {author} {\bibinfo {author} {\bibfnamefont {K.}~\bibnamefont
  {Goeke}}, \bibinfo {author} {\bibfnamefont {M.~V.}\ \bibnamefont {Polyakov}},
  \ and\ \bibinfo {author} {\bibfnamefont {M.}~\bibnamefont {Vanderhaeghen}},\
  }\href {\doibase 10.1016/S0146-6410(01)00158-2} {\bibfield  {journal}
  {\bibinfo  {journal} {Prog. Part. Nucl. Phys.}\ }\textbf {\bibinfo {volume}
  {47}},\ \bibinfo {pages} {401} (\bibinfo {year} {2001})},\ \Eprint
  {http://arxiv.org/abs/hep-ph/0106012} {arXiv:hep-ph/0106012} \BibitemShut
  {NoStop}%
\bibitem [{\citenamefont {Belitsky}\ \emph {et~al.}(2002)\citenamefont
  {Belitsky}, \citenamefont {Mueller},\ and\ \citenamefont
  {Kirchner}}]{Belitsky:2001ns}%
  \BibitemOpen
  \bibfield  {author} {\bibinfo {author} {\bibfnamefont {A.~V.}\ \bibnamefont
  {Belitsky}}, \bibinfo {author} {\bibfnamefont {D.}~\bibnamefont {Mueller}}, \
  and\ \bibinfo {author} {\bibfnamefont {A.}~\bibnamefont {Kirchner}},\ }\href
  {\doibase 10.1016/S0550-3213(02)00144-X} {\bibfield  {journal} {\bibinfo
  {journal} {Nucl. Phys.}\ }\textbf {\bibinfo {volume} {B629}},\ \bibinfo
  {pages} {323} (\bibinfo {year} {2002})}\BibitemShut {NoStop}%
\bibitem [{\citenamefont {Diehl}(2003)}]{Diehl:2003ny}%
  \BibitemOpen
  \bibfield  {author} {\bibinfo {author} {\bibfnamefont {M.}~\bibnamefont
  {Diehl}},\ }\href {\doibase 10.1016/j.physrep.2003.08.002} {\bibfield
  {journal} {\bibinfo  {journal} {Phys. Rept}\ }\textbf {\bibinfo {volume}
  {388}},\ \bibinfo {pages} {41} (\bibinfo {year} {2003})},\ \Eprint
  {http://arxiv.org/abs/hep-ph/0307382} {arXiv:hep-ph/0307382} \BibitemShut
  {NoStop}%
\bibitem [{\citenamefont {Tu}\ \emph {et~al.}(2020)\citenamefont {Tu},
  \citenamefont {Jentsch}, \citenamefont {Baker}, \citenamefont {Zheng},
  \citenamefont {Lee}, \citenamefont {Venugopalan}, \citenamefont {Hen},
  \citenamefont {Higinbotham}, \citenamefont {Aschenauer},\ and\ \citenamefont
  {Ullrich}}]{Tu:2020ymk}%
  \BibitemOpen
  \bibfield  {author} {\bibinfo {author} {\bibfnamefont {Z.}~\bibnamefont
  {Tu}}, \bibinfo {author} {\bibfnamefont {A.}~\bibnamefont {Jentsch}},
  \bibinfo {author} {\bibfnamefont {M.}~\bibnamefont {Baker}}, \bibinfo
  {author} {\bibfnamefont {L.}~\bibnamefont {Zheng}}, \bibinfo {author}
  {\bibfnamefont {J.-H.}\ \bibnamefont {Lee}}, \bibinfo {author} {\bibfnamefont
  {R.}~\bibnamefont {Venugopalan}}, \bibinfo {author} {\bibfnamefont
  {O.}~\bibnamefont {Hen}}, \bibinfo {author} {\bibfnamefont {D.}~\bibnamefont
  {Higinbotham}}, \bibinfo {author} {\bibfnamefont {E.-C.}\ \bibnamefont
  {Aschenauer}}, \ and\ \bibinfo {author} {\bibfnamefont {T.}~\bibnamefont
  {Ullrich}},\ }\href@noop {} {\  (\bibinfo {year} {2020})},\ \Eprint
  {http://arxiv.org/abs/2005.14706} {arXiv:2005.14706 [nucl-ex]} \BibitemShut
  {NoStop}%
\end{thebibliography}%
\end{document}